\documentclass[preprint,3p,times,onecolumn]{elsarticle} 
\pdfoutput=1
\usepackage{etoolbox}
\usepackage[utf8]{inputenc}
\usepackage[colorlinks]{hyperref}
\usepackage{graphicx}
\usepackage[sectionbib]{bibunits}
\usepackage{tabularx}
\usepackage{wrapfig}
\usepackage{amsmath}
\usepackage{float}
\usepackage{subfigure}
\usepackage{gensymb}
\usepackage{hyperref}
\usepackage{cleveref}
\usepackage{chngcntr}
\usepackage{multirow}
\usepackage{makecell}
\usepackage{array}
\usepackage{booktabs}
\usepackage{enumitem}
\usepackage{ragged2e}
\usepackage{amssymb}
\usepackage[section]{placeins}
\usepackage{siunitx}
\usepackage{epigraph}
\usepackage{lineno}
\usepackage{braket}
\usepackage{xparse,nameref}
\usepackage[usestackEOL]{stackengine}
\usepackage{xcolor}
\usepackage{mathtools}
\usepackage{multirow}
\usepackage{makecell}
\usepackage{array}
\usepackage{booktabs}
\usepackage{textcase}
\usepackage[toc,page]{appendix}
\usepackage{pdfpages}
\usepackage{titlesec}

\titleformat{\subsection} {\normalfont\bfseries}{\thesubsection}{1em}{}
\titleformat{\paragraph} {\normalfont\bfseries\itshape}{\paragraph}{1em}{}
\usepackage{dcolumn}   
\newcommand\Tstrut{\rule{0pt}{2.4ex}}       
\newcommand\Bstrut{\rule[-1.3ex]{0pt}{0pt}} 

\newcommand\TstrutLarge{\rule{0pt}{3.6ex}}       
\newcommand\BstrutLarge{\rule[-2.3ex]{0pt}{0pt}} 

\usepackage[tablename=TABLE, labelsep=period, figurename=FIG., skip=0pt, font=small]{caption}

\usepackage{etoolbox}
\makeatletter
    \patchcmd{\tnotemark}{\ding{73}}{\dag}{}{\@latex@error{Failed to path \string\tnotemark\space for \string\ding{73}}}
    \patchcmd{\tnotemark}{\ding{73}\ding{73}}{\dag\dag}{}{\@latex@error{Failed to path \string\tnotemark\space for \string\ding{73}\string\ding{73}}}
    \patchcmd{\tnotetext}{\ding{73}}{\dag}{}{\@latex@error{Failed to path \string\tnotetext\space for \string\ding{73}}}
    \patchcmd{\tnotetext}{\ding{73}\ding{73}}{\dag\dag}{}{\@latex@error{Failed to path \string\tnotetext\space for \string\ding{73}\string\ding{73}}}
\makeatother

\begin{document}
  \begin{frontmatter}
     \title{\textbf{Comprehensive geoneutrino analysis with Borexino}}   
     \author[Munchen]{M.~Agostini}
    \author[Munchen]{K.~Altenm\"{u}ller}
    \author[Munchen]{S.~Appel}
    \author[Kurchatov]{V.~Atroshchenko}
    \author[Juelich]{Z.~Bagdasarian}
    \author[Milano]{D.~Basilico}
    \author[Milano]{G.~Bellini}
    \author[PrincetonChemEng]{J.~Benziger}
    \author[Hamburg]{D.~Bick}
    \author[LNGS]{G.~Bonfini}
% ...
    
    \fntext[Roma]{Present address: Dipartimento di Fisica, Sapienza Universit\`a di Roma e INFN, 00185 Roma, Italy}
    \fntext[Napoli]{Present address: Dipartimento di Fisica, Universit\`a degli Studi Federico II e INFN, 80126 Napoli, Italy}
    \fntext[Madrid]{Present address: Universidad Autónoma de Madrid, Ciudad Universitaria de Cantoblanco, 28049 Madrid, Spain}
    \fntext[Fermi]{Present address: Fermilab National Accelerator Laboratory (FNAL), Batavia, IL 60510, USA}
    \fntext[LNGSG]{Present address: INFN Laboratori Nazionali del Gran Sasso, 67010 Assergi (AQ), Italy}
    \fntext[Mainz]{Institute of Physics and Excellence Cluster PRISMA$^+$, Johannes Gutenberg-Universit\"at Mainz, 55099 Mainz, Germany}
    \fntext[KFKI]{Also at: MTA-Wigner Research Centre for Physics, Department of Space Physics and Space Technology, Konkoly-Thege Miklós út 29-33, 1121 Budapest, Hungary}
    \tnotetext[fn1]{Deceased in August 2019}
    
    \address{The Borexino Collaboration}

% ...
\address[Kurchatov]{National Research Centre Kurchatov Institute, 123182 Moscow, Russia}
% ...
\address[LNGS]{INFN Laboratori Nazionali del Gran Sasso, 67010 Assergi (AQ), Italy}
    \address[Milano]{Dipartimento di Fisica, Universit\`a degli Studi e INFN, 20133 Milano, Italy}
% ...
\address[PrincetonChemEng]{Chemical Engineering Department, Princeton University, Princeton, NJ 08544, USA}
% ...
\address[Munchen]{Physik-Department and Excellence Cluster Universe, Technische Universit\"at  M\"unchen, 85748 Garching, Germany}
% ...
\address[Juelich]{Institut f\"ur Kernphysik, Forschungszentrum J\"ulich, 52425 J\"ulich, Germany}
% ...
\address[Hamburg]{Institut f\"ur Experimentalphysik, Universit\"at Hamburg, 22761 Hamburg, Germany}
% ...

     \begin{abstract}
      This paper presents a comprehensive geoneutrino measurement using the Borexino detector, located at Laboratori Nazionali del Gran Sasso (LNGS) in Italy. The analysis is the result of 3262.74\,days of data between December 2007 and April 2019. The paper describes improved analysis techniques and optimized data selection, which includes enlarged fiducial volume and sophisticated cosmogenic veto. The reported exposure of (1.29 $\pm$ 0.05)$ \times 10^{32}$ protons $\times$ year represents an increase by a factor of two over a previous Borexino analysis reported in 2015. By observing $52.6 ^{+9.4}_{-8.6}\,({\rm stat}) ^{+2.7}_{-2.1}\,({\rm sys})$ geoneutrinos (68\% interval) from $^{238}$U and $^{232}$Th, a geoneutrino signal of $47.0^{+8.4}_{-7.7}\,({\rm stat)}^{+2.4}_{-1.9}\,({\rm sys})$\,TNU with $^{+18.3}_{-17.2}$\% total precision was obtained. This result assumes the same Th/U mass ratio as found in chondritic CI meteorites but compatible results were found when contributions from $^{238}$U and $^{232}$Th were both fit as free parameters. 
      Antineutrino background from reactors is fit unconstrained and found compatible with the expectations. The null-hypothesis of observing a geoneutrino signal from the mantle is excluded at a 99.0\% C.L. when exploiting detailed knowledge of the local crust near the experimental site. Measured mantle signal of $21.2 ^{+9.5}_{-9.0}\,({\rm stat}) ^{+1.1}_{-0.9}\,({\rm sys})$\,TNU corresponds to the production of a radiogenic heat of $24.6 ^{+11.1}_{-10.4}$\,TW (68\% interval) from $^{238}$U and $^{232}$Th in the mantle. Assuming 18\% contribution of $^{40}$K in the mantle and $8.1^{+1.9}_{-1.4}$\,TW of total radiogenic heat of the lithosphere, the Borexino estimate of the total radiogenic heat of the Earth is $38.2 ^{+13.6}_{-12.7}$\,TW, which corresponds to the convective Urey ratio of 0.78$^{+0.41}_{-0.28}$. These values are compatible with different geological predictions, however there is a $\sim$2.4$\sigma$ tension with those Earth models which predict the lowest concentration of heat-producing elements in the mantle. In addition, by constraining the number of expected reactor antineutrino events, the existence of a hypothetical georeactor at the center of the Earth having power greater than 2.4\,TW is excluded at 95\% C.L. Particular attention is given to the description of all analysis details which should be of interest for the next generation of geoneutrino measurements using liquid scintillator detectors.
     \end{abstract}
    \end{frontmatter}
    \twocolumn 
      \tableofcontents
   
    \section{INTRODUCTION}
    \label{sec:intro}
 
 Neutrinos, the most abundant massive particles in the universe, are produced by a multitude of different processes. They interact only by the weak and gravitational interactions, and so are able to penetrate enormous distances through matter without absorption or deflection. Thus, they represent a unique tool to probe otherwise inaccessible objects, such as distant stars, the Sun, as well as the interior of the Earth. 

The present availability of large neutrino detectors has opened a new window to study the deep Earth's interior, complementary to more conventional direct methods used in seismology and geochemistry. For example, atmospheric neutrinos can be used as probe of the Earth's structure~\cite{Winter:2015zwx}. This absorption tomography is based on the fact that the Earth begins to become opaque to neutrinos with energies above $\sim$10\,TeV. Thus, the attenuation of the neutrino flux, as measured by the signals in large Cherenkov detectors, provides information about the nucleon matter density of the Earth. Recently, IceCube determined the mass of the Earth and its core, its moment of inertia and verified that the core is denser than the mantle using data obtained from atmospheric neutrinos~\cite{Donini:2018tsg}.  A complementary information about the electron density could, in principle, be inferred by exploiting the flavor oscillations of atmospheric neutrinos in the energy range from MeV to GeV~\cite{Bourret:2017tkw}.

An independent method to study the matter composition deep within the Earth, can be provided by geoneutrinos, i.e. (anti)neutrinos emitted by the Earth's radioactive elements. Their detection allows to assess the Earth's heat budget, specifically the heat emitted in the radioactive decays. The latter, the so-called {\it radiogenic heat} of the present Earth, arises mainly from the decays of isotopes with half-lives comparable to, or longer than Earth's age (4.543\,$\cdot$\,10$^{9}$\,years): $^{232}$Th ($T_{1/2}$ = 1.40\,$\cdot$\,10$^{10}$\,years), $^{238}$U ($T_{1/2}$ = 4.468\,$\cdot$\,10$^{9}$\,years), $^{235}$U ($T_{1/2}$ = 7.040\,$\cdot$\,10$^{8}$\,years), and $^{40}$K ($T_{1/2}$ = 1.248\,$\cdot$\,10$^{9}$\,years)~\cite{IAEA}.
All these isotopes are labeled as {\it heat-producing elements} (HPEs). The natural Thorium is fully composed of $^{232}$Th, while the natural isotopic abundances of $^{238}$U, $^{235}$U, and $^{40}$K are 0.992742, 0.007204, and $1.17 \times 10^{-4}$, respectively. In each decay, the emitted radiogenic heat is in a well-known ratio\footnote{The energy expressed in the following equations is the total energy, from which the released geoneutrinos take away about 5\% as their kinetic energy.} to the number of emitted geoneutrinos~\cite{Fiorentini:2007te}:
\begin{align}
^{238}\mathrm{U} \rightarrow& \, ^{206}\mathrm{Pb} + 8\alpha + 8 e^{-} + 6 \bar{\nu}_e + 51.7\,\mathrm{MeV}\label{Eq:geo1}\\
^{235}\mathrm{U} \rightarrow& \, ^{207}\mathrm{Pb} + 7\alpha + 4 e^{-} + 4 \bar{\nu}_e + 46.4\,\mathrm{MeV} \label{Eq:geo2}\\
^{232}\mathrm{Th} \rightarrow& \, ^{208}\mathrm{Pb} + 6\alpha + 4 e^{-} + 4 \bar{\nu}_e + 42.7\,\mathrm{MeV} \label{Eq:geo3} \\
^{40}\mathrm{K} \rightarrow& \, ^{40}\mathrm{Ca}  +  e^{-} +  \bar{\nu}_e + 1.31\,\mathrm{MeV}~\mathrm{(89.3\%)} \label{Eq:geo4}\\
^{40}\mathrm{K} + e^{-} \rightarrow& \, ^{40}\mathrm{Ar}  +  \nu _{e} +  1.505\,\mathrm{MeV}~\mathrm{(10.7\%)} \label{Eq:geo5}
\end{align}
Obviously, the total amount of emitted geoneutrinos scales with the total mass of HPEs inside the Earth. Hence, geoneutrinos' detection provides us a way of measuring this radiogenic heat.

This idea was first discussed by~G.~Marx and N.~Menyh\'ard~\cite{Marx1960}, G.~Eder~\cite{EDER1966657}, and G.~Marx~\cite{Marx1969} in the 1960's. It was further developed by M.\,L.~Krauss, S.\,L.~Glashow, and D. N.~Schramm~\cite{KraussGlashowSchramm} in 1984. Finally, the potential to measure geoneutrinos with liquid scintillator detectors was suggested in the '90s by C.\,G.~Rothschild, M.\,C.~Chen, and F.\,P.~Calaprice~\cite{Rothschild:1997dd} and independently by R.~Raghavan et al.~\cite{Raghavan:1997gw}. 

It took several decades to prove these ideas feasible. Currently, large-volume liquid-scintillator neutrino experiments KamLAND~\cite{Araki:2005qa, Abe:2008aa, Gando:1900zz, Gando:2013nba} and Borexino~\cite{Bellini:2010geo, Bellini:2013geo, Agostini:2015cba} have demonstrated the capability to efficiently detect a geoneutrino signal.
These detectors are thus offering a unique insight into 200 years long discussion about the origin of the Earth's internal heat sources.

The Borexino detector, located in hall-C of Laboratori Nazionali del Gran Sasso in Italy (LNGS), was originally designed to measure $^7$Be solar neutrinos. However thanks to the unprecedented levels of radiopurity, Borexino has surpassed its original goal and has now measured all\footnote{The upper limit was placed for $hep$ solar neutrinos, the flux of which is expected to be about 3 orders of magnitude smaller than that of $^8$B solar neutrinos~\cite{Vinyoles2016}.} the $pp$-chain neutrinos~\cite{Bellini:2013lnn, Bellini:2014uqa,Agostini:2018uly}. We report here a comprehensive geoneutrino measurement based on the Borexino data acquired during 3262.74\,days (December 2007 to April 2019). Thanks to an improved analysis with optimized data selection cuts, an enlarged fiducial volume, and a sophisticated cosmogenic veto, the exposure of (1.29 $\pm$ 0.05)$ \times 10^{32}$ protons $\times$ year represents a factor 2 increase with respect to the previous Borexino analysis~\cite{Agostini:2015cba}.

A detailed description of all the steps in the analysis is reported, and should be important to new experiments measuring geoneutrinos, e.g. SNO+~\cite{Chen:2005zza}, JUNO~\cite{An:2015jdp}, and Jinping~\cite{Wan:2016nhe}. Hanohano~\cite{Dye:2011mc} is an interesting, additional proposal to use a movable 5\,kton detector resting on the ocean floor. As the oceanic crust is particularly thin and relatively depleted in HPEs, this experiment could provide the most direct information about the mantle. Finally, it is anticipated that using antineutrinos to study the Earth's interior will increase in the future based on the availability of new detectors and the continuous development of analysis techniques.

This paper is structured as follows: Section~\ref{sec:geo} introduces the fundamental insights on what the geoneutrino studies can bring to the comprehension of the Earth's inner structure and thermal budget. Section~\ref{sec:det} details a description of the Borexino detector and the structure of its data.  In Section~\ref{sec:IBD}, the ${\bar{\nu}_e}$ detection reaction - the Inverse Beta Decay on free proton, that will be abbreviated as IBD through the text - is illustrated.
It is shown that only geoneutrinos above 1.8\,MeV kinematic threshold can be detected, leaving $^{40}$K and $^{235}$U geoneutrinos completely unreachable with present-day detection techniques. Section~\ref{sec:antinu} deals with the estimation of the expected antineutrino signal from geoneutrinos, through background from reactor and atmospheric neutrinos, up to a hypothetical natural georeactor in the deep Earth. Section~\ref{sec:bgr} describes the non-antineutrino backgrounds, e.g. cosmogenic or natural radioactive nuclei whose decays could mimic IBD. The criteria to selectively identify the best candidates in the data, are discussed in Sec.~\ref{sec:data_sel}, which involves the optimization of the signal-to-background ratio.  Section~\ref{sec:mc} shows how the signal and background spectral shapes, expressed in the experimental energy estimator (normalized charge), were constructed and how the detection efficiency is calculated. Both procedures are based on Borexino Monte Carlo (MC)~\cite{Agostini:2017aaa}, that was tuned on independent calibration data. Section~\ref{sec:sig_bgr_est} introduces the analyzed data set and discusses the number of expected signal and background events passing the optimized cuts, based on Secs.~\ref{sec:antinu} and \ref{sec:bgr}. In Section~\ref{sec:sensitivity}, the Borexino sensitivity to extract geoneutrino signals is illustrated. Finally, Sec.~\ref{sec:results} discusses our results. The golden IBD candidate sample is presented (Sec.~\ref{subsec:golden_candidates}) together with the spectral analysis (Sec.~\ref{subsec:ngeo}) and sources of systematic uncertainty (Sec.~\ref{subsec:syst}). The measured geoneutrino signal at LNGS is compared to the expectations of different geological models in Sec.~\ref{subsec:geo_lngs}. The extraction of the mantle signal using knowledge of the signal from the bulk lithosphere is discussed in Sec.~\ref{subsec:mantle}. The consequences of the new geoneutrino measurement with respect to the Earth's radiogenic heat are discussed in Sec.~\ref{subsec:radiogenic}. Placing limits on the power of a hypothetical natural georeactor, located at different positions inside the Earth, is discussed in Sec.~\ref{subsec:georeactor-results}. Final summary and conclusions are reported in Sec.~\ref{sec:conclusion}. The acronyms used within the text are listed in alphabetical order in the Appendix.

   \section{WHY STUDY GEONEUTRINOS}
    \label{sec:geo}

Our Earth is unique among the terrestrial planets\footnote{Mercury, Venus, Earth, and Mars} of the solar system. It has the strongest magnetic field, the highest surface heat flow, the most intense tectonic activity, and it is the only one to have continents composed of a silicate crust~\cite{RN1100}. Understanding the thermal, geodynamical, and geological evolution of our planet is one of the most fundamental questions in Earth Sciences~\cite{national2008origin}. 

\begin{figure}[t]
        \centering
        \includegraphics[width = 0.45\textwidth]{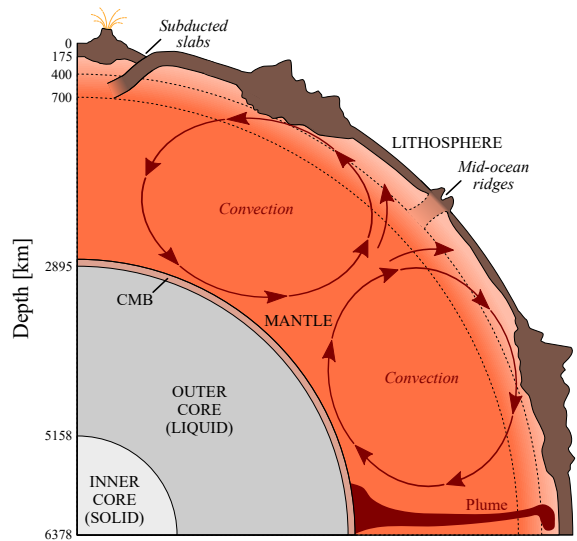}
        \vspace{3mm}
        \caption{Schematic cross-section of the Earth. The Earth has a concentrically layered structure with an equatorial radius of 6378\,km. The metallic {\it core} includes an inner solid portion (1220\,km radius) and an outer liquid portion which extends to a depth of 2895\,km, where the core is isolated from the silicate {\it mantle} by the {\it core-mantle boundary} (CMB). Seismic tomography suggests a convection through the whole depth of the viscose mantle, that is driving the movement of the {\it lithospheric tectonic plates}. The lithosphere, subjected to brittle deformations, is composed of the {\it crust} and {\it continental lithospheric mantle}. The {\it mantle transition zone}, extending from a depth of 400 to 700\,km, is affected by partial melting along the mid-oceanic ridges where the {\it oceanic crust} is formed. The {\it continetal crust} is more complex and thicker than the oceanic crust.}
        \label{fig:earth}
    \end{figure}

The Earth was created in the process of accretion from undifferentiated material of solar nebula~\cite{RN1398, RN607}. The bodies with a sufficient mass undergo the process of differentiation, i.e. a transformation from a homogeneous object into a body with a layered structure. The geophysical picture of a concentrically layered internal structure of the present Earth (Fig.~\ref{fig:earth}), with mass $M_{\mathrm{E}}$ = 5.97 $ \times 10^{24}$\,kg, is relatively well established from its density profile, which is obtained by precise measurements of seismic waves on its surface.

 During the first differentiation, metallic segregation occurred and the {\it core} ($\sim$32.3\% of $M_{\mathrm{E}}$) separated from the silicate {\it Primitive Mantle} or {\it Bulk Silicate Earth} (BSE). The latter further differentiated into the present {\it mantle} ($\sim$67.2\% of $M_{\mathrm{E}}$) and {\it crust} ($\sim$0.5\% of $M_{\mathrm{E}}$). The metallic core has Fe--Ni chemical composition and is expected to reach temperatures up to about 6000\,K in its central parts. The inner core ($\sim$1220\,km radius) is solid due to high pressure, while the 2263\,km thick outer core is liquid~\cite{BELLINI20131}. The outer core has an approximate 10\% admixture of lighter elements and plays a key role in the geodynamo process generating the Earth's magnetic field. The {\it core-mantle boundary} (CMB) seismic discontinuity divides the core from the mantle. The mantle reaches temperature of about 3700\,K at its bottom, while being solid but viscose on long time scales, so the mantle convection can occur. The latter drives the movement of tectonic plates at the speed of few cm per year. A whole mantle convection is supported by high resolution seismic tomography~\cite{RN1391}, which proves existence of material exchange across the mantle, as in the zones of deeply subducted lithosperic slabs and mantle plumes rooted close to the CMB. At a depth of [400 - 700]\,km, the mantle is characterized by a {\it transition zone}, where a weak seismic-velocity heterogeneity is measured. The upper portion of the mantle contains the viscose {\it asthenosphere} on which the {\it lithospheric tectonic plates} are floating. These comprise the uppermost, rigid part of the mantle (i.e. the {\it continental lithospheric mantle} (CLM)) and the two types of crust: {\it oceanic crust} (OC) and {\it continental crust} (CC). The CLM is a portion of the mantle underlying the CC included between the Moho discontinuity\footnote{The Moho (Mohorovi\v{c}i\'{c}) discontinuity is the boundary between the crust and the mantle, characterized by a jump in seismic compressional waves velocities from $\sim$7 to $\sim$8 km/s occurring beneath the CC at typical depth of $\sim$35\,km.} and a seismic and electromagnetic transition at a typical depth of $\sim$175\,km \cite{RN367}. The CC with a thickness of (34 $\pm$ 4)\,km~\cite{Huang2013} has the most complex history being the most differentiated and heterogeneous layer. It consists of igneous, metamorphic, and sedimentary rocks. The OC with (8 $\pm$ 3)\,km thickness~\cite{Huang2013} is created along the mid-oceanic ridges, where the basaltic magma differentiates from the partially melting mantle up-welling towards the ocean floor.

Traditionally, direct methods to obtain information about the deep Earth's layers, from where there are few or no direct rock samples, are limited to seismology. Seismology provides relatively precise information about the density profile of the deep Earth~\cite{RN1394}, but it lacks direct information about the chemical composition and radiogenic heat production. Geoneutrinos come into play here: their small interaction cross-section ($\sim 10^{-42}$\,cm$^{2}$ at MeV energy for the IBD, Sec.~\ref{sec:IBD}), on one hand, limits our ability to detect them, on the other hand, it makes them a unique probe of inaccessible innermost parts of the Earth. In the radioactive decays of HPEs, the amount of released geoneutrinos and radiogenic heat are in a well-known ratio (Eqs.~\ref{Eq:geo1} to~\ref{Eq:geo5}). Thus, a direct measurement of the geoneutrino flux provides useful information about the composition of the Earth's interior~\cite{Fiorentini:2007te}. Consequently, it also provides an insight into the radiogenic heat contribution to the measured Earth's surface heat flux. 

\begin{table}[t]
	\centering	
	\caption{\label{tab:heat_flux} Integrated terrestrial surface heat fluxes $H_{\mathrm{tot}}$ estimated by different authors. The lower limit estimation~\cite{RN1384} is due to the approach based only on direct heat flow measurements in contrast with the thermal model of {\it half space cooling} adopted by the remaining references.} \vskip 2pt
	{\small	
	\begin{tabular*}{\columnwidth}{l@{\extracolsep{\fill}}c}
		\hline
		\hline
		Reference  &   Earth’s heat flux [TW]  \Tstrut\Bstrut \\ 
		\hline
		Williams \& Von Herzen (1980)	\cite{RN1386}	&	43	\Tstrut\\ 
		Davies (1980)	\cite{RN1385}	&	41	\\				
		Sclater et al. (1980)	\cite{RN1381}	&	42	\\				
		Pollack et al. (1993) 	\cite{RN1382}	&	44 $\pm$ 1	\\				
		Hofmeister et al (2005)	\cite{RN1384}	&	31 $\pm$ 1	\\				
		Jaupart et al. (2007)	\cite{RN1387}	&	46 $\pm$ 3	\\				
		Davies \& Davies (2010)	\cite{RN593}	&	47 $\pm$ 2	\Bstrut \\ 
		\hline
		\hline
	\end{tabular*}
	}
\end{table}

The heat flow from the Earth's surface to space results from a large temperature gradient across the Earth~\cite{RN404}. Table~\ref{tab:heat_flux} shows estimations of this heat flow, $H_{\mathrm{tot}}$, integrated over the whole Earth's surface. Different studies of this flux are based on several thousands of inhomogeneously distributed measurements of the thermal conductivity of rocks and the temperature gradients within deep bore holes. The existence of perturbations produced by volcanic activity and hydrothermal circulations, especially along the mid-ocean ridges (where the data are sparse), requires the application of energy-loss models~\cite{RN1395}. Except for Ref.~\cite{RN1384}, the papers account for the hydrothermal circulation in the young oceanic crust by utilizing the {\it half-space cooling model}, which describes ocean depths and heat flow as a function of the oceanic lithosphere age. The latter is unequivocally correlated with the distance to mid-ocean ridges, where the oceanic crust is created~\cite{RN1397} and from where the older crust is pushed away by a newly created crust. This approach leads to an $H_{\mathrm{tot}}$ estimation between 41 and 47\,TW, with the oceans releasing $\sim$70\% of the total escaping heat. The most recent models~\cite{RN1387,RN593} are in excellent agreement and provide a value of (46-47)\,TW with (2-3)\,TW error.  However, Ref.~\cite{RN1384}, based only on direct measurements and not applying the half-space cooling model, provides a much lower $H_{\mathrm{tot}}$ of (31 $\pm$ 1)\,TW. We assume a $H_{\mathrm{tot}}$ = (47 $\pm$ 2)\,TW as the best current estimation. 

Neglecting the small contribution ($<$0.5\,TW) from tidal dissipation and gravitational potential energy released by the differentiation of crust from the mantle, the $H_{\mathrm{tot}}$ is typically expected to originate from two main processes: ({\it i}) {\it secular cooling} $H_{\mathrm {SC}}$ of the Earth, i.e. cooling from the time of the Earth's formation when gravitational binding energy was released due to matter accretion, and ({\it ii}) {\it radiogenic heat} $H_{\mathrm {rad}}$ from HPEs' radioactive decays in the Earth. The relative contribution of radiogenic heat to the $H_{\mathrm{tot}}$ is crucial in understanding the thermal conditions occurring during the formation of the Earth and the energy now available to drive the dynamical processes such as the mantle and outer-core convection. The {\it Convective Urey Ratio} ($UR_{\mathrm{CV}}$) quantifies the ratio of internal heat generation in the mantle over the mantle heat flux, as the following ratio~\cite{RN404}:
\begin{equation}
UR_{\mathrm{CV}} = \frac {H_{\mathrm {rad}} - H_{\mathrm {rad}}^{\mathrm{CC}}} {H_{\mathrm{tot}} - H_{\mathrm {rad}}^{\mathrm{CC}}},
\label{eq:URCV}
\end{equation}
where $H_{\mathrm {rad}}^{\mathrm{CC}}$ is the radiogenic heat produced in the continental crust. The secular cooling of the core is expected in the range of [5 - 11]\,TW~\cite{RN1395}, while no radiogenic heat is expected to be produced in the core.

\begin{table*}[t]
	\centering
	\caption{\label{tab:BSE} Masses $M$ and abundances $a$ of HPEs in the Bulk Silicate Earth (M$_{\mathrm{BSE}}$ = 4.04 $\times$ 10$^{24}$\,kg~\cite{BELLINI20131}) predicted by different models: J: Javoy et al., 2010~\cite{RN367}, L \& K: Lyubetskaya \& Korenaga, 2007~\cite{RN747}, T: Taylor, 1980~\cite{RN1378}, M \& S: McDonough \& Sun, 1995~\cite{RN1380}, A: Anderson, 2007~\cite{RN372}, W: Wang et al., 2018~\cite{RN1319}, P \& O: Palme and O'Neil, 2003~\cite{RN400}, T \& S: Turcotte \& Schubert, 2002~\cite{RN368}. The {\it Cosmochemical} (CC), {\it Geochemical} (GC), and {\it Geodynamical} (GD) BSE models correspond to the estimates reported in~\cite{RN630}; the {\it Fully Radiogenic} (FR) model is defined adopting the approach of~\cite{RN356}, assuming that the total heat $H_{\mathrm {tot}}$ = (47 $\pm$ 2)\,TW is due to only the radiogenic heat production $H_{\mathrm{rad}}$(U+Th+K). The CC and GD models correspond to the estimates based on predictions made by Javoy et al., 2010~\cite{RN367} and Turcotte \& Schubert, 2002~\cite{RN368}, respectively. The GC is based on estimates reported by McDonough \& Sun, 1995~\cite{RN1380} with K abundances corrected following~\cite{RN361}. The radiogenic heat $H_{\mathrm{rad}}$ released in the radioactive decays of HPEs is calculated adopting the element specific heat generation $h$ ($H_{\mathrm{rad}}$ = $h$ $\times$ $M$ with $h$(U) = 98.5\,$\mu$W/kg, $h$(Th) = 26.3\,$\mu$W/kg, and $h$(K) = 3.33 $\times$ 10$^{-3}$\,$\mu$W/kg taken from~\cite{RN356}). It is assumed that the uncertainties on U, Th, and K abundances are fully correlated.}	\vskip 2pt
	{\small
	\begin{tabular*}{\textwidth}{c@{\hskip 9pt}c@{\hskip 8pt}c@{\hskip 8pt}c@{\hskip 9pt}c@{\hskip 8pt}c@{\hskip 8pt}c@{\hskip 9pt}c@{\hskip 8pt}c@{\hskip 8pt}c@{\hskip 8pt}c}
		\hline
		\hline
		Model  &	
		\thead{{\small $a$(U)} \\ {\small $[$ng/g$]$}}  &
		\thead{{\small $a$(Th)} \\ {\small $[$ng/g$]$}} & 
		\thead{{\small $a$(K)} \\ {\small $[\mu$g/g$]$}} & 
		\thead{{\small $M$(U)} \\ {\small $[10^{16}$\,kg$]$}} & 
		\thead{{\small $M$(Th)} \\ {\small $[10^{16}$\,kg$]$}} &
		\thead{{\small $M$(K)} \\ {\small $[10^{19}$\,kg$]$}}  &
		\thead{{\small $H_{\mathrm{rad}}$(U)} \\ {\small $[$TW$]$}} & 
		\thead{{\small $H_{\mathrm{rad}}$(Th)} \\ {\small $[$TW$]$}} &
		\thead{{\small $H_{\mathrm{rad}}$(K)} \\ {\small $[$TW$]$}}  &
		\thead{{\small $H_{\mathrm{rad}}$(U+Th+K)} \\ {\small $[$TW$]$}}   \Tstrut\Bstrut \\ 
		\hline
J       &	12	&	43	&	146	&	4.85	&	17.4	&	59		&	4.8	&	4.6		&	2.0	&	11.3 \Tstrut \\      
L \& K 	&	17	&	63	&	190	&	6.87	&	25.5	&	76.8	&	6.8	&	6.7		&	2.6	&	16		 \\
T       &	18	&	70	&	180	&	7.28	&	28.3	&	72.8	&	7.2	&	7.5		&	2.4	&	17		 \\ 
M \& S  &	20	&	80	&	240	&	8.09	&	32.4	&	97.1	&	8.0	&	8.5		&	3.2	&	19.7	 \\
A       & 	20	&	77	&	151	&	8.09	&	31.1	&	61.1	&	8.0	&	8.2		&	2.0	&	18.2	 \\	 
W       &	20	&	75	&	237	&	8.09	&	30.3	&	95.8	&	8.0	&	8.0		&	3.2	&	19.1	 \\    
P \& O  &	22	&	83	&	260	&	8.9		&	33.6	&	105.1	&	8.8	&	8.9		&	3.5	&	21.1	 \\	
T \& S  & 	35	&	140	&	350	&	14.2	&	56.6	&	141.5	&	13.9	&	14.9	&	4.7	&	33.5 \Bstrut \\ 
\hline
$CC$    &	12 $\pm$ 2	&	43  $\pm$ 4	&	146  $\pm$ 29	&	5  $\pm$ 1	&	17  $\pm$ 2	&	59  $\pm$ 12	&	4.8  $\pm$ 0.8	&	4.6  $\pm$ 0.4	&	2.0  $\pm$ 0.4	&	11.3  $\pm$ 1.6  \Tstrut \\
$GC$    &	20 $\pm$ 4	&	80  $\pm$ 13	&	280  $\pm$ 60	&	8 $ \pm$ 2	&	32  $\pm$ 5	&	113  $\pm$ 24	&	8.0  $\pm$ 1.6	&	8.5  $\pm$ 1.4	&	3.8  $\pm$ 0.8	&	20.2  $\pm$ 3.8 \\
$GD$    & 	35  $\pm$ 4	&	140  $\pm$ 14	&	350  $\pm$ 35	&	14  $\pm$ 2	&	57 $\pm$ 6	&	142  $\pm$ 14	&	13.9 $\pm$ 1.6	&	14.9  $\pm$ 1.5	&	4.7  $\pm$ 0.5	&	33.5  $\pm$ 3.6	\\
$FR$    & 	49 $\pm$ 2	&	189 $\pm$ 8 &	554 $\pm$ 24 	&	20 $\pm$ 1 	&	77 $\pm$ 3 	&	224 $\pm$ 10  	&	19.4 $\pm$ 0.8 	&	20.2 $\pm$ 0.8  &	7.5 $\pm$ 0.3  &	47 $\pm$ 2 \Bstrut \\
		\hline
		\hline
	\end{tabular*}
	}
\end{table*}
 
Preventing dramatically high temperatures during the initial stages of Earth  formation, the present-day $UR_{\mathrm{CV}}$ must be in the range between 0.12 to 0.49~\cite{RN1395}. Additionally, HPEs' abundances, and thus $H_{\mathrm {rad}}$ of Eq.~\ref{eq:URCV}, are globally representative of {\it BSE models}, defining the original chemical composition of the Primitive Mantle. The elemental composition of BSE is obtained assuming a common origin for celestial bodies in the solar system. It is supported, for example, by the strong correlation observed between the relative (to Silicon) isotopical abundances in the solar photosphere and in the CI chondrites (Fig.~2 in ~\cite{BELLINI20131}). Such correlations can be then assumed also for the material from which the Earth was created. The BSE models agree in the prediction of major elemental abundances (e.g. O, Si, Mg, Fe) within 10\%~\cite{RN1380}. Uranium and Thorium are {\it refractory} (condensate at high temperatures) and {\it lithophile} (preferring to bind with silicates over metals) elements. The relative abundances of the refractory lithophile elements are expected to be stable to volatile loss or core formation during the early stage of the Earth~\cite{RN1396}. The content of refractory lithophile elements (e.g. U and Th), which are excluded from the core\footnote{Recent speculations~\cite{RN1399} about possible partitioning of some lithophile elements (including U and Th) into the metallic core are still debated~\cite{RN1400,RN1401}. This would explain the anomalous Sm/Nd ratio observed in the silicate Earth and would represent an additional radiogenic heat source for the geodynamo process.}, are assumed based on relative abundances in chondrites, and dramatically differ between different models. In Table~\ref{tab:BSE} global masses of HPEs and their corresponding radiogenic heat are reported, covering a wide spectrum of BSE compositional models. The contributions to the radiogenic heat of U, Th, and K vary in the range of [39 - 44]\%, [40 - 45]\%, and [11 - 17]\%, respectively.

Three classes of BSE models are adopted in this work: the {\it Cosmochemical}, {\it Geochemical}, and {\it Geodynamical} models, as defined in~\cite{BELLINI20131,RN630}. The {\it Cosmochemical} (CC) model~\cite{RN367} is characterized by a relatively low amount of U and Th producing a total $H_{\mathrm{rad}}$ = (11 $\pm$ 2)\,TW. This model bases the Earth's composition on enstatite chondrites. The {\it Geochemical} (GC) model class predicts intermediate HPEs'abundances for primordial Earth. It adopts the relative abundances of refractory lithophile elements as in CI chondrites, while the absolute abundances are constrained by terrestrial samples~\cite{RN1380, RN361}. The {\it Geodynamical} (GD) model shows relatively high U and Th abundances. It is based on the energetics of mantle convection and the observed surface heat loss~\cite{RN368}. Additionally, an extreme model can be obtained following the approach described in~\cite{RN356}, where the terrestrial heat $H_{\mathrm {tot}}$ of 47\,TW is assumed to be fully accounted for by radiogenic production $H_{\mathrm {rad}}$. When keeping the HPEs' abundance ratios fixed to chondritic values and rescaling the mass of each HPE component accordingly, one obtains estimates for {\it Fully Radiogenic} (FR) model (Table~\ref{tab:BSE}). 

A global assessment of the Th/U mass ratio of the Primitive Mantle could hinge on the early evolution of the Earth and its differentiation. The most precise estimate of the planetary Th/U mass ratio reference, having a direct application in geoneutrino analysis, has been refined to a value of $M_{\mathrm{Th}}$/$M_{\mathrm {U}}$ = (3.876 $\pm$ 0.016)~\cite{RN1377}. Recent studies~\cite{RN1328}, based on measured molar $^{232}$Th/$^{238}$U values and their time integrated Pb isotopic values, are in agreement estimating $M_{\mathrm{Th}}$/$M_{\mathrm {U}}$ = $3.90^{\mathrm{+0.13}}_{\mathrm{-0.08}}$. Significant deviations from this average value can be found locally, especially in the heterogeneous continental crust. This fact is attributable to many different lithotypes, which can be found surrounding the individual geoneutrino detectors~\cite{RN1290, RN872}. In the local reference model for the area surrounding the Borexino detector (see also Sec.~\ref{subsec:geo}), the reservoirs of the sedimentary cover, which account for 30\% of the geoneutrino signal from the regional crust, are characterized by a Th/U mass ratio ranging from $\sim$0.8 (carbonatic rocks) to $\sim$3.7 (terrigenous sediments)~\cite{coltorti}.

The determination of the radiogenic component of Earth's internal heat budget has proven to be a difficult task, since an exhaustive theory is required to satisfy geochemical, cosmochemical, geophysical, and thermal constraints, often based on indirect arguments. In this puzzle, direct U and Th geoneutrino measurements are candidates to play a starring role. Geoneutrinos have also the potential to determine the mantle radiogenic heat, the key unknown parameter. This can be done by constraining the relatively-well known litospheric contribution, as we show in Sec.~\ref{subsec:mantle}. The lithospheric contribution would be particularly small and easily determined on a thin, HPEs depleted oceanic crust. This would make the ocean floor an ideal environment for geoneutrino detection. Geoneutrino measurements can also contribute to the discussion about possible additional heat sources, which have been proposed by some authors. For example, stringent limits (Sec.~\ref{subsec:georeactor-results}) can be set on the power of a hypothetical Uranium natural georeactor suggested in~\cite{herndon1993feasibility,herndon1996substructure,rusov2007geoantineutrino,de2008feasibility} and discussed in Sec.~\ref{subsec:georeactor}. In future, by combining measurements from several experiments placed in distant locations and in distinct geological environments, one could test whether the mantle is laterally homogeneous or not~\cite{RN630}, as suggested, for example, by the Large Shear Velocity Provinces observed at the mantle base below Africa and Pacific ocean~\cite{wang}. 

In future, detection of $^{40}$K geoneutrinos might be possible~\cite{Szczerbinska2011,OpaqueDetector}. This would be extremely important, since Potassium is the only semi-volatile HPE. Our planet seems to show $\sim$1/3~\cite{RN1380} to $\sim$1/8~\cite{RN367} Potassium when compared to chondrites, making its expected bulk mass span of a factor $\sim$2 across different Earth's models. Two theories on the fate of the mysterious ``missing K" include loss to space during accretion~\cite{RN1380} or segregation into the core~\cite{Goettel1976}, but no experimental evidence has been able to confirm or rule out any of the hypotheses, yet. As a consequence, the different BSE class models predict a K/U ratio in the mantle in a relatively wide range from 9700 to 16000~\cite{RN630}. According to these ratios, the Potassium radiogenic heat of the mantle varies in the range [2.6 – 4.3]\,TW, which translates to an average contribution of 18\% to the mantle radiogenic power. We will use this value in the evaluation of the total Earth radiogenic heat from the Borexino geoneutrino measurement (Sec.~\ref{subsec:radiogenic}).

    \section{THE BOREXINO DETECTOR}
    \label{sec:det}
   
    \begin{figure}[t]
        \centering
        \includegraphics[width = 0.45\textwidth]{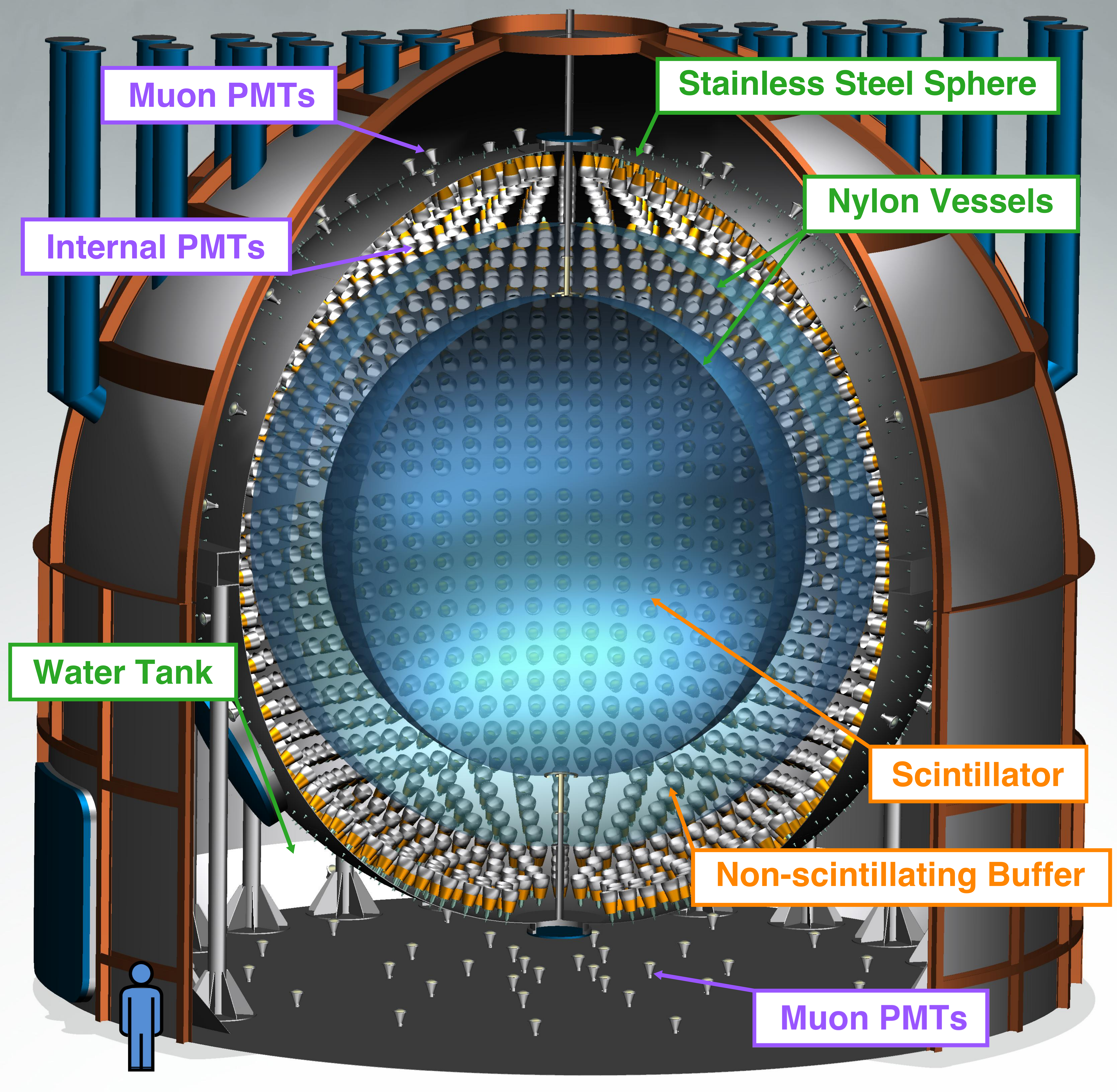}
    \vspace{2mm}
        \caption{Scheme of the Borexino detector.}
        \label{fig:detector}
    \end{figure}
    
    	Borexino is an ultra-pure liquid scintillator detector~\cite{Alimonti:2008gc} operating in real-time mode. It is located in the hall-C of the Gran Sasso National Laboratory in central Italy at a depth of some 3800\,m w.e. (meter water equivalent). The rock above the detector provides shielding against cosmogenic backgrounds such that the muon flux is decreased to $(3.432 \pm 0.003) \cdot 10^{-4}$\,m$^{-2}$\,s$^{-1}$~\cite{Agostini:2018fnx}. The general scheme of the Borexino detector is shown in Fig.~\ref{fig:detector}. The detector has a concentric multi-layer structure. The outer layer ({\it Outer Detector} (OD)) serves as a passive shield against external radiation as well as an active Cherenkov veto of cosmogenic muons. It consists of a steel {\it Water Tank} (WT) of 9\,m base radius and 16.9\,m height filled with approximately 1\,kt of ultra-pure water. Cherenkov light in the water is registered in 208 8" photo-multiplier tubes (PMTs) placed on the floor and outer surface of a {\it Stainless Steel Sphere} (SSS, 6.85\,m radius), which is contained within the WT. The {\it Inner Detector} (ID) within the SSS comprises three layers and it is equipped with 2212 8" PMTs mounted on the inner surface of the SSS. Over time, the number of working PMTs in the ID has decreased, from 1931 in December 2007 to 1183 by the end of April 2019. The three ID layers are formed by the insertion of the two 125\,$\mu$m thick nylon ``balloons",  the {\it Inner Vessel} (IV) and the {\it Outer Vessel} (OV) with the radii 4.25\,m and 5.50\,m, respectively. The two layers between the SSS and the IV, separated by the OV, form the {\it Outer Buffer} (OB) and the {\it Inner Buffer} (IB). 
 
       The antineutrino target is an organic {\it liquid scintillator} (LS) confined by the IV. The scintillator is composed of pseudocumene (PC, 1,2,4-trimethylbenzene, C$_6$H$_3$(CH$_3$)$_3$) solvent doped with a fluorescent dye PPO (2,5-diphenyloxazole, C$_{15}$H$_{11}$NO) in concentration of 1.5\,g/l. The scintillator density is $(0.878 \pm  0.004)$\,g\,cm$^{-3}$, where the error considers the changes due to the temperature instabilities over the whole data acquisition period. The nominal total mass of the target is 278\,ton and the proton density is $(6.007 \pm 0.001) \times 10^{28}$ per 1\,ton. A careful selection of detector materials, accurate assembling, and a complex radio-purification of the liquid scintillator guaranteed extremely low contamination levels of $^{238}$U and $^{232}$Th. After the additional LS purification in 2011, they achieved $< 9.4 \times 10^{-20}$\,g/g ($95\%$ C.L.) and $< 5.7 \times 10^{-19}$\,g/g ($95\%$ C.L.), respectively.    
    
The buffer liquid, consisting of a solution of the dimethylphthalate (DMP, C$_6$H$_4$(COOCH$_3$)$_2$) light quencher in PC, shields the core of the detector against external $\gamma$s and neutron radiation. The OV and IV themselves block the inward transfer of Radon emanated from the internal PMTs and SSS. The quencher concentration has been varied twice, changing it from the initial 5\,g/l to 3\,g/l and then to 2\,g/l. These operations have reduced the density difference between the buffer and the scintillator in order to minimize the scintillator leak (appeared in April 2008) from the central volume through the small hole in the IV to the IB as much as possible. This campaign was mostly successful, but the IV shape has become non-spherical and changing in time. We are able to reconstruct the IV shape from the data itself, as it will be described in Sec.~\ref{subsec:IV}.     

	Borexino has a main data acquisition system ({\it the main DAQ}) and a semi-independent {\it Fast Wave Form Digitizer} (FWFD) or {\it Flash Analog-to-Digital Converter} (FADC) sub-system designed for energies above 1\,MeV. Both systems process signals from both the ID and the OD PMTs, but in different ways. Every ID PMT is AC-coupled to an electronic chain made by an analogue front end (so-called {\it FE boards}, FEBs) followed by a digital circuit (so-called {\it Laben boards}\footnote{The boards were designed and built in collaboration with Laben s.p.a.}). While the main DAQ treats every PMT individually, the FADC sub-system receives as input the sums of up to 24 analogue FEB outputs. More details about the Borexino data structure are given in Sec.~\ref{subsec:data}.

		The effective light yield in Borexino is approximately 500\,detected {\it photoelectrons} (p.e.) per 1\,MeV of deposited energy. This results in the $5\%/\sqrt{E~(\text{MeV})}$ energy resolution. Borexino is a position sensitive detector. For point-like events, the vertex is reconstructed based on the time-of-flight technique~\cite{Bellini:2013lnn} with $\sim$10\,cm at 1\,MeV resolution at the center of the detector. For other positions with larger radii, the resolution decreases on average by a few centimeters. 
		
		A comprehensive calibration campaign~\cite{Back:2012awa} was performed in 2009. It served as a base for understanding of the detector's performance and for tuning a custom, Geant4 (release 4.10.5)-based, Monte Carlo (MC) code called {\it G4Bx2}. This MC simulates all processes after the interaction of a particle in the detector, including all known characteristics of the apparatus~\cite{Agostini:2017aaa}. Since the Borexino MC chain results in data files with the same format as real data, the same software can be applied to both of them. During the calibration, radioactive sources were employed in approximately 250 points through the IV scintillator volume. Using seven CCD cameras mounted inside the detector, the positions of the sources could be determined with a precision better than 2\,cm. Several gamma sources with energies between 0.12 to 1.46\,MeV were used for studying the energy scale. $^{222}$Rn source, emitting alpha particles characterized by point-like interactions, was applied to study the homogeneity of the detector's response as well as position reconstruction. For geoneutrino studies, employment of the $^{241}$Am-$^9$Be source is of particular interest, since the emitted neutrons closely represent the delayed signal of an Inverse Beta Decay (Sec.~\ref{sec:IBD}), which is the interaction used to detect geoneutrinos. In addition, a $^{228}$Th source emitting 2.615\,MeV gammas was placed in 9 detector inlets, constructed in a way that the sources were practically positioned at the SSS. This calibration was fundamental in the optimization of the biasing technique used in the MC simulation of the external background.  
	
	Along with the special calibration campaign at the beginning of data acquisition, there are constant offline checks of the detector's stability and regular online PMTs' calibration. The time equalization among PMTs is performed once a week with a special laser ($\lambda$ = 394\,nm, 50\,ps wide peak) calibration run. This procedure is of utmost importance for position and muon track reconstructions, as well as for the $\alpha$/$\beta$ discrimination (Sec.~\ref{subsec:ab}), based on their different fluorescence time profiles. The charge calibration of the single photoelectron response of each PMT is preformed typically 4 times a day. It is based on plentiful $\beta ^{-}$ decays of $^{14}$C ($Q$ = 156\,keV), inevitably present in each organic liquid scintillator and dominating the triggering rate, which varied between 20-30\,s$^{-1}$ (above roughly 50\,keV threshold) during the analyzed period. A new {\it Borexino Trigger Board} (BTB) was installed in May 2016, in place of the old module which began to have failures. The thorough tests have proven unbiased performance and further improved stability of the detector.

    \subsection{Borexino data structure}
    \label{subsec:data}

Borexino electronics must handle about 10$^{6}$ events per day, which are dominated by $^{14}$C decays. The residual flux of cosmogenic muons results in the detection of approximately 4300 per day {\it internal muons} which cross IV scintillator and/or the buffer, and approximately the same number of {\it external muons} which cross only the OD but not the ID~\cite{Bellini:2011yd}. The details of the read-out system, electronics, and trigger can be found in~\cite{Alimonti:2008gc}. The key features relevant for the geoneutrino analysis are presented in this Section.

 \paragraph{The main DAQ}
 
 The main DAQ reads individually all channels from both the ID and the OD when the BTB issues a global trigger. A global trigger condition occurs when at least one of the two sub-detectors has a trigger. 

 The ID trigger threshold, set to 25-20 PMTs triggered in a selected time window (typically 90\,ns wide),  corresponds to a deposited energy of approximately 50\,keV. The trigger threshold in the {\it Muon Trigger Board} (MTB) is 6 hits in a 150\,ns time window. The information of the OD trigger is stored in the $2^2$ bit of the trigger word, and is referenced as the {\it BTB4 condition}, or {\it Muon Trigger Flag} (MTF). This is described in Sec.~\ref{subsec:muon}. In addition, the BTB processes special calibration triggers such as random, electronic pulse, and timing-laser triggers. These triggers are used to monitor the detector status and are regularly generated, typically every few seconds. Service interrupts, used in the generation of calibration triggers and synchronized with 20\,MHz base clock, can raise other bit fields of the trigger word. Each event is associated with an absolute time read from a GPS receiver. Listed below are different {\it Trigger Types} ($TT$), their names and main purpose. Each trigger (or event) has a 16\,$\mu$s long DAQ gate, with the exception of a 1.6\,ms long {\it TT128} associated with the detection of cosmogenic neutrons.
 
\begin{itemize}
    \item{\it TT1 \& BTB0} -  {\it point-like events} - the main physics trigger type for ID events, when OD did not see a signal. No bit of the trigger word is raised. After each event, there is approximately a 2-3\,$\mu$s dead time window, during which no trigger can be issued.
    \item{\it TT1 \& BTB4} - {\it internal muons} - the main category of internal muons which did trigger both the OD as well as the ID. 
    \item{\it TT128} - {\it neutron trigger} - special 1.6\,ms long trigger issued shortly (order 100\,ns) after every {\it TT1 \& BTB4} event to guarantee a high detection efficiency of cosmogenic neutrons, sometimes created with very high multiplicity. The duration of this trigger type corresponds to about six times the neutron capture time~\cite{Bellini:2011yd}.
    \item{\it TT2} - {\it external muons} - the main category of external muons detected by the OD only. 
    \item {\it TT8} - {\it laser} - calibration trigger of the timing laser used to monitor the quality of the laser pulse.
    \item {\it TT32} - {\it pulser} - calibration trigger with the electronics pulse, used to monitor the number of working electronic channels. 
    \item {\it TT64} - {\it random} - forced trigger, used to monitor the dark noise and the $^{14}$C-dominated energy spectrum below the BTB threshold.
    \end{itemize}

             \begin{figure*}
        \centering
         \includegraphics[width = 0.88 \textwidth]{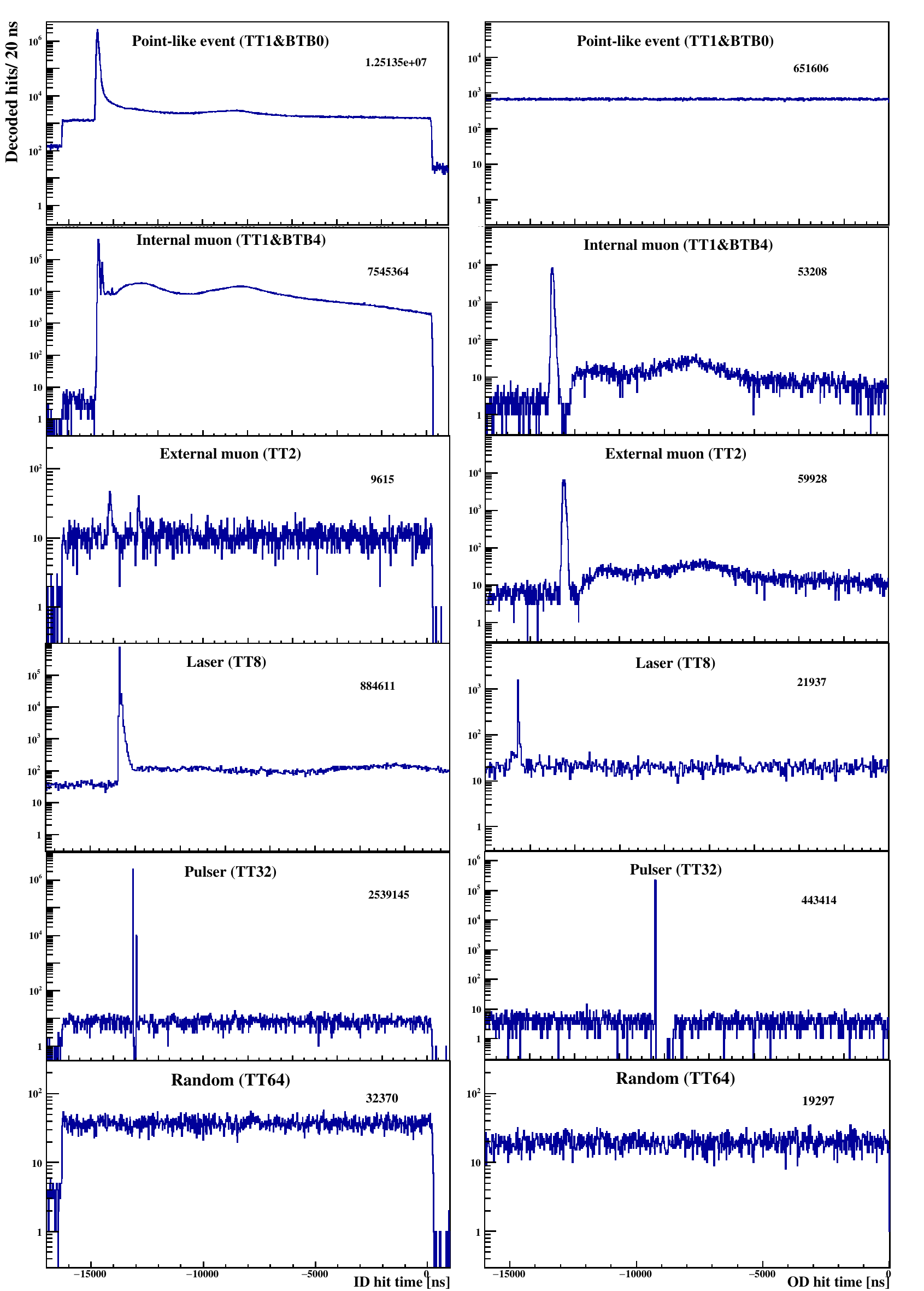}
         \vspace{2mm}
        \caption{Event structure, i.e. hit time distribution in a 16\,$\mu$s DAQ gate with respect to the trigger time, for different event types as acquired during a typical 6 hour run by the ID (left column) and the OD (right column). The hit times are negative, since the hits are detected before the trigger decision is made. Different rows represent, from top to bottom, the different trigger types: point-like events in the ID ({\it TT1 \& BTB0}); internal muons ({\it TT1 \& BTB4}); external muons ({\it TT2}); and the three types of calibration triggers: laser ({\it TT8}), pulser ({\it TT32}), and random ({\it TT64}). The total number of decoded hits detected for each trigger type is shown on the top right corner of the respective pads.}
        \label{fig:event_run}
        \end{figure*}
       
         \begin{figure*}
        \centering
         \includegraphics[width = 0.88 \textwidth]{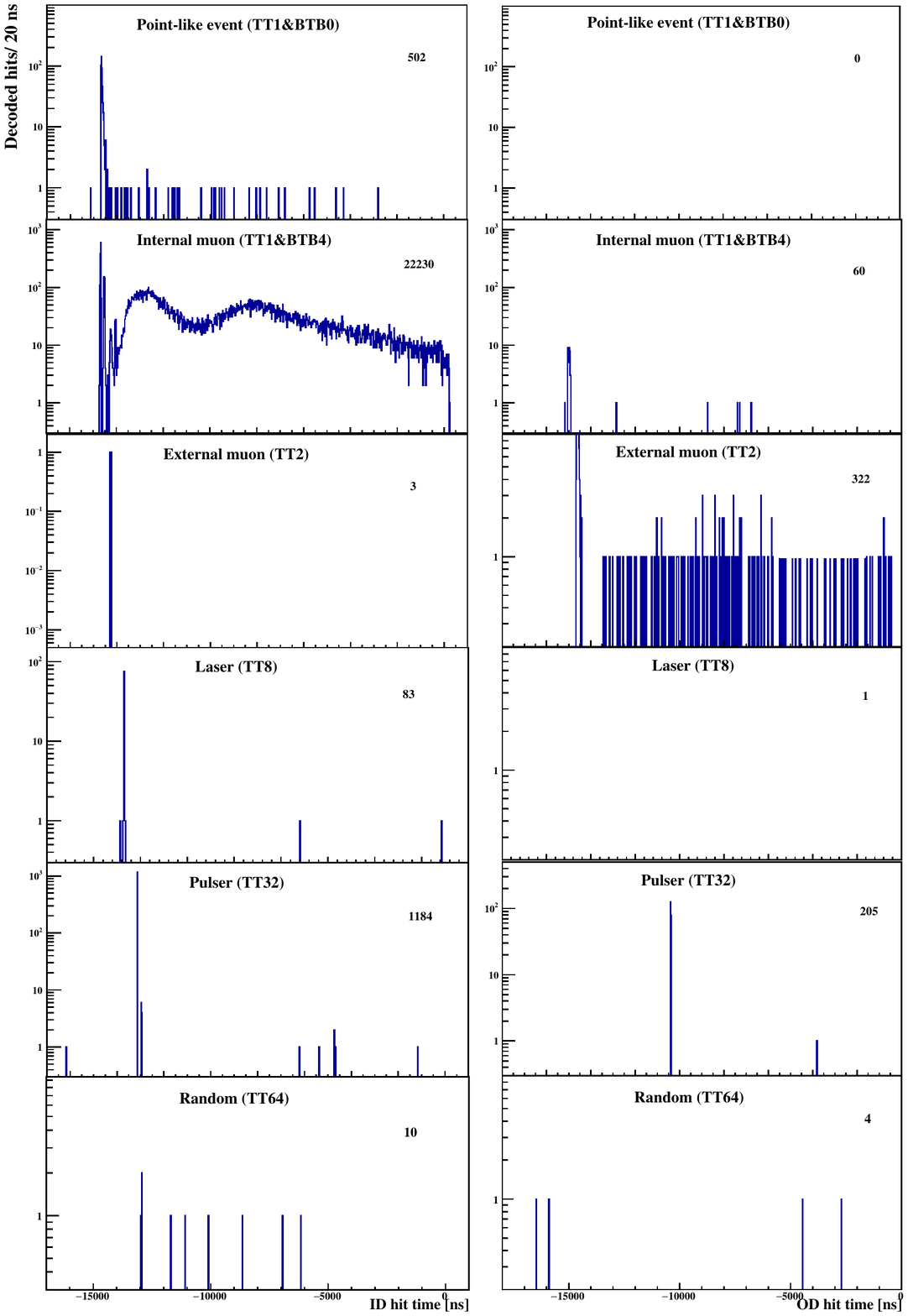}
         \vspace{2mm}
        \caption{Analogy of the Fig.~\ref{fig:event_run}, but the event structure is shown for individual events of each trigger type.}
        \label{fig:event_single}
        \end{figure*}

        \begin{figure}[t]
        \centering
        \includegraphics[width = 0.49\textwidth]{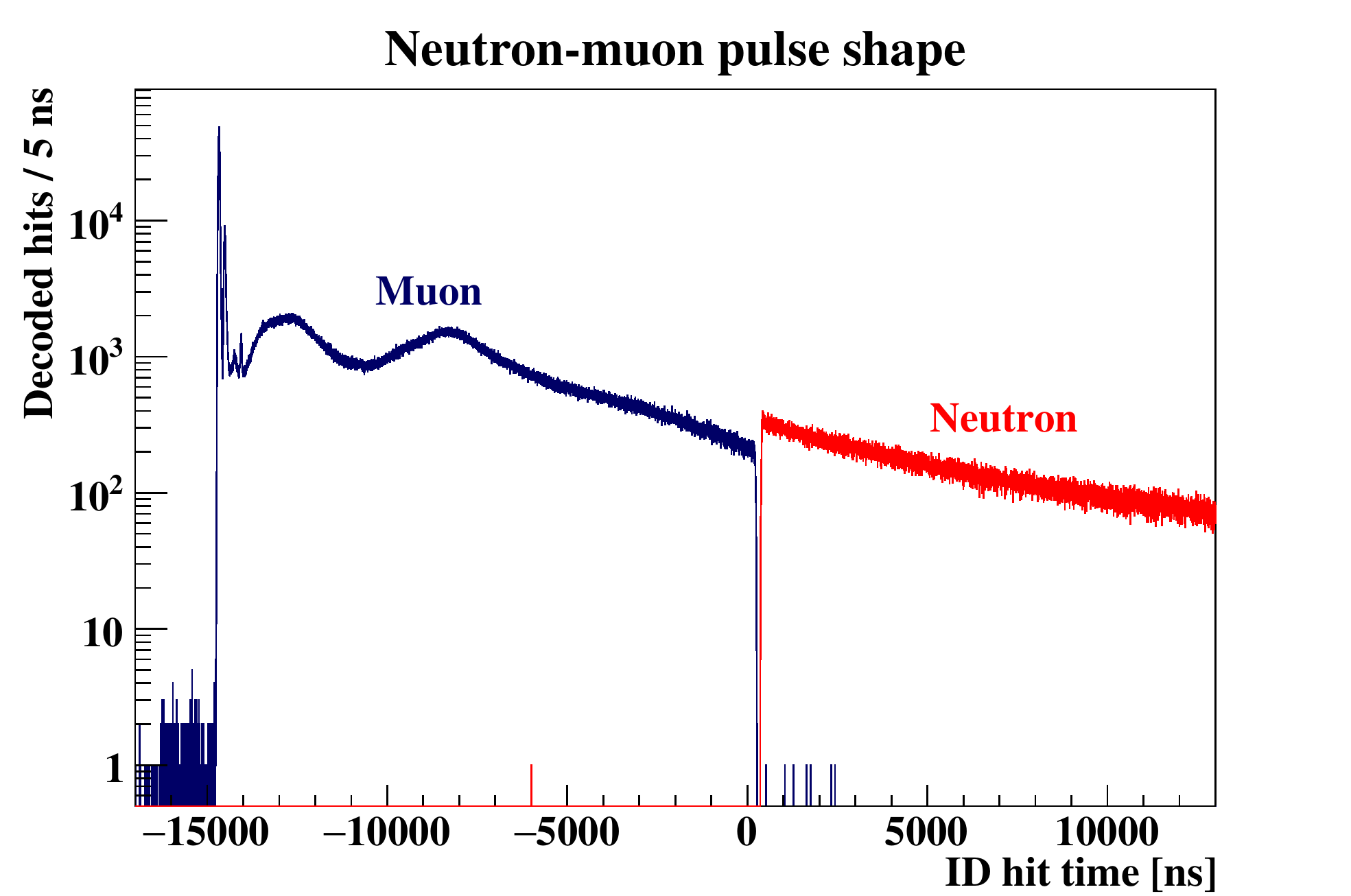}
       \includegraphics[width = 0.49\textwidth]{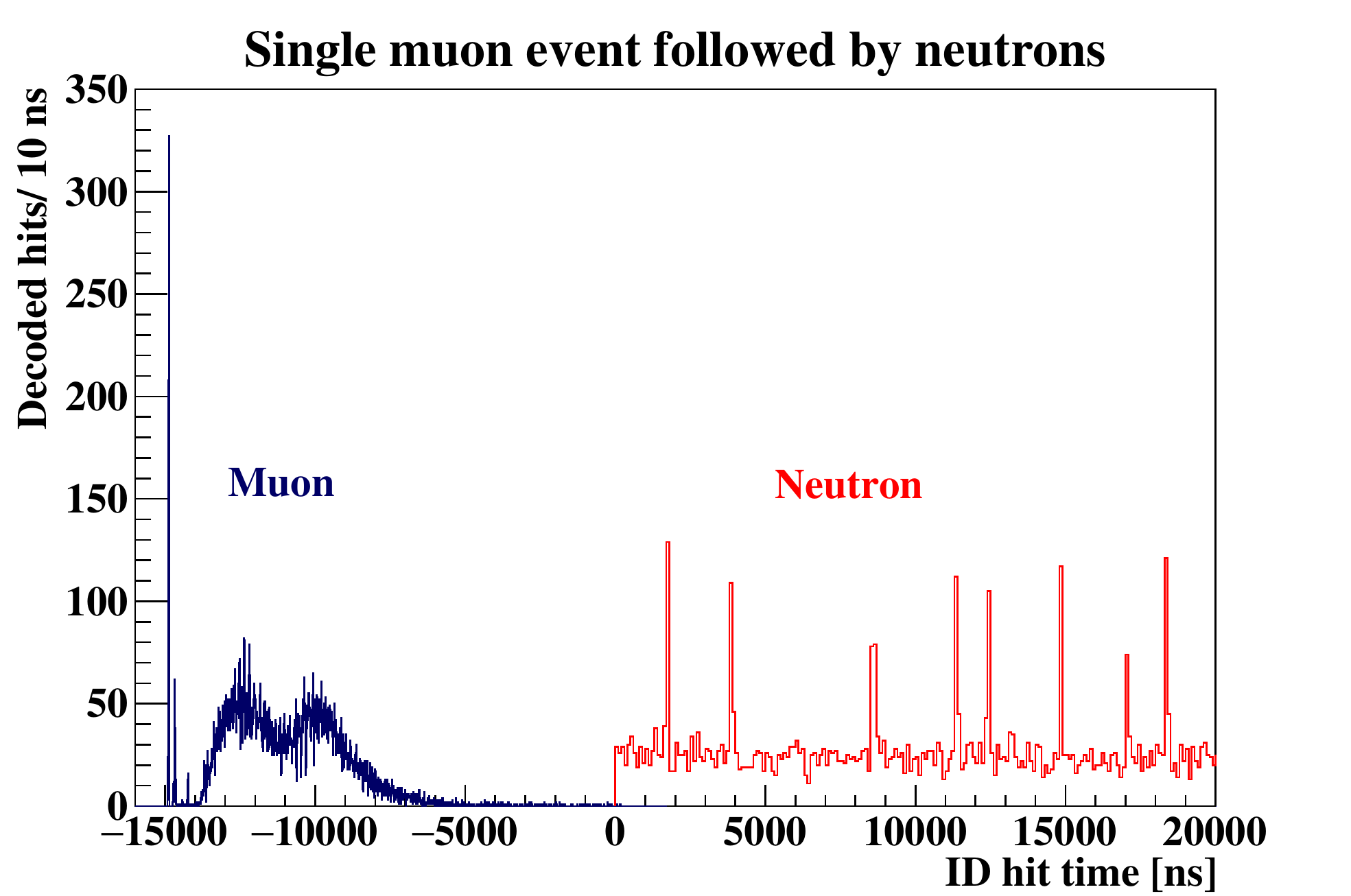}
       \vspace{1mm}
        \caption{Top: Event structure of {\it TT1 \& BTB4} muons (blue) and the start of the follow-up 1.6\,ms-long {\it TT128} neutron gates (red) collected in a typical 6\,hour run. Due to the large amount of light for muon events, the detected hits at the end of {\it TT1 \& BTB4} events are pre-scaled, causing the step between the two trigger types. Bottom: example of a {\it TT1 \& BTB4} muon (blue) followed by a {\it TT128} event (red) with high neutron multiplicity.}
        \label{fig:tt128}
        \end{figure}
        
          \begin{figure} [t]
    \centering  
    \subfigure[]{\includegraphics[width = 0.49\textwidth]{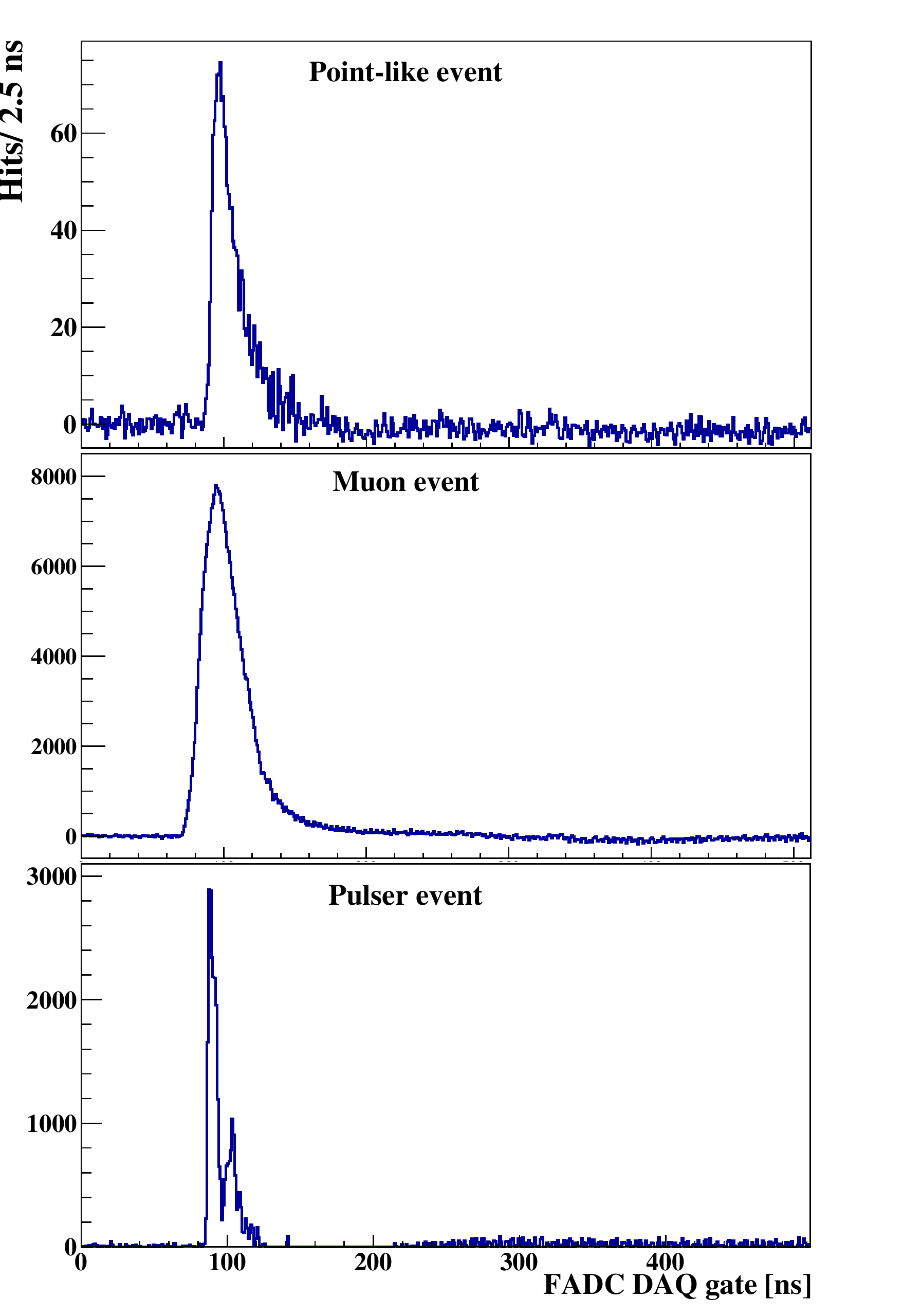}}
        \caption{Examples of the FADC 1.28\,$\mu$s long waveforms with 2.5\,ns binning: 3.9\,MeV point-like event (top), muon with 90\,MeV deposited energy (middle), calibration pulser event (bottom).}
        \label{fig:FADC_pulse_shapes}
	\end{figure}
	
During each trigger, the {\it raw hits} are the actual hits recorded during the event, while the {\it decoded hits} are all valid raw hits, i.e. the hits that are not accompanied by possible error messages from the Laben digital boards. A {\it cluster} is defined as an aggregation of decoded hits in the DAQ gate, well above the random dark-noise coincidence (typically, about 1 dark noise hit per 1\,$\mu$s in the whole detector is observed). Each cluster represents a physical event and its visible energy is parametrized by the energy estimators, each normalized to 2000 working channels:
   %
    \begin{itemize}
        \item $N_{P}$ - number of triggered PMTs;
        \item $N_h$ - number of hits within the cluster, including possible multiple hits from the same PMT;
        \item $N_{pe}$ - number of photoelectrons, calculated as a sum of charges of all individual hits contributing to $N_h$.
    \end{itemize} 

   The {\it event structure}, i.e. the depiction of the time distribution of the decoded hits in the DAQ gate, is shown in Fig.~\ref{fig:event_run} for different trigger types and for both the ID and the OD, as an integral of many events acquired during a typical 6\,hours run. Instead, Fig.~\ref{fig:event_single} shows in an analogous way the typical structure of individual events. The structure of trigger {\it TT128} is shown in Fig.~\ref{fig:tt128}. 

\paragraph {The FADC DAQ sub-system}
      
        There is an auxiliary data acquisition system based on 34 FWFD (also known as FADC), 400 MHz, 8 bit VME boards with 3 input channels on each board. This additional read-out system was created to extend the Borexino energy range to $\sim$50\,MeV, which is important to detect supernova neutrinos. The FADC DAQ energy threshold is $\sim$1\,MeV. The system has been essentially continuously operational since it was started in December 2009, with the exception in 2014 when the system had technical problems and was not operating properly for about 5\,months. 
      
        Each FADC channel receives a summed signal from up to 24 PMTs. The system works independently from the main DAQ if the PMTs and FEBs are operating. The FADC DAQ has a separate trigger, implemented through the programmable FPGA unit. The FADC event time window (DAQ gate) is 1.28\,$\mu$s long. The trigger module receives additional trigger flags such as Muon Trigger Flag (MTF, will be explained in Sec.~\ref{subsec:muon}), the output of the OD analog sum discriminator, the calibration pulser, and laser flags. The trigger unit produces the permissions and prohibitions of signals allowing recording of physical events and moderating the rate of the calibration signals. Figure~\ref{fig:FADC_pulse_shapes} shows examples of typical waveforms, in particular for a point-like event, a muon, and a calibration pulser signal. As it will be discussed in Sec.~\ref{subsec:muon}, the FADC system allows for an accurate pulse shape discrimination which is of paramount importance to achieve high muon detection efficiency.
        
        The main and FADC DAQ systems are synchronized and merged offline, based on the GPS time of each trigger, using a special software utility. Typically, FADC waveforms are extended up to $16\,\mu$s (or 1.6\,ms for $TT128$ events) with a simple unperturbed baseline. When multiple FADC events correspond to different clusters of the same main DAQ event, they are merged together and eventual gaps are substituted with simple baselines.

        \subsection{Muon detection}
        \label{subsec:muon}
 
         The overall muon detection efficiency with the main DAQ system has been evaluated on 2008 - 2009 data to be at least 99.992\%~\cite{Bellini:2011yd}. When using the additional FADC system for muon detection, this efficiency increases to 99.9969\%. The high performance of muon tagging over the 10 years period was demonstrated in a recent study of the seasonal modulation of the muon signal~\cite{Agostini:2018fnx}. In this Section we review the muon tagging methods in Borexino, and also provide updated analysis of the overall muon tagging efficiency using the main DAQ and evaluate its stability over the analyzed period. 
         The principal muon tagging is performed by a specifically designed Water-Cherenkov OD. Additionally, ID pulse-shape and several special muon flags have been designed particularly for the geoneutrino analysis, where undetected single muons could become an important background among the approximate 15 antineutrino candidates per year. Below is a summary of different categories of muon detection in Borexino.
         
        \paragraph{Muon Trigger Flag (MTF)}
        MTF muons trigger the OD and set the bit BTB4 of the trigger board, as described above in Sec.~\ref{subsec:data}.
        
        \paragraph{Muon Cluster Flag (MCF)}

        The MCF flag is set by a software reconstruction algorithm using the hits acquired from the OD. It considers separately two subsets of OD PMTs: those mounted on the SSS and those on the floor of the Water Tank. The MCF condition is met if 4 PMTs of either subset are fired within 150\,ns. 
        
        \paragraph{Inner Detector Flag (IDF)}
        
        The IDF identifies muons based on different time profiles of hits originating from muon tracks with respect to those from point-like events. These time profiles are characterized by the {\it peak time} and the {\it mean time} of the cluster of hits from the ID. The mean time is the mean value of the times of decoded hits that belong to the same cluster, while the peak time is the time at which most of the decoded hits are deposited. The IDF is optimized in three different energy ranges. For muons, the $N_h$ energy estimator is proportional to the track length across the ID rather than to the muon energy. Events with $N_h > 2100$ and with mean time greater than 100\,ns are considered as muons. In the lower energy interval $80 < N_h < 2100$, events with peak time $>$30(40)\,ns are tagged as muons above (below) $N_h$ = 900, respectively. In addition, the Gatti $\alpha$/$\beta$ discrimination parameter (see Sec.~\ref{subsec:pulse_shape}) is used in IDF to reject electronics noise, re-triggering of muons, as well as scintillation pulses coming from the buffer. In the very low energy interval $N_h < 80$ dominated by $^{14}$C background, IDF is not applied due to the limited performance of pulse-shape identification. A detailed explanation of the IDF flag can be found in~\cite{Bellini:2011yd}.

    \begin{figure}[t]
	\includegraphics[width = 0.5\textwidth]{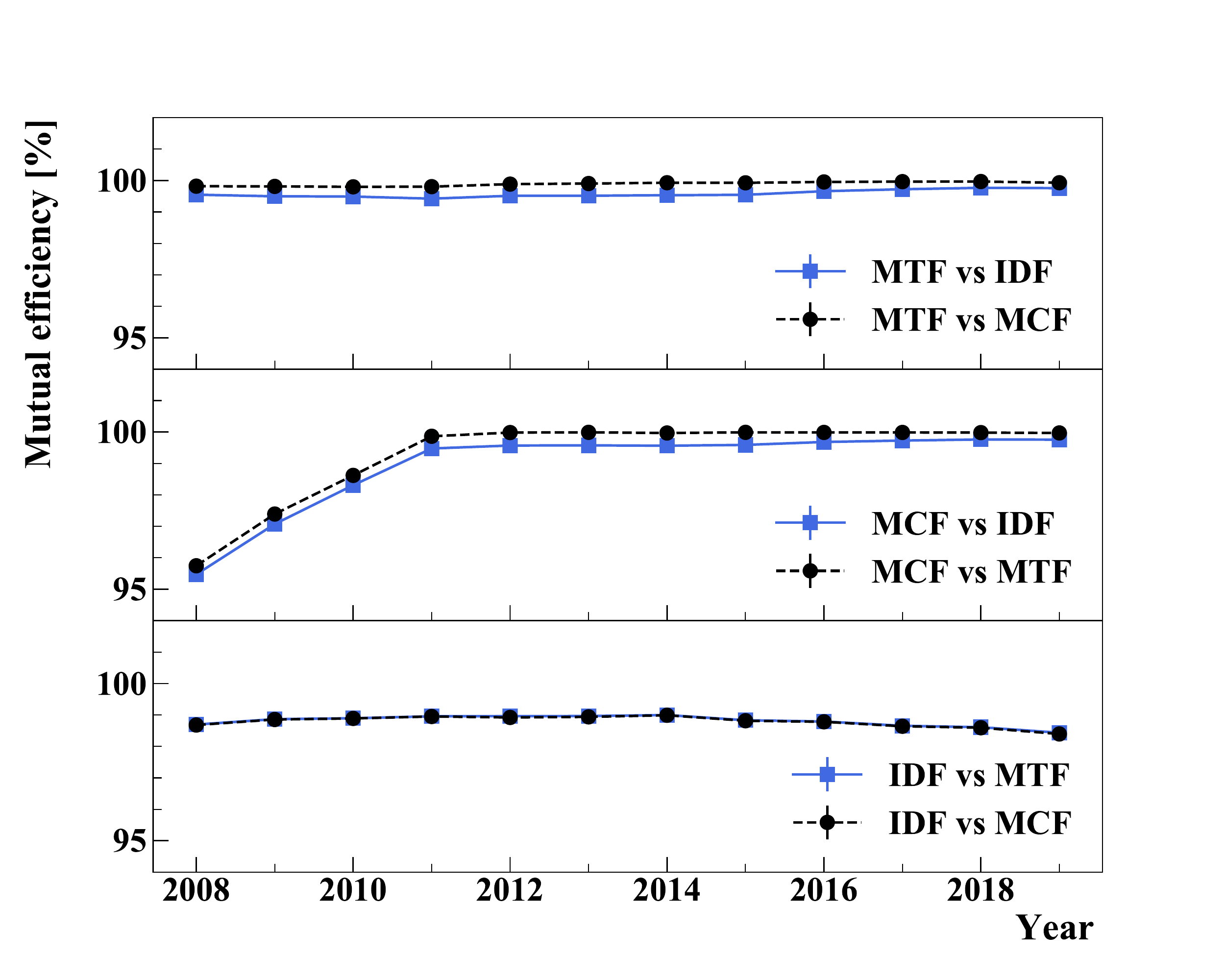}
		\caption{Mutual efficiencies of the three strict muon flags MTF (top), MCF (middle), and IDF (bottom) for events with $N_h > 80$ calculated for calendar years (2019 contains data up to end of April).}
		\label{fig:mut_eff}
	\end{figure}
	%

        \paragraph{External Muon Flag}
  
  External muons are those that did not deposit energy in the ID and passed only through the OD. For all of them we require that they do not have a cluster in the ID-data. Most of these muons are $TT2$ events, very few also $TT1$ triggers, that satisfy MTF or MCF (slightly modified) conditions.

        \paragraph{Strict Internal Muon Flag}
  
  The three above mentioned muon flags MTF, MCF, IDF are optimized in order to maximize the muon tagging efficiency while keeping the muon sample as clean as possible. Events that deposited energy in the ID, i.e. those that have a cluster of hits in the ID data, and are tagged by any of these three tags, are identified as {\it Strict Internal Muons}. These events are largely dominated by $TT1$ events. However, we include in this category also the rare $TT2$ events, which did not trigger the ID, but have a cluster with more than 80 hits in the ID-data.
  
\begin{table*}
	\centering
	\caption{\label{tab:mut_eff_tab} Mutual efficiencies for the three muon flags MTF, MCF, and IDF are given for the period December 2007 -- April 2019 used in the geoneutrino analysis. In the three last columns, the muon reference sample was defined, from left to right, by IDF, MTF, and MCF flags, respectively. In the last 4 rows, the IDF efficiency is shown for different energy ranges as well.} \vskip 2pt
	\begin{tabular*}{14 cm}{c @{\hskip 24pt} c @{\hskip 24pt} c @{\hskip 23pt} c @{\hskip 23pt} c}
		\hline
		\hline
		\thead{{\normalsize Visible energy} \\ {\normalsize (N$_{h}$)}} & \thead{{\normalsize  Mutual} \\  {\normalsize efficiency $\varepsilon$}} & vs IDF & vs MTF & vs MCF \Tstrut\Bstrut \\ 
		\hline
		$\ge$80 & $\varepsilon_{\mathrm{MTF}}$ & 0.9957 $\pm$ 0.0004 & x & 0.9989 $\pm$ 0.0004 \Tstrut \\
		$\ge$80 & $\varepsilon_{\mathrm{MCF}}$ & 0.9894 $\pm$ 0.0004 & 0.9927 $\pm$ 0.0004 & x  \\ 
		$\ge$80 & 	$\varepsilon_{\mathrm{IDF}}$ & x & 0.9882 $\pm$ 0.0004 & 0.9881 $\pm$ 0.0004  \Bstrut \\
		\hline
		 80-900 & 	$\varepsilon_{\mathrm{IDF}}$  & x & 0.7941 $\pm$ 0.0014 & 0.7921 $\pm$ 0.0014 \Tstrut \\ 
	   	900-2100 & $\varepsilon_{\mathrm{IDF}}$  & x & 0.9988 $\pm$ 0.0008 & 0.9984 $\pm$ 0.0008\\ 
        $\ge$2100 & $\varepsilon_{\mathrm{IDF}}$  & x & 1.0000 $\pm$ 0.0005 & 1.0000 $\pm$ 0.0005 \Bstrut \\ 
		\hline
		\hline
	\end{tabular*}
\end{table*}

In the lack of a pure muon sample, we introduce the concept of mutual efficiencies. We first define the muon reference sample with one flag, with respect to which we express the efficiency of the other two flags. Such mutual efficiencies of the three strict muon flags for muons with $N_h > 80$ were studied over the years, as shown in Fig.~\ref{fig:mut_eff}. These are crucial in order to estimate the number of untagged muons which affect the geoneutrino candidate sample (Sec.~\ref{subsec:cosmogenic_est}).

The MTF efficiency has been mostly stable over the years. The MCF efficiency was slowly increasing since 2008 and has reached its optimal stable performance in 2011. The lowered MCF efficiency during the first years was due to occasional instability of the OD in the main DAQ data stream, that however did not influence neither the trigger system nor the MTF flag. In this case the OD data was not acquired. The IDF efficiency has been slowly decreasing since 2014 due to the decreasing number of active PMTs in the ID. This efficiency decrease is limited only to low energy ranges.

The average mutual efficiencies over the whole analyzed period are shown in Table~\ref{tab:mut_eff_tab}.
The highest mutual inefficiency is 1.19$\%$ (the inefficiency of the IDF flag with respect to the MCF flag) and the highest mutual efficiency is 99.89$\%$ (the efficiency of the MTF flag with respect to the MCF flag). This can be used to calculate the overall inefficiency of the three muon flags as 0.0119 $\times$ (1 - 0.9989) = (1.31 $\pm$ 0.5)$\times 10^{-5}$.

        \paragraph{Special Muon Flags}
  
\begin{figure}[t]
	\includegraphics[width = 0.49\textwidth]{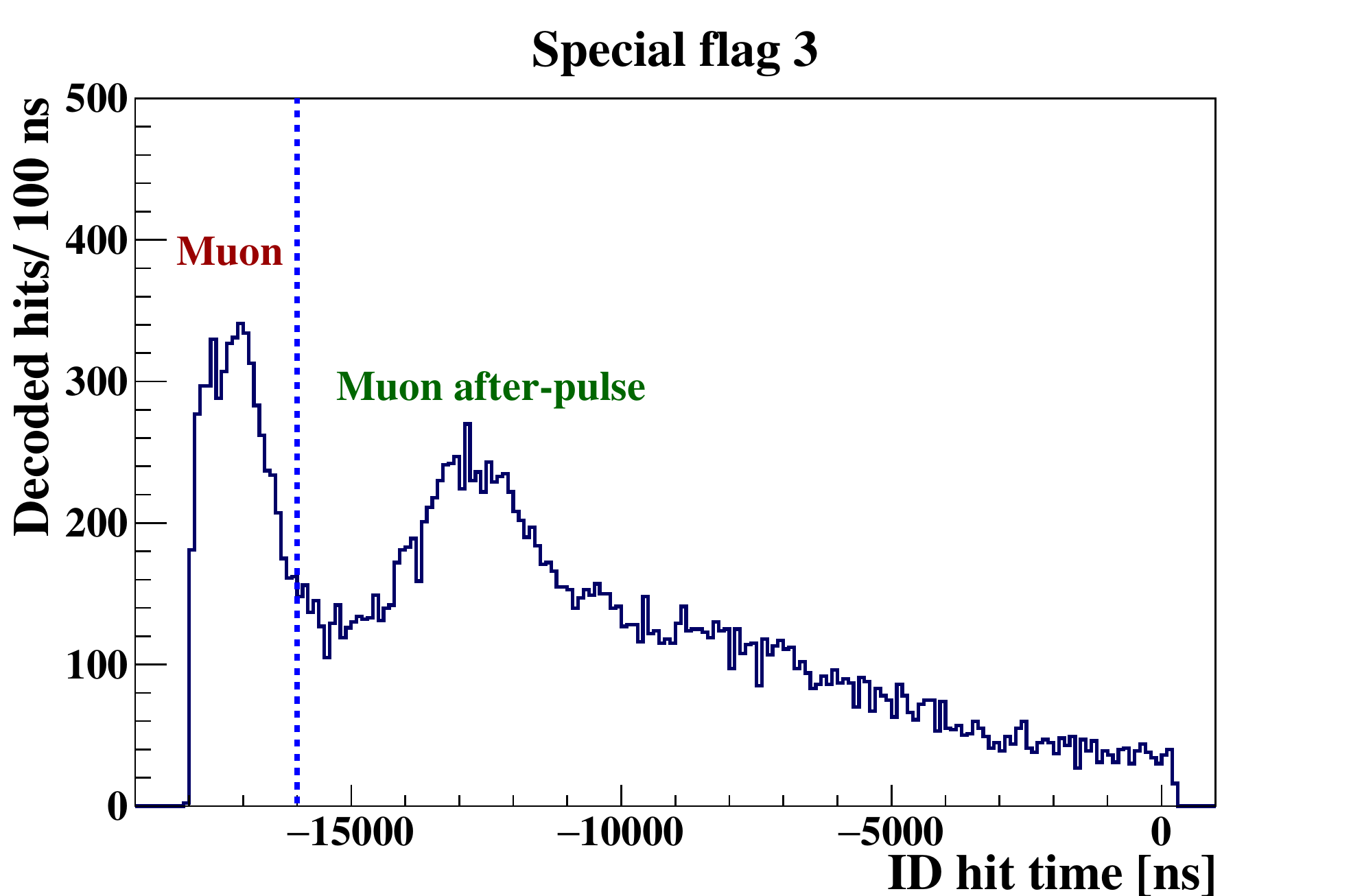}	\includegraphics[width =0.49\textwidth]{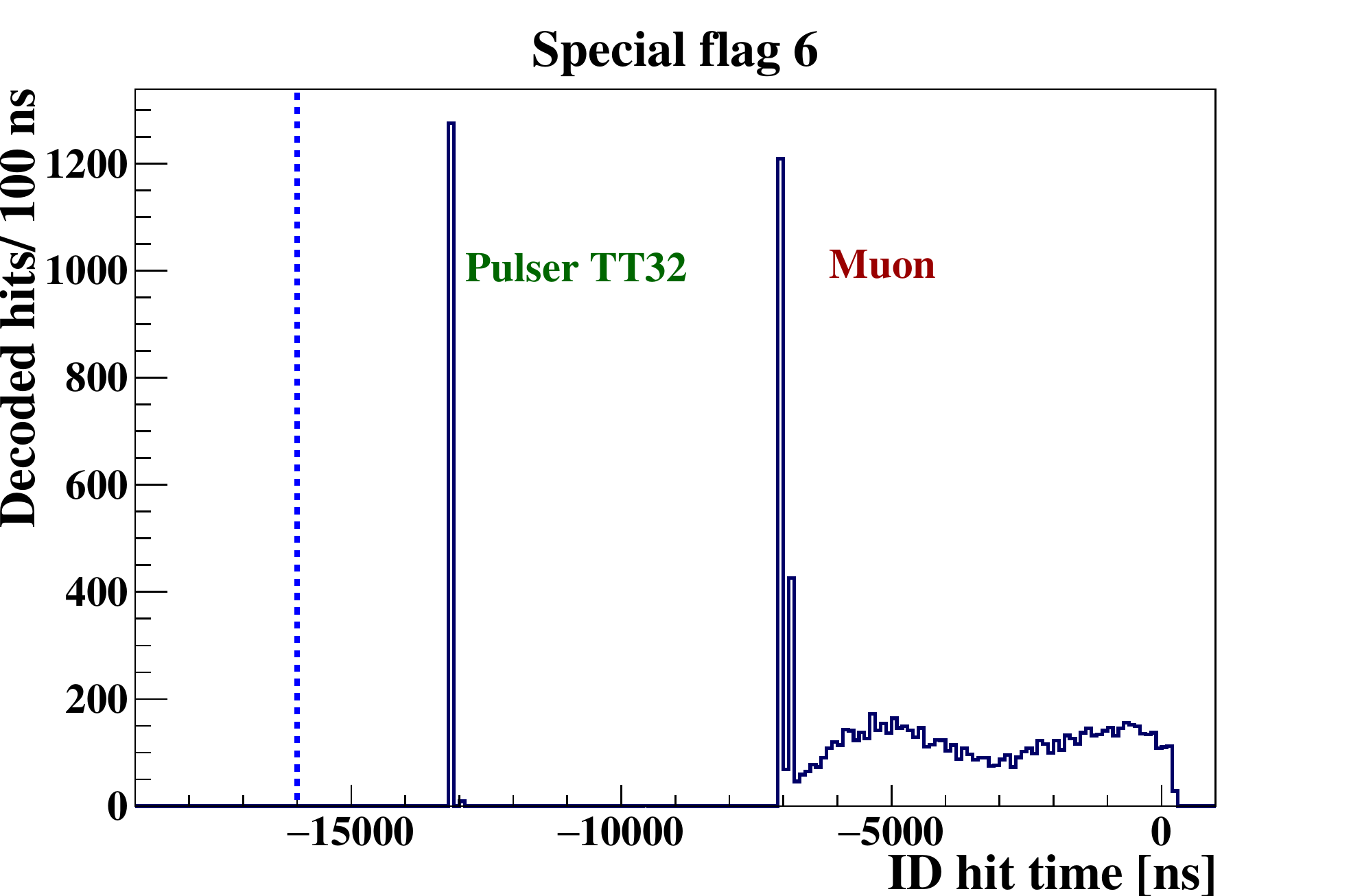}
		\caption{Top: Muon type {\it special flag 3} showing the muon which crossed the detector during the dead-time after the previous trigger. While the main peak is out of gate, the muon was detected anyway, triggering on the after-pulse. The main muon peak is out of the gate. Bottom: Muon {\it special flag 6} type showing the pulser trigger $TT32$ with a muon occurring during the same gate. The vertical dashed lines represent the start of the DAQ gate at -16\,$\mu$s.}
		\label{fig:sp_muon}
	\end{figure}

       In addition to the Strict Internal Muon Flags, there are also six kinds of {\it Special Muon Flags}. These are designed to tag the small, remaining fraction of muons at the cost of decreased purity of the muon sample. In fact, special tags mark as muons also noise events. In the antineutrino analysis, all special muons are conservatively treated as internal muons (see Sec.~\ref{subsec:cosmogenic}), regardless of the fact whether they have a cluster of hits in the ID-data.  
        \begin{itemize}
            \item  \textit{Special flag 1} This flag was introduced to tag muons when the electronics was too saturated to correctly read out all photons, i.e. there was sufficiently high amount of raw hits ($NR$) but very few decoded hits ($ND$). Therefore, this flag tags events with $NR > 200$ when only less than 5\% of them are decoded ($ND/NR < 0.05$). These events can have trigger type $TT1$ or $TT2$, without any condition regarding the BTB trigger word.
            \item \textit{Special flag 2} This flag tags events of trigger type $TT1$ or $TT2$ with $ND > 100$ that have the $2^2$ bit of the trigger word raised, but the OD triggered in coincidence with the service interrupt which generated service triggers. In this case, additional bits of the trigger word can be raised. For these events, only part of the muon data can be present in the DAQ gate or additional calibration pulses are possibly present in the event. 
           \item \textit{Special flag 3} Events with a cluster start time out of the DAQ gate, as shown in Fig.~\ref{fig:sp_muon}, are tagged as special muons. These events typically tag muons that pass through the detector during the 2-3\,$\mu$s dead time after {\it TT1 \& BTB0} events. The muon itself does not generate a trigger, but the hits from the PMT after-pulses do. Even if the detection efficiency for out-of-gate hits is $<$100\%, the main muon pulse is detected and positioned before the start of the DAQ gate.   
           \item \textit{Special flag 4} The point-like events are expected to have cluster mean time smaller than 200\,ns. Conservatively, all events with a cluster mean time greater than 200\,ns are tagged as special muons.
           \item \textit{Special flag 5} There is an extremely small number of $TT1$ events, that by definition triggered the ID, but have zero clusters in the ID-data. Conservatively, if they are tagged by the MTF or MCF, they are considered as special muons.
           \item \textit{Special flag 6} The events of trigger type $TT8$, $TT32$, or $TT64$ (examples of these events are shown in the three lower rows of Fig.~\ref{fig:event_single}) that have unexpectedly high number of $ND$ hits can be due to a muon occurring inside a service event. This is graphically shown in Fig.~\ref{fig:sp_muon}. 
         \end{itemize}
         
  \begin{figure} [t]
\centering
\includegraphics[width = 0.44\textwidth]{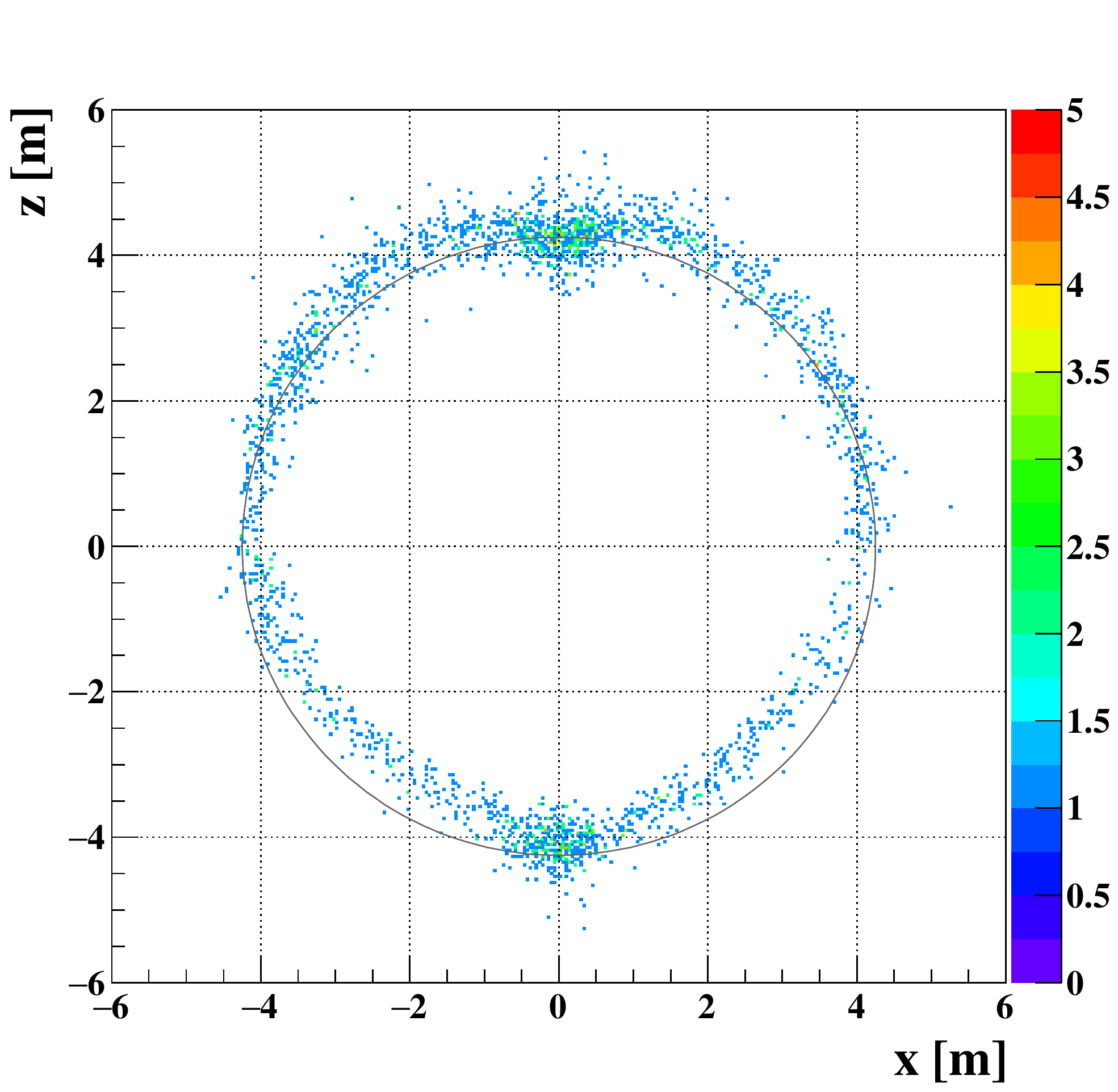}
\caption{A cross section ($z$ - $x$ plane, $|y| < 0.5$\,m) view of the distribution of 2478 events, acquired during a 3 week period, and selected for the IV shape reconstruction. The color axis represents the number of events per 0.0016\,m$^3$, in a pixel of 0.04\,m $\times$ 1.00\,m $\times$ 0.04\,m ($x \times y \times z$). This distribution reveals the IV shift and deformation with respect to its nominal spherical position shown in solid black line.}
\label{fig:Bi210_vessel}
\end{figure}

        \paragraph{FADC muon identification} 
    
  The FADC DAQ improves the efficiency of muon detection in the ID. The advantage of the FADC system over the main electronics is the availability of detailed pulse shape information of the event. The selection of muons is obtained using a special algorithm which includes 8 different independent tests to classify a muon event. In addition to checking information about the triggers of  and the data from the OD, four different classifiers are deployed. Three of the classifiers are based on machine learning and one of them considers formal and simple quantitative characteristics of the pulse shape, as the rise of the leading edge and the pulse amplitude. The following approaches are used for machine learning classifiers: a neural network based on the {\it Multi-Layer Perceptron} (MLP)~\cite{Voss:2007jxm}, a {\it Support Vector Machine} (SVM)~\cite{Cortes:1995}, and a {\it Boosted Decision Tree} (BDT)~\cite{Freund:1997xna}. These were implemented to make a decision using the {\it Toolkit for Multivariate Data Analysis} (TMVA) with ROOT~\cite{Voss:2007jxm}. The classifiers use the event pulse-shape and additional parameters that determine the distribution of the digitized waveform. Different tests have different tagging efficiencies. There are three levels of reliability of tagging, defined according to the number of classifiers which tag an event as a muon. The lowest level of reliability is suitable for the antineutrino analysis. In this case, the maximum tagging efficiency is achieved with the price of the greatest over-efficiency. The combined muon tagging efficiency of the main and the FADC DAQ systems is 99.9969\%~\cite{Lukyanchenko:2017}.

 \paragraph{Internal Large Muon Flag}
     
The term {\it Internal Large Muon Flag} is used for the muon category that considers as internal muons not only the strict internal flag muons, but includes also special and FADC muon flags. This approach is conservative, since internal muons, contrary to external muons, are able to create cosmogenic background other than neutrons and thus, generally, require longer veto. This will be discussed in Sec.~\ref{subsec:vetoes}.

        \subsection{Inner Vessel shape reconstruction}
        \label{subsec:IV}

The shape of the thin nylon IV holding the Borexino scintillator changes with time, deviating from a spherical shape. This deformation is a consequence of a small leak in the IV, with a location estimated as $26^{\circ} < \theta < 37^{\circ}$ and $225^{\circ} < \phi < 270^{\circ}$~\cite{Bellini:2013lnn}. The leak developed approximately in April 2008 and was detected based on a large amount of events reconstructed outside the IV. In order to minimize the buoyant force between the buffer and the scintillator liquids, their density difference was reduced by partial removal of DMP from the buffer by distillation, with negligible consequences on the buffer's optical behaviour. The evolution of the IV shape needs to be monitored and is also crucial for the antineutrino analysis. 

In the geoneutrino analysis, the so-called {\it Dynamical Fiducial Volume} (DFV) (Sec.\ref{subsec:DFV}) is defined along the time-dependent reconstructed IV shape. The reconstruction method, introduced in~\cite{Bellini:2013lnn}, is based on events in the 800 - 900\,keV energy range ($N_{pe}$ = 290 - 350\,p.e.) reconstructed on the IV surface (Fig.~\ref{fig:Bi210_vessel}). These events originate in the radioactive contamination of the nylon and are dominated by $^{210}$Bi, $^{40}$K, and $^{208}$Tl. The reconstructed position of selected events is fit assuming uniform azimuthal symmetry ($x$ - $y$ plane) so that the $\theta$-dependence of the vessel radius $R$ can be determined. Three weeks of data provide sufficient statistics for this analysis.

 \begin{figure}[H]
\centering
\subfigure[]{\includegraphics[width = 0.49\textwidth]{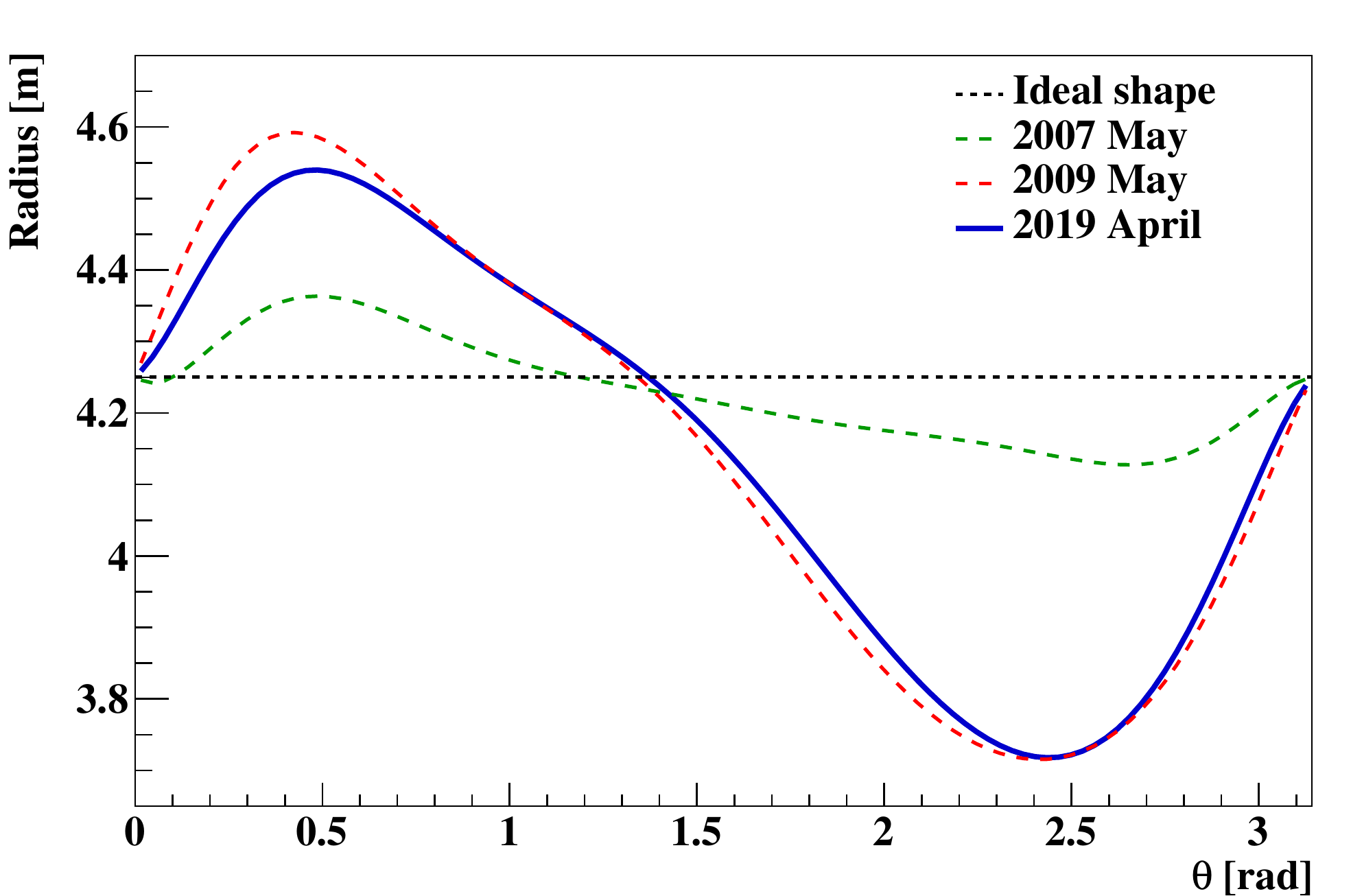}\label{fig:VShape}}
\subfigure[]{\includegraphics[width = 0.49\textwidth]{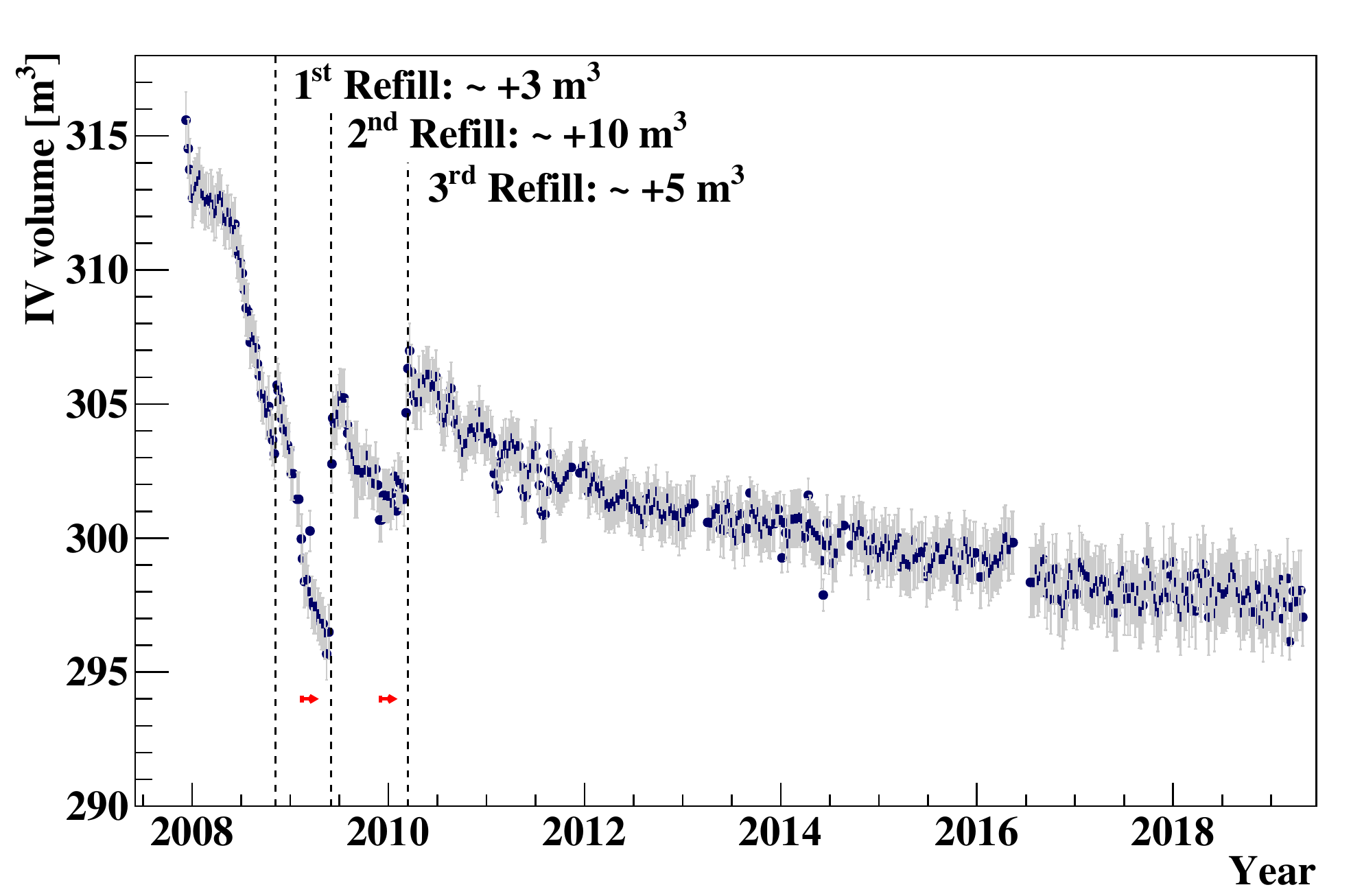}\label{fig:VolumeTrend}}
\caption{(a) Examples of the reconstructed IV shapes, i.e. reconstructed IV radius $R$ as a function of $\theta$, resulting from fitting of the ($R, \theta$) distributions of the selected events originating in the IV contamination. The solid blue line shows one recent vessel shape. The dashed lines represent the least (green) and the most (red) deformed IV shapes registered. The dashed black line shows the ideal sphere with 4.25\,m radius. (b) Trend of the reconstructed IV volumes as a function of time (1 week bin), starting from 2007, December 09 up to 2019, April 28. Each point represents 3 weeks of data. The error bars reflect the goodness of the fit of ($R, \theta$) distributions. The three dashed vertical lines represent the LS refill into the IV: we observe that the increase in the reconstructed volume corresponds to the amount of inserted LS. During the two periods (small red arrows in the lower part of the plot), the concentration of the DMP in the buffer was decreased from the original 5.0\,g/l to 3.0 and then to 2.0\,g/l. Thanks to the resulting better match between the densities of the LS and the buffer liquid, the rate of the leak was minimized, but not fully stopped.  }
\label{VshapeVolumeTrend}
\end{figure}

A combination of a high-order polynomial, a Fourier series, and a Gaussian distribution was used as the function to fit the 2-dimensional distribution ($R, \theta$). Figure~\ref{fig:VShape} shows examples of the reconstructed IV shapes, in particular the reconstructed radius $R$ as a function of $\theta$. Fixed parameters of the fit are the two end points ($\theta=0, \theta=\pi$) of the distribution where the vessel radius is imposed to be 4.25\,m, since the IV is fixed to the end caps that are held in place by rigid support. 
This procedure was cross-checked and calibrated over several ID pictures with an internal CCD camera system, which were taken throughout the data collection. The precision of this method is found to be $\sim$1\% ($\pm$5\,cm). 
An estimation of the active volume of scintillator is then calculated by revolving the ($R, \theta$) function around the z-axis. Figure~\ref{fig:VolumeTrend} shows the results of the rotational integration as a function of time. This evolution is very important to monitor the status and the stability of the Borexino IV. The errors in the plot represent the goodness of the 2D fit only. The effect of the IV reconstruction precision is included in the systematic uncertainty of the geoneutrino measurement and will be discussed in Sec.~\ref{subsec:syst}.
        \subsection{$\alpha$ / $\beta$ discrimination}
        \label{subsec:ab}

        The time distribution of the photons emitted by the scintillator depends on the details of the energy loss, and consequently on the particle type that produced the scintillation. For example, $\alpha$ particles have high specific energy loss due to their higher charge and mass. The energy deposition of a particle provides a way to characterize the pulse shape which can be used for particle identification~\cite{Bellini:2013lnn}.
        Thus Borexino has different pulse-shape discrimination parameters which aid in the distinction of $\alpha$-like and $\beta$-like interactions, and even more generally, to discriminate highly ionizing particles ($\alpha$, proton) from particles with lower specific ionization ($\beta^-$, $\beta^+$, $\gamma$).
        These parameters were tuned using the Radon-correlated $^{214}$Bi($\beta^{-}$) - $^{214}$Po($\alpha$) coincidence sample that was present in the detector during the WE-cycles. This occurred between June 2010 and August 2011 as a part of the scintillator purification process. The $\alpha$/$\beta$ discrimination parameters are important in the geoneutrino analysis since they help in distinguishing the nature of the delayed signals, as it will be shown in Sec.~\ref{subsec:pulse_shape}.

        \paragraph{Gatti Optimal Filter}
        
\begin{figure}[H]
            \centering
             \subfigure[]{\includegraphics[width = 0.49\textwidth]{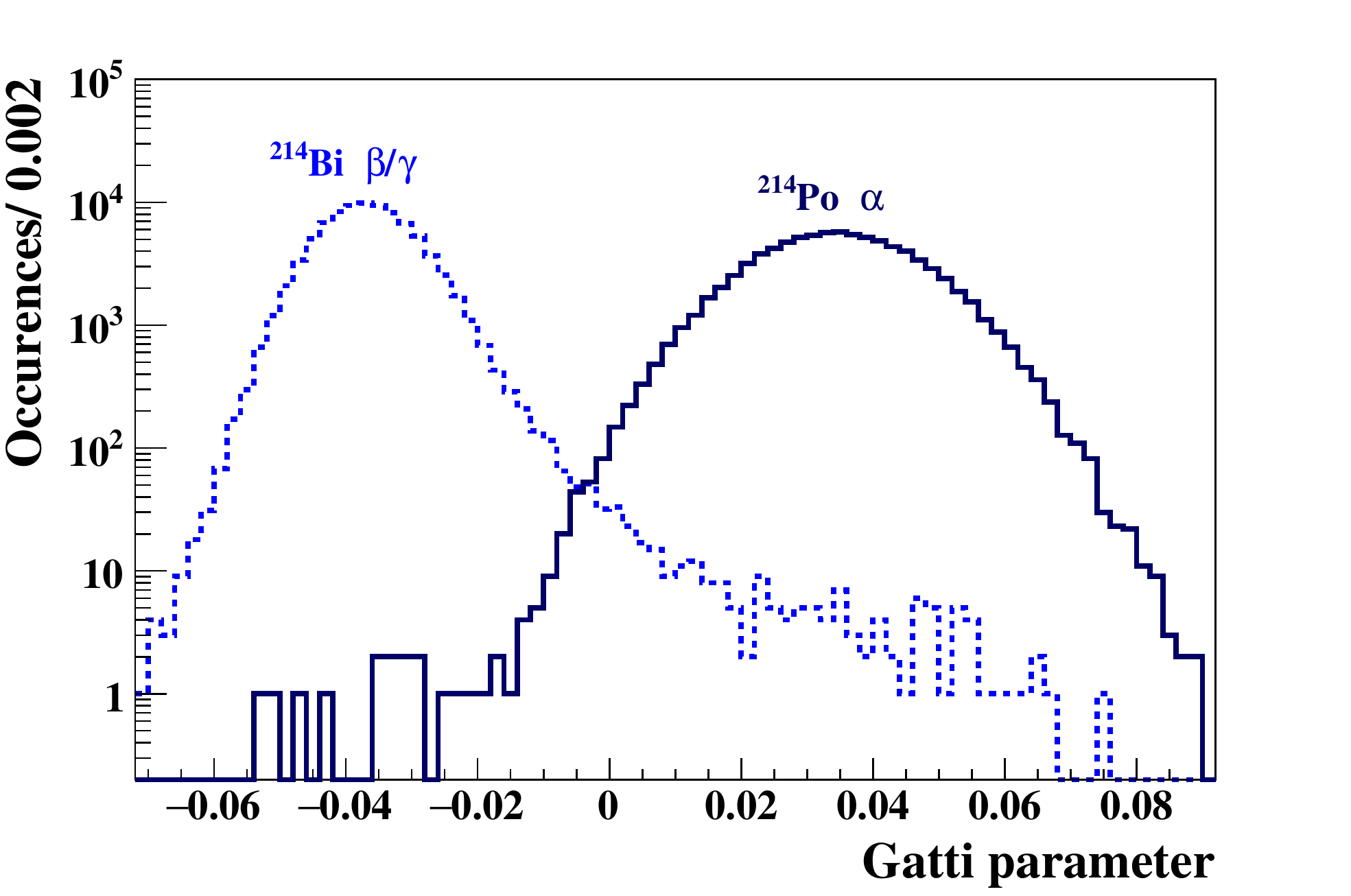}\label{fig:gatti_pic}}               \subfigure[]{\includegraphics[width = 0.49\textwidth]{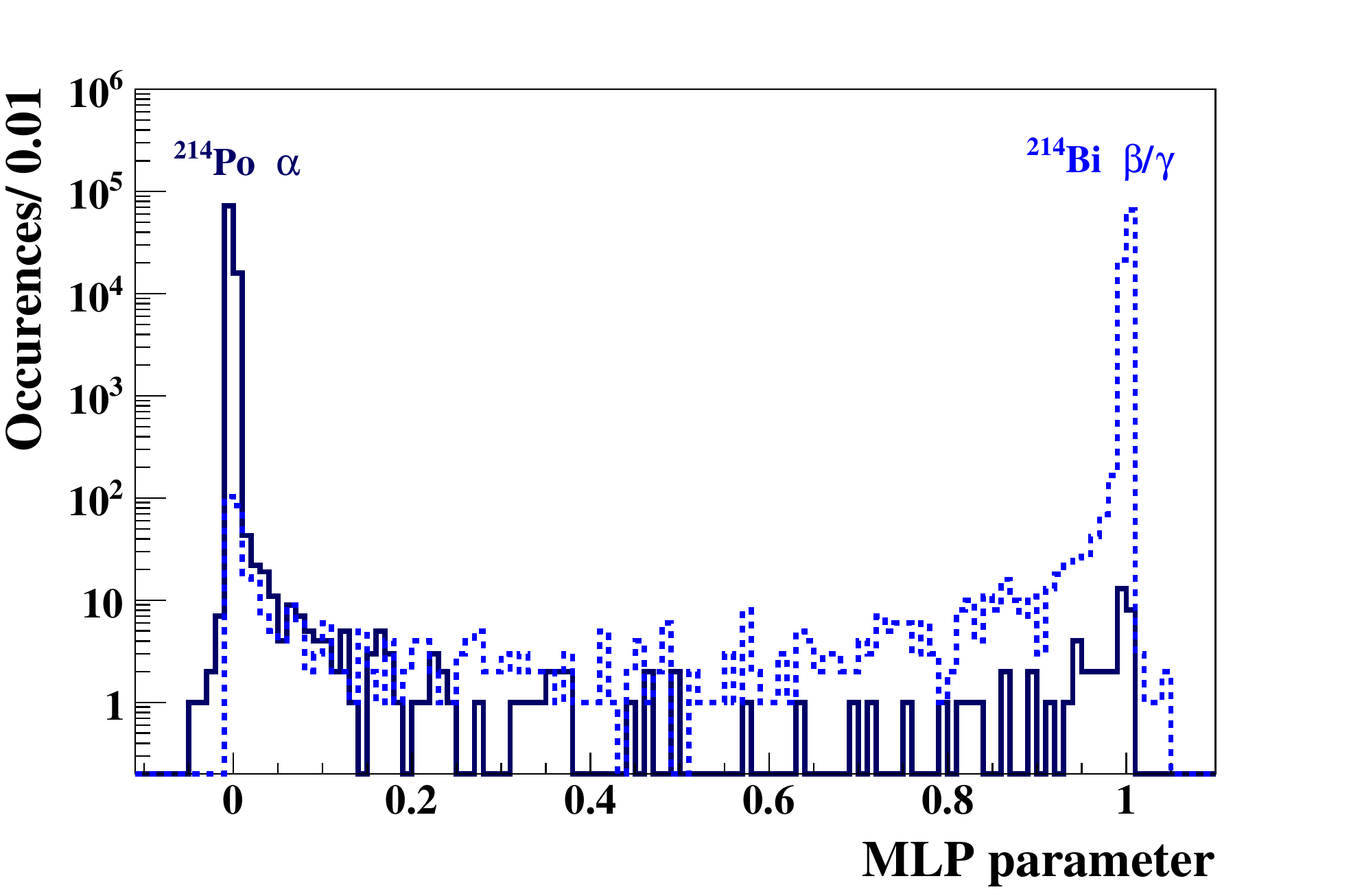}\label{fig:mlp_pic}}
          \caption{Distributions of the Gatti ($G$) (a) and the Multi-Layer Perceptron (MLP) (b) $\alpha / \beta$ discrimination parameters for $^{214}$Bi($\beta^{-}$) (dashed line) and $^{214}$Po($\alpha$) (solid line) events.}
        \end{figure}

            The {\it Gatti optimal filter} ($G$) is a linear discrimination technique, which allows to separate two classes of events with different time distributions~\cite{Back:2008,Bellini:2013lnn}. First, using the typical $\beta$ and $\alpha$ time profiles after the time-of-flight subtraction, the so-called {\it weights} $w(t_{n})$ are defined for time bins $t_n$:
                \begin{equation}
                \centering
                w(t_{n}) \equiv \frac{P_{\alpha}(t_{n}) - P_{\beta}(t_{n})} {P_{\alpha}(t_{n}) + P_{\beta}(t_{n})},
                \label{eq:gatti_w}
                \end{equation}
            where $P_{\alpha}(t_n)$ and $P_{\beta}(t_n)$ are the probabilities that a photoelectron is detected at the time bin $t_n$ for $\alpha$ and $\beta$ events, respectively. The Gatti parameter $G$ for an event with the hit time profile $f(t_n)$, after the time-of-flight subtraction, is then defined as:
                \begin{equation}
                \centering
                G = \sum_{n}^{}{f(t_{n})w(t_{n}) }.
                \label{eq:gatti_eq}
                \end{equation}
            Figure~\ref{fig:gatti_pic} shows the distributions of the Gatti parameter for $\beta$ particles from $^{214}$Bi ($G < 0$) and $\alpha$ particles from $^{214}$Po ($G > 0$).
             %
            
    \paragraph{Multi-Layer Perceptron}  
    
The {\it Multi-Layer Perceptron} (MLP) is a non-linear technique developed using deep learning for supervising binary classifiers, {\it i.e.} functions that can decide whether an input (represented by a vector of numbers) belongs to one class or another. In Borexino this technique was applied for $\alpha$/$\beta$ discrimination~\cite{Agostini:Jun2017}, and uses several pulse-shape variables, parametrizing the event hit-time profile, as input. Among these variables are, for example, tail-to-total ratio for different time bins $t_n$, mean time of the hits in the cluster, their variance, skewness,  kurtosis, and so on. The $\alpha$-like events tend to have an MLP value of 0, while the $\beta$-like events tend to have an MLP value of 1, as it can be seen in Fig.~\ref{fig:mlp_pic}.
    
    \section{ANTINEUTRINO DETECTION}
    \label{sec:IBD}
     \begin{figure}
        	\centering
            \includegraphics[width = 0.46\textwidth]{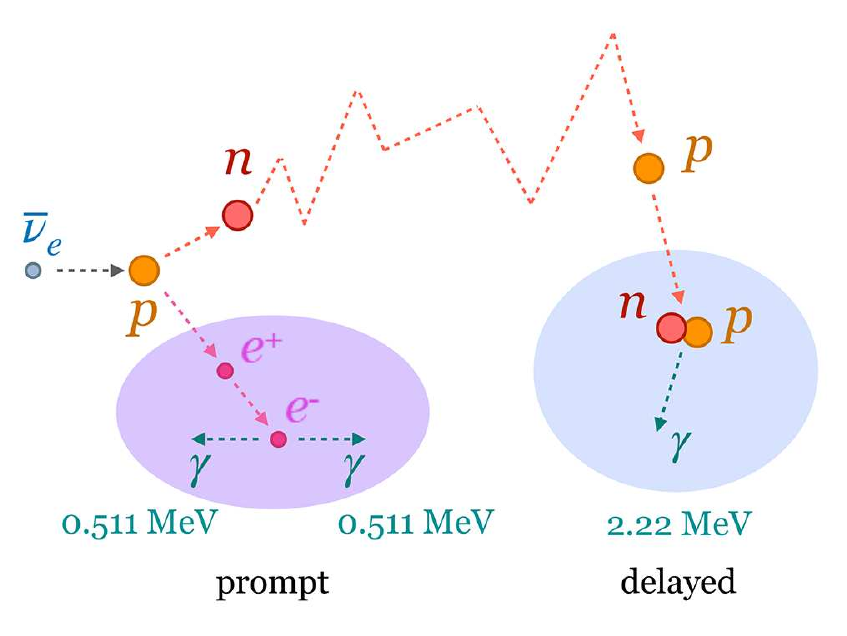}
            \caption{Schematic of the proton Inverse Beta Decay interaction, used to detect geoneutrinos, showing the origin of the prompt (violet area) and the delayed (blue area) signals. The visible energy of the prompt signal includes the contribution from the kinetic energy of the positron as well as from its annihilation. The neutron thermalizes and scatters until it is captured on a free proton. The 2.2\,MeV de-excitation gamma of the deuteron represents the delayed signal.}
        	\label{fig:ibd_int}
        \end{figure}

    Antineutrinos are detected in liquid scintillator detectors through the Inverse Beta Decay (IBD) reaction illustrated in Fig.~\ref{fig:ibd_int}:
         \begin{equation}
            \bar{\nu}_{e} + p \rightarrow e^{+} + n,
            \label{feq:ibd_int}
        \end{equation}

\noindent in which the free protons in hydrogen nuclei, that are copiously present in hydrocarbon (C$_{n}$H$_{2n}$) molecules of organic liquid scintillators, act as target. IBD is a charge-current interaction which proceeds only for electron flavoured antineutrinos. Since the produced neutron is heavier than the target proton, the IBD interaction has a kinematic threshold of 1.806\,MeV. The cross section of the IBD interaction can be calculated precisely with an uncertainty of 0.4\%~\cite{strumia2003precise}. In this process, a positron and a neutron are emitted as reaction products. The positron promptly comes to rest and annihilates emitting two 511\,keV $\gamma$-rays, yielding a {\it prompt} signal, with a visible energy $E_{p}$, which is directly correlated with the incident antineutrino energy $E_{\bar{\nu}_e}$: 
\begin{equation}
E_{p} \sim E_{\bar{\nu}_e}- 0.784\,\, \mathrm {MeV}.
\label{eq:Epro}
\end{equation}
The offset results mostly from the difference between the 1.806\,MeV, absorbed from $E_{\bar{\nu}_e}$ in order to make the IBD kinematically possible, and the 1.022\,MeV energy released during the positron annihilation.
The emitted neutron initially retains the information about the ${\bar{\nu}_e}$ direction. However, the neutron is detected only indirectly, after it is thermalized and captured, mostly on a proton. Such a capture leads to an emission of a 2.22\,MeV $\gamma$-ray, which interacts typically through several Compton scatterings. These scattered Compton electrons then produce scintillation light that is detected in a single coincident {\it delayed} signal. In Borexino, the neutron capture time was measured with the $^{241}$Am-$^{9}$Be calibration source to be $(254.5 \pm1.8)$\,$\mu$s~\cite{Bellini:2011yd}. During this time, the directional memory is lost in many scattering collisions.

\begin{figure}[t]
     \centering  
    \includegraphics[width = 0.50\textwidth]{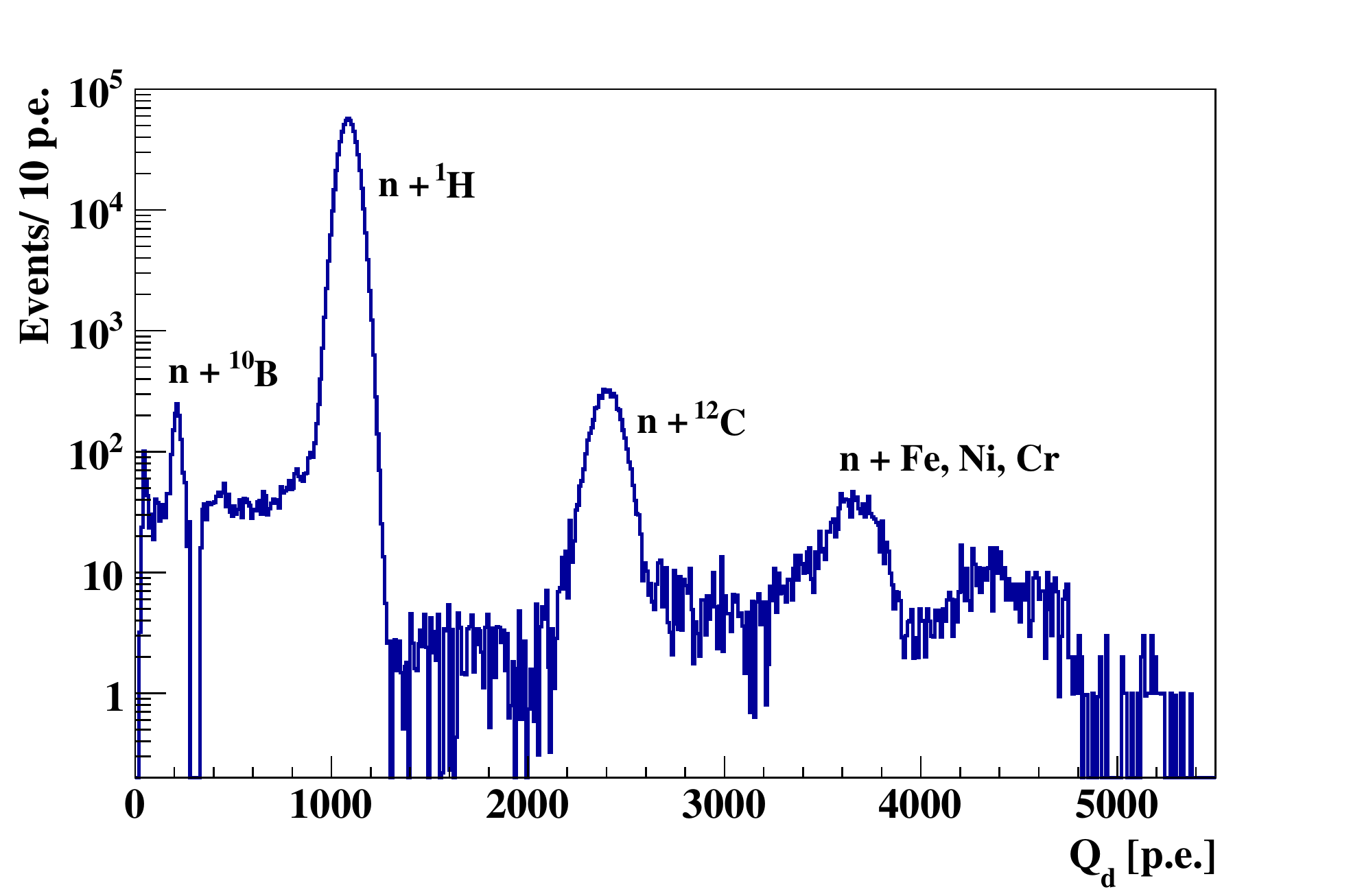} 
    \caption{The $N_{pe}$ charge spectrum of delayed signals, expressed in the number of detected photoelectrons, due to the gammas from captures of neutrons emitted from the $^{241}$Am-$^{9}$Be calibration source placed in the center of the detector. The clearly visible peaks of 2.22\,MeV and 4.95\,MeV gammas from the neutron captures on proton and $^{12}$C are positioned at 1090\,p.e. and 2400\,p.e., respectively. The other peaks are from neutron captures on nuclei of stainless steel used in the source and its insertion system construction: at $>$7\,MeV energies due to captures on (Fe, Ni, Cr) and at 477.6\,keV on $^{10}$B.}
    \label{fig:AmBeCenterDelayed} 
    \end{figure}

Figure~\ref{fig:AmBeCenterDelayed} shows the $N_{pe}$ spectrum of delayed signals due to the gammas from captures of neutrons emitted from the $^{241}$Am-$^{9}$Be calibration source placed in the center of the detector. In addition to the main 2.22\,MeV peak due to the neutron captures on protons, higher energy peaks are clearly visible. The 4.95\,MeV $\gamma$s originate from the neutron captures on $^{12}$C present in LS which occurs with about 1.1\% probability. The higher energy peaks are from neutron captures on stainless steel nuclei (Fe, Ni, Cr) used in the source construction.

The pairs of time and spatial coincidences between the prompt and the delayed signals offer a clean signature of $\bar{\nu}_e$ interactions, which strongly suppresses backgrounds. In the following sections, we refer to these signals as {\it prompt} and {\it delayed}, respectively.

    \section{EXPECTED ANTINEUTRINO SIGNAL}
    \label{sec:antinu}

This section describes the expected antineutrino signals at the LNGS location ($\vec{r}$ = $\ang{42.4540}$N, $\ang{13.5755}$E).
We express them in {\it Terrestrial Neutrino Units} (TNU). This unit eases the conversion of antineutrino fluxes to the number of expected events: 1\,TNU corresponds to 1 antineutrino event detected via IBD (Sec.~\ref{sec:IBD}) over 1\,year by a detector with 100\% detection efficiency containing 10$^{32}$ free target protons (roughly corresponds to 1\,kton of LS).

Since we detect only electron flavour of the total antineutrino flux (Sec.~\ref{sec:IBD}), neutrino oscillations affect the expected signal expressed in TNU. Thus, the neutrino oscillations and the adopted parameters are discussed in Sec.~\ref{subsec:oscil}. The evaluation of the expected geoneutrino signal from the Earth's crust and mantle is described in Sec.~\ref{subsec:geo}. Section~\ref{subsec:rea} details the estimation of the signal from antineutrinos from the world reactors, the most important background for geoneutrino measurements. Atmospheric neutrinos, discussed in Sec.~\ref{subsec:atm}, also represent a potential background source for geoneutrinos. The existence of a {\it georeactor}, a naturally occurring Uranium fission in the deep Earth, was suggested by some authors. We present this idea as well as the expected signal from such a hypothetical source in Sec.~\ref{subsec:georeactor}. The final number of the expected events from each of these sources, expected in our data set (Sec.~\ref{subsec:data}) and passing all optimized selection cuts (Sec.~\ref{subsec:cuts}) will be presented in Sec.~\ref{subsec:antinu_est_ev}.
    \begin{figure*}[h]
     \centering  
    \subfigure[]{\includegraphics[width = 0.49\textwidth]{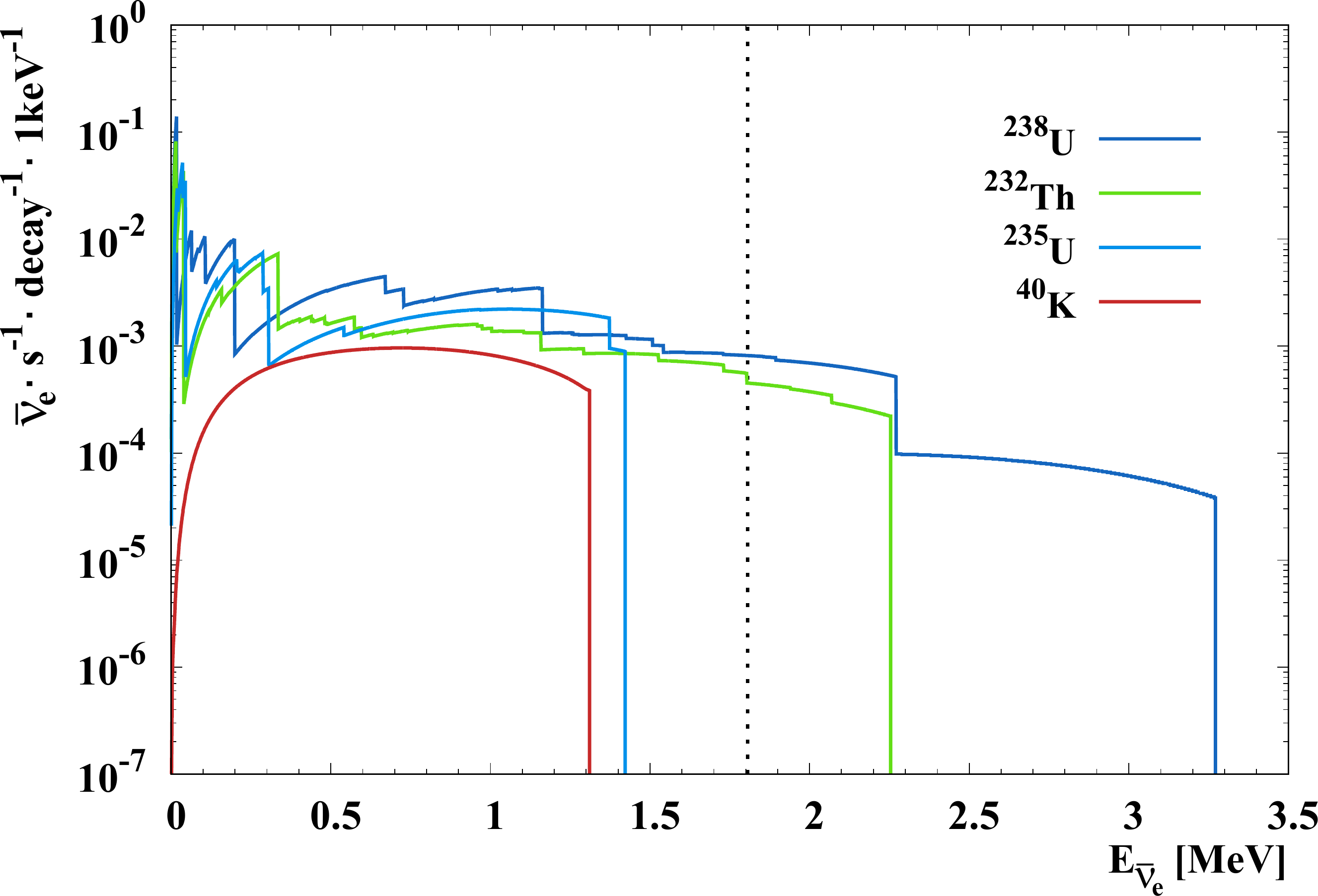}\label{fig:GeoNuDecay} }
    \subfigure[]{\includegraphics[width = 0.49\textwidth]{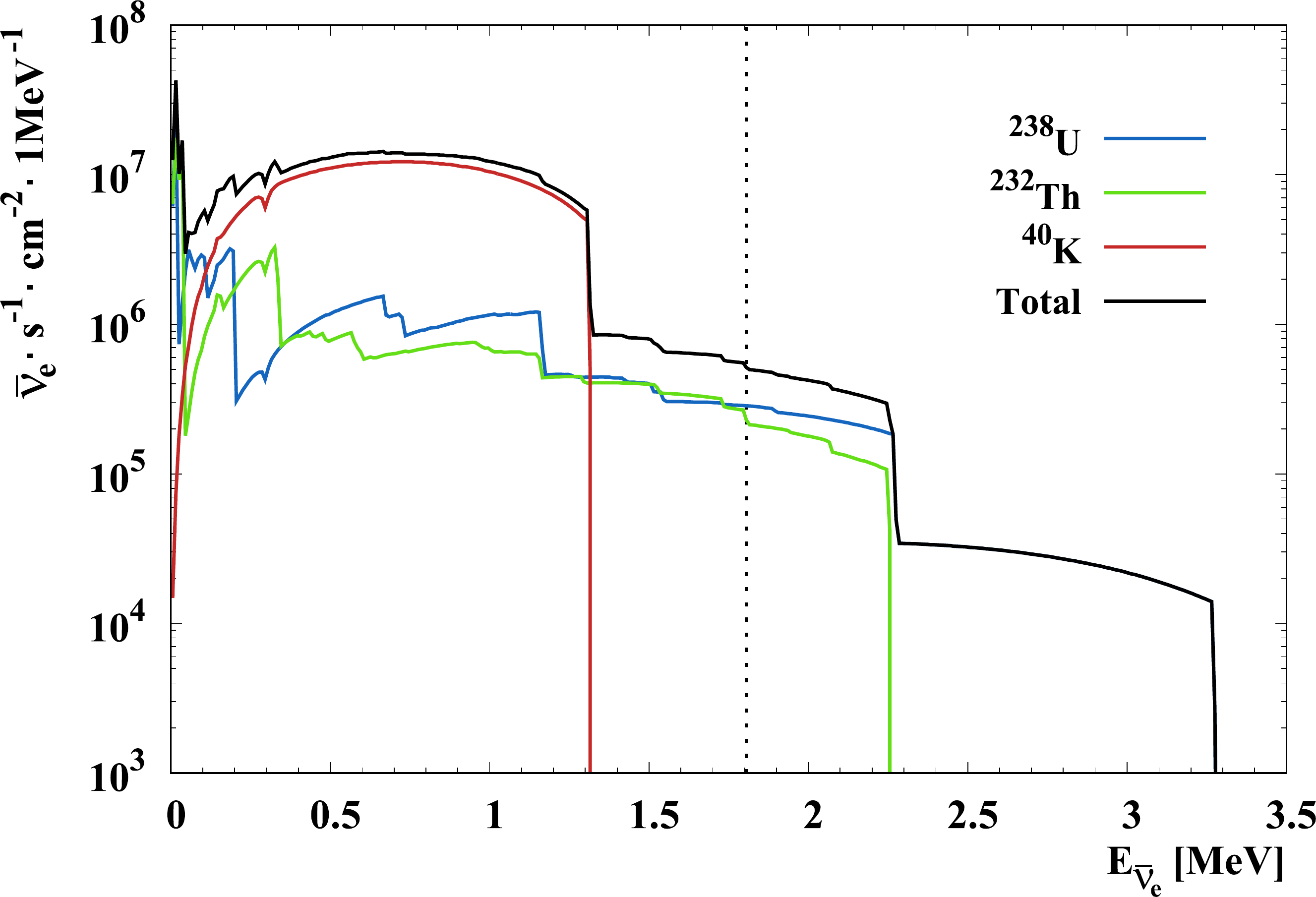}\label{fig:GeoNuFlux}}
    \vspace{3mm}
    \caption{(a) Geoneutrino energy spectra from the decays of $^{40}$K and of the $^{238}$U, $^{235}$U, and $^{232}$Th chains. All spectra are normalized to one decay of the head element of the chain. The integral from zero to the end point of the total spectrum is 6 for $^{238}$U, 4 for $^{235}$U and $^{232}$Th, and 0.89 for $^{40}$K. Data are from~\cite{Enomoto}. (b) Geoneutrino fluxes from different isotopes and their sum at LNGS as a function of geoneutrino energies calculated adopting geophysical and geochemical inputs from~\cite{Huang2013} for the far-field lithosphere and from ~\cite{coltorti} for the local crust. The flux from the mantle is calculated assuming a two-layer distribution (Fig.~\ref{fig:MantleScenarios}b) and adopting HPEs' abundances in BSE according to the GC model. The vertical dashed lines in both plots represent the kinematic threshold of the IBD interaction.}
      
    \end{figure*} 
 \begin{figure*}[h]
     \centering  
    \includegraphics[width = 0.90\textwidth]{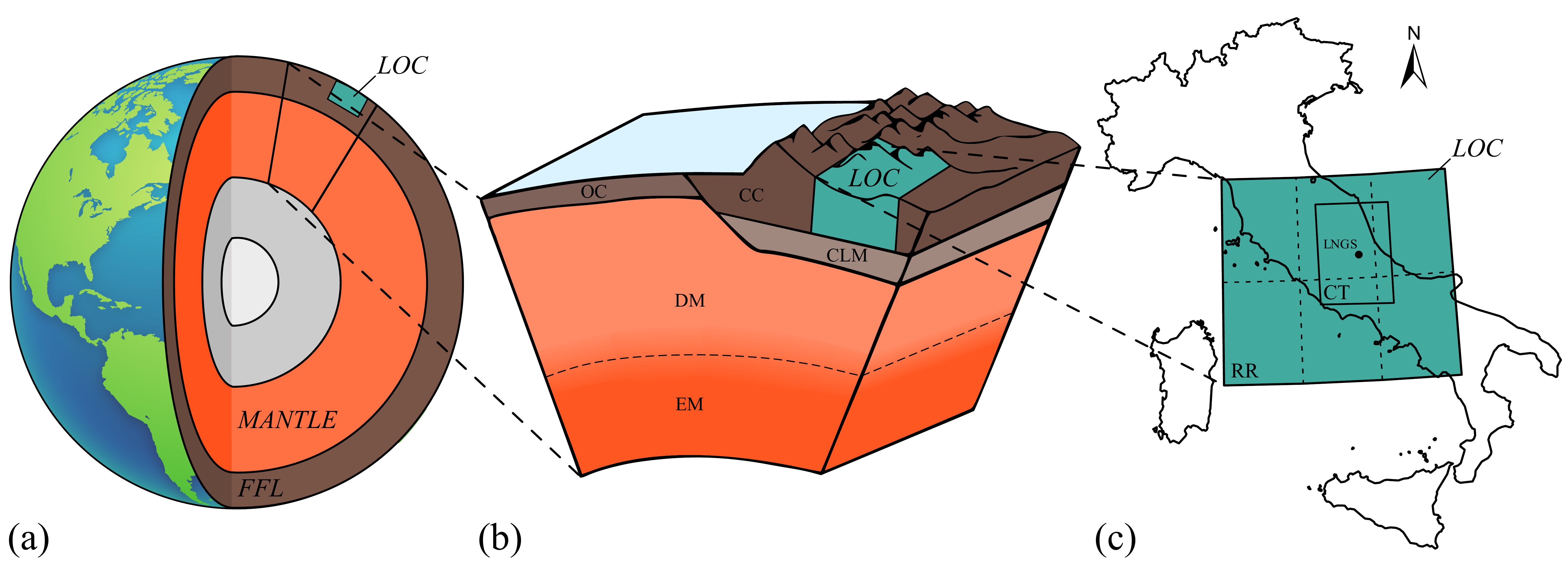}
    \vspace{3mm}
        \caption{(a) Schematic drawing of the Earth’s structure showing the three units contributing to the expected geoneutrino signal at LNGS: (i) the {\it local crust} (LOC), (ii) the {\it far field lithosphere} (FFL), and (ii) the {\it mantle}. The inner and outer portions of the {\it core} (in grey) do not contribute to the geoneutrino signal. Not to scale. (b) Schematic section detailing the components of the BSE. The {\it lithosphere} includes the LOC and the FFL. The latter comprises the rest of the {\it continental crust} (CC), the {\it oceanic crust} (OC), and the {\it continental lithospheric mantle} (CLM). In the mantle, two portions can be distinguished: a lower {\it enriched mantle} (EM) and an upper {\it depleted mantle} (DM). Not to scale. (c) Simplified map of the LOC. The {\it central tile} (CT) of the $\ang{2}$ $\times$ $\ang{2}$ centered at LNGS is modelled separately from the remaining six tiles which represent the {\it rest of the region} (RR).}
       \label{fig:GeolContrib} 
    \end{figure*} 
    
 \subsection{Neutrino oscillations} 
 \label{subsec:oscil}

The presently accepted Standard Model of elementary particles describes neutrinos existing in three flavours (electron, muon, and tau) with masses smaller than 1/2 of the $Z^0$ boson mass. The experiments with solar, atmospheric, as well as reactor antineutrinos observed that the neutrino flavour can change during the travel between the source and the detector. The process of {\it neutrino oscillations} has been established and confirmed that neutrinos have a non-zero rest mass.
At present, most experimental results on neutrino flavor oscillation agree with a three neutrino scenario, where the weak neutrino eigenstates, i.e. flavor eigenstates $(\nu_{e}, \nu_{\mu}, \nu_{\tau})$ mix with the mass eigenstates $(\nu_{1}, \nu_{2}, \nu_{3})$ via the Pontecorvo–Maki–Nakagawa–Sakata (PMNS) matrix, parametrized with the three mixing angles $(\theta_{12}, \theta_{13} , \theta_{23})$ and possible CP-violating and Majorana phases. 

Therefore, to establish the expected electron antineutrino flux at a given site, it is necessary to consider the survival probability $P_{ee}$ of the electron flavoured neutrinos, which depends on the PMNS mixing matrix, as well as on the differences between the squared masses of the mass eigenstates, neutrino energy $E_{\bar {\nu}}$, and the travelled baseline $L$. In the calculation of $P_{ee}$ for MeV antineutrinos, Eq.~(37) from~\cite{capozzi2014neutrino} is adopted, using the neutrino oscillation parameters as in Table~\ref{tab:oscillation_param}, obtained by NU-FIT 3.2 (2018)~\cite{NUFIT2018} from a global fit to data provided by different experiments:
\begin{equation}
    \begin{split}
    \label{eq:pee}
    P_{ee} (L, E_{\bar{\nu}_e}) = &1 - 4 c^4_{13}s^2_{12}c^2_{12} \sin^2 \delta  \\
                                  &- 4 s^2_{13}c^2_{13}c^2_{12} \sin^2(\Delta+\delta/2) \\
    &- s^2_{13} c^2_{13} s^2_{12} \sin^2(-\Delta+\delta/2),
    \end{split}
\end{equation}
where
\begin{equation}
    \begin{aligned}
         \delta = \frac{\delta m^{2} L}{4E_{\bar{\nu}_e}} & \quad \Delta = \frac{\Delta m^{2}L}{4E_{\bar{\nu}_e}} \\ 
        \delta m^2= \Delta m_{12}^2 &  \quad
         \Delta m^2 = \frac{1}{2} | \Delta m_{31}^2   +  \Delta m_{32}^2 |  \\
          \Delta m_{i,j} & = m_i^2 - m_j^2  \\
        c_{12}=\cos\theta_{12}   & \quad  s_{12}=\sin\theta_{12}  \\
        c_{13}=\cos\theta_{13}   & \quad  s_{13}=\sin\theta_{13} 
    \end{aligned}
\end{equation}
and $L$ and $E_{\bar{\nu}_e}$ are expressed in natural units ($\hbar$ = $c$ = 1). We assume Normal Hierarchy for neutrino mass eigenstates $(m_{1} < m_{2} < m_{3})$ and neutrino oscillations in vacuum. In addition  we assume
$\Delta m^2 = \Delta m_{31}^2  $
 since 
$ |\Delta m_{31}^2 | \approx   |\Delta m_{32}^2 |$ and in 
Normal Hierarchy both difference squared masses are positive.

The size of matter effects on the $P_{ee}$, when the neutrinos cross the Earth, is discussed individually for each antineutrino source in the following subsections. It depends on the baseline lenght in matter, the matter electron density $N_e$, as well as on the antineutrino energy. Equation~62 of~\cite{capozzi2014neutrino} is adopted in this calculation:
\begin{equation}
    \begin{split}
    \label{eq:pee_matter}
    P_{ee}^{\mathrm {matter}} (L, E_{\bar{\nu}_e}, N_e) =  & c^4_{13} ( 1 - 4 \tilde{s}^2_{12} \tilde{c}^2_{12} \, \sin^2\tilde{\delta} )+ s^4_{13},\\
    \end{split}
\end{equation}
where ``tilde" denotes the mixing parameters $(\tilde{\theta}_{12}, \tilde{\delta})$ in matter, related with the vacuum oscillation parameters $(\theta_{12}, \delta)$ through relations:
\begin{equation}
    \begin{aligned}
    \label{eq:param_matter}
    \sin2\tilde{\theta}_{12}= & \sin2\theta_{12} \, (1 - \mu_{12}\, \cos2\theta_{12})      \quad  \\
    \tilde{\delta} = &\delta(1+\mu_{12}\, \cos 2\theta_{12})  \quad    \\
    \end{aligned}
\end{equation}
with
\begin{equation}
    \mu_{12} = \frac{2\sqrt{2} G_F N_e E_{\bar{\nu}_e}}{\delta m^2},
    \label{eq:mu12}
\end{equation}
where $G_F$ if the Fermi coupling constant.

\begin{table}[h]
	\centering
	\caption{\label{tab:oscillation_param} The 3$\nu$ parameters, taken from NU-FIT 3.2 (2018)~\cite{NUFIT2018}, entering the calculation of the survival probability $P_{ee}$ for MeV electron antineutrinos.} \vskip 2pt
	\begin{tabular*}{\columnwidth}{c @{\hskip 30pt} c}
		\hline
		\hline
		Oscillation parameter & Value \Tstrut\Bstrut \\
		\hline 
		$\delta m^{2}$ [eV$^{2}$]& (7.40$^{+ 0.21}_{-0.20} )\cdot 10^{-5} $ \Tstrut \\ [4pt]
		$\Delta m^{2}$ [eV$^{2}$]& (2.494$^{+ 0.033}_{-0.031} )\cdot 10^{-3}$  \\ [4pt]
		sin$^{2}\theta_{12}$  & 0.307$^{+ 0.013}_{-0.012} $ \\ [4pt]
		sin$^{2}\theta_{13}$  & 0.02206$^{+ 0.00075}_{-0.00075} $ \Bstrut \\ 
		\hline
		\hline
	\end{tabular*}
\end{table}

 \subsection{Geoneutrinos} 
 \label{subsec:geo}
 
The Earth is a planet shining essentially in a flux of antineutrinos with a luminosity $L\sim 10^{25}$\,s$^{-1}$. For a detector placed on the continental crust, the expected U and Th geoneutrino flux is of the order of $10^6$\,cm$^{-2}$\,s$^{-1}$ and is typically dominated by the crustal contribution. The differential flux of geoneutrinos emitted from isotope $i$ = ($^{238}$U, $^{232}$Th) and expected at LNGS location $\vec{r}$ is calculated using the following expression:
 \begin{eqnarray}
\frac{d\Phi (i;E_{\bar{\nu}},\vec{r})}{dE_{\bar{\nu}}} &=&  \varepsilon_{\nu} (i) \frac{dn(i;E_{\bar{\nu}})}{dE_{\bar{\nu}}} \\
\nonumber
    &\times&  \int_{V}d\vec{r \prime} P_{ee} \left( E_{\bar{\nu}}, \lvert\vec{r} - \vec{r \prime}\rvert\right) \frac{a(i; \vec{r \prime}) \cdot \rho (\vec{r \prime})}{4 \pi \lvert \vec{r} - \vec{r \prime}\rvert^2},
 \label{eq:geonuS}
 \end{eqnarray}
where $\varepsilon_{\nu} (i)$ is the {\it specific antineutrino production rate} for isotope $i$ per 1\,kg of naturally occurring element ($7.41 \cdot 10^{7}$\,kg$^{-1}$\,s$^{-1}$ for $^{238}$U and $1.62 \cdot 10^{7}$\,kg$^{-1}$\,s$^{-1}$ for $^{232}$Th). $E_{\bar{\nu}}$ is geoneutrino energy. The {\it geoneutrino energy spectra} $\frac{dn(i;E_{\bar{\nu}})}{dE_{\bar{\nu}}}$, discussed in Sec.~\ref{subsubsec:geo-nucl}, are normalized to one. The electron-flavour survival probability $P_{ee}$ after the propagation of geoneutrinos from a geological reservoir located at $\vec{r \prime}$ to the detector is calculated
considering oscillations in vacuum (Eq.~\ref{eq:pee}). The matter effect (Sec.~\ref{subsec:oscil}) is estimated to be of the order of 1\%~\cite{baldoncini2015reference}, i.e. much less than other uncertainties involved in the geoneutrino signal prediction. The average survival probability $\left \langle P_{ee} \right \rangle$ = 0.55. The $\rho(\vec{r \prime})$ is the density of the voxel emitting geoneutrinos and it is taken from geophysical models of lithosphere~\cite{Huang2013} and mantle~\cite{Dziewonski:1981xy}. The abundances $a(i; \vec{r \prime})$ of isotope $i$ are expressed per mass unit of rock. The integration is done over the whole volume of the Earth, considering geological constraints of the main HPEs reservoirs, as discussed in Sec.~\ref{subsubsec:geoinput}.

 To convert the differential geoneutrino flux $\frac{d\Phi (i;E_{\bar{\nu}},\vec{r})}{dE_{\bar{\nu}}}$ to geoneutrino signal $S (i)$ expressed in TNU (given in Sec.~\ref{subsubsec:geoinput}), it is necessary to account for the detection process via IBD on free protons and to perform integration over the geoneutrino energy spectra:
\begin{equation}
    \begin{aligned}
        S(i) = N_{p} t \int dE_{\bar{\nu}}\frac{d\Phi (i;E_{\bar{\nu}},\vec{r})}{dE_{\bar{\nu}}} \sigma (E_{\bar{\nu}}),
    \end{aligned}
\end{equation}
where $N_p=10^{32}$ target protons, $t$ is 1\,year measuring time, and $\sigma (E_{\bar{\nu}})$ is the IBD cross section~\cite{strumia2003precise}. Note that for a reference oscillated flux of $10^6$\,cm$^{-2}$\,s$^{-1}$, the geoneutrino signals from U and Th are $S$(U) = 12.8\,TNU and $S$(Th) = 4.04\,TNU, respectively. Considering the specific antineutrino production rates $\varepsilon_{\nu} (\mathrm {U, Th})$, one can calculate the signal ratio $R_S$ for a homogeneous reservoir characterized by a fixed $a$(Th)/$a$(U) ratio:
\begin{equation}
    \begin{aligned}
        R_s=\frac{S(\mathrm{Th})}{S(\mathrm{U})}=0.069\frac{a(\mathrm{Th})}{a(\mathrm{U})}.
    \end{aligned}
    \label{eq:RS}
\end{equation}
This signal ratio thus depends on the composition of the reservoir.
Adopting the CI chondrites $a$(Th)/$a$(U) = 3.9 for the bulk Earth, we get $R_s$ = 0.27.
Geophysical and geochemical observations of the lihtosphere constrain the $a$(Th)/$a$(U) = 4.3 (Table~\ref{tab:MH_litho}), implying a signal ratio of 0.29 (Table~\ref{tab:S_litho}). As a consequence, maintaining the global chondritic ratio of 3.9 for the bulk Earth, the inferred mantle ratio $a$(Th)/$a$(U) results to be 3.7, which corresponds to a $R_S$ = 0.26.

 \subsubsection{Geoneutrino energy spectra} 
 \label{subsubsec:geo-nucl}
 
The expected geoneutrino signal depends on the shape and rates of the individual decays. The HPEs, i.e. $^{238}$U, $^{235}$U, $^{232}$Th, and $^{40}$K release geoneutrinos with different energy spectra reported in Fig.~\ref{fig:GeoNuDecay}) for one decay of the head element of the chain. The number of emitted antineutrinos per decay is 6 for $^{238}$U, 4 for $^{235}$U and $^{232}$Th, and 0.89 for $^{40}$K. Note that the maximal energy of both $^{40}$K and $^{235}$U antineutrinos is below the IBD threshold (Sec.~\ref{sec:IBD}), while 0.38 and 0.15 antineutrinos per one decay are above this threshold for $^{238}$U and $^{232}$Th, respectively. The effective transitions producing detectable antineutrinos are given by $^{234\mathrm{m}}$Pa and $^{214}$Bi in $^{238}$U decay chain and $^{228}$Ac and $^{212}$Bi in the $^{232}$Th decay chain. Neglecting $^{210}$Tl, having branching probability $<$0.1\%, the $^{238}$U and $^{232}$Th antineutrino maximal energies are 3.27\,MeV and 2.25\,MeV, produced from $^{214}$Bi and $^{212}$Bi, respectively. In the energy distribution of U and Th antineutrinos reported in Fig.~\ref{fig:GeoNuDecay} and Fig.~\ref{fig:GeoNuFlux}, the spectral structures of $\beta$ and $(\beta+\gamma)$ decays are clearly visible. Note that only $^{214}$Bi decay spectral shape has been studied on the basis of experimental measurements~\cite{RN885}. The other energy distributions of antineutrinos are given with unknown uncertainties, since they are generally calculated assuming a well-known universal shape distribution. Figure~\ref{fig:GeoNuFlux}) shows the oscillated geoneutrino spectra expected at LNGS considering the geophysical and geochemical inputs as discussed in the following Sec.~\ref{subsubsec:geoinput}. 

  \subsubsection{Geological inputs}
  \label{subsubsec:geoinput}

The geoneutrino signal together with its uncertainty can be calculated considering the observational data concerning U and Th abundances in the lithosphere~\cite{Huang2013}, the density profile of the Earth~\cite{Dziewonski:1981xy}, and the BSE constraints on the global amounts of HPEs (Table~\ref{tab:BSE}). In Table~\ref{tab:MH_litho} the masses of the main reservoirs of the lithosphere are reported together with the HPEs' masses and the released radiogenic heat. Note that, although the mass of the bulk crust is less than 1\% of the BSEs mass, it contains $\sim$35\% of U and Th masses predicted by the GC model (Table~\ref{tab:BSE}). The HPEs' radiogenic heat of the whole lithosphere is $8.1^{+1.9}_{-1.4}$\,TW.

\begin{table*}[h]
	\centering
	\caption{\label{tab:MH_litho} Total masses, HPEs’ masses M, and total radiogenic heat $H_{\mathrm{rad}}$ of the CC, OC, CLM, bulk crust (CC + OC), and bulk lithosphere (bulk crust + CLM). The error propagation assumes no correlation among different lithospheric units and is performed via a Monte Carlo sampling of HPEs abundances according to their probability density function in order to propagate the asymmetrical uncertainties of the non-Gaussian distributions. The median values and the $1\sigma$ uncertainties are shown. Due to the positive asymmetry of the distributions, in some cases the median of the output matrices is not coincident with the sum of the medians of the individual components.}	\vskip 2pt
	\begin{tabular*}{14 cm}{l @{\hskip 16pt} c @{\hskip 14pt} c @{\hskip 14pt} c @{\hskip 14pt} c @{\hskip 14pt} c}
		\hline
		\hline
   	 	& Mass  & $M$(U) & $M$(Th)   & $M$(K)  & $H_{\mathrm {rad}}$(U+Th+K) \Tstrut \\
   		& [10$^{21}$\,kg] & [10$^{16}$\,kg] & [10$^{16}$\,kg]& [10$^{19}$\,kg]  & [TW] \Bstrut\\
   		\hline     
	    Continental crust (CC) 		& 20.6 $\pm$ 2.5       & 2.7$^{+0.6}_{-0.5}$ & 11.9$^{+3.2}_{-2.1}$ & 31.9$^{+6.4}_{-4.9}$ & 6.8$^{+1.4}_{-1.1}$ \Tstrut \\ [4pt]
		Oceanic crust (OC)  	& 6.7 $\pm$ 2.3          &  0.10 $\pm$ 0.03 & 0.4 $\pm$ 0.1  & 1.0 $\pm$ 0.3  & 0.2 $\pm$ 0.1 \\ [4pt]   
		Bulk crust (CC + OC)  & 27.3 $\pm$ 4.8       & 2.8$^{+0.6}_{-0.5}$ & 12.3$^{+3.2}_{-2.1}$ & 33.0$^{+6.5}_{-4.9}$ & 7.0$^{+1.4}_{-1.1}$ \\ [4pt]
		CLM         & 97 $\pm$ 47          & 0.3$^{+0.5}_{-0.2}$ & 1.5$^{+2.9}_{-0.9}$  & 3.1$^{+4.7}_{-1.8}$  & 0.8$^{+1.1}_{-0.6}$ \Bstrut\\
		\hline
		Bulk lithosphere  & 124 $\pm$ 47         & 3.3$^{+0.8}_{-0.6}$ & 14.3$^{+4.8}_{-2.8}$ & 36.9$^{+8.4}_{-6.0}$ & 8.1$^{+1.9}_{-1.4}$ \Tstrut \\
		(Bulk crust + CLM) & & & & & \Bstrut \\
		\hline
		\hline  
	\end{tabular*}
\end{table*}

Following the scheme reported in Fig.~\ref{fig:GeolContrib}, the expected geoneutrino signal in Borexino $S$(U+Th) can be expressed as the sum of three components:  
\begin{itemize}
    \item $S_{\mathrm{LOC}}$(U+Th), the {\it local crust} (LOC) signal produced from the \ang{6} $\times$ \ang{4} crustal area surrounding LNGS, 
    \item $S_{\mathrm{FFL}}$(U+Th), the signal from the {\it far field lithosphere} (FFL), which includes the continental lithospheric mantle (CLM), i.e. the brittle portion of the mantle underlying the CC, and the remaining crust obtained after the removal of the LOC.
    \item $S_{\mathrm{mantle}}$(U+Th), the signal from the mantle.
\end{itemize}
The signal expected from the bulk lithosphere, as the sum of LOC and FFL contributions, is given in Table~\ref{tab:S_litho}, while the mantle signals, using in inputs different BSE models (Table~\ref{tab:BSE}), in Table~\ref{tab:S_mantle}. 

\begin{table*}[t]
    \centering
    \caption{\label{tab:S_litho} Geoneutrino signals $S$ (median and $1\sigma$ uncertainties) and the ratio $R_s$ = $S$(Th)/$S$(U) expected at Borexino originated from U and Th in the LOC and FFL. The bulk lithosphere signal is obtained summing the FFL and LOC contributions as linearly independent. The total geoneutrino signal $S$(U+Th) of each reservoir is obtained assuming $S$(U) and $S$(Th) are fully positive correlated. The asymmetrical uncertainties of the non-Gaussian distributions are propagated via a Monte Carlo sampling performed according to the signal probability density functions of each component.} \vskip 2pt
    \begin{tabular*}{14 cm}{c @{\hskip 36pt} c @{\hskip 35pt} c @{\hskip 35pt} c @{\hskip 35pt} c @{\hskip 36pt} c}
        \hline
        \hline
	    & $S$(U) & $S$(Th) 	& $S$(U+Th) & $R_s$ = $S$(Th)/$S$(U) \Tstrut \\
	    & [TNU]  & [TNU] & [TNU]  & \Bstrut \\ 
	    \hline
        LOC &	7.4 $\pm$ 1.0	& 1.8 $\pm$ 0.3 &	9.2 $\pm$ 1.2	& 0.24 \Tstrut \\ [2pt]
        FFL	& 12.4$^{+3.5}_{-2.7}$ &	4.0$^{+1.4}_{-1.0}$ &	16.3$^{+4.8}_{-3.7}$ &	0.33  \Bstrut\\ 
        \hline
        Bulk lithosphere &	19.8$^{+3.6}_{-2.9}$ &	5.8$^{+1.4}_{-1.1}$ &	25.9$^{+4.9}_{-4.1}$	& 0.29 \Tstrut \\ 
        (Bulk crust + CLM) & & & & \Bstrut \\
        \hline
        \hline
    \end{tabular*}
\end{table*}

\paragraph{Local crust contribution}

The $S_{\mathrm{LOC}}$(U+Th) is estimated adopting the local refined model based on specific geophysical and geochemical data described in~\cite{coltorti}. The 492\,km $\times$ 444\,km  region of continental crust surrounding the LNGS is divided in a Central Tile (CT) and the Rest of the Region (RR) (Fig.~\ref{fig:GeolContrib}c).
For the CT, which includes the crustal portion within $\sim$100\,km from the Borexino detector, a 3D model with a typical resolution of (2.0\, km $\times$ 2.0\, km $\times$ 0.5\,km) is built. The crustal structure of the CT is based on a simplified tectonic model that includes the main crustal thrusts and near vertical reflection seismic profiles of the CROP project~\cite{RN572}. The $\sim$35\,km thick crust has a layered structure typical of Central Apennines, characterized by thick sedimentary cover ($\sim$13\,km) which is not reported in any global crustal model. It is constituted by three Permo-Mesozoic carbonatic successions and a unit of the Cenozoic terrigenous sediments. Since the local seismic sections do not highlight any evidence of {\it middle crust}, the crystalline basement is subdivided into {\it upper crust} ($\sim$13\,km) and {\it lower crust} ($\sim$9\,km). The U and Th mass abundances are obtained by ICP-MS and gamma spectroscopy measurements of the rock samples collected within 200\,km from the LNGS and from representative outcrops of upper and lower crust of the south Alpine basement. It's relevant to note that $\sim$75\% of the sedimentary cover volume of CT is constituted by Mesozoic carbonates particularly poor of U and Th. It implies that the overall U and Th abundances of sediments are $a(\mathrm{U})$ = (0.8 $\pm$ 0.2)\,$\mu$g/g and $a(\mathrm{Th})$ = (2.0 $\pm$ 0.5)\,$\mu$g/g to compare with $a(\mathrm{U})$ = (1.73 $\pm$ 0.09)\, $\mu$g/g and $a(\mathrm{Th})$ = (8.10 $\pm$ 0.59)\, $\mu$g/g~\cite{RN392} used for the global crustal estimations. A geophysical model with a lower spatial resolution ($\ang{0.25}$ $\times$ $\ang{0.25}$) is built for the RR, which treats the sedimentary cover as a single and homogeneous layer with the same U and Th abundances of CT sediments. The geoneutrino signal of the LOC is $S_{\mathrm{LOC}}$(U+Th) = (9.2 $\pm$ 1.2)\,TNU\footnote{The difference of $\sim$0.8\,TNU  with respect to the value reported in~\cite{coltorti} is the result of the neutrino survival probability function calculated from each cell using the updated oscillation parameters. The oscillation amplifies the reduction of the signal due to the presence of surrounding carbonatic rocks poor in Th and U.} (Table~\ref{tab:S_litho}) where 77\% of the signal originates from U and Th distributed in the CT. The maximal and minimal excursions of various input values and uncertainties reported in~\cite{coltorti} are taken as the $\pm$3$\sigma$ error range. The U and Th signal errors are conservatively considered fully positively correlated. Note that the reduction of $\sim$6\,TNU with respect to the estimations of the global reference model~\cite{Huang2013} is due to presence of thick sedimentary deposits composed primarily of U- and Th-poor carbonate rocks. The signal ratio $R_s$ (Eq.~\ref{eq:RS}) for the local crustal contribution is 0.24 (Table~\ref{tab:S_litho}).

\paragraph{Far field lithosphere contribution}

The FFL includes the CLM and the remaining crust after subtracting the LOC (Fig.~\ref{fig:GeolContrib}). The geoneutrino signal $S_{\mathrm{FFL}}$ is calculated adopting the $\ang{1}$ $\times$ $\ang{1}$ geophysically based, 3D global reference model~\cite{Huang2013}, which provides the abundances and distributions of HPEs in the lithosphere, together with their uncertainties. 

The crust is subdivided in 64,800 cells labeled with their thickness, density, and velocity of compressional and shear waves for eight layers (ice, water, three sediment layers, upper, middle, and lower crust). The total crustal thickness and the associated uncertainty correspond, respectively, to the mean and the half range of three crustal models:
\begin{itemize}
    \item CRUST 2.0~\cite{bassin2000current, RN896}, a global crustal model with ($2^{\circ} \times 2^{\circ}$) resolution based on refraction and reflection seismic  experiments and on extrapolations from geological and tectonic settings for regions lacking field measurements.
    \item CUB 2.0~\cite{RN345}, a ($2^{\circ} \times 2^{\circ}$) resolution model, provided with crustal thickness uncertainties, obtained by applying a Monte Carlo multi-step process with a priori constraints to invert surface wave dispersion data.
    \item GEMMA~\cite{RN1096}, a high-resolution ($0.5^{\circ} \times 0.5^{\circ}$) map of Moho depth obtained by inverting satellite gravity field data collected by GOCE. Additional external information (e.g. topography, bathymetry, and ice sheet models) and prior hypotheses on crustal density relative to the main geological provinces are taken also into account. 
\end{itemize}

The relative thickness of the crustal layers are incorporated from CRUST 2.0, while the information about the sedimentary cover is adopted from~\cite{RN895}. The HPEs' abundances in the sediments, OC, and upper crust layers are taken from published reference values reporting the uncertainties~\cite{Huang2013}, while U and Th abundances in the deep crust are inferred using seismic velocity arguments. The distinctive ultrasonic velocities reported in geophysical databases can be related to acidity (SiO$_2$ content) of igneous rocks (Fig.\,3 in~\cite{Huang2013}), which is generally correlated with the U and Th abundances.

The CLM is geophysically and geochemically distinct from the rest of the mantle (the sublithospheric mantle), and it is characterized by abundances of $a$(U) = 0.03$^{+0.05}_{-0.02}$\,$\mu$g/g and $a$(Th) = 0.15$^{+0.28}_{-0.10}$\,$\mu$g/g taken from a database of $\sim$500 xenolithic peridotite samples, representing the typical rock types of the CLM. The CLM geoneutrino signal is $S_{\mathrm{CLM}}$(U+Th) = 2.3$^{+3.1}_{-1.3}$\,TNU, corresponding to $\sim$14\% of the $S_{\mathrm{FFL}}$(U+Th).

The geoneutrino signal of the FFL is $S_{\mathrm{FFL}}$(U+Th) = 16.3$^{+4.8}_{-3.7}$\,TNU  (Table~\ref{tab:S_litho}) and it constitutes the 63\% of the signal of the bulk lithosphere.

\begin{table*}[t]
		\centering
		\caption{\label{tab:S_mantle}Ranges of HPEs' masses and of the radiogenic heat in the mantle derived from different BSE models (Table~\ref{tab:BSE}): the low and high mantle values are obtained by subtracting the 1$\sigma$ high and low values of the lithosphere (Table \ref{tab:MH_litho}), respectively. The range of the expected mantle geoneutrino signal is then obtained by distributing the remaining HPEs’ masses according to
		the {\it low scenario} (minimal value) and the {\it high scenario} (maximal value). The {\it convective Urey ratio} $UR_{\mathrm{CV}}$ values are obtained assuming a total heat flux $H_{\mathrm{tot}}$(U+Th+K) = 47\,TW and taking into account the radiogenic heat produced by the continental crust,  $H_{\mathrm{rad}}^{\mathrm{CC}}$ = 6.8$^{+1.4}_{-1.1}$\,TW (Table~\ref{tab:MH_litho}).} \vskip 2pt
		{\small
	\begin{tabular*}{\textwidth}{c @{\hskip 16pt} c @{\hskip 16pt} c @{\hskip 16pt} c @{\hskip 16pt} c @{\hskip 16pt} c @{\hskip 16pt} c @{\hskip 16pt} c}
		\hline
		\hline
		Model  &
		\thead{$M_{\mathrm{mantle}}$(U) \\ $[10^{16}$\,kg$]$} &
		\thead{$M_{\mathrm{mantle}}$(Th)\\ $[10^{16}$\,kg$]$} & 
		\thead{$M_{\mathrm{mantle}}$(K) \\ $[10^{19}$\,kg$]$} &
		\thead{$H_{\mathrm{rad}}^{\mathrm{mantle}}$(U+Th) \\ $[$TW$]$} &
		\thead{$H_{\mathrm{rad}}^{\mathrm{mantle}}$(U+Th+K) \\ $[$TW$]$} &
		\thead{$S_{\mathrm{mantle}}$(U+Th) \\ $[$TNU$]$} &
		$UR_{\mathrm{CV}}$ \Tstrut\Bstrut \\
		\hline
		J      & 0.8 - 2.2    & 0.0 - 5.9     & 13.8 - 28.1   & 0.7 - 3.8   & 1.2 - 4.7    		& 0.9 - 4.1   	            &	0.02 - 0.20	\Tstrut \\
		L \& K & 2.8 - 4.3    & 6.5 - 14.0    & 31.6 - 45.9   & 4.5 - 7.9   & 5.5 - 9.4    		& 3.9 - 8.0   	    &	0.14 - 0.32	\\ 
		T      &  3.2 - 4.7    & 9.3 - 16.8    & 27.5 - 41.9   & 5.6 - 9.0   & 6.5 - 10.4    		& 4.7 - 8.9  	        &	0.16 - 0.34	\\
		M \& S & 4.0 - 5.5    & 13.3 - 20.9   & 51.8 - 66.2   & 7.5 - 10.9  & 9.2 - 13.1   		& 6.0 - 10.6 	        &	0.23 - 0.41	\\
		A      & 4.0 - 5.5    & 12.1 - 19.7   & 15.8 - 30.2   & 7.1 - 10.6  & 7.7 - 11.6   		& 5.9 - 10.5  	            &	0.19 - 0.37	\\
		W      & 4.0 - 5.5    & 11.3 - 18.9   & 50.6 - 64.9   & 6.9 - 10.4  & 8.6 - 12.5   	& 5.8 - 10.4 	            &	0.21 - 0.39	\\
		P \& O & 4.8 - 6.3    & 14.6 - 22.1   & 59.9 - 74.2   & 8.6 - 12.0  & 10.6 - 14.5  		& 7.1 - 12.0  	        &	0.26 - 0.44	\\ 
		T \& S & 10.1 - 11.5  & 37.6 - 45.2   & 96.3 - 110.6  & 19.8 - 23.3 & 23.5 - 26.9  		& 15.7 - 22.4 	    &	0.57 - 0.75	\Bstrut \\
		\hline
		CC   &  0.8 - 2.2    & 0.0 - 5.9     & 13.8 - 28.1   & 0.7 - 3.8   & 1.2 - 4.7    			& 0.9 - 4.1   	&	0.02 - 0.20	\Tstrut \\
		GC   & 4.0 - 5.5    & 13.3 - 20.9   & 68.0 - 82.3   & 7.5 - 10.9  & 9.7 - 13.6  			& 6.0 - 10.6  	&	0.24 - 0.42	\\
		GD   & 10.1 - 11.5  & 37.6 - 45.2   & 96.3 - 110.6  & 19.8 - 23.3 & 23.0 - 26.9  			& 15.7 - 22.4 	&	0.57 - 0.75	\\
		FR   & 15.6 - 17.1  & 57.6 - 65.2   & 178.7 - 193.1 & 30.5 - 34.0 & 36.5 - 39.8  			& 24.2 - 33.0 	&	0.85 - 1.15	\Bstrut \\ 
		\hline
		\hline
	\end{tabular*}
	}
\end{table*}

\begin{figure*}[h]
     \centering  
    \includegraphics[width = 0.70\textwidth]{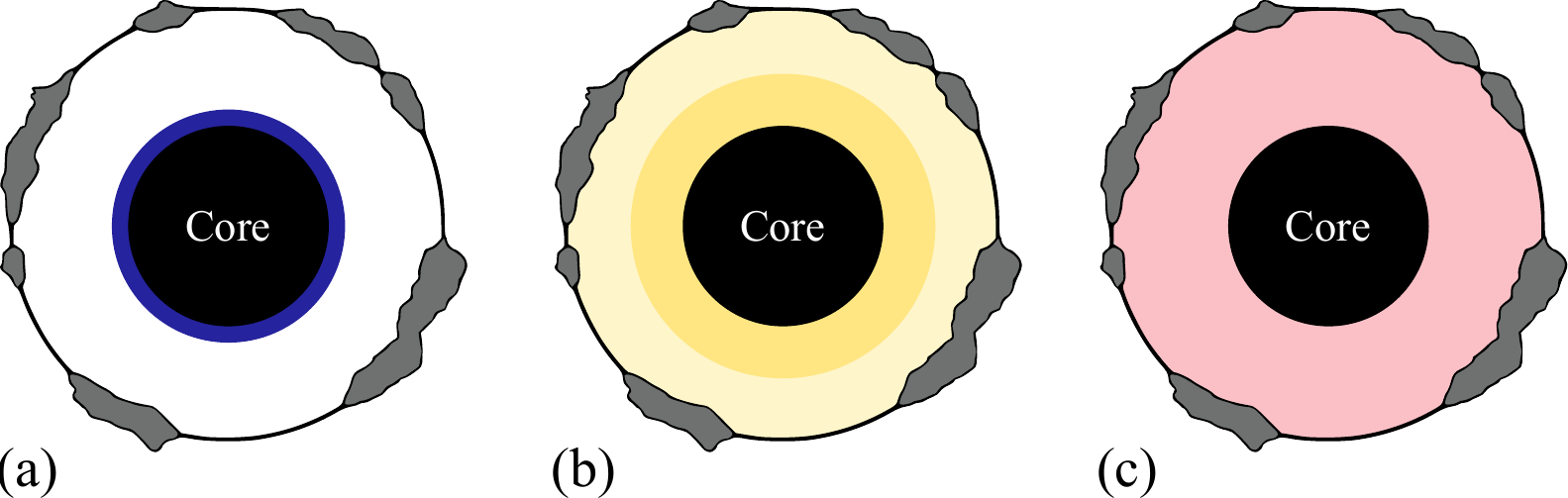}
    \vspace{3mm}
        \caption{Cartoons of the distributions of HPEs’ masses in the mantle predicted according to three different scenarios. (a) {\it Low scenario}: the HPEs are placed in a thin layer (in blue) above the core-mantle boundary. (b) {\it Intermediate scenario}: the HPEs are distributed differently in the upper depleted (in light yellow) and in the lower enriched (in dark yellow) layer. (c) {\it High scenario}: the HPEs are distributed homogenously (in red) in the mantle.}
       \label{fig:MantleScenarios} 
\end{figure*} 

\paragraph{Mantle contribution}

Earth scientists have debated the picture of the mantle convection over the last decades. Some geochemical arguments support a two-layer convection, while geophysical reasoning affirms a whole-mantle convection. The main arguments for a layered mantle are based on (i) chemical and isotopic differences between {\it Mid-Ocean-Ridge Basalts} (rocks differentiated from mantle transition zone depleted in incompatible elements) and {\it Ocean Island Basalts} (rocks melted out from the deeper enriched mantle), (ii) isotope variations between continental and oceanic crust, and (iii) the missing radiogenic heat source paradox~\cite{RN361}.

Beyond this controversy, all models agree that U and Th abundances are basically spherically distributed and non-decreasing with depth. This is an important point which permits us to keep the masses of HPEs in the unexplored mantle as free parameters and to provide constraints on the mantle contribution to the geoneutrino signal. For fixed HPEs' masses in a mantle having PREM density profile~\cite{Dziewonski:1981xy}, three different predictions for the mantle geoneutrino signal can be calculated varying their distribution in the mantle according to:
\begin{itemize}
   \item {\it Low scenario (LSc)} (Fig.~\ref{fig:MantleScenarios}a): the HPEs' masses are placed in a layer just above CMB;
   \item{\it Intermediate scenario (ISc)} (Fig.~\ref{fig:MantleScenarios}b): the HPEs' masses are distributed in two layers, a {\it lower enriched mantle} (EM) and an upper {\it depleted mantle} (DM) separated at 2180\,km of depth (for more details, see Sec.~2.4 of~\cite{Huang2013}).
   \item {\it High scenario (HSc)} (Fig.~\ref{fig:MantleScenarios}c): the HPEs’ masses are homogeneously distributed in the mantle.
\end{itemize}
Adopting these low and high scenarios, the predicted mantle geoneutrino signals $S_{\mathrm{mantle}}$, for each of the BSE models (Table~\ref{tab:BSE}), are reported in Table~\ref{tab:S_mantle}, together with the corresponding HPEs' masses ($M_{\mathrm{mantle}}$), the radiogenic heat power ($H_{\mathrm{rad}}^{\mathrm{mantle}}$), and the convective Urey ratio ($UR_{\mathrm{CV}}$, Eq.~\ref{eq:URCV}). Since the BSE model which is based on enstatitic chondrites composition~\cite{RN356} is poor in HPEs, the Th mass calculated after lithosphere subtraction shows slightly negative values which have been set to zero.

\paragraph{Total geoneutrino signal at Borexino}

The total geoneutrino signal from the bulk lithosphere expected at Borexino is a crucial piece of information for extracting the mantle signal from the Borexino measurement (Sec.~\ref{subsec:mantle}). Since LOC and FFL are modelled independently, for each element the signal contributions are summed as linearly independent. The obtained {\it bulk lithosphere signal} is $S_{\mathrm{LSp}}$(Th+U) = 25.9$^{+4.9}_{-4.1}$\,TNU (Table~\ref{tab:S_litho}).

Considering the intermediate scenario (Fig.~\ref{fig:MantleScenarios}b) for each mantle model (Table~\ref{tab:S_mantle}), the total expected geoneutrino signal can cover a wide range from $S^{\mathrm{CC}}_{\mathrm{tot}}$ = 28.5$^{+5.5}_{-4.8}$\,TNU to  
$S^\mathrm{GD}_{\mathrm{tot}}$ = 45.6$^{+5.6}_{-4.9}$\,TNU passing through $S^{\mathrm{GC}}_{\mathrm{tot}}$ = 34.6$^{+5.5}_{-4.8}$\,TNU (Table~\ref{tab:antinu-signals-expected}). The highest signal $S^{\mathrm{FR}}_{\mathrm{tot}}$ = 55.3$^{+5.7}_{-5.0}$\,TNU is given by a Fully Radiogenic Earth. The estimated 1$\sigma$ error of the mantle signals in Table~\ref{tab:antinu-signals-expected} corresponds to [$S^{\mathrm{HSc}}_{\mathrm{mantle}} - S^{\mathrm{LSc}}_{\mathrm{mantle}}]/6$ and is conservatively summed to lithospheric uncertainty as fully positive correlated.  

\begin{figure}[t]
\centering  
\includegraphics[width = 0.47\textwidth]{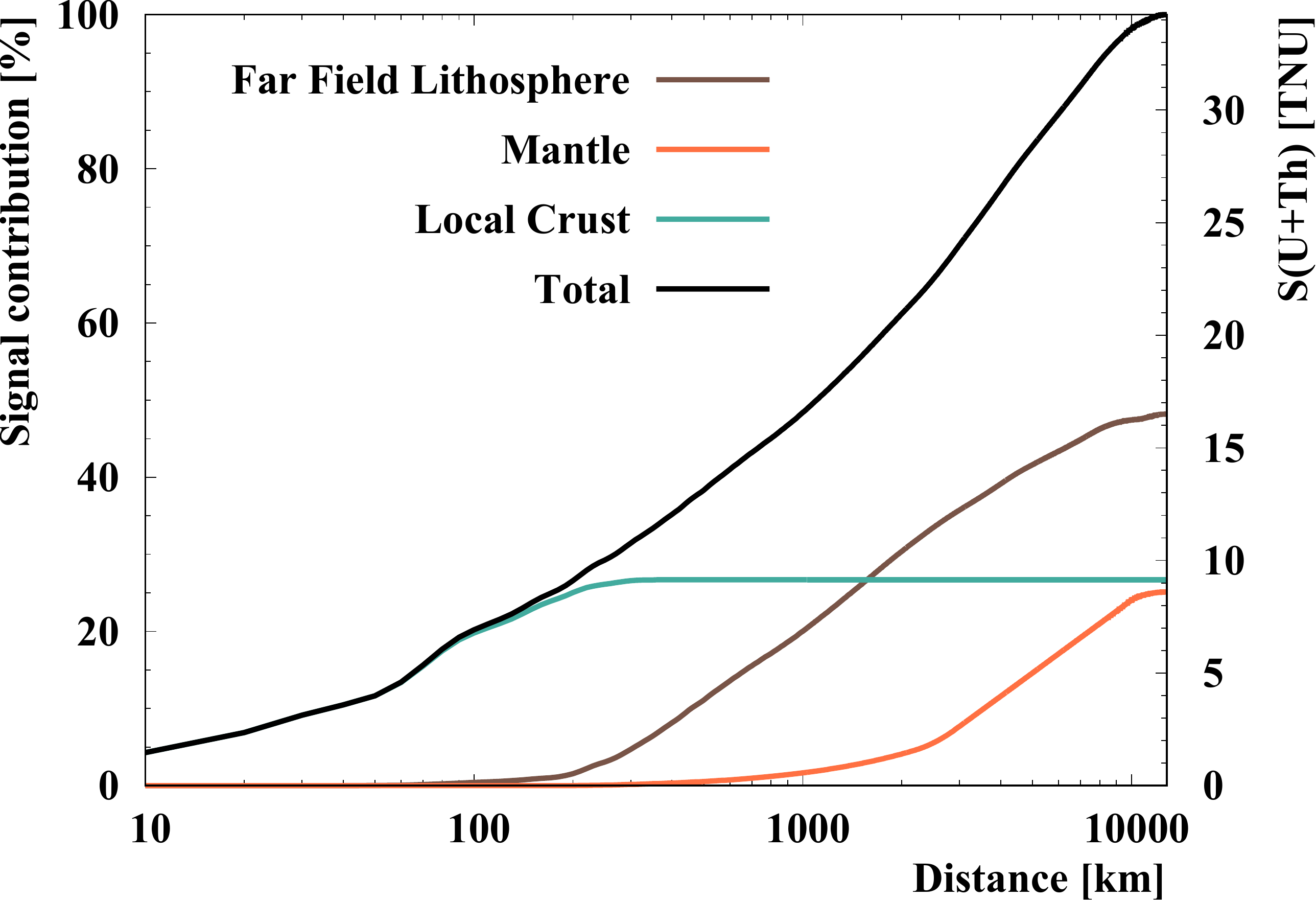}
\vspace{0.5mm}
\caption{The cumulative geoneutrino signal and percentage signal contribution of the Far Field Lithosphere (FFL), mantle, LOCal crust (LOC) and their sum (total) are represented as function of the distance from LNGS. Geoneutrino signals are calculated by adopting geophysical and geochemical inputs from~\cite{Huang2013} for the FFL and from\cite{coltorti} for the LOC.  The signal from the mantle is calculated by assuming a two-layer distribution and by adopting HPEs’ abundances in the BSE from the GC model.}
\label{fig:SGeoCumul} 
\end{figure}

Plotting the cumulative geoneutrino signal, as a function of the distance from Borexino (Fig.~\ref{fig:SGeoCumul}), we observe that 40\% of the total signal comes from U and Th in the regional crust that lies within 550\,km of the detector. Up to a distance of $\sim$150\,km from Borexino, 100\% of the geoneutrino signal is generated from the LOC.

\begin{figure*}[h]
    \centering     
       \subfigure[] {\includegraphics[width = 0.48\textwidth]{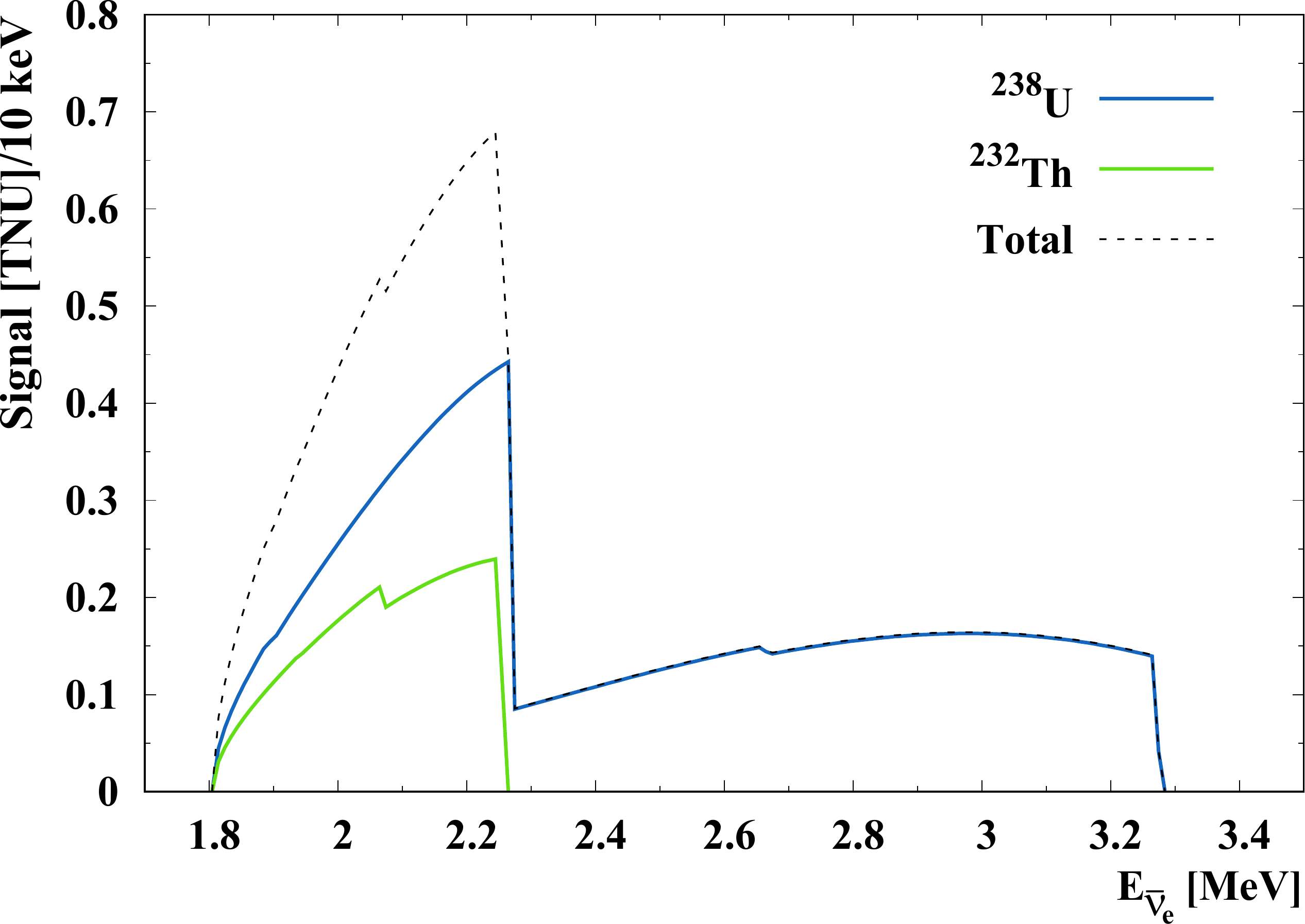}}
       \hspace{2mm}
    \subfigure[]{\includegraphics[width = 0.48\textwidth]{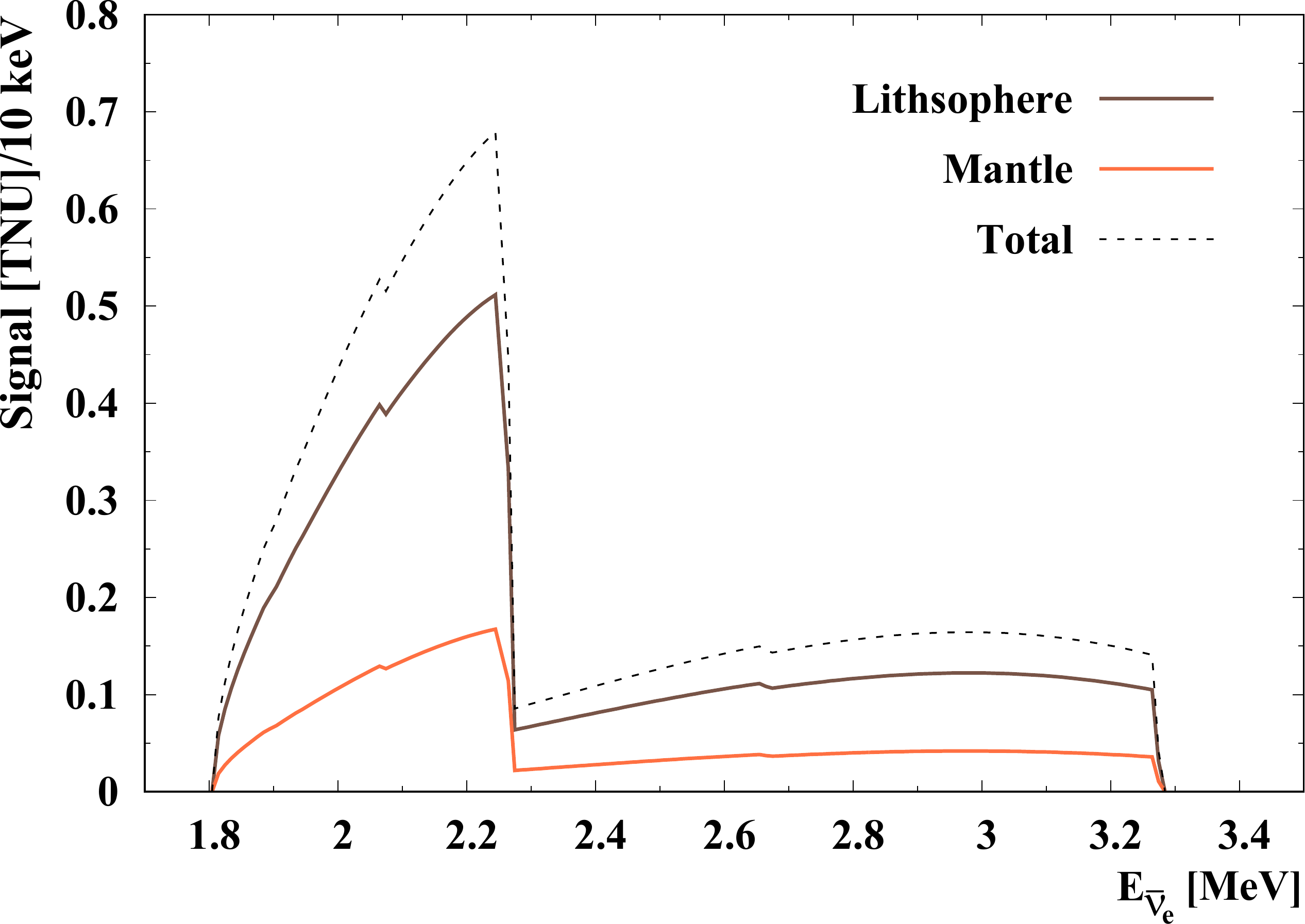}}
    \vspace{1mm}
        \caption{The total geoneutrino energy spectrum expected to be detected at LNGS via IBD interaction on free proton (black dashed lines). The geological considerations are the same as in Fig.~\ref{fig:GeoNuFlux}. Left: The contributions from $^{238}$U (green solid line) and $^{232}$Th (blue solid line) components are shown. Right: The contributions from the lithosphere (brown solid line) and from the mantle (orange solid line) are reported separately.}
         \label{Fig:GeoNuSignal}
    \end{figure*}

The geoneutrino spectrum expected at LNGS is presented in Fig.~\ref{fig:GeoNuFlux}. Figure~\ref{Fig:GeoNuSignal} shows instead the geoneutrino spectrum as expected to be detected via the IBD interaction, showing explicitly the contributions from $^{238}$U and $^{232}$Th, as well as those from the bulk lithosphere and the mantle.

\subsection{Reactor antineutrinos}
\label{subsec:rea}
 
The main source of background in geoneutrino detection is the production of electron antineutrinos by nuclear power plants, the strongest man-made antineutrino source. Many nuclei, produced in the fission process of nuclear fuel decay through $\beta$-processes with the consequent emission of electron antineutrinos, the so-called {\it reactor antineutrinos}. Their energy spectrum extends up to $\simeq$10\,MeV, 
well beyond the end point of the geoneutrino spectrum (3.27\,MeV). As a consequence, in the geoneutrino energy window (1.8 - 3.27\,MeV), there is an overlap between geoneutrino and reactor antineutrino signals.

At present, there are approximately 440 nuclear power reactors in the world, providing, nominally, a total amount of about 1200\,GW thermal power, corresponding to approximately 400\,GW of electrical power. 
With $\sim$200\,MeV average energy released per fission and 6 $\bar{\nu}_e$ produced along the $\beta$-decay chains of the neutron-rich unstable fission products, a reactor with a typical thermal power of 3\,GW emits $5.6 \times 10^{20} \,  \bar{\nu}_e$ s$^{-1}$.

An accurate determination of the expected signal and spectrum of reactor antineutrinos requires a wide set of information, spanning from the characteristics of nuclear cores to neutrino properties. The spectrum of the reactor antineutrino events expected to be measured during the acquisition time $t$ by a detector with efficiency $\varepsilon$ and $N_p$ target protons is:
\begin{eqnarray}\label{Eq:ReactorFlux}
\frac{dN_{\mathrm{rea}}}{dE_{\bar{\nu}_e}} &=& \varepsilon \,  N_p \,  t \,
 \sum_{r=1}^{N_{\mathrm{rea}}}
\frac { P_{r}}{4 \pi L_{r}^{2}} <LF>_r      \\
\nonumber
&\times&
 \sum_{i=1}^4 \frac {p_{ri}}{Q_{i}} \frac{\phi_{i}(E_{\bar{\nu}_e})}{dE_{\bar{\nu}_e}} 
 \sigma(E_{\bar{\nu}_e})
P_{ee}(L_r, E_{\bar{\nu}_e}),
\end{eqnarray}
where the index $r$ cycles over $N_{\mathrm{rea}}$ reactors considered: $P_{r}$ is its nominal thermal power, $L_{r}$ is the reactor-to-detector distance (the core positions are taken from~\cite{coresposition} and we assume a spherical Earth with radius $R$ = 6371\,km), $<LF>_r$ indicates the weighted average of monthly thermal {\it load factors} ($LF_{mr}$). The index $i$ stands for the different components of nuclear fuel ($^{235}$U, $^{238}$U, $^{239}$Pu, and $^{241}$Pu), $p_{i}$ is the {\it power fraction} of the component $i$, $Q_i$ is the {\it energy released per fission} of the component $i$ taken from~\cite{ma2013improved} with a 0.2\% quoted uncertainty, $\phi_i(E_{\bar{\nu}_e})$ is the antineutrino spectrum originating from the fission of the $i^{\rm th}$ component, 
$\sigma(E_{\bar{\nu}_e})$ is the IBD cross section~\cite{strumia2003precise}, 
and $P_{ee}$ is the survival probability use in Eq.~\ref{eq:pee}.

The $<LF>_r$ is calculated for each reactor $r$:
\begin{equation}\label{LFmedio}
<LF>_r= \frac  { \sum_{m=1}^{137} LF_{mr}  \mathcal{E}_m } { \sum \mathcal{E}_m},
\end{equation}
where $\mathcal{E}_m$ are the monthly Borexino exposures in kton $\times$ year during the 137 months period from December 2007 to April 2019 (Sec.~\ref{subsec:exposure}). 

 In our calculation the nominal thermal power $P_r$ and the monthly load factors $L_{mr}$ originate from the {\it Power Reactor Information System} (PRIS), developed and maintained by the International Atomic Energy Agency (IAEA)~\cite{iaeaLF}. Each year in summer, the PRIS produces documents containing information about the nuclear power reactor performance
 relative to the previous year\footnote{For year 2019 the data are not yet available, so we use the data of 2018.}.
The $L_{mr}$ reported are defined as the ratio 
between the net electrical energy produced during a reference period (after subtracting the electrical energy taken by auxiliary units) and the net electrical energy that would have been supplied to the grid if the unit were operated continuously at the nominal power during the whole reference period. In our calculation we assume that such {\it electrical load factors} are equal to the {\it thermal ones}, which are not available at present. We also consider an uncertainty on thermal power $P_{r}$ of the order of 2\%.

Concerning power fractions $p_{ri}$, one has to take into account that throughout the years, several technologies in building nuclear power plants have been developed. Different core types are characterized by different fuel compositions which give rise to different isotope contributions to total thermal power. In addition, during the power cycle of a nuclear reactor, the composition of the fuel changes since Pu isotopes are bred and U is consumed. Thus, the power fractions $p_{ri}$ are in principle 
quantities which depend on the reactor type and time. 
At the moment, we do not know the exact time-dependent fuel composition in each core operating in the world, so,
as in~\cite{baldoncini2015reference}, we assume some representative values for the power fractions. Pressurized Water Reactors, Boiled Water Reactors, Light Water Graphite Reactors, and Gas Cooled Reactors are assumed to adopt an enriched Uranium composition with power fractions 
$\mbox{$^{235}$U}:\mbox{$^{238}$U}:\mbox{$^{239}$Pu}:\mbox{$^{241}$Pu} =0.567:0.075:0.307:0.054$. For about thirty Pressurized Water Reactors, mainly located in Europe, using MOX fuel (i.e. plutonium recovered from spent nuclear fuel, reprocessed 
Uranium or depleted Uranium), we assume that 30\% of their thermal power was originated with power fractions $\mbox{$^{235}$U}  :\mbox{$^{238}$U}:\mbox{$^{239}$Pu}:\mbox{$^{241}$Pu}  = 0.00 :  0.080 : 0.708 :  0.212$ and the remaining 70\% of the thermal power originated by the previous composition. For Pressurized Heavy Water Reactors, we adopt $\mbox{$^{235}$U}  :\mbox{$^{238}$U}:\mbox{$^{239}$Pu}:\mbox{$^{241}$Pu}  = \, 0.543: 0.024 : 0.411 : 0.022$. The range of variations of the different power fractions available in the literature reported in~\cite{baldoncini2015reference} is considered as the uncertainty on $p_{ri}$.

\begin{figure}[t]
     \centering
     \vspace{-2.5mm}
 \includegraphics[width = 0.48\textwidth]{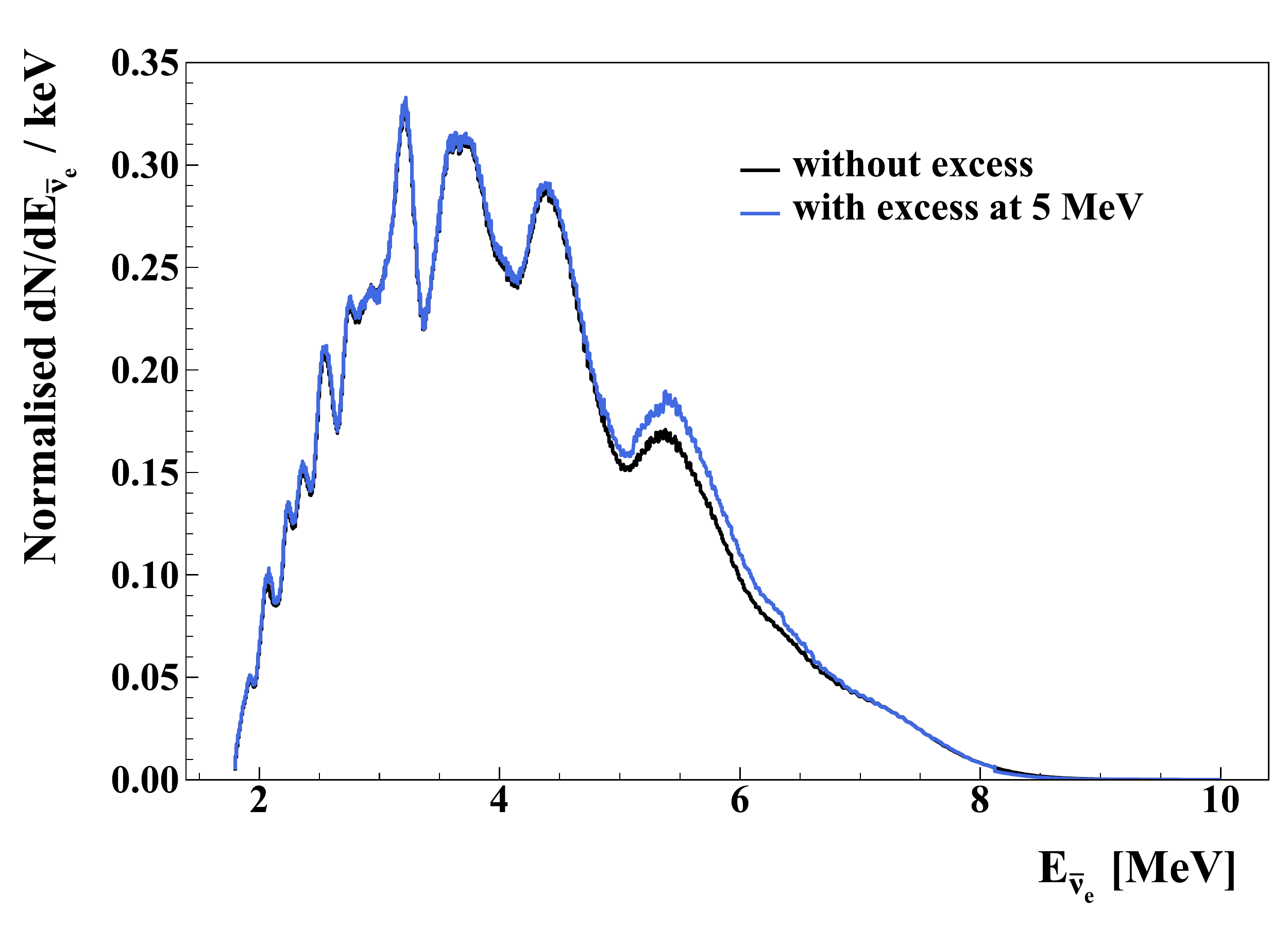}
\caption{Comparison of the spectral shapes ``with and without the 5\,MeV excess" for reactor antineutrinos expected to be detected in Borexino via the IBD interaction for the period December 2007 to April 2019. The spectra are normalized to one. We note that the expected number of events predicted by the spectrum ``with 5\,MeV excess" diminishes by about 6\%
with respect to the one "without 5 MeV excess", consistently with the
normalization factor $R$ = 0.946 found in~\cite{an2016measurement}.}
\label{fig:reactor_antinu}
\end{figure}

The antineutrino energy spectra $\phi_i(E_{\bar{\nu}_e})$ deserve particular attention. Experimental results from the reactor antineutrino experiments Daya Bay~\cite{an2016measurement}, Double CHooz~\cite{abe2014improved}, RENO~\cite{reno2018nufact}, NEOS~\cite{neos2017} coherently show that the measured IBD positron energy spectrum deviates significantly from the spectral predictions of Mueller et al. 2011~\cite{mueller2011improved} in the energy range between 4 - 6 MeV: the so-called  {\it 5\,MeV excess}. In addition, an overall deficit is observed with respect to the prediction of ~\cite{mueller2011improved}. The origin of this effect is being still debated~\cite{PhysRevLett.118.251801,MENTION2017307,Ejiri2019}. In order to take into account this effect, we proceed in the following way: first, we calculate the neutrino spectra corresponding to all four isotopes according to the parametrization from~\cite{mueller2011improved}. Then we multiply the total spectrum by an energy-dependent correction factor based on the Daya Bay high precision measurement (extracted from lower panel of Fig.~3 in~\cite{an2016measurement}). The shapes of the antineutrino spectra expected in Borexino in the period from December 2007 to April 2019, without and with the 5\,MeV excess, are compared in Fig.~\ref{fig:reactor_antinu}. The effect of this shape difference on the precision of the geoneutrino measurement is discussed in Sec.~\ref{subsec:syst}.

Finally, the expected signal from reactor antineutrinos $S_{\mathrm{rea}}$, expressed in TNU, can be obtained by integrating Eq.~\ref{Eq:ReactorFlux} and assuming 100\% detection efficiency for a detector containing 
 $N_p=10^{32}$ target protons and operating continuously for $t$ = 1\,year. The $S_{\mathrm{rea}}$ for the reactor spectra without and with the 5\,MeV excess, are $84.5 ^{+1.5}_{-1.4}$\,TNU and $79.6 ^{+1.4}_{-1.3}$\,TNU, respectively. In the geoneutrino energy window, the respective signals are $22.4 ^{+0.4}_{-0.4} $\,TNU and $20.7 ^{+0.4}_{-0.4}$\,TNU. The quoted errors are at $1\sigma$ level, where the uncertainties related to reactor antineutrino production, propagation, and detection processes are estimated using a Monte Carlo-based approach discussed in~\cite{baldoncini2015reference}. 
 As expected, by introducing the ``5\,MeV excess", the predicted signal varies by about 6\% consistently with the normalization factor $R$ = 0.946 found in~\cite{an2016measurement}.

In~\cite{baldoncini2015reference} the authors investigated the matter effects concerning the antineutrino propagation from the reactor to 
several experimental sites including LNGS. Since the size of the effect is proportional to the baseline travelled in matter, a maximum effect is found for the location in Hawaii, far away from all the reactors, and amounts to 0.7\%. For Borexino, it can be considered negligible with respect to the overall uncertainties on the reactor antineutrino signal.  

        \subsection{Atmospheric neutrinos} 
        \label{subsec:atm}

\begin{figure}
     \centering  
    \includegraphics[width = 0.48\textwidth]{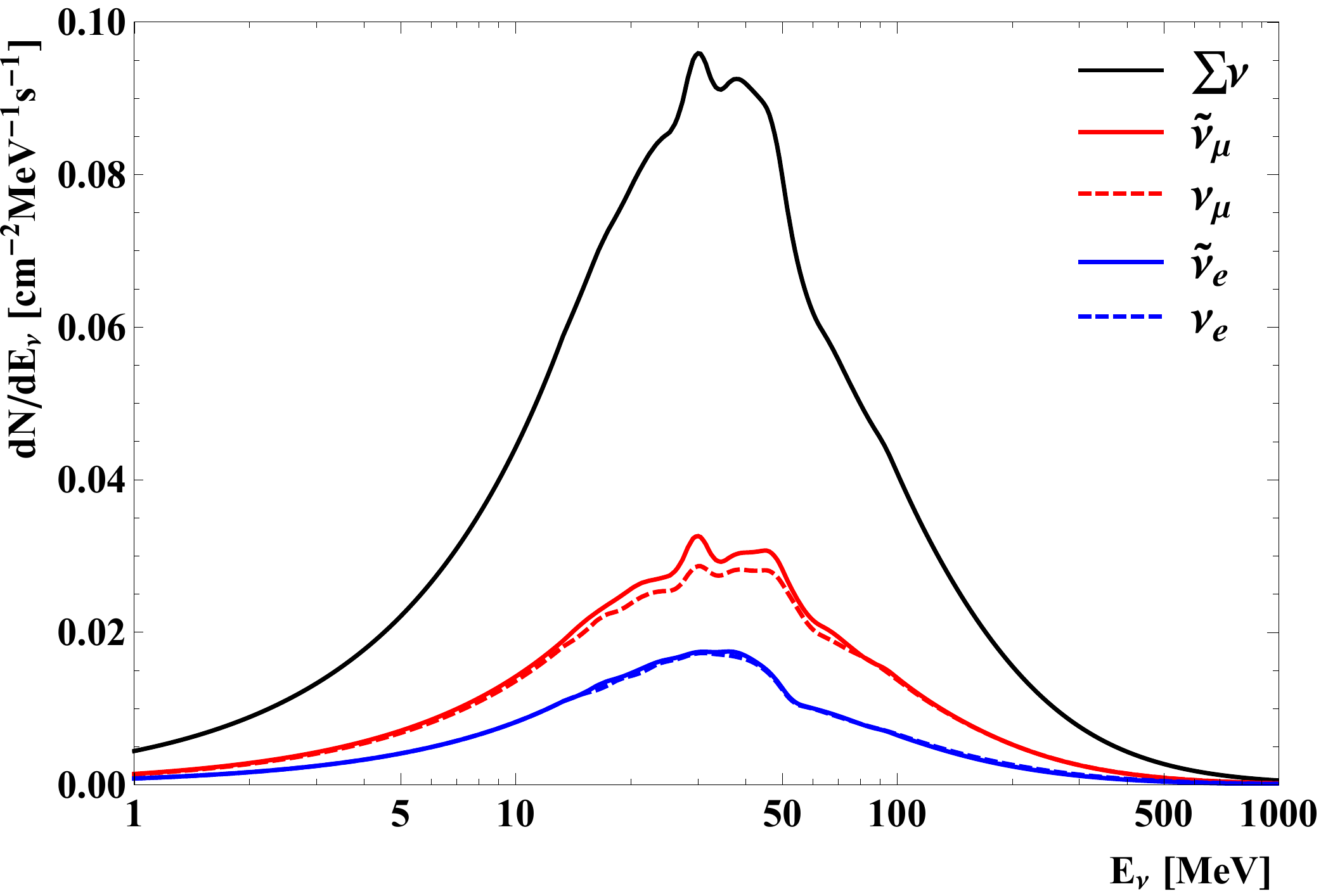}
        \caption{Energy spectra of atmospheric neutrinos as created in the atmosphere by cosmic rays, obtained with the HKKM2014 and FLUKA simulations.}
        \label{fig:AtmoNuSpectrum}
    \end{figure}

We have studied atmospheric neutrinos as a potential background source for the geoneutrino measurement. Atmospheric neutrinos originate in
sequential decays of $\pi(K)^{\pm}$ mesons and $\mu^{\pm}$ muons produced in cosmic rays' interactions with atmospheric nuclei. The flux of atmospheric neutrinos, the energy spectrum of which is shown in Fig.~\ref{fig:AtmoNuSpectrum}, contains both neutrinos and antineutrinos, and the muon flavour is roughly twice abundant than the electron flavour. The process of neutrino oscillations then alters the flavour composition of the neutrino flux passing through the detector.

Atmospheric neutrinos interact in many ways with the nuclei constituting the Borexino scintillator. The most copious isotopes in the Borexino scintillator are $^1$H ($6.00\times 10^{31}$/kton), $^{12}$C ($4.46\times 10^{31}$/kton), and $^{13}$C ($5.00\times 10^{29}$/kton). Besides the IBD reaction itself, there are many reactions with $^{12}$C and $^{13}$C atoms that may, in some cases, mimic the IBD interactions. They have the form of $\nu + A \to \nu(l) + n + \dots + A'$, where $A$ is the target nucleus, $A'$ is the nuclear remnant, $l$ is charged lepton produced in CC processes, $n$ is the neutron, and dots are for other produced particles like nucleons (including additional neutrons) and mesons (mostly $\pi$ and $K$ mesons). A dedicated simulation code was developed to precisely calculate this background in Borexino. 

For energies above 100\,MeV, the atmospheric neutrino fluxes are taken from the HKKM2014 model~\cite{Honda:2015fha}, while below 100\,MeV the fluxes from the FLUKA code~\cite{Battistoni:2005pd} are adopted. The resulted energy spectrum is the one shown in Fig.~\ref{fig:AtmoNuSpectrum}. We consider the neutrino fluxes averaged from all directions. We calculated the flavour oscillations during neutrino propagation through the Earth, including the matter effects, with the modified Prob3++ software~\cite{bib:ProbPP} that comprises 1\,km wide constant-density layers according to the PREM Earth's model~\cite{Dziewonski:1981xy}. The neutrino interactions with $^{12}$C, $^{13}$C, and $^{1}$H nuclei were simulated with the GENIE Neutrino Monte Carlo Generator (version 3.0.0)~\cite{Andreopoulos:2009rq}.
GENIE output final state particles are used as primary particles for the {\it G4Bx2} Borexino MC code~\cite{Agostini:2017aaa}. The number of expected IBD-like interactions due to atmospheric neutrinos passing all the optimized selection cuts (Sec.~\ref{subsec:cuts}) will be given in Sec.~\ref{subsec:antinu_est_ev}, Table~\ref{tab:antinu-events-expected}.

 \subsection{Georeactor} 
 \label{subsec:georeactor}
   
      \begin{figure} [t]
    	\centering    
          \subfigure[]{ \includegraphics[width = 0.25\textwidth]{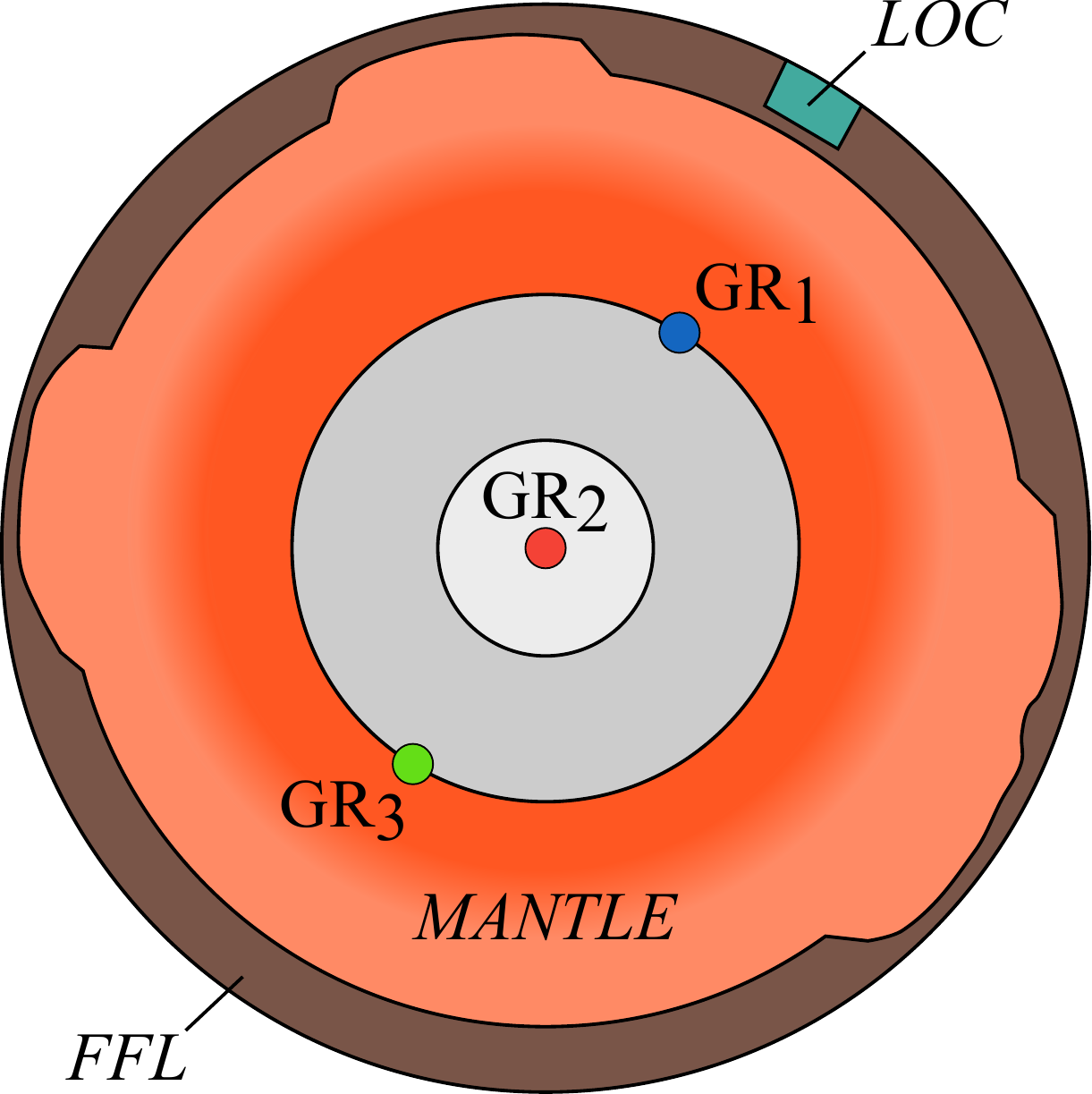}
          \label{fig:GeoReactorPosition}}
          \subfigure[]{\includegraphics[width = 0.45\textwidth]{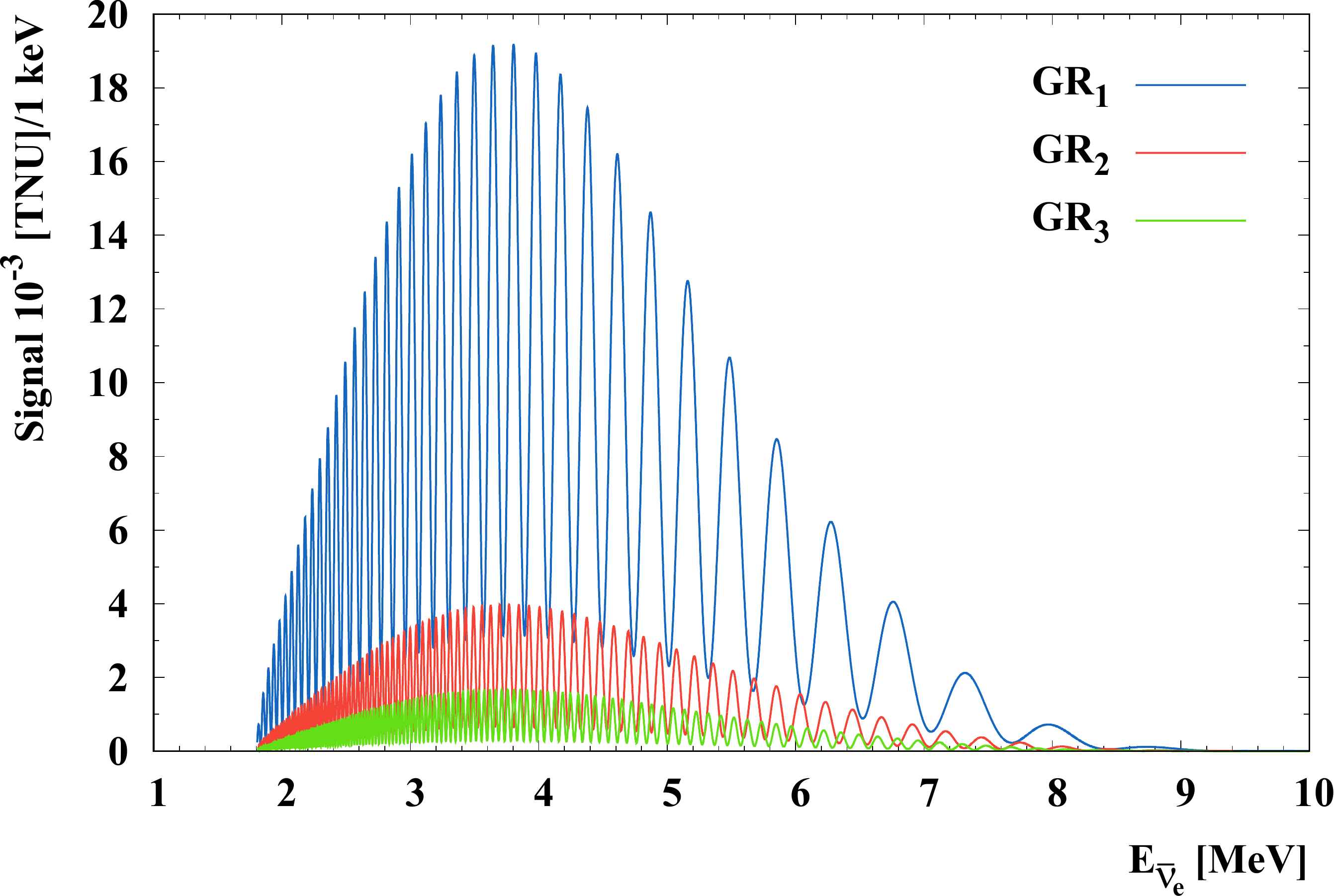}
          \label{fig:GeoReactorSpectrum}}  \caption{Hypothetical georeactor: (a) schematics of the three studied locations: the Earth's center (GR2) and the core-mantle-boundary (GR1 and GR3). The green area labeled LOC shows the position of the local crust around LNGS. Not to scale. (b) The oscillated antineutrino spectra expected at LNGS from 1\,TW georeactor positioned at the three locations.}
        \label{fig:GeoReactor}
    \end{figure}

A possible existence of georeactor, i.e. natural nuclear fission reactor in the Earth interior, was first suggested by Herndon in 1993~\cite{ herndon1993feasibility}. Since then, several authors have discussed its possible existence and the characteristics. Different models suggest the existence of natural nuclear reactors at different depths: at the center of the core~\cite{herndon1996substructure}, at the inner core boundary~\cite{rusov2007geoantineutrino}, and the core-mantle boundary~\cite{de2008feasibility}. These models predict that the georeactor output power sufficient to explain terrestrial heat flow measurements (Table~\ref{tab:heat_flux}) and helium isotope ratios in oceanic basalts~\cite{farley1998noble}.

In principle, characteristics of the antineutrino spectrum observed at the Earth’s surface could specify the location and the power of a georeactor, discriminating among these models~\cite{dye2009neutrino}. Some authors~\cite{rusov2010kamland} even suggested an alternative interpretation of the KamLAND data through the existence of a $\sim$30\,TW georeactor at the boundary of the liquid and solid phases of the Earth’s core. Borexino 2013 results~\cite{Bellini:2013geo} set a 4.5\,TW upper bound at 95\% C.L. for a georeactor in the Earth's center. KamLAND has also studied this hypothesis and obtained an upper limit of 3.7\,TW at 95\% C.L.~\cite{Gando:2013nba}.

In this paper, we set the new upper bounds on the power of a potential georeactor in Sec.~\ref{subsec:georeactor-results}. In order to be able to set such limits for different hypothetical locations of the georeactor, we have calculated the expected antineutrino spectra of a 1\,TW point-like georeactor, operating continuously during the data taking period (December 2007 - April 2019) with the power fractions of fuel components as suggested in~\cite{herndon2005background} ($^{235}$U : $^{238}$U $\approx$ 0.76 : 0.23). The energy released and the antineutrino spectra per fission are as in Sec.~\ref{subsec:rea}. Using only the flux parametrisation of~\cite{mueller2011improved}, we neglect the question of 5\,MeV excess. As shown in Fig.~\ref{fig:GeoReactorPosition}, we consider three different depths as extreme georeactor locations: 1) GR2: the Earth center ($d$ = $R_{\mathrm{Earth}}$), 2) GR1: the CMB placed just below the LNGS site ($d$ = 2900\,km), and 3) GR3: the CMB on the opposite hemisphere ($d$ = 2$R_{\mathrm{Earth}}-2900$ = 9842\,km).

In the calculation of the survival probability (Sec.~\ref{subsec:oscil}), the matter effect (Eq.~\ref{eq:pee_matter}) is taken into account by assuming an Earth constant density $\rho=5.5$\,g/cm$^3$. With respect to oscillations in vacuum, we observe 1.4\% increase of the signal.

The expected oscillated spectra for 1\,TW georeactor at the three locations GR1, GR2, and GR3 are shown in Fig.~\ref{fig:GeoReactorSpectrum}. Since the oscillation length of MeV neutrinos is much smaller with respect to the studied baselines, we observe very fast oscillations in all three spectra. As it will be discussed in Sec.~\ref{sec:mc}, Fig.~\ref{fig:PDFs-georeactor}, Borexino energy resolution does not allow to distinguish the spectral differences due to oscillations. The total expected signal $S_{\mathrm{geo}}$ is reported in Table~\ref{tab:antinu-signals-expected}.

Concerning error estimation, the same procedure as in Sec.~\ref{subsec:rea} in the determination of the contribution due to oscillation parameters, IBD cross section, and energy released per fission were adopted. Uncertainties due to power fractions were not taken into account, since the georeactor is assumed not to have a significant change of isotopic composition on a scale of few years. The systematic uncertainties due to our simplified treatment of neutrino oscillations in matter were estimated: the density variation in the range between 2.2\,g cm$^{-3}$ (crust) and 13\,g cm$^{-3}$ (core) causes about 3\% (1\%) variation  of the expected signal in the total (geoneutrino) energy window. The errors reported in Table~\ref{tab:antinu-signals-expected} have been calculated by adding in quadrature the errors discussed above. 
        \subsection{Summary of antineutrino signals}
        \label{subsec:antinu_est}

  Table~\ref{tab:antinu-signals-expected} summarizes the expected antineutrino signals from geoneutrinos, reactor antineutrinos, and from the georeactor, as they were discussed above. For geoneutrinos, we show the predictions for Cosmochemical (CC), Geochemical (GC), Geodynamical (GD) BSE models, as well as for the Fully Radiogenic (FR) model, as defined in Table~\ref{tab:BSE}. The contribution from each component is specified explicitly in the geoneutrino energy window (1.8 - 3.3\,MeV) as well as in the reactor antineutrino window (1.8 - 8.0\,MeV).

 \begin{table}[H]
	\centering
	\caption{	\label{tab:antinu-signals-expected} Summary of the antineutrino signals expected at LNGS in different energy windows from geoneutrinos (for the whole Earth and from the mantle separately), reactor antineutrinos, and from 1\,TW georeactor located in three different positions. The geoneutrino signals are calculated summing the lithospheric contribution $S_{\mathrm{LSp}}$(Th + U) = 25.9$^{+4.9}_{-4.1}$\,TNU and the mantle signal $S_{\mathrm{mantle}}$. The latter is predicted according to Cosmochemical (CC), Geochemical (GC), Geodynamical (GD), and Fully radiogenic (FR) models (Table~\ref{tab:S_mantle}). The central value represents the intermediate scenario estimates (Fig.~\ref{fig:MantleScenarios}), while the 1$\sigma$ uncertainties corresponds to [$S^{\mathrm{HSc}}_{\mathrm{mantle}} - S^{\mathrm{LSc}}_{\mathrm{mantle}}]/6$.} \vskip 2pt
	\begin{tabular*}{\columnwidth}{l @{\hskip 18pt} c @{\hskip 18pt} c}
		\hline
		\hline
		Source & Energy & Signal \Tstrut \\ 
		       & [MeV] & [TNU] \Bstrut \\ 
		\hline
		\multicolumn{3}{c}{Geoneutrinos} \Tstrut\Bstrut \\
		\hline
		Bulk lithosphere & 1.8 - 3.3 & 25.9$^{+4.9}_{-4.1}$ \Tstrut  \\ [8pt]  
	        CC BSE (total)  & 1.8 - 3.3 & 28.5$^{+5.5}_{-4.8}$ \\ [4pt] 
	        CC BSE (mantle)   & 1.8 - 3.3 & 2.5 $\pm$ 0.5 \\ [8pt] 
	        GC BSE (total) & 1.8 - 3.3 & 34.6$^{+5.5}_{-4.8}$ \\ [4pt] 
	        GC BSE (mantle)  & 1.8 - 3.3 & 8.7 $\pm$ 0.8 \\ [8pt] 
	        GD BSE (total) & 1.8 - 3.3 & 45.6$^{+5.6}_{-4.9}$ \\ [4pt] 
	        GD BSE (mantle) & 1.8 - 3.3 & 19.6 $\pm$ 1.1 \\ [8pt]
	        FR (total) & 1.8 - 3.3 & 55.3$^{+5.7}_{-5.0}$ \\ [4pt] 
	        FR (mantle)  & 1.8 - 3.3 & 29.4 $\pm$ 1.5 \Bstrut \\
	        \hline
	        \multicolumn{3}{c}{Reactor antineutrinos} \Tstrut\Bstrut \\
	        \hline
			without                                             & 1.8 - 3.3  & 22.4$^{+0.4}_{-0.4}$ \Tstrut \\ [4pt]
		    ``5\,MeV excess"         & 1.8 - 8.0 & 84.5$^{+1.5}_{-1.4}$ \\ [8pt]
		    with                                                & 1.8 - 3.3  & 20.7$^{+0.4}_{-0.4}$ \\ [4pt]
		    ``5\,MeV excess"         & 1.8 - 8.0 & 79.6$^{+1.4}_{-1.3}$ \Bstrut \\
		    \hline
	  	    \multicolumn{3}{c}{1\,TW Georeactor} \Tstrut\Bstrut \\
	 	    \hline
		  GR2: Earth's center & 1.8 - 3.3 & 1.87 $\pm$ 0.05 \Tstrut \\
			              & 1.8 - 8.0 & 7.73 $\pm$ 0.23 \\ [4pt]
          GR1: CMB at 2900\,km & 1.8 - 3.3 & 9.0  $\pm$ 0.30 \\
			              & 1.8 - 8.0 & 37.3 $\pm$ 1.12 \\ [4pt]
		  GR3: CMB at 9842\,km & 1.8 - 3.3 & 0.78 $\pm$ 0.02\\
			              & 1.8 - 8.0 & 3.24 $\pm$ 0.10 \Bstrut \\
	    \hline
	    \hline
	\end{tabular*}
\end{table}

\section{NON-ANTINEUTRINO BACKGROUNDS}
\label{sec:bgr}

In this section we describe the origin of non-antineutrino backgrounds for geoneutrino measurements. In particular, the cosmogenic background in Sec.~\ref{subsec:cosmogenic}, background due to accidental coincidences in Sec.~\ref{subsec:acc},
due to the $(\alpha,$ n) and $(\gamma,$ n) reactions in Sec.~\ref{subsec:alpha_n} and Sec.~\ref{subsec:gamma_n}, respectively, while the radon-correlated background in Sec.~\ref{subsec:radon} and $^{212}$Bi-$^{212}$Po coincidences in Sec.~\ref{subsec:212BiPo}. Typically, the rates of these backgrounds depend on the selection cuts for IBD events, whose optimization is described in Sec.~\ref{sec:data_sel}. Therefore, the final evaluation of the non-antineutrino background levels is given in the following Sec.~\ref{sec:sig_bgr_est}.

        \subsection{Cosmogenic background}
        \label{subsec:cosmogenic}
  
        One of the most important non-antineutrino backgrounds for geoneutrino detection is the cosmogenic background due to the residual muon flux. It is necessary to eliminate muons along with their spallation daughters as they can imitate the IBD signals. The various cosmogenic backgrounds relevant in geoneutrino measurement are explained in this section. The muon detection methods and efficiencies were already explained in Sec.~\ref{subsec:muon}. The actual evaluation of the cosmogenic background passing the IBD selection cuts is explained in detail in Sec.~\ref{subsec:cosmogenic_est}.

            \paragraph{Hadronic background}
 
\begin{table*}
	\centering
	\caption{\label{tab:isotopes_muon} The lifetimes ($\tau$), $Q$-values, decay modes with respective branching ratios, and the isotope production rates in Borexino of the hadronic spallation products relevant for geoneutrino measurement~\cite{Bellini:2013cosmo,lightnuclei}.} \vskip 2pt
    \begin{tabular*}{\textwidth}{c @{\hskip 24pt} c @{\hskip 24pt} c @{\hskip 24pt} c @{\hskip 24pt} c @{\hskip 24pt} c}
        \hline
    		\hline
        Isotope  & $\tau$ & $Q$   & Decay mode & Branching ratio & Production rate \Tstrut \\ 
                 & [ms]   & [MeV] &            & [\%]            & [(day $\times$ 100\,ton)$^{-1}$] \Bstrut \\ 
        \hline
        $^{9}$Li  & 257.2 & 13.6 & $\beta^{-} + n$  & 51 & 0.083 $\pm$ 0.009 (stat.) $\pm$ 0.001 (syst.) \Tstrut \\ [4pt]
        $^{8}$He & 171.7 & 10.7 & $\beta^{-} + n $  & 16 & $<$0.042 (3$\sigma$ CL) \\ [4pt]
        $^{12}$B & 29.1 & 13.4 & $\beta^-$  & 98.3 & 1.62 $\pm$ 0.07 (stat.) $\pm$ 0.06 (syst.) \Bstrut \\
        \hline
        \hline
	\end{tabular*}
\end{table*}

      Table~\ref{tab:isotopes_muon} summarizes the lifetimes ($\tau$), $Q$-values, branching ratios of the relevant decays, and the production rates of the three hadronic spallation products relevant for geoneutrino measurement.
      $^{9}$Li and $^{8}$He isotopes are the most important cosmogenic background, since they can produce $(\beta^- + n$) pairs during their decays:
                \begin{equation}
                \centering
      	        ^{9}\text{Li} \rightarrow e^{-} + \bar{\nu}_{e} + n + 2\alpha
      	         \label{eq:Li9}
                \end{equation}
                \begin{equation}
                \centering
                ^{8}\text{He} \rightarrow e^{-} + \bar{\nu}_{e} + n + ^{7}\text{Li},
                \label{eq:He8}
                \end{equation}
        indistinguishable from IBD's products. Generally speaking, 1 to 2\,s veto after all internal muons can sufficiently suppress this kind of background. Due to the very small residual muon flux at LNGS, this kind of approach was applied in the previous Borexino geoneutrino analyses and is further improved in this work, as will be shown in Sec.~\ref{subsec:vetoes}. Borexino has studied the production of hadronic cosmogenic backgrounds~\cite{Bellini:2013cosmo}. For the $^{8}$He production rate, only an upper limit was set (Table~\ref{tab:isotopes_muon}). When taking the 3$\sigma$ CL limit as the production rate itself, the ratio of products [production rate $\times$ branching ratio] for $^{9}$Li:$^{8}$He is 6:1 in Borexino. Considering also the fact that in the far detector of Double Chooz, where the $^{9}$Li production rate is 575 times higher than in Borexino, the $^{8}$He production rate is observed to be compatible with zero~\cite{deKerret:2018fqd}, we conclude that the $(\beta^- + n)$ decaying hadronic background is largely dominated by $^{9}$Li. 
  
       In addition, two $\beta^{-}$s from two decays of cosmogenic $^{12}$B can imitate IBD signals:
                \begin{equation}
                \label{eq:B12}
                \centering
	            ^{12}\text{B} \rightarrow e^{-} + \bar{\nu}_{e} + \mathrm{^{12}C}.
                \end{equation}
      \noindent  The second $^{12}$B decay should happen within the IBD coincidence time window (1.28\,ms, Sec.~\ref{sec:data_sel}) in order to be a background for geoneutrino analysis. Since this time window is much smaller when compared to $^{12}$B lifetime (29.1\,ms), only [1- exp(-1.28/29.1)] = 4.3\% fraction of $^{12}$B-$^{12}$B coincidences can contribute to this background. In addition, the 13.4\,MeV $Q$ value of $^{12}$B decay largely extends over the end-point of geoneutrino as well as reactor antineutrino spectra, which further suppresses the importance of this background for geoneutrinos.

        \paragraph{Untagged muons}

The small amount of muons that go undetected can cause muon-related background in the geoneutrino sample. This kind of background can consist of:
       \begin{itemize}
                \item {\it $\mu$ + $\mu$}: two untagged muons close in time. In particular muons with short tracks inside the buffer, where the scintillation light yield is strongly suppressed, could fall within the IBD selection cuts. 
                \item {\it $\mu$ + $n$}: an untagged muon, again especially a buffer muon, can mimic a prompt, while muon-generated spallation neutron is indistinguishable from an IBD delayed signal. Also, the capture time of cosmogenic and IBD-produced neutrons is the same.
                \item {\it Muon daughters}: obviously, after undetected or unrecognized muons, no vetoes are applied. Thus, their spallation products can mimic IBDs. They can be, for example, neutron - neutron pairs or other hadronic backgrounds or fast neutrons described below.
            \end{itemize} 
            %

            \paragraph{Fast cosmogenic neutrons}
            
            Cosmic muons can produce neutrons that have a very hard energy spectrum extending to several GeV. These so-called {\it fast neutrons} can penetrate significant depth both through the surrounding rock as well as detector shielding materials. Neutrons can reach the LS, where they can scatter off protons. The proton mimics the IBD-prompt signal, while the neutron itself, after being captured, provides a delayed signal. Thus, cosmogenic neutrons are potentially a dangerous background for the geoneutrino analysis. In principle, some level of pulse-shape discrimination between the scattered proton and the positron signal, prompt signal of a real IBD, is possible. 
            
            In Borexino, 2\,ms veto is sufficient to suppress this background after muons that cross the detector and are detected either as internal or external muons. Possible background of this kind can arise from undetected muons crossing the water tank and from the muons producing spallation products in interactions outside the detector, e.g. in the surrounding rock. The details of the estimation of both categories of this kind of background is explained in Sec.~\ref{subsec:cosmogenic_est}.

        \subsection{Accidental coincidences}
        \label{subsec:acc}
  
        In Borexino, the rate of prompt- and delayed-like events is 0.03\,s$^{-1}$ and 0.01\,s$^{-1}$, respectively, in the whole scintillator volume. The accidental coincidences of these events in time and space can imitate IBD signals. These events are dominated by the external background, thus their reconstructed positions are mostly at large radii and close to the IV. This means that increasing the Fiducial Volume (FV) for the geoneutrino analysis necessarily increases also the rate of accidental coincidences. These coincidences are searched in an extended off-time window of 2\,ms - 20\,s and scaled back to the geoneutrino search time window of 1.28\,ms for a certain set of selection cuts. The evaluation of this background is explained in detail in Sec.~\ref{subsec:acc_est}.

        \subsection{($\alpha$, n) background}
        \label{subsec:alpha_n}
  
  The decays along the chains of $^{238}$U, $^{235}$U, and $^{232}$Th generate $\alpha$ particles that can initiate ($\alpha, n$) reactions, a possible background for geoneutrinos. Thanks to the ultra-radiopurity levels achieved in Borexino,
  the only relevant source of $\alpha$-particles in this terms is $^{210}$Po ($\tau = 199.6\,\text{days}$), a product of $^{222}$Rn, found fully out of equilibrium with the rest of the $^{238}$U chain. $^{210}$Po decays into a stable $^{206}$Pb emitting $\alpha$ with the energy of 5.3\,MeV:
        \begin{equation}
	        ^{210}\textrm{Po} \rightarrow \: ^{206}\textrm{Pb} + \alpha.
        \end{equation}  
 Thanks to the $\alpha/\beta$ discrimination techniques developed in Borexino (Sec.~\ref{subsec:ab}), it is possible to count $^{210}$Po decays via an event-by-event basis, as it will be shown in Sec.~\ref{subsec:alpha_n_est}, along with the estimation of this background passing the IBD selection cuts.

\begin{figure}
     \centering  
  \includegraphics[width = 0.48\textwidth]{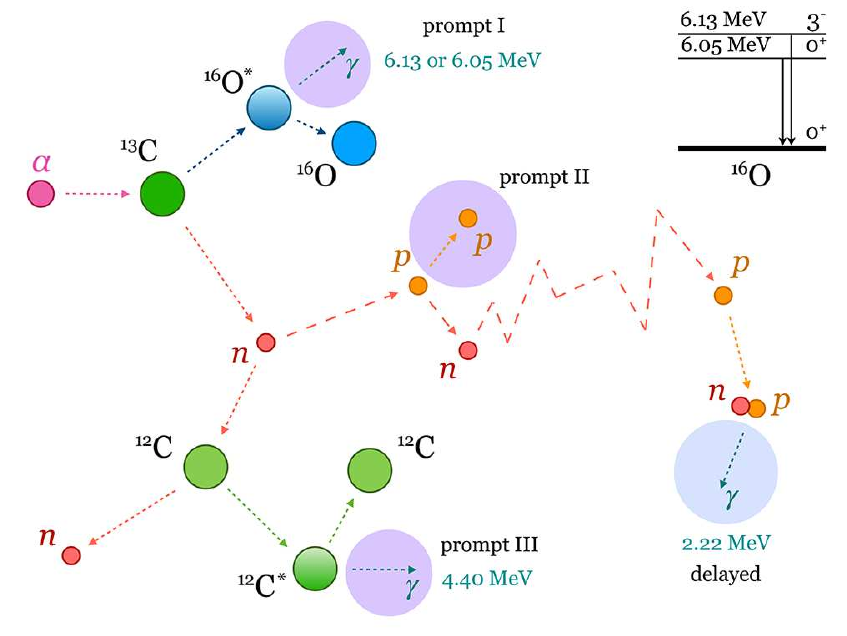}

  \caption{Scheme of the ($\alpha$, n) interaction on $^{13}$C. The three processes that can generate the three kinds of the prompt signals, {\it prompt I}, {\it II}, and {\it III}, as described in text, are shown in violet areas. The blue area indicates the delayed signal from the neutron capture.}
   \label{fig:alphaNBgr}
    \end{figure}

The ($\alpha, n$) reactions can occur on different nuclides, but it takes place mostly on $^{13}$C in the Borexino scintillator:
   \begin{equation}
    ^{13}\text{C} + \alpha \longrightarrow \: ^{16}\text{O} + n.
    \label{eq:alpha_n}
    \end{equation}	
According to the recent revision~\cite{Mohr:2018alphaNBgr} of the previous data~\cite{Harissopulos:2005alphaNBgr, Koning:2012zqy}, the respective cross section equals to $200$\,mb. The produced neutron, with energies up to 7.3\,MeV, is indistinguishable from the neutron that originates in the IBD interaction (Sec.~\ref{sec:IBD}). The relevance of this interaction for the antineutrino measurement, first studied by KamLAND~\cite{Abe:2008aa}, arises from the fact that there are also three possibilities for the generation of the prompt, as it is schematically shown in Fig.~\ref{fig:alphaNBgr}:
    \begin{itemize}
    	\item \textit{prompt I:} $\gamma$-emission with energy of 6.13\,MeV or 6.05\,MeV, as a result of $^{16}$O$^*$ de-excitation, if $^{16}$O is produced in such an excited state;
	    \item \textit{prompt II:} recoil proton appearing after scattering of the fast neutron on proton; 
	    \item \textit{prompt III:} 4.4\,MeV $\gamma$-ray that is a product of the two-stage process: First, $^{12}$C is excited into $^{12}$C$^*$ in an inelastic scattering off fast neutron. Then, $^{12}$C$^*$ transits to the ground state, accompanied by the $\gamma$ emission:
	    \begin{align}
		    n + \: ^{12}\text{C} &\longrightarrow \: ^{12}\text{C}^* + n, \\
		    ^{12}\text{C}^* &\longrightarrow \: ^{12}\text{C} + \gamma \: (4.4\,\text{MeV}).
		    \label{eq:inelastic_scattering}
	    \end{align}
    \end{itemize}

        \subsection{($\gamma$, n) interactions and fission in PMTs}
        \label{subsec:gamma_n}
        
Energetic $\gamma$ rays are produced by neutron capture reactions in the detector materials and in the surrounding rocks or in natural radioactive decays. These $\gamma$'s may in turn give ($\gamma$, n) reactions with the nuclei of the LS or the buffer in Borexino. In particular, if the $\gamma$-ray makes a Compton scattering before being absorbed, its interaction and the following neutron capture gives a coincidence that almost perfectly mimics the IBD signal and in such a case, the pulse shape discrimination is not effective. Table~\ref{tab:gamman} shows the ($\gamma$, n) reactions' thresholds for the most abundant isotopes in the Borexino scintillator.

\begin{table}[h]
	\centering
	\caption{\label{tab:gamman} Reaction thresholds for the ($\gamma$, n) interactions on the isotopes present in the Borexino scintillator. The middle column gives the isotopic abundances. The most relevant isotopes are those present in the PC molecule (in bold).} \vskip 2pt
	\begin{tabular*}{\columnwidth}{c @{\hskip 50pt} c @{\hskip 50pt} c}     
		\hline
		\hline
		Target & Abundance & Threshold \Tstrut \\
		      & [\%] & [MeV] \Bstrut \\
		\hline
		{\bf $^{2}$H } & {\bf 0.015} &   {\bf 2.22} \Tstrut  \\ [8 pt]
		{\bf $^{12}$C} &  {\bf 98.9} &   {\bf 18.7} \\
		{\bf $^{13}$C} & {\bf 1.10} &   {\bf 4.95} \\ [8 pt]
		$^{14}$N &  99.634 &   10.6 \\
		$^{15}$N &  0.366 &   10.8 \\[8 pt]
		$^{16}$O & 99.762 &   15.7 \\
		$^{17}$O & 0.038 &   4.14 \\
		$^{18}$O & 0.200 &   8.04 \Bstrut \\
		\hline
		\hline
	\end{tabular*}
\end{table}

The ($\gamma$, n) interaction having the lowest energy threshold (2.22\,MeV) is the one on $^{2}$H. Taking into account that the IBD's prompt energy starts at 1\,MeV (Eq.~\ref{eq:Epro}) and considering the energy resolution, we conclude that only gammas with energies higher than 3\,MeV could first give a prompt IBD-like signal and then induce photo-dissociation process. Although there are some uncertainties in the branching ratios listed in the isotopes' tables, the natural U and Th chains do not emit sizeable $\gamma$ lines at energies above 3\,MeV. Thus, the intrinsic scintillator and vessel contaminations are not an important source of background, while  muon and neutron induced gammas have to be carefully studied. 

Natural radioactivity of detector materials can be a source of an additional type of background. Spontaneous fission can generate neutrons up to several MeV. These can mimic antineutrino signals in the target mass. $^{238}$U has by far the largest fission probability among the nuclei we can consider and equals to $5.45 \cdot 10^{-7}$. The PMTs can be considered as the most important source of this background.

The final evaluation of the background due to ($\gamma$, n) interactions as well as to spontaneous fission in PMTs will be given in Sec.~\ref{subsec:gamma_n_est}.

        \subsection{Radon background}
        \label{subsec:radon}
    
In 2010-11, the Borexino scintillator was subjected to several purifications with water extraction (WE) procedures followed by nitrogen stripping. These operations were mainly aimed to remove the $^{85}$Kr contamination and to lower as much as possible the $^{210}$Bi one. During the purification campaigns, some radon entered the detector due to the emanation in the purification system. The $^{222}$Rn has a lifetime of $\tau$ = 5.52 days and, after the operations, the correlated backgrounds typically disappear in a time window of a couple of weeks, leaving the corresponding amount of $^{210}$Pb in the detector. An easy approach to precisely evaluate the Rn contamination level during the operations is to observe the $^{214}$Bi($\beta^-$,$Q$ = 3.272\,MeV)-$^{214}$Po($\alpha$, $Q$ = 7.686\,MeV) coincidence, among the radon daughters. By following this approach, we have estimated a contamination factor 100-1000 larger than the typical amount in the Borexino scintillator, during the WE period. For this reason, these transition periods are not used in solar neutrino studies, but, with some care can be used for $\bar{\nu}_e$ analysis. 

It is precisely the $^{214}$Bi-$^{214}$Po coincidence that can become a potential source of background for the IBD selection. This coincidence has a time constant of $\tau$ = $(236.0 \pm 0.5)$\,$\mu$s ($^{214}$Po lifetime~\cite{Bellini2013}), very close to the neutron capture time in PC. The $\alpha$-particles emitted by $^{214}$Po usually have a visible energy well below the neutron capture energy window. However, in 1.04 $\times$ $10^{-4}$ and in 6 $\times$ $10^{-7}$ of cases, the $^{214}$Po decays to excited states of $^{210}$Pb and the $\alpha$ is accompanied by the emission of prompt gammas of 799.7\,keV and of 1097.7\,keV, respectively (see Table~\ref{tab:214Po}). In liquid scintillators, the $\gamma$ of the same energy produces more light with respect to an $\alpha$-particle. Therefore, for these ($\alpha$ + $\gamma$) decay branches, the observed light yield is higher with respect to pure $\alpha$-decays and is very close to the neutron capture energy window. As already mentioned in Sec.~\ref{subsec:ab}, the Borexino liquid scintillator offers a possibility to discriminate highly ionizing particles ($\alpha$, proton) from particles with lower specific ionization ($\beta^-$, $\beta^+$, $\gamma$) by means of pulse shape analysis. In the case of these rare branches, the scintillation pulses from the $\alpha$ and $\gamma$ decays are so close in time that they practically overlap and result to be partially $\alpha$-like. In order to suppress this background to negligible levels, during the purification periods we have increased the lower limit on the charge of delayed signal and applied a slight pulse shape cut as described below in Sec.~\ref{sec:data_sel}. The final evaluation of the radon correlated background passing the optimized IBD selection cuts will be given in Sec.~\ref{subsec:radon_est}.

\begin{table}[h]
	\centering
	\caption{\label{tab:214Po} Decay modes of $^{214}$Po.} \vskip 2pt
	\begin{tabular*}{\columnwidth}{c @{\hskip 20pt} c @{\hskip 20pt} c }   
		\hline
		\hline
		 Decay mode & Branching ratio & Energy \Tstrut \\
		                      & [\%] & [keV] \Bstrut \\
		\hline
		$\alpha$ & 99.99  &  E$_{\alpha}$ = 7833.46 \Tstrut \\[ 4pt]
		$\alpha$+$\gamma$ & 1.04 $\times$ $10^{-4}$  &  E$_{\alpha}$ = 7033.66  \\ 
		  &   &  E$_{\gamma}$ = 799.7  \\[4pt]
				$\alpha$+$\gamma$ & 6.0 $\times$ 10$^{-7}$  &  E$_{\alpha}$ = 6735.76  \\
		  &   &  E$_{\gamma}$ = 1097.7 \Bstrut \\
		\hline
		\hline
	\end{tabular*}
\end{table}

        \subsection{$^{212}$Bi - $^{212}$Po background}
            \label{subsec:212BiPo}
  
During the WE periods, an increased contamination of $^{220}$Rn ($^{232}$Th chain) was also observed. Among the $^{220}$Rn daughters, there is the fast decay sequence of $^{212}$Bi($\beta^-$) and $^{212}$Po($\alpha$):
\begin{equation}
{^{212}}\rm{Bi} \rightarrow {^{212}}\rm{Po} + e^- + \bar{\nu} _e
\end{equation}
\begin{equation}
{^{212}}\rm{Po} \rightarrow {^{208}}\rm{Pb}+\alpha,  
\label{eq:Po212}
\end{equation}
characterized with $\tau$ = $(425.1 \pm 1.5)$\,ns of the $^{212}$Po decay~\cite{Bellini2013}. The $^{212}$Bi is a $\beta^-$ emitter with $Q$ = 2.252\,MeV, while the $\alpha$ of $^{212}$Po decay has 8.955\,MeV energy. Given the short time coincidence, they could be a potential source of background for the IBD candidates, searched among the double cluster events, when both the prompt and delayed fall within one 16\,$\mu$s DAQ gate (Sec.~\ref{subsec:data}). This kind of events was included in the geoneutrino analysis for the first time (Sec.~\ref{subsec:dt}). Fortunately, the $^{212}$Po is not giving any ($\alpha$ + $\gamma$) decay branch, so its effective energy distribution is below the neutron capture peak. Moreover, being a pure $\alpha$ decay, it can be easily recognised and rejected with a proper pulse shape analysis. The final evaluation of this background passing the optimized selection cuts will be given in Sec.~\ref{subsec:212BiPo_est}. 

        \section{DATA SELECTION CUTS}
        \label{sec:data_sel}

 In this section we describe the cuts for the selection of antineutrino candidates and the process of their optimization. The vetoes applied after different muon categories are described in Sec.~\ref{subsec:vetoes}. Sections~\ref{subsec:dt} and \ref{subsec:dR} deal with the definitions of the time and spatial correlation windows between the prompt and delayed IBD candidates. The application of the $\alpha/\beta$ discrimination techniques (Sec.~\ref{subsec:ab}) on the delayed, in order to suppress the radon background (Sec.~\ref{subsec:radon}), is shown in Sec.~\ref{subsec:pulse_shape}. Optimization of the energy cuts for the prompt and delayed signals is shown in Sec.~\ref{subsec:energy}, while the selection of the Dynamic Fiducial Volume in Sec.~\ref{subsec:DFV}. The so-called {\it multiplicity cut} to suppress the background due to undetected muons with multiple neutrons and some noise is explained in  Sec.~\ref{subsec:multiplicity}. Finally, Sec~\ref{subsec:cuts} summarizes all the optimized values for all cuts. 

        \begin{figure}
         \hspace{-4mm}	\includegraphics[width =  0.53\textwidth]{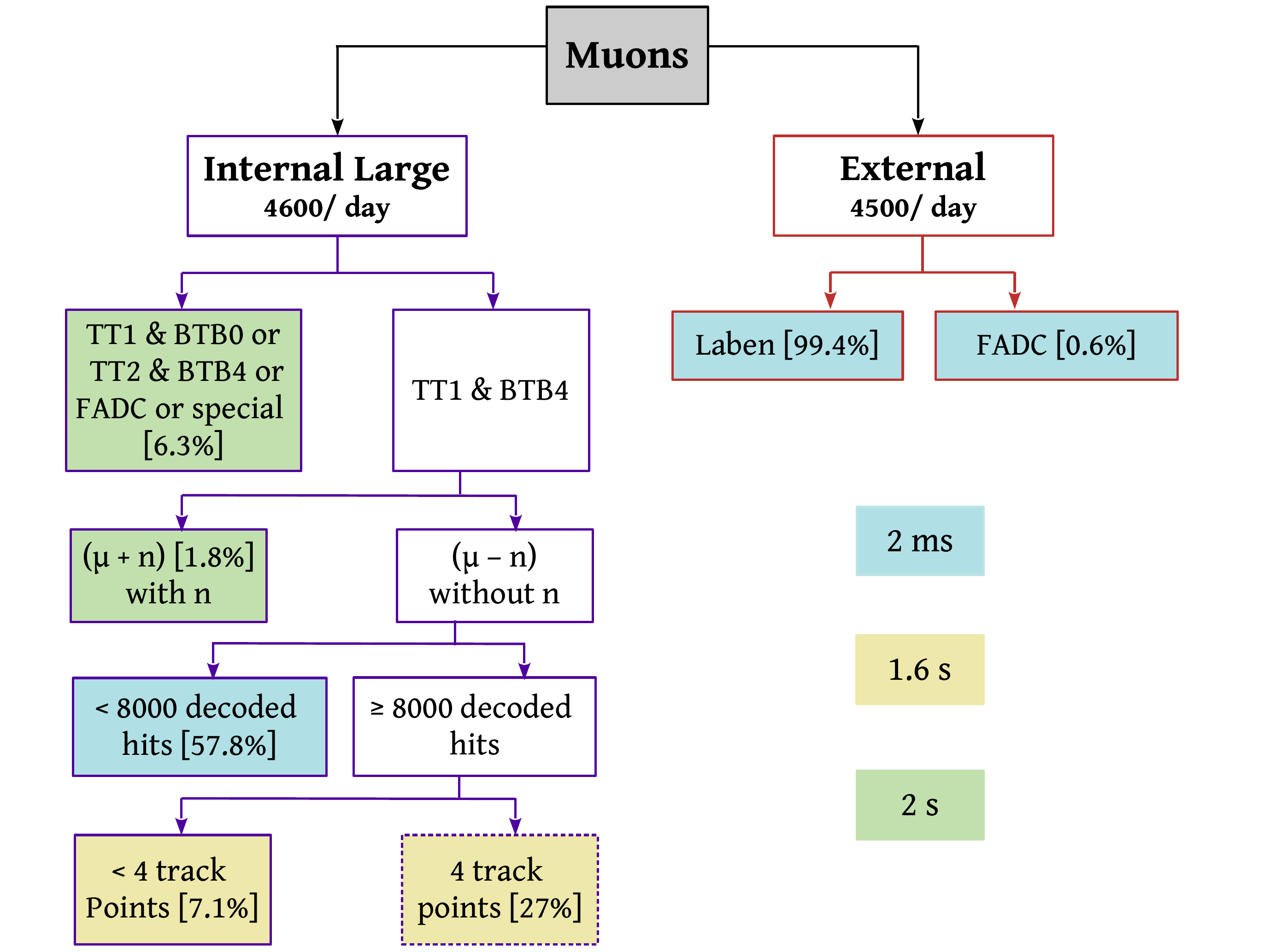}
        	\caption{Scheme of different types of vetoes applied after various muon categories. The relative fraction of each muon category is given by the numbers in brackets. Different colours represent different durations of the time veto. The dashed line around the box at the bottom of the plot represents the only muon category when the cylindrical veto around the muon track is applied. For other categories, the whole detector is vetoed.}
        	\label{fig:vetoes_scheme}
        \end{figure}
    %
 
        \subsection{Muon vetoes}
        \label{subsec:vetoes}
 
        Muons and related spallation products represent an important background for geoneutrino measurement, as described in Sec.~\ref{subsec:cosmogenic}. In the data selection, we first remove all categories of detected muons, as defined in Sec.~\ref{subsec:muon}. After different types of detected muons, different kinds of vetoes are applied. They are described in this section and are schematically shown in Fig.~\ref{fig:vetoes_scheme}.

        \paragraph{Veto after external muons}
 
 Among different kinds of spallation products of external muons, only cosmogenic neutrons can penetrate inside the scintillator and thus present a background for geoneutrino analysis. The neutron capture time in Borexino is $(254.5 \pm1.8)$\,$\mu$s~\cite{Bellini:2011yd}. Therefore, a 2\,ms veto, i.e. about 8 times the neutron capture time, is applied after all external muons detected by the OD.
 
 \paragraph{Veto after internal muons}
  
 For internal muons, in addition to fast neutrons, $^{9}$Li, $^{8}$He, and $^{12}$B isotopes (Table~\ref{tab:isotopes_muon}) are potential background sources for IBD selection as well. In the past analyses, a 2\,s veto of the whole detector has been applied after all categories of internal strict and special muons. Since 2\,s is nearly 8 times the lifetime of the longest-lived isotope ($^{9}$Li), this background was effectively eliminated for the price of about 10-11\% loss of exposure. In this analysis, we reduce the exposure loss to 2.2\% by introducing different categories of vetoes.

   \begin{itemize}
     \item {\it 2\,s veto of the whole detector}
     
  We apply a conservative 2\,s veto of the whole detector after internal strict muons, special muons, and FADC muons that did not trigger the OD (not tagged by MTF flag). These represent 6.3\% of all internal large muon category, which includes some noise events as well (Sec.~\ref{subsec:muon}). In addition, we apply this kind of veto after the so-called ($\mu$ + n) muons, i.e. those internal muons that triggered the OD ({\it TT1 \& BTB4}, Sec.~\ref{subsec:data}) and were followed by at least one neutron observed in the following event of {\it TT128}. These muons represent only 1.8\% of all internal muons and have a higher probability to produce $^{9}$Li events with a detectable decay neutron. The ($\mu$ + n) muon sample was used to characterize the veto parameters after the independent sample of ($\mu$ - n) muons, i.e. MTF/BTB4 internal muons that are not followed by any neutron, as described below.

  \begin{figure} [t]
        	\centering
        	\includegraphics[width = 0.5\textwidth]{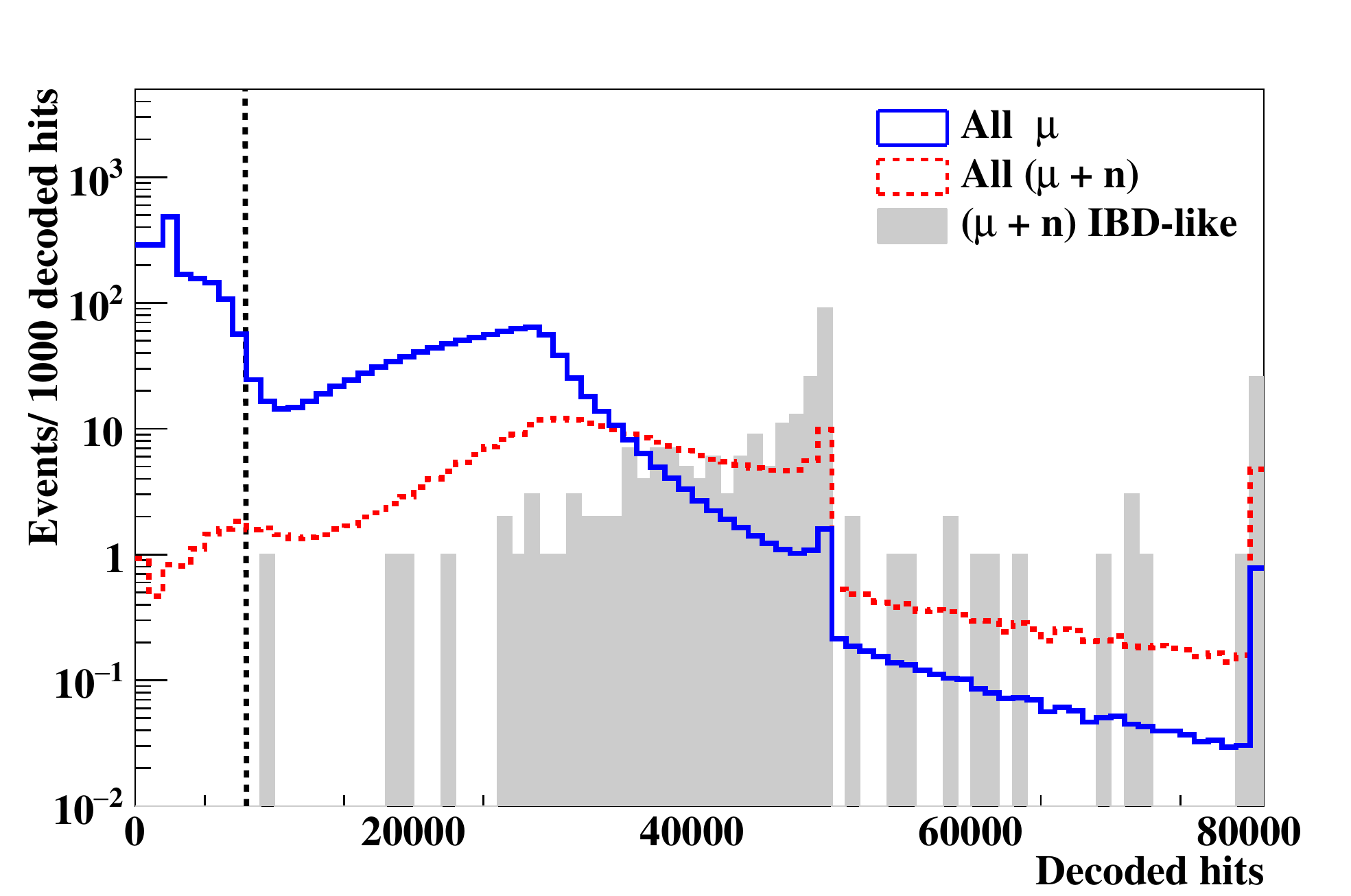}
        	\caption{Comparison of the decoded hits spectra of all the muons (solid blue line), all ($\mu$ + n) muons (dotted red line), and those ($\mu$ + n) muons, which produce IBD-like candidates due to hadronic background within 2\,s after muons (filled grey area). The dashed vertical line shows 8000 decoded hits threshold used in the definition of the veto duration. Spectra of all and ($\mu$ + n) muons are normalized to the number of ($\mu$ + n) muons producing IBD-like candidates.}
        	\label{fig:vetoes_dechits}
        \end{figure}
\item {\it 2\,ms veto of the whole detector}

 It was observed that ($\mu$ + n) muons that also produce IBD-like hadronic background, always have more than 8000 decoded hits (Fig.~\ref{fig:vetoes_dechits}). In the following text, we apply the notation ($\mu$ + n)$_{\ge 8000}$ for this and similar muon types.
 Muons producing less than 8000 decoded hits are typically not crossing the scintillator and are passing only through the buffer, where neutrons cannot be effectively detected due to low light yield.  Thus, ($\mu$ - n)$_{<8000}$ have little chance to produce $^{9}$Li events with a detectable decay neutron. We apply a conservative 2\,ms dead time  after them, suppressing potential fast neutrons with negligible exposure loss. These muons represent 57.8\% of all internal muons.

\item {\it 1.6\,s veto of the whole detector}

Muons passing through the scintillator, i.e. with $\ge$8000 decoded hits, have high probability that neutron from a potential $^{9}$Li decay would be detected. Thus, for these muons, a 2\,ms veto is not sufficient.
For the ($\mu$ - n)$_{\ge 8000}$ muon category that represents 34.1\% of all internal muons, we apply a veto reduced from 2\,s to 1.6\,s, after which only 0.2\% of $^{9}$Li candidates survive. In addition, only 10\% of the observed $^{9}$Li background is produced due to ($\mu$ - n)$_{\ge 8000}$ muons, as it will be explained in detail in Sec.~\ref{subsec:cosmogenic_est}.

Muons in Borexino can be tracked based on their reconstructed entry and exit points in the OD and ID~\cite{Bellini:2011yd}. We consider the muon track to be reliable only when all the four track points are reconstructed. When this is not the case, the 1.6\,s veto is applied to the whole detector, for 7.1\% of all internal muons.

 \begin{figure} [t]
        	\centering    
          \subfigure[]{ \includegraphics[width = 0.27\textwidth]{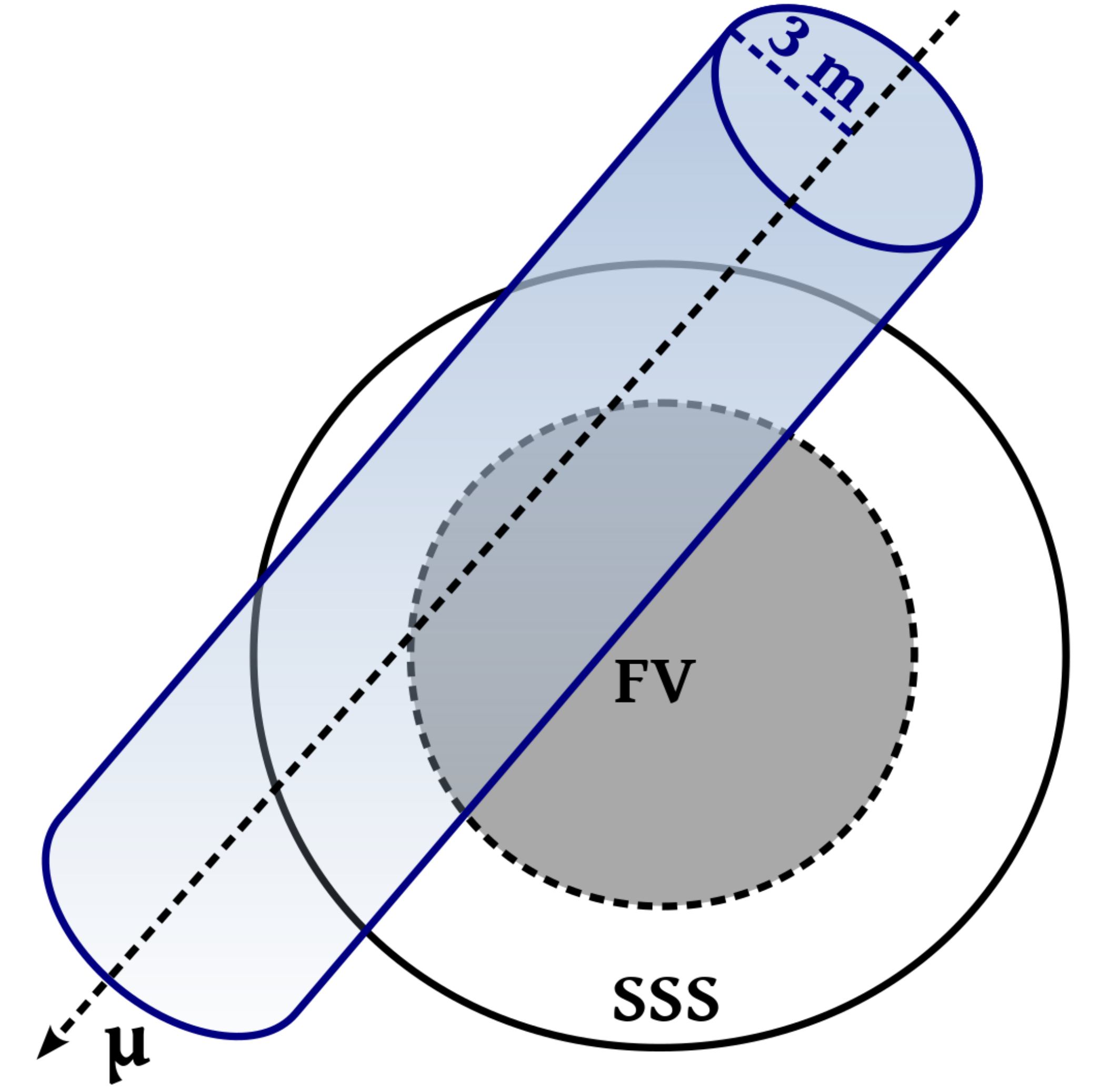}
          \label{fig:cylinder}}
          \subfigure[]{\includegraphics[width =0.5\textwidth]{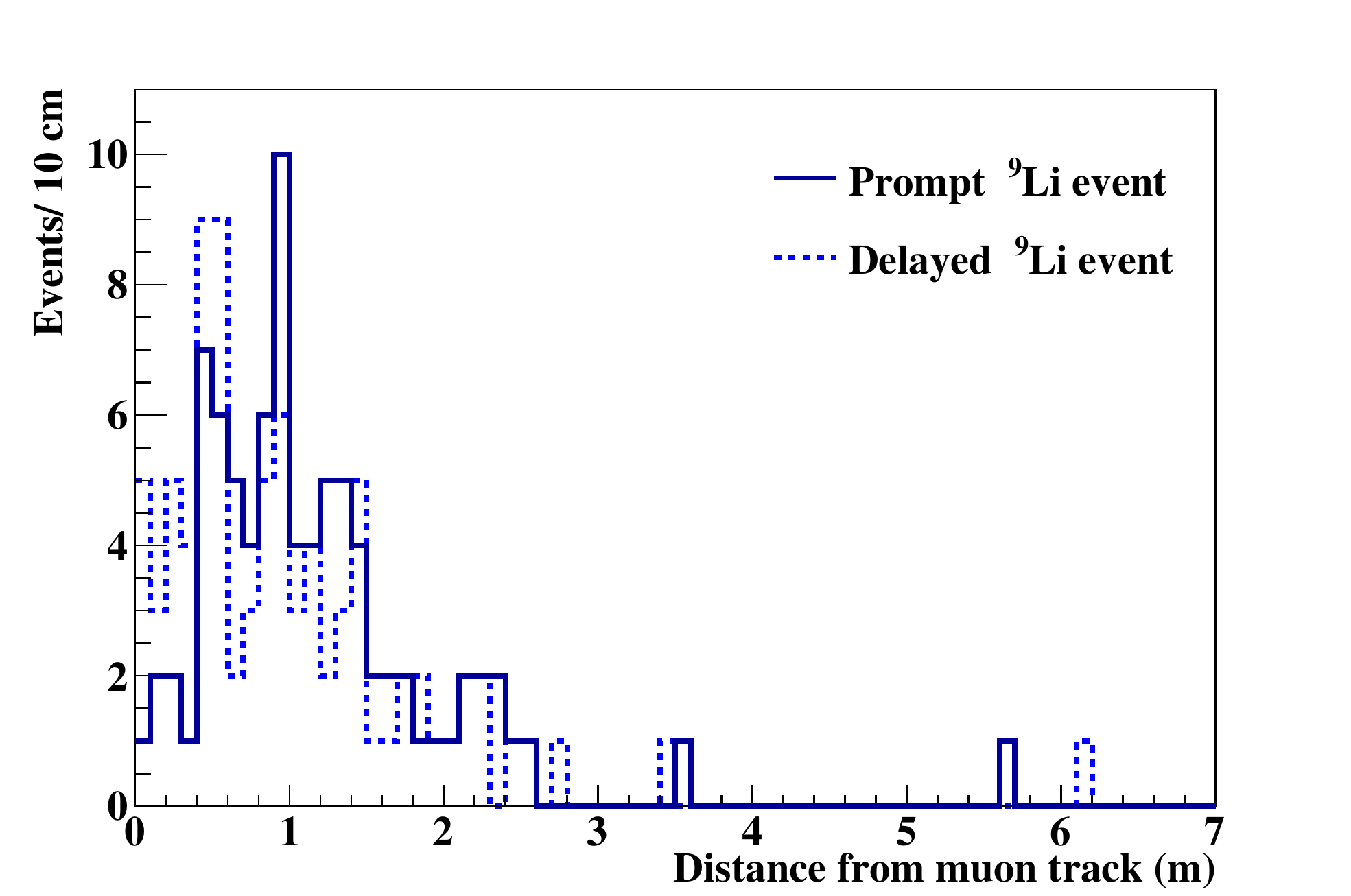}
          \label{fig:dist}}
            \caption{Top (a): A schematic representation of the cylindrical veto applied around the muon track for 1.6\,s after the ($\mu$ - n)$_{\ge 8000}$ muons with reliable track reconstruction. Bottom (b): Distributions of the distances of 85 $^9$Li IBD-like events from reliably reconstructed tracks of ($\mu$ + n) muons: solid and dashed lines represent the prompt and delayed candidates, respectively.}
         \end{figure}

    \begin{figure} [t]
    	\centering	\includegraphics[width=0.49\textwidth]{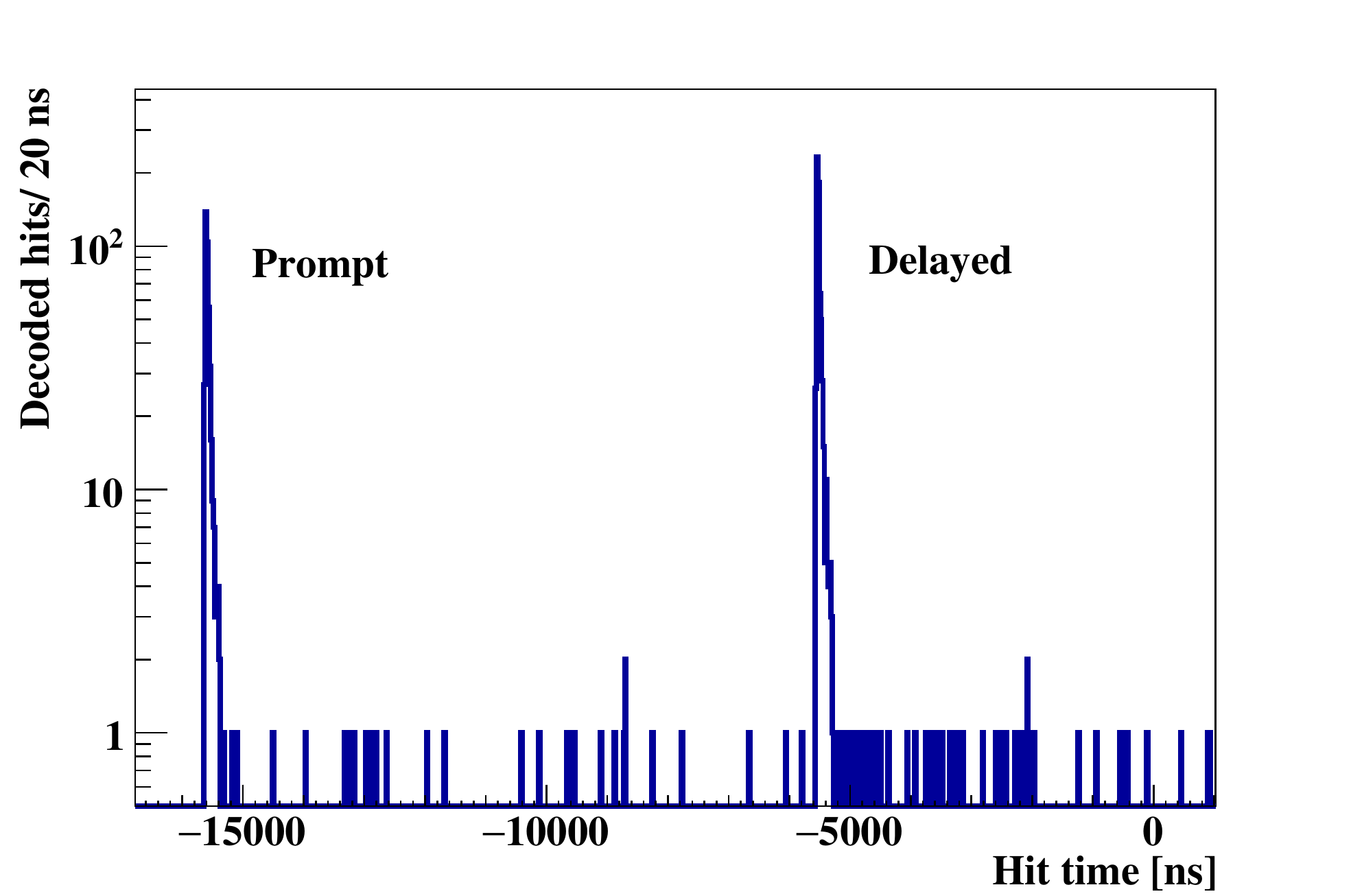}
        \caption{Example of an IBD candidate, when both the prompt and the delayed are individual clusters in a single event of 16\,$\mu$s DAQ time window. The negative hit times are expressed with respect to the trigger reference time.}
  	\label{fig:dcl_fig}
   \end{figure}

\item {\it 1.6\,s cylindrical veto}

The application of a cylindrical veto around the muon track, as schematically shown in Fig.~\ref{fig:cylinder}, instead of vetoing the whole detector, can further increase the exposure for geoneutrino analysis. This kind of veto is applied to ($\mu$ - n)$_{\ge 8000}$ muons for which all the four muon track points are reconstructed.   

The radius of the cylindrical veto is set by studying the lateral distance between the muon and IBD-like prompt and delayed observed within 2\,s after the passage of a ($\mu$ + n) muon with four track points (Fig.~\ref{fig:dist}). Within the observed statistics, this distance is very similar for the prompt and the delayed. 97.7$\%$ of the prompts (on which the DFV cut is applied) lie within a 3\,m radius from the muon track in the IBD selection (Sec.~\ref{subsec:DFV}). Since the lateral distribution of the muon daughters is expected to be the same for ($\mu$ - n) muons as well, a cylindrical veto of 1.6\,s duration with 3\,m radius is applied for ($\mu$ - n)$_{\ge 8000}$ muons, which constitute 27\% of all internal muons. 

The resulting exposure loss of 2.2\% after all muon vetos was calculated using a Monte Carlo simulation. In total, 74 million point-like events were generated homogeneously in the dynamical fiducial volume for this study (Sec.~\ref{subsec:DFV}), following the changing shape of the IV (Sec.~\ref{subsec:IV}). After considering the GPS times of all the muons and the track geometry reconstructed for ($\mu$ - n)$_{\ge 8000}$ muons, the relative exposure loss was calculated as the fraction of the events removed by all vetoes.
 \end{itemize}

        \subsection{Time coincidence}
        \label{subsec:dt}

    The coincidence time window between the prompt and the delayed ($dt$) is an important background-suppressing cut. It is implemented based on the neutron capture time that was measured during the calibration campaign with the  $^{241}$Am-$^{9}$Be neutron source to be (254.5 $\pm$ 1.8)\,$\mu$s~\cite{Bellini:2011yd}.
    Considering the 16\,$\mu$s DAQ window, followed by an electronics dead time of 2-3\,$\mu$s (Sec.~\ref{subsec:data}), one has to consider separately the case when the prompt and delayed are either two separate events/triggers with single cluster each (similar to the event in top left of Fig.~\ref{fig:event_single}), or are represented by two clusters in a single event, as shown in Fig.~\ref{fig:dcl_fig}.

  \begin{figure}[t]
    	\includegraphics[width = 0.47\textwidth]{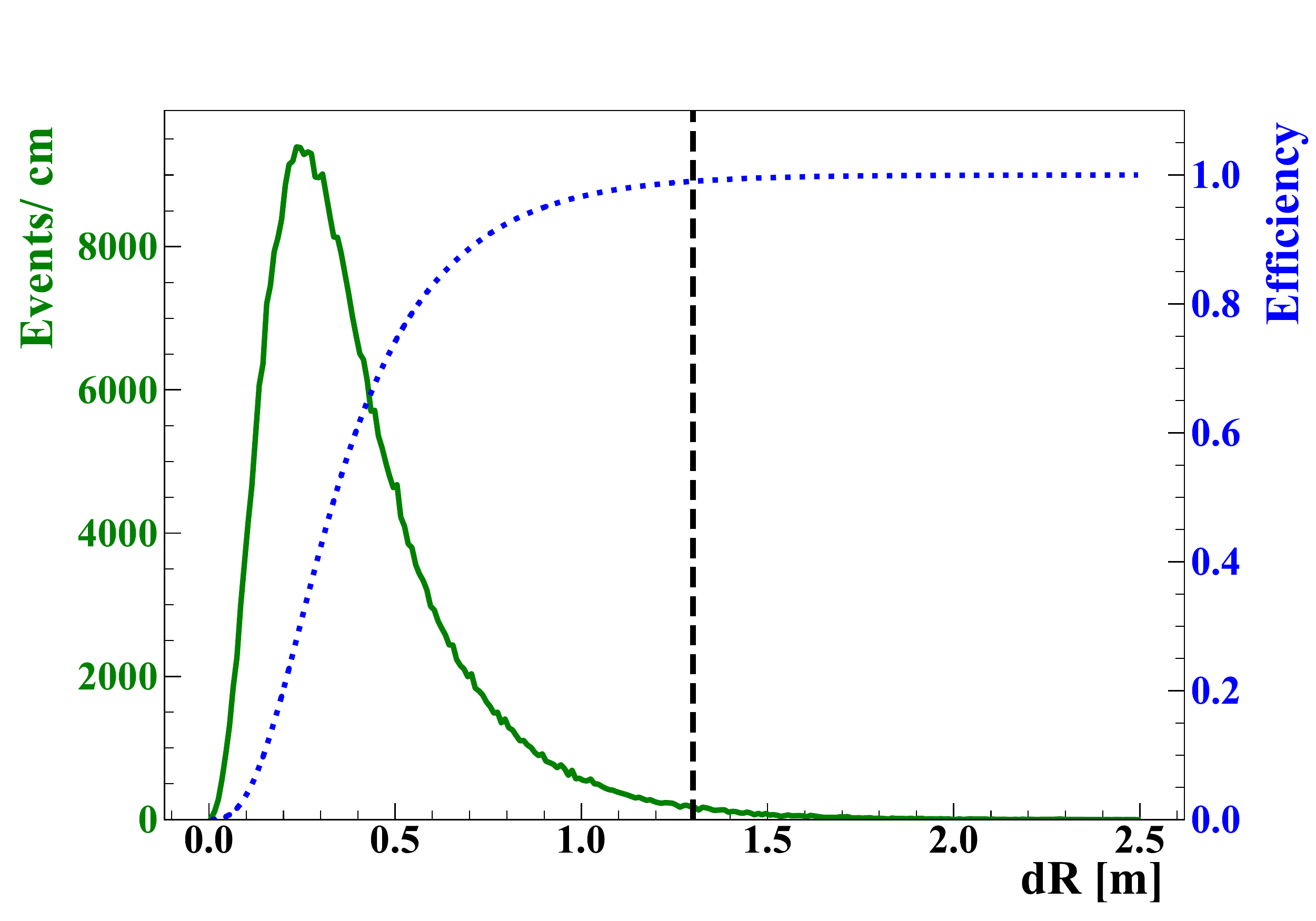}
    	\caption{Distribution of the prompt-delayed reconstructed distance $dR$ (solid green line) shown for the geoneutrino MC sample and the corresponding efficiency (blue dottted line). The vertical dashed line shows the optimized $dR$ cut at 1.3\,m.}
    	\label{fig:dR_MC}
    \end{figure}
            
    \paragraph{Single cluster events}
             
     For this category of IBD candidates, the coincidence time window is between $dt_{\mathrm{min}}$ = 20\,$\mu$s and $dt_{\mathrm{max}}$ = 1280\,$\mu$s. The lower threshold guarantees that the delayed signal can trigger after the dead-time of the prompt trigger. The $dt_{\mathrm{max}}$
     corresponds to about five times the measured neutron capture time. This time window covers 91.8\% of all IBD interactions. 
            
    \paragraph{Double cluster events}
            
    This category of IBD candidates is included in the present analysis for the first time. In this case, the coincidence time window is between $dt_{\mathrm{min}}$ = 2.5\,$\mu$s and $dt_{\mathrm{max}}$ = 12.5\,$\mu$s. The inclusion of double cluster events led to a 3.8\% increase in the IBD tagging efficiency.

    The lower threshold was optimized by studying the cluster duration of prompts from the MC of reactor antineutrinos (Sec.~\ref{sec:mc}), which spectrum extends up to about 10\,MeV. It guarantees that even for the highest energy prompt, after the $dt_{\mathrm{min}}$, there is no light that could alter the hit pattern of the delayed. 
        
    The $dt_{\mathrm{max}}$ was set considering the variable position of the trigger-generating cluster (prompt) inside the DAQ window, that occurred during the analyzed period due to the changes in the trigger system. 
    At the same time, it guarantees that the delayed can always have cluster duration of up to 2.5 $\mu$s before the end of the DAQ window.
    \begin{figure}[t]
    	\centering
    	\includegraphics[width = 0.47\textwidth]{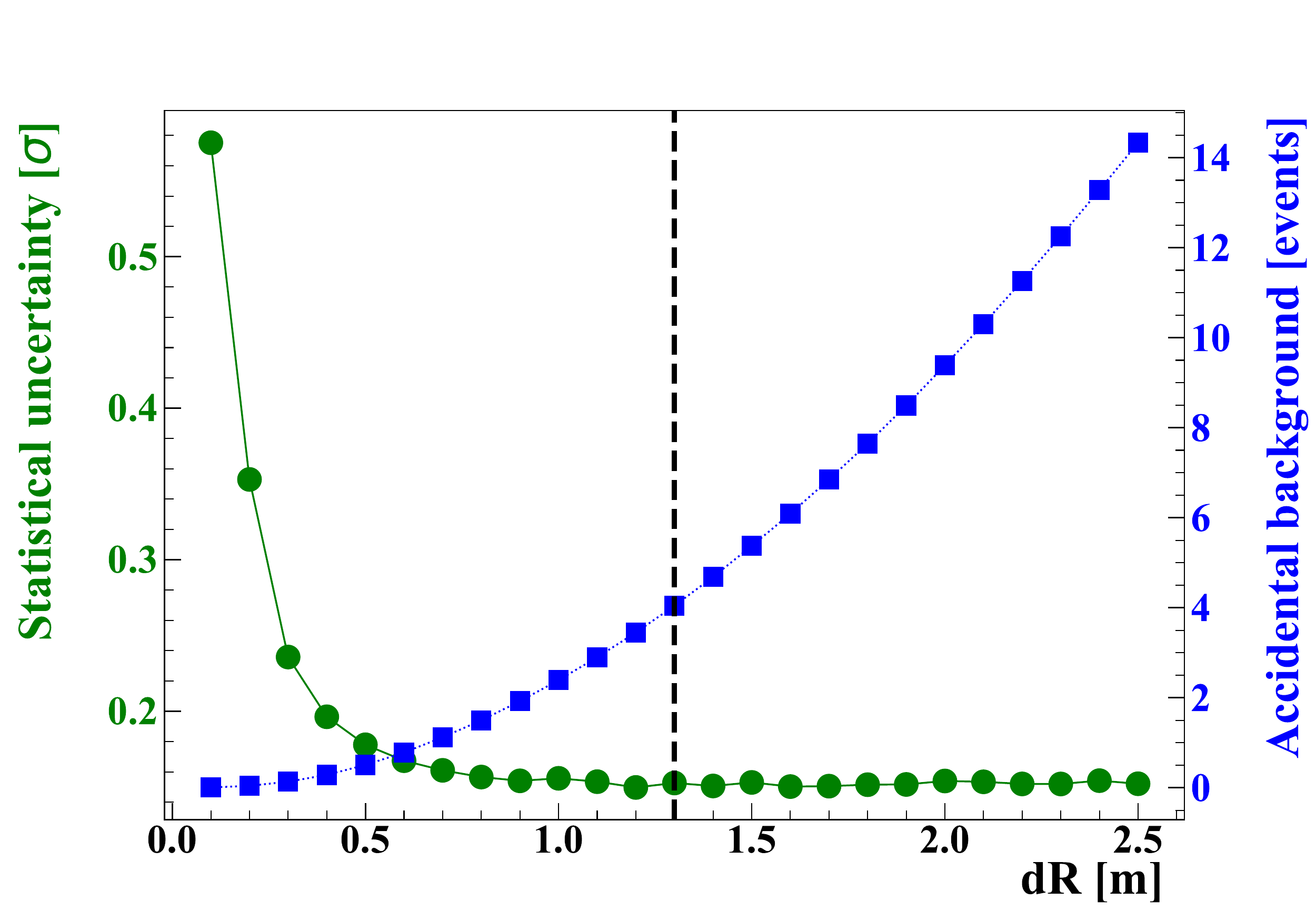}
    	\caption{The expected statistical uncertainty of the geoneutrino measurement using the MC study (filled green circles) and the accidental background (filled blue squares) for different values of $dR$ cut, while the other cuts were set to the optmized values. The vertical dashed line shows the selected value for the $dR$ cut at 1.3\,m.}
    	\label{fig:sens_dR}
    \end{figure}
    %
        \subsection{Space correlation}
        \label{subsec:dR}

     Similar to $dt$, the spatial distance between the prompt and the delayed ($dR$) is also an important background-suppressing cut. In the previous analyses, $dR$ = 1\,m was used. The reconstructed distance between the prompt and the delayed is larger with respect to the distance between their respective points of production. This is caused predominantly by these effects:
     \begin{itemize}
         \item {\it Interaction of gammas:} the gammas, from the positron annihilation and the neutron capture, interact in the LS mostly through several Compton scatterings. Thus, their interaction is intrinsically not point-like and the barycenter of the cloud of these Compton electrons is not identical to the point of the generation of the gammas.
         \item  {\it Position reconstruction:} the position reconstruction (10\,cm at 1\,MeV) further smears the reconstructed positions. 
     \end{itemize}

    In the optimization of the $dR$ cut, two main aspects have to be considered. On one hand, in the prompt-delayed reconstructed distance, shown in Fig.~\ref{fig:dR_MC} for the MC sample of geoneutrinos (Sec.~\ref{sec:mc}), the efficiency is quickly dropping below 1\,m. On the other hand, with the increasing $dR$, the accidental background is also strongly increasing (Fig.~\ref{fig:sens_dR}). In order to find the optimized value, we have generated thousands of MC pseudo-experiments as described in Sec.~\ref{sec:sensitivity}. The cuts were set to optimized values, while the $dR$ cut was varied. The $dR$ cut was then set to 1.3\,m, within the interval, where the variation in the expected statistical uncertainty of the geoneutrino measurement is small, as shown in Fig.~\ref{fig:sens_dR}. We note that the procedure of the so-called ``sensitivity study'' used to estimate the expected statistical uncertainty is described in Sec.~\ref{subsec:senstool}.

\begin{figure} [t]
    \centering
    \includegraphics[width = 0.5\textwidth]{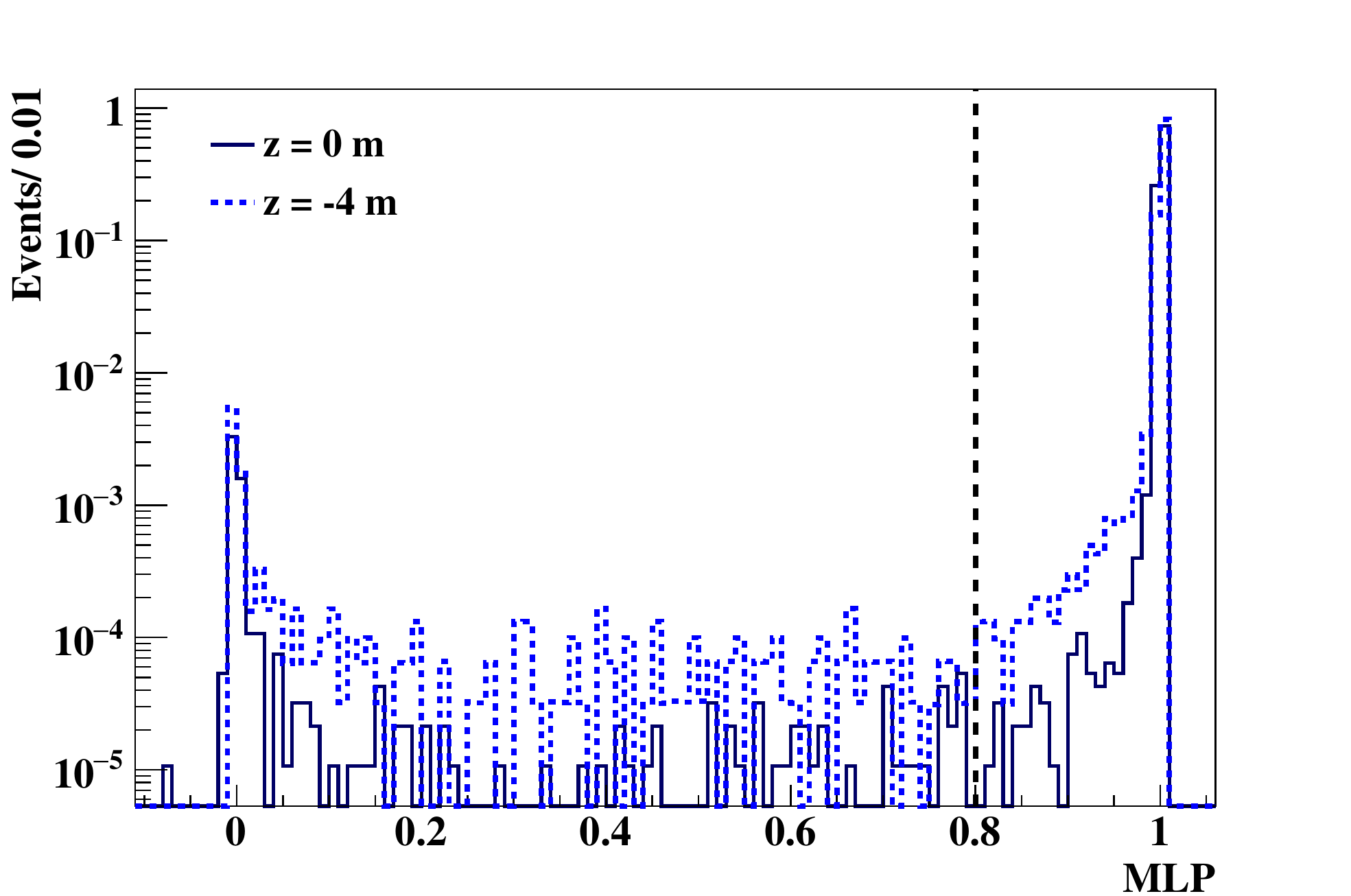}
    \caption{Distribution of the MLP $\alpha$/$\beta$ discriminator (Sec.~\ref{subsec:ab}) for the gammas from the capture of neutrons from $^{241}$Am-$^{9}$Be calibration source placed at the detector center (x,y,z) = (0,0,0)\,m, as well as at (0,0,-4)\,m, where the observed distribution is the broadest. The dashed line shows the  threshold set at 0.8 for the delayed in IBD search.}
    \label{fig:ambe_mlp}
\end{figure}

        \subsection{Pulse shape discrimination}
        \label{subsec:pulse_shape}
             
In the previous geoneutrino analyses, a Gatti cut $G < 0.015$ was applied on the delayed for efficient rejection of $\alpha$-like events from radon-correlated background (Sec.~\ref{subsec:radon}). The cut value was set using gammas from neutron-captures from the $^{241}$Am-$^{9}$Be calibration source and cross checked with the MC. In this analysis, an MLP cut $>$ 0.8 was applied to the delayed using the better $\alpha$/$\beta$ discrimination power of the MLP (Sec.~\ref{subsec:ab}) when compared to the Gatti. The cut threshold was chosen based on the $^{241}$Am-$^{9}$Be calibration data as shown in Fig.~\ref{fig:ambe_mlp}. It was found that only a 5.4 (8.1)$\cdot$10$^{-3}$ fraction of the neutron-capture $\gamma$s remains below the MLP threshold of 0.8 for the source position in the detector's center (close to IV border).”

            \begin{figure}[t]
     \centering  
    \includegraphics[width = 0.5\textwidth]{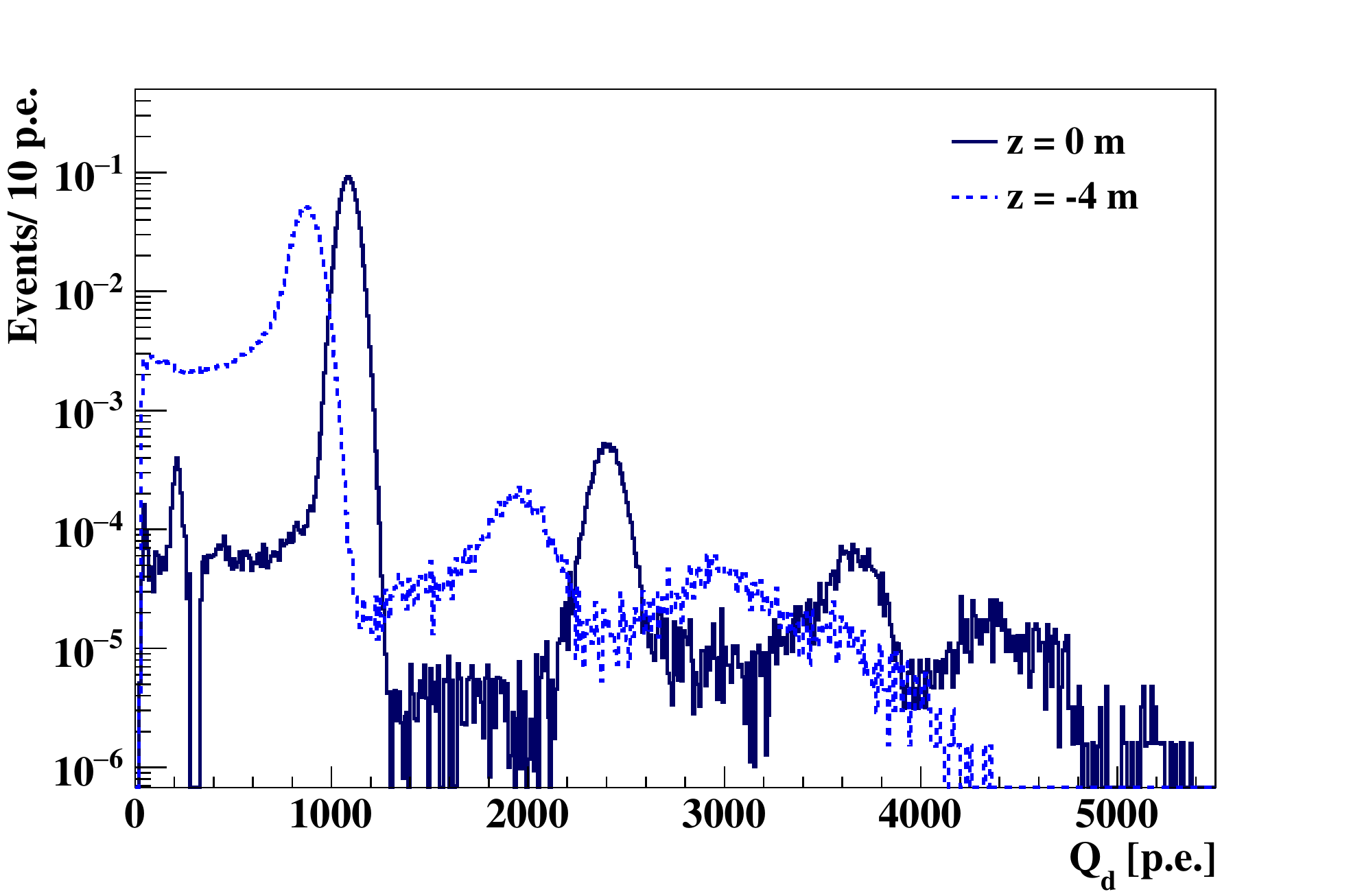}
        \caption{The $N_{pe}$ spectrum of delayed candidates from the $^{241}$Am-$^{9}$Be calibration source placed at (x,y,z) = (0,0,-4)\,m inside the detector (dashed line) compared to the spectrum for the source placed at the detector's center (solid line). The two peaks from the left correspond to 2.2\,MeV and 4.95\,MeV gammas from neutron captures on $^1$H and $^{12}$C nuclei, respectively, while the higher energy peaks are from neutron captures on stainless steel nuclei (Fe, Ni, Cr) used in the source construction.
        All gamma peaks in the off-center spectrum are shifted to lower energies and develop tails due to the partial energy deposits in the buffer. The spectra are normalised to one.}
       \label{fig:AmBeSouthDelayed} 
    \end{figure}  
        \subsection{Energy cuts}
        \label{subsec:energy}

        Energy cuts were applied to the the prompt and delayed to aid the identification of IBD signals. The analysis was performed using the charge energy estimator $N_{pe}$ (Sec.~\ref{subsec:data}). The intervals in the respective charges $Q_{\rm{p}}$ and $Q_{\rm{d}}$ are optimized as explained in this Section.

 \paragraph{Charge of prompt signal}
        The energy spectrum of the prompt $E_{\rm {p}}$ starts at $\sim$1\,MeV, which corresponds only to the two 511\,keV annihilation gammas (Eq.~\ref{eq:Epro}). Therefore, the threshold on the prompt charge was set to $Q_{\rm{p}}^{\rm{min}}$ = 408\,p.e. This threshold value corresponds to approximately 0.8\,MeV and remains unchanged from previous analyses. No upper limit is set on the charge of the prompt candidate.

  \paragraph{Charge of delayed}
            
The delayed signal can be either due to a 2.22\,MeV ($n$-capture on $^{1}$H), or a 4.95\,MeV gamma ($n$-capture on $^{12}$C) with about 1.1\% probability, as described in Sec.~\ref{sec:IBD}. The corresponding values in the $N_{pe}$ variable are 1090\,p.e. and 2400\,p.e., respectively, as measured in the detector center and shown in Fig.~\ref{fig:AmBeCenterDelayed}. However, at large radii, gammas can partially deposit their energy in the buffer, which decreases the visible energy. Consequently, the $\gamma$-peak develops a low-energy tail and even the peak position can shift to lower values (Fig.~\ref{fig:AmBeSouthDelayed}).

In the last geoneutrino analysis~\cite{Agostini:2015cba}, $Q_{\rm{d}}^{\rm{min}}$ was set to 860\,p.e., a conservatively large value because of the radon-correlated background, particularly due to $^{214}$Po($\alpha$ + $\gamma$) decays. This was discussed in Sec.~\ref{subsec:radon}. In this analysis, $Q_{\rm{d}}^{\rm{min}}$ was decreased to 700\,p.e. based on the improved performance of $\alpha$/$\beta$ separation with MLP (Sec.~\ref{subsec:pulse_shape}), which improved the ability to suppress $^{214}$Po($\alpha$ + $\gamma$) decays. This cut was applied to all data with the exception of the water-extraction period, which had an increased radon contamination. In this case the $Q_{\rm{d}}^{\rm{min}}$ 860\,p.e. was retained. The $Q_{\rm{d}}^{\rm{min}}$ cut was not decreased below 700\,p.e. for the following reasons:  
\begin{enumerate}
    \item The $N_{pe}$ spectrum of accidental background increases at lower energies, as shown in Fig.~\ref{fig:acc_charge}. 
    \item The end point of the $\alpha$ peak from the main $^{214}$Po decay in the radon-correlated background is $\sim$600\,p.e., as described in Sec.~\ref{subsec:radon_est}.
\end{enumerate}

            \noindent In this analysis, we include the neutron captures on $^{12}$C (4.95\,MeV) as well. Consequently, $Q_{\rm{d}}^{\rm{max}}$ was increased from 1300\,p.e. (2.6\,MeV) to 3000\,p.e. ($\approx$5.5\,MeV). 

\begin{figure}[t]
\vspace{-2mm}
	\centering  
   \includegraphics[width =0.46\textwidth]{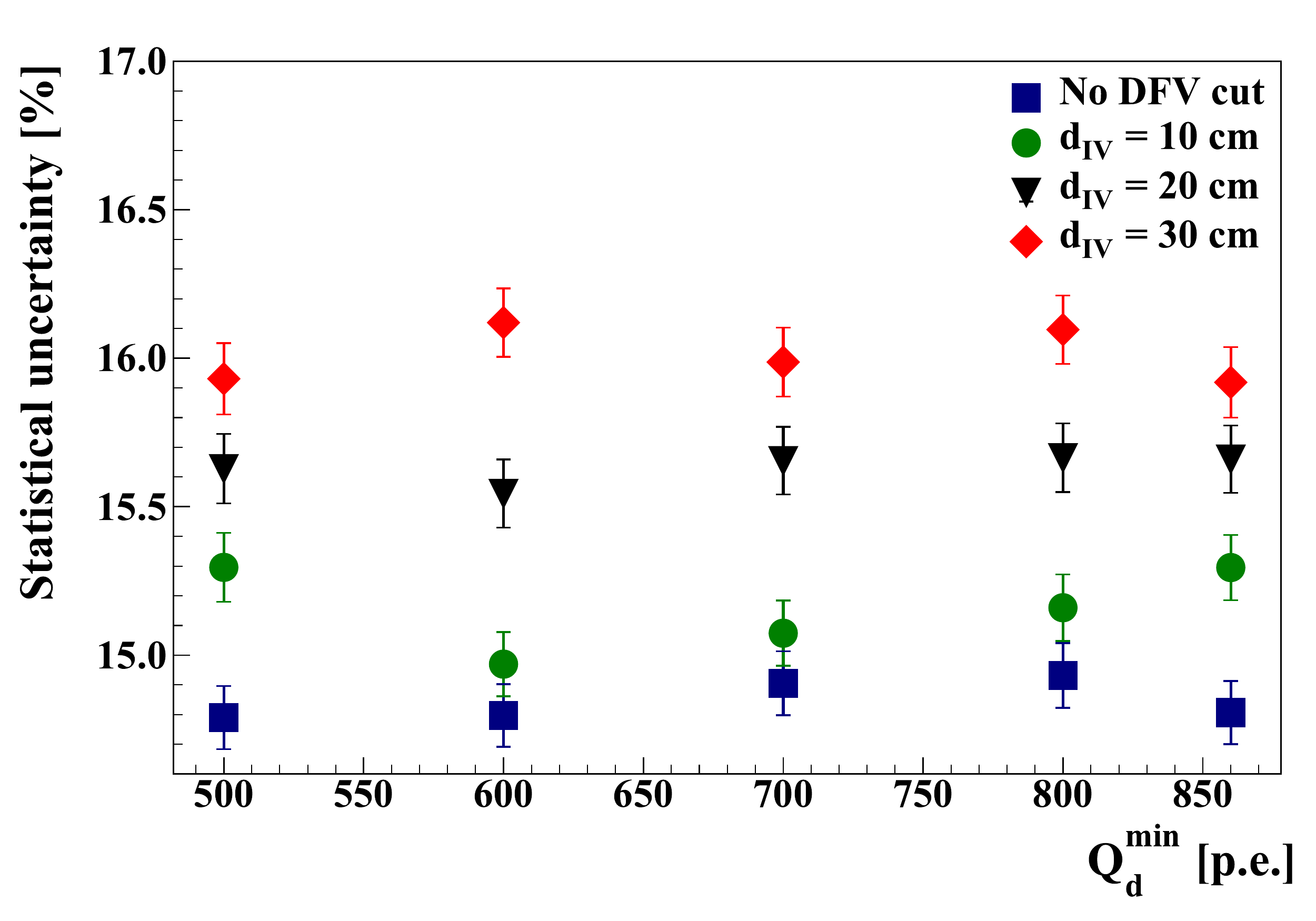}
   \caption{The expected precision of the geoneutrino measurement for different DFV cuts. For a given DFV cut, the choice of $Q_{\rm d}^{\rm{min}}$, shown on the $x$-axis, does not strongly influence the geoneutrino expected precision. Also note, that for $d_{\mathrm {IV}} = 10$\,cm (final choice) and no DFV cut, the sensitivity to geoneutrinos is nearly the same.}
   \label{fig:unc_dfv}
\end{figure}

    \subsection{Dynamical fiducial volume cut}
    \label{subsec:DFV}

    The shape of the Borexino IV, that is changing due to the presence of a small leak, can be periodically reconstructed by using the data (Sec.~\ref{subsec:IV}). A DFV cut, {\it i.e.} a requirement of some minimal distance of the prompt from the IV, $d_{\mathrm {IV}}$, is applied along the reconstructed IV shape. In the previous analysis~\cite{Agostini:2015cba} a conservative cut of $d_{\mathrm {IV}} = 30$\,cm has been applied to account for the uncertainty of the IV shape reconstruction and the potential background coincidences near the IV. In this analysis, the DFV was increased by using $d_{\mathrm {IV}} = 10$\,cm, which leads to a 15.8\% relative increase in exposure. This choice is justified below.

  The geoneutrino sensitivity studies (Sec.~\ref{sec:sensitivity}) were performed for different combinations of the DFV cut and the $Q_{\rm {min}}^{\rm{d}}$, as shown in Fig.~\ref{fig:unc_dfv}. The choice of $Q_{\rm d}^{\rm{min}}$ does not have a big impact on the expected precision, for a given DFV cut. Order of 5\% improvement in the statistical uncertainty of the geoneutrino measurement is expected when the cut is lowered to $d_{\mathrm {IV}}$ = 10\,cm, while there is no further improvement when no DFV cut is applied. A $d_{\mathrm {IV}}$ = 10\,cm DFV cut is also sufficient to account for the precision of the IV shape reconstruction (Sec.~\ref{subsec:IV}). In addition, we have verified that no excess of the IBD candidates is observed at large radii (close to the IV), as it will be shown in Sec.~\ref{subsec:golden_candidates}.

\begin{table}[t]
    	\centering
    	\caption{\label{tab:sel_cuts} Summary of the optimized selection cuts for IBD candidates: charge cut on the prompt, $Q_{\mathrm{p}}$ and delayed, $Q_{\mathrm{d}}$, time and space correlation $dt$ and $dR$, respectively, distance to the IV, $d_{\mathrm {IV}}$, MLP $\alpha/\beta$ Particle IDentification parameter cut on delayed, and multiplicity cut. The scheme indicating the duration and geometry of the different muon vetoes is shown in Fig.~\ref{fig:vetoes_scheme}. }  \vskip 2pt
    \begin{tabular*}{\columnwidth}{l @{\hskip 40pt} l}
    		\hline
    		\hline
    		Cut   &  Condition \Tstrut\Bstrut \\
        \hline
       	$Q_{\mathrm{p}}$  & $>$ 408\,p.e. \Tstrut \\ 
		$Q_{\mathrm{d}}$  & (700 - 3000)\,p.e. \\ 
		                  & (860 - 3000)\,p.e. (WE period) \\ [4pt]
		$d$t             & double cluster: (2.5 - 12.5)\,$\mu$s \\ 
		 & single cluster: (20 - 1280)\,$\mu$s \\ [4pt]
		$dR$              & 1.3\,m \\ [4pt]
		Muon veto         & 2\,s or 1.6\,s or 2\,ms    (internal $\mu$)  \\ 
			              & 2\,ms (external $\mu$) \\ [4pt]
		$d_{\mathrm {IV}}$  &  10\,cm (prompt) \\ [4pt]
	    PID ($\alpha$/$\beta$)  &  MLP$_{\rm d}$ $>$ 0.8 \\ [4pt]
		Multiplicity      &  no $N_{pe} >$ 400\,p.e. event \\
			              & $\pm 2$\,ms around prompt/delayed  \Bstrut \\ 
        \hline
        \hline
	\end{tabular*}
\end{table}
 \subsection{Multiplicity cut}
 \label{subsec:multiplicity}

The {\it multiplicity cut} requires that no additional ``high-energy" ($N_{pe} > 400$\,p.e.) event is observed within $\pm 2$\,ms around either the prompt or the delayed candidate. This cut is designed to suppress the background from undetected muons, for example, neutron-neutron or buffer muon-neutron pairs. This justifies the selected time window which is nearly 8 times the neutron capture time. The charge cut was lowered to account for those neutrons that are depositing their energy partially in the buffer. Thanks to  a high radiopurity of the LS, the corresponding exposure loss due to accidental coincidences of IBD candidates with $N_{pe} > 400$\,p.e. events within 2\,ms is of the order of 0.01\%, which is negligible.

        \subsection{Summary of the selection cuts}
        \label{subsec:cuts}
        
    The summary of all the optimized selection cuts is listed in Table~\ref{tab:sel_cuts}.
        %

\section{MONTE CARLO OF SIGNAL AND BACKGROUNDS}
 \label{sec:mc}

The spectral fit of the prompt charge $Q_p$ (Sec.~\ref{subsec:nutshell}), relies extensively on the use of MC-constructed probability distribution functions (PDFs) which represent the shapes of geoneutrino signal and, with the exception of accidental coincidences (Sec.~\ref{subsec:acc_est}), all backgrounds. The construction of these PDFs is described in Sec.~\ref{subsec:MCspectra}. The Geant4 based MC of the Borexino detector was tuned on independent data acquired during an extensive calibration campaign with radioactive sources~\cite{Back:2012awa} and is described in detail in~\cite{Agostini:2017aaa}. For the antineutrino analysis, the calibration with $^{241}$Am-$^{9}$Be neutron source is of particular importance, since the delayed IBD (Sec.~\ref{sec:IBD}) signal is represented by a neutron. The comparison of the neutron spectra from the $^{241}$Am-$^{9}$Be calibration source at the detector center and at (0, 0, -4)\,m position inside the detector is shown in Fig.~\ref{fig:AmBeSouthDelayed}.

 \begin{figure*}[h]
     \centering  
    \subfigure[]{\includegraphics[width =0.49\textwidth]{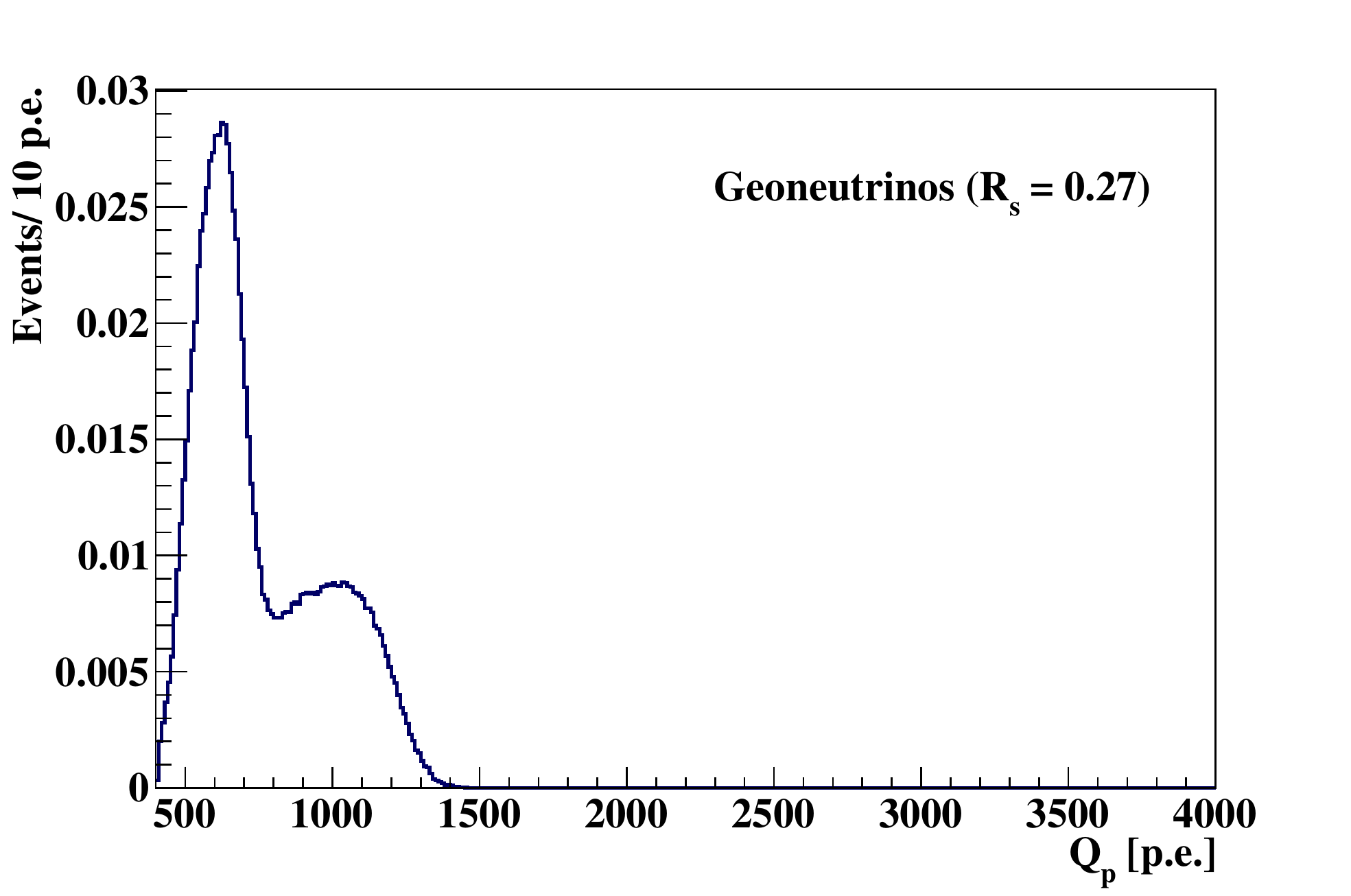}
    \label{fig:PDF_geo}}
     \subfigure[]{\includegraphics[width =0.49\textwidth]{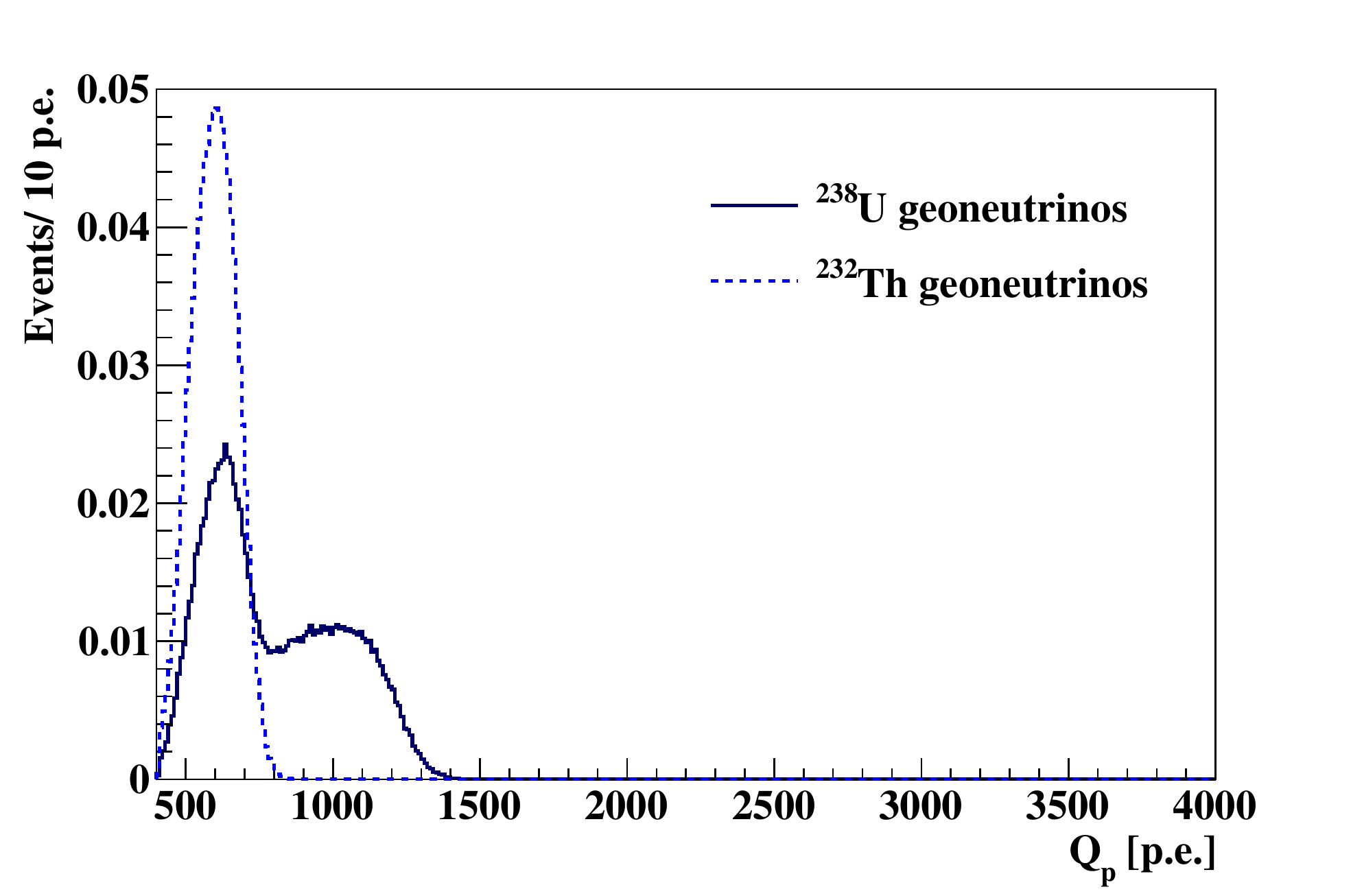}
     \label{fig:PDF_UTh}}
     \subfigure[]{\includegraphics[width =0.49\textwidth]{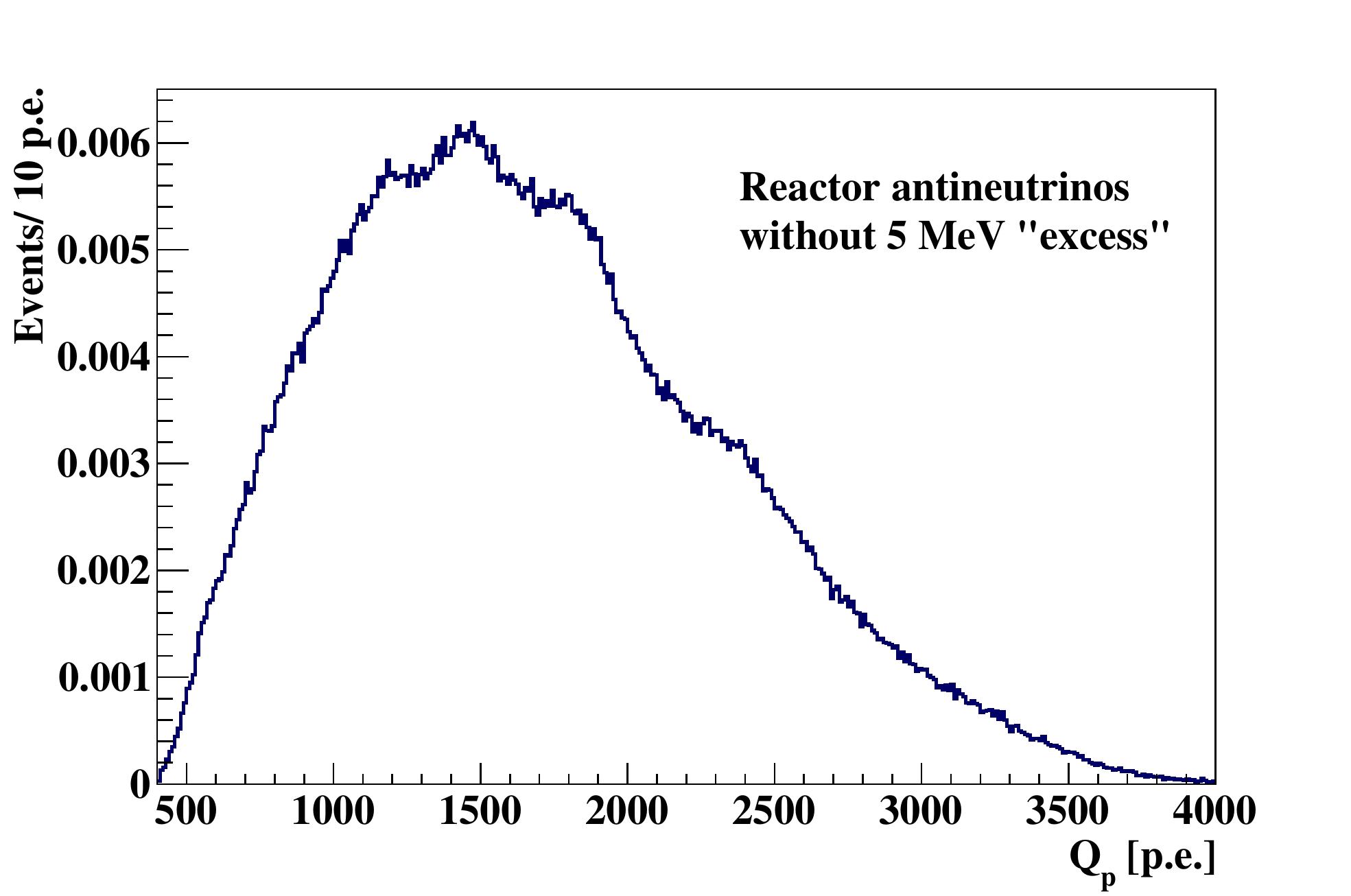}
     \label{fig:PDF_rea}}
    \subfigure[]{\includegraphics[width =0.49\textwidth]{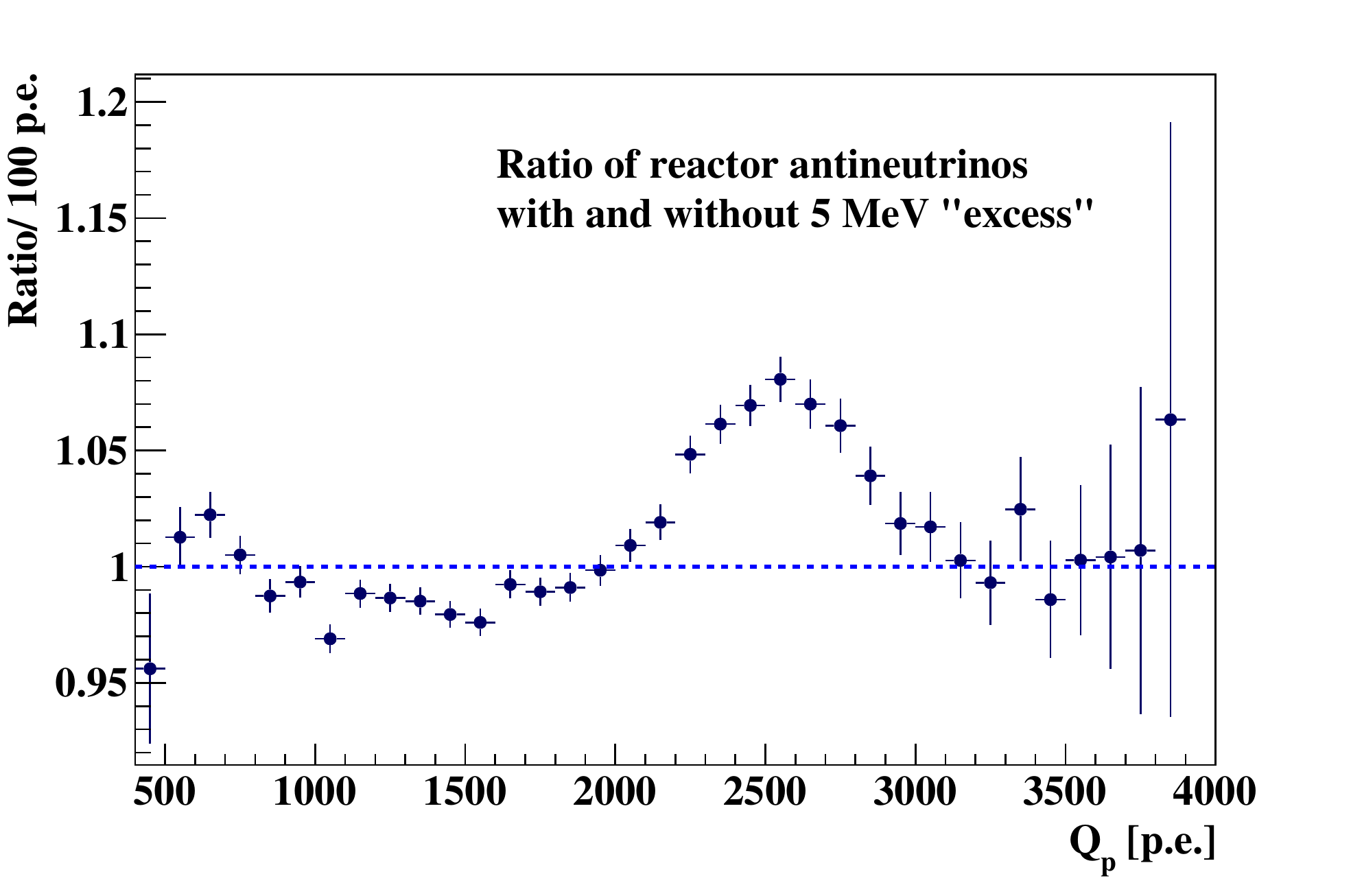}
    \label{fig:PDF_bump}}
        \caption{The MC-based PDFs normalized to one, i.e. the expected shapes of prompts for geoneutrinos and reactor antineutrinos, after optimized geoneutrino selection cuts, that include the detector response. Top-left: geoneutrinos with Th/U ratio fixed to the chondritic value ($R_{\mathrm{S}} = 0.27$). Top-right: $^{238}$U and $^{232}$Th PDFs shown separately. Bottom-left: reactor antineutrinos ``without 5\,MeV excess". Bottom-right: ratio of reactor antineutrino spectra ``with/without 5\,MeV excess" (Sec.~\ref{subsec:rea}), normalized to the same number of events each, in order to demonstrate the difference in shape only. }
        \label{fig:PDFs-Geo-Rea}
    \end{figure*}

        \begin{figure}[t]
     \centering  
    \subfigure[]{\includegraphics[width =0.5\textwidth]{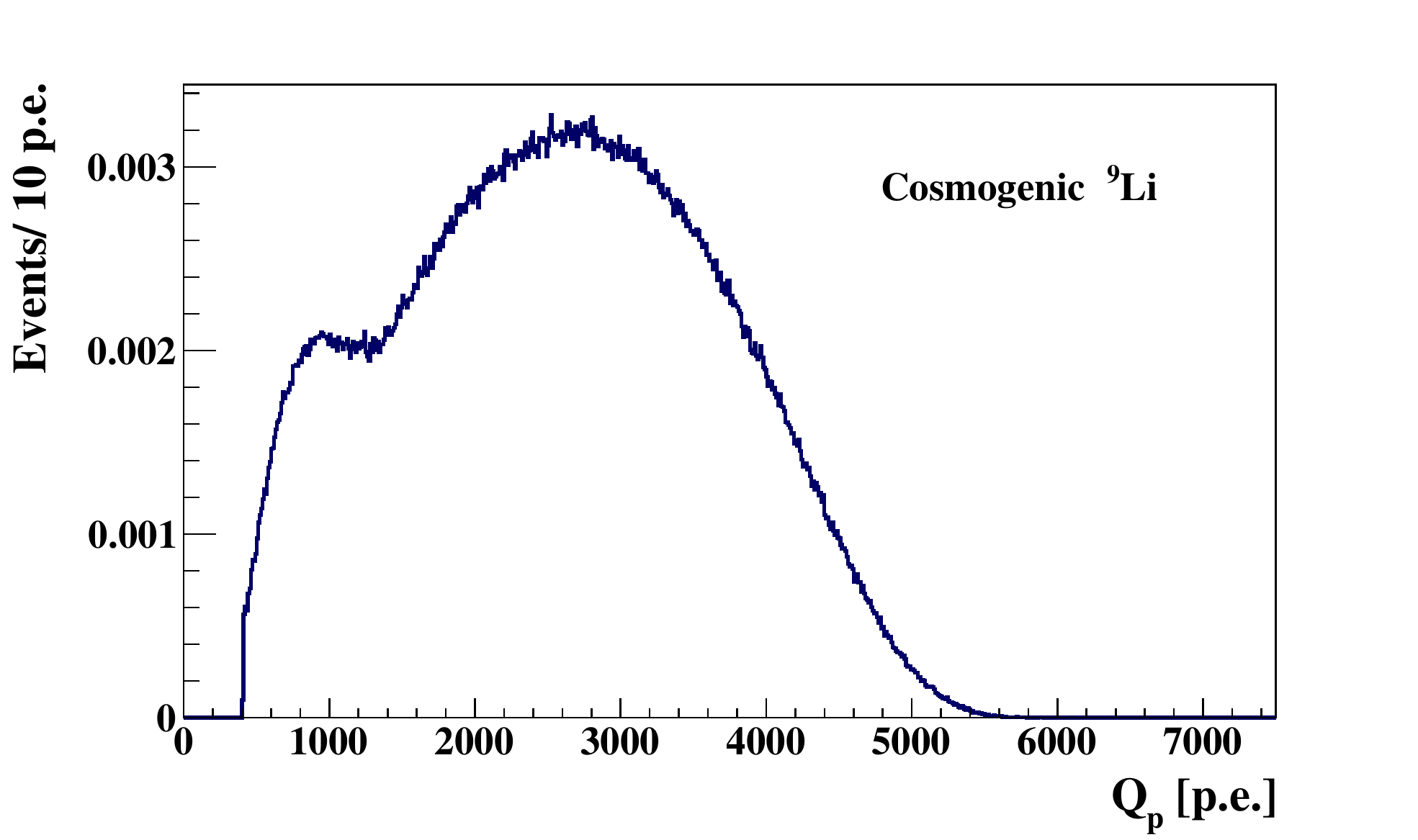}}
     \subfigure[]{\includegraphics[width =0.5\textwidth]{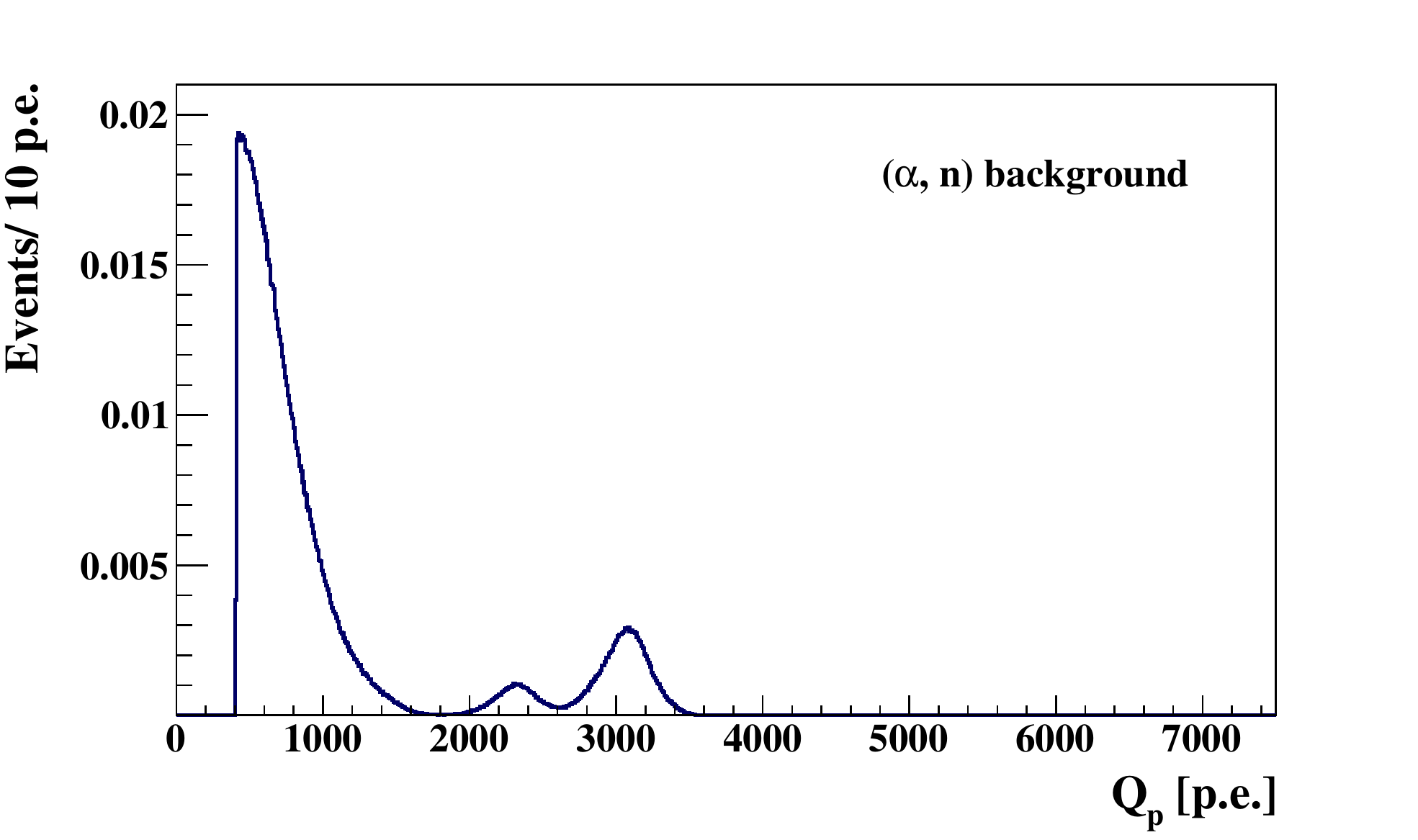}}
        \subfigure[]{ \includegraphics[width =0.5\textwidth]{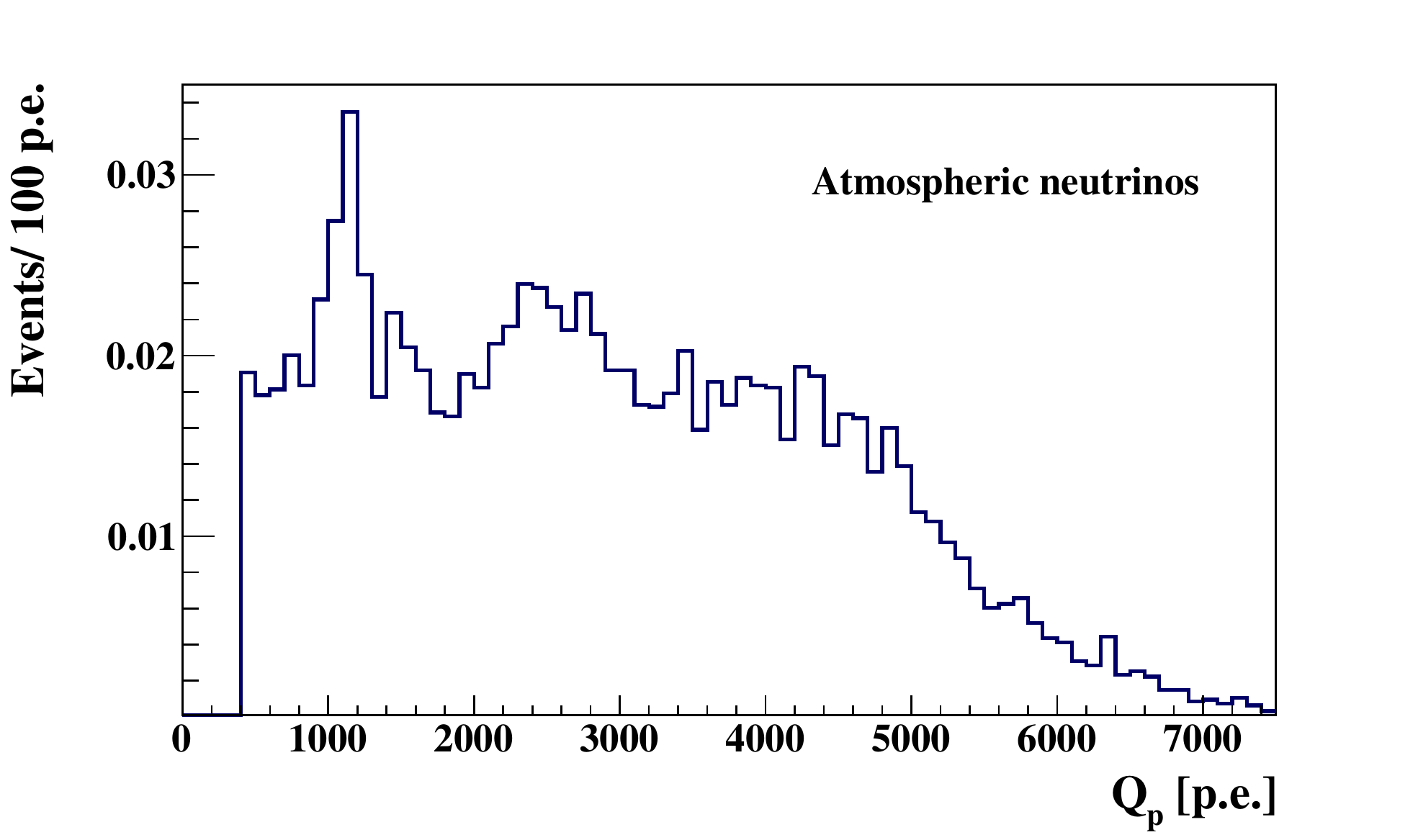}}
        \caption{MC-based PDFs of prompts for different backgrounds after optimized geoneutrino selection cuts, normalized to one. Top: cosmogenic $^9$Li background. Middle: ($\alpha$, n) background. Bottom: atmospheric neutrino background.}
        \label{fig:PDFs-backgrounds}
    \end{figure}  

    \subsection{Monte Carlo spectral shapes}
    \label{subsec:MCspectra}

Once the full {\it G4Bx2} MC code is working reliably and the origin of the signal and backgounds is known, it is, in principle, easy to simulate the PDFs that incorporate the detector response and that can be used directly in the spectral fit (Sec.~\ref{subsec:nutshell}).

The simulated signal and backgrounds follow the same experimental conditions as observed in real data, including the number of working channels, the shape of the IV, and the dark noise, as described in~\cite{Agostini:2017aaa}. Each run of the complete data set from December 2007 to April 2019 is simulated individually. After the simulation, the optimized geoneutrino selection cuts (Sec.~\ref{subsec:cuts}) are applied as in the real data.

For antineutrinos, pairs of positrons and neutrons were simulated. The neutron energy spectrum is taken from~\cite{PhysRevD.60.053003}, while for the positrons the energy spectra as discussed in Sec.~\ref{sec:antinu} are used. The antineutrino energy spectra are transformed to positron energy spectra following Eq.~\ref{eq:Epro}. In particular, for geoneutrinos, the energy spectra as in Fig.~\ref{Fig:GeoNuSignal} are used. Individual spectra from $^{232}$Th and $^{238}$U chains were also simulated, so that they can be weighted according to the expected $R_s$ ratio (Eq.~\ref{eq:RS}) for different geological contributions. For reactor antineutrinos, calculated energy spectra ``with and without 5\,MeV excess" as in Fig.~\ref{fig:reactor_antinu} are used as MC input. The resulting PDFs are shown in Fig.~\ref{fig:PDFs-Geo-Rea}.

The MC-based PDFs for non-antineutrino backgrounds are shown in Fig.~\ref{fig:PDFs-backgrounds}. A dedicated code is developed within the {\it G4Bx2} simulation framework for the generation of $^{9}$Li events, based on Nuclear Data Tables and literature data~\cite{PREZADO200543}. The input for ($\alpha, n)$ background simulation is discussed in Sec.~\ref{subsec:alpha_n}, while for atmospheric neutrinos in Sec.~\ref{subsec:atm}.

\begin{figure} [t]
     \centering  
   \includegraphics[width = 0.49\textwidth]{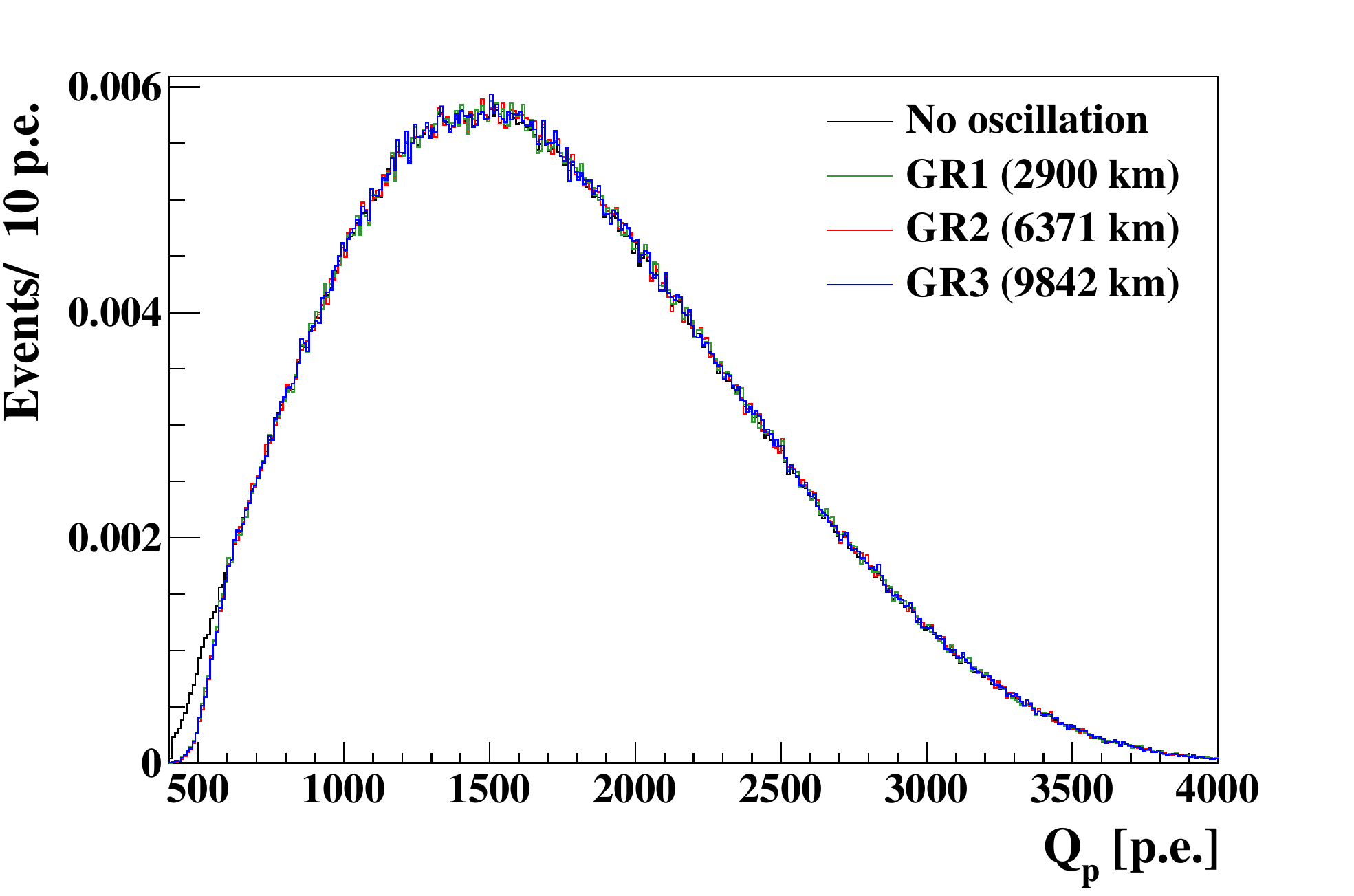}
        \caption{MC-based PDFs of prompts for a georeactor (Sec.~\ref{subsec:georeactor}) after optimized IBD selection cuts, normalized to one. We show both the case of non-oscillated spectrum, as well as the oscillated spectra for the georeactor placed in three positions at different depths: GR1 ($d$ = 2900\,km), GR2 ($d$ = 6371\,km), and GR3 ($d$ = 9842\,km), defined in Fig.~\ref{fig:GeoReactorPosition}.}
        \label{fig:PDFs-georeactor}
    \end{figure}

The PDFs of prompts due to antineutrinos from a hypothetical georeactor (Sec.~\ref{subsec:georeactor}) are shown in Fig.~\ref{fig:PDFs-georeactor}. We compare the shapes (in all cases normalized to one) for the non-oscillated spectrum and for the oscillated cases with the georeactor placed at three different positions GR1, GR2, and GR3 as shown in Fig.~\ref{fig:GeoReactorPosition}. As it can be seen, the shapes of the  prompt energy spectra are almost identical.
\begin{table} [t]
	\centering
	\caption{\label{tab:eff} Detection efficiencies after the optimized selection cuts for geoneutrinos from $^{238}$U and $^{232}$Th chains individually and for their summed contribution (according to the chondritic Th/U mass ratio), as well as for antineutrinos from reactors and from a hypothetical georeactor. The efficiencies were determined based on the MC simulation of each component. The error is estimated using the calibration data.} \vskip 2pt
	\begin{tabular*}{\columnwidth}{ l @{\hskip 42pt} l }
			\hline
			\hline
			Source & Efficiency \Tstrut \\
			       & [\%] \Bstrut \\
			\hline
			$^{238}$U geoneutrinos & 87.6 $\pm$ 1.5 \Tstrut \\
	        $^{232}$Th geoneutrinos & 84.8 $\pm$ 1.5  \\
	        Geoneutrinos ($R_s$ = 0.27)& 87.0 $\pm$ 1.5  \\ [4 pt]
			Reactor antineutrinos & 89.5 $\pm$ 1.5 \\ [4 pt]
			Georeactor & 89.6 $\pm$ 1.5 \Bstrut \\
			\hline
			\hline
		\end{tabular*}
\end{table}

    \subsection{Detection efficiency}
    \label{subsec:efficiency}

The detection efficiencies for geoneutrinos ($\varepsilon_{\mathrm{geo}}$, $\varepsilon_{\mathrm{Th}}$, $\varepsilon_{\mathrm{U}}$), reactor antineutrinos ($\varepsilon_{\mathrm{rea}})$, and antineutrinos from a hypothetical georeactor ($\varepsilon_{\mathrm{georea}})$ are summarized in Table~\ref{tab:eff}. They represent a fraction of MC events passing all the optimized data selection cuts (Sec.~\ref{subsec:cuts}) from those generated in the FV of this analysis (10\,cm DFV cut). The errors due to the FV definition and the position reconstruction resolution are included in the calculation of the systematic uncertainty, as it will be discussed in Sec.~\ref{subsec:syst}. The error on the detection efficiency was estimated based on the comparison of the calibration data (most importantly $^{241}$Am-$^{9}$Be neutron source data) with MC simulation. The major contribution comes from the uncertainties of the detector response close to the edge of IV.

    \section{EVALUATION OF THE EXPECTED SIGNAL AND BACKGROUNDS WITH OPTIMIZED CUTS}
    \label{sec:sig_bgr_est}
     
    In this Section the evaluation of the number of expected antineutrino signal and background events after the optimized selection cuts (Table~\ref{tab:sel_cuts}) is described. The data set and the total exposure are presented in Sec.~\ref{subsec:exposure}. The number of expected antineutrino events from different sources, based on the estimated antineutrino signals as in Table~\ref{tab:antinu-signals-expected}, is presented in Sec.~\ref{subsec:antinu_est_ev}.
    The next sections treat the non-antineutrino background, following the structure of Sec.~\ref{sec:bgr}, where the physics of each of these background categories is described. In particular, cosmogenic background is discussed in Sec.~\ref{subsec:cosmogenic_est}, accidental background in Sec.~\ref{subsec:acc_est}, ($alpha$, n) interactions in Sec.~\ref{subsec:alpha_n_est}, ($\gamma$, n) and fission in PMTs in Sec.~\ref{subsec:gamma_n_est}, radon correlated background in Sec.~\ref{subsec:radon_est}, and finally $^{212}$Bi-$^{212}$Po background in Sec.~\ref{subsec:212BiPo_est}. Section~\ref{subsec:bckg_sum_est} summarizes the expected total number of non-antineutrino background events.
    
        \subsection{Data set and exposure}
        \label{subsec:exposure}
 In this analysis the data taken between December 9, 2007 and April 28, 2019, corresponding to $t_{\mathrm{DAQ}}$ = 3262.74\,days of data taking, are considered. The average life-time weighted IV volume and the FV used in this analysis (DFV cut $d_{\mathrm {IV}} = 10$\,cm) are $\overline{V}_{\mathrm{IV}}$ = (301.3 $\pm$ 10.9)\,m$^{3}$ and $\overline{V}_{\mathrm{FV}}$ = (280.1 $\pm$ 10.1)\,m$^{3}$, respectively, after taking the changing shape of the IV into account (Sec.~\ref{subsec:IV}). These correspond to the average IV and FV mass of $\overline{m}_{\mathrm{IV}}$ = (264.5 $\pm$ 9.6)\,ton and $\overline{m}_{\mathrm{FV}}$ = (245.8 $\pm$ 8.7)\,ton, respectively, considering the scintillator density of $\rho_{\mathrm{LS}}$ = $(0.878 \pm 0.004)$\,g\,cm$^{-3}$. The total exposure after the cosmogenic veto (Sec.~\ref{subsec:vetoes}) and in the FV is $\mathcal{E}$ = (2145.8 $\pm$ 82.1)\,ton $\times$ yr, considering the systematic uncertainty on position reconstruction (Sec.~\ref{subsec:syst}). This can be expressed as $\mathcal{E}_p$ = (1.29 $\pm$ 0.05) $\times 10^{32}$\, protons $\times$ yr, using the proton density in Borexino LS of $N_p$ = $(6.007 \pm 0.001) \times 10^{28}$ protons ton$^{-1}$. Applying the geoneutrino detection efficiency described in Sec.~\ref{subsec:efficiency}, the effective  exposure for the geoneutrino detection reduces to $\mathcal{E'}$ = (1866.4 $\pm$ 78.4)\,ton $\times$ yr and $\mathcal{E'}_p$ = (1.12 $\pm$ 0.05) $\times 10^{32}$\,protons $\times$ yr. 
 
\begin{table} 
	\centering
	\caption{\label{tab:antinu-events-expected} Summary of the expected number of antineutrino events with optimized selection cuts in the geoneutrino (408 - 1500)\,p.e and reactor antineutrino (408 - 4000)\,p.e energy ranges. The errors include uncertainties on the predicted signal only. The reference exposure $\mathcal{E}_p$ = (1.29 $\pm$ 0.05) $\times 10^{32}$\, protons $\times$ yr corresponds to the analyzed period.}	\vskip 2pt
	\begin{tabular*}{\columnwidth}{l @{\hskip 10 pt} c @{\hskip 10 pt} c}
		\hline
		\hline
		Model & Energy range & Signal \Tstrut \\
			   & [p.e.] & [Events] \Bstrut \\
		\hline
		\multicolumn{3}{c}{Geoneutrinos} \Tstrut\Bstrut \\
		\hline
		Bulk lithosphere & 408 - 1500 & 28.8 $^{+5.5}_{-4.6}$ \Tstrut \\ [4pt]
	    CC BSE (total)  & 408 - 1500 & 31.9$^{+6.2}_{-5.4}$ \\
	    CC BSE (mantle)   & 408 - 1500& 2.8 $\pm$ 0.6 \\ [4pt]
	    GC BSE (total) & 408 - 1500 & 38.8$^{+6.2}_{-5.4}$ \\ 
	    GC BSE (mantle)  & 408 - 1500& 9.8 $\pm$ 0.9\\ [4pt]
	    GD BSE (total) & 408 - 1500 & 51.1$^{+6.3}_{-5.5}$\\
	    GD BSE (mantle) & 408 - 1500 & 22.0 $\pm$ 1.2\\ [4pt]
	    FR (total)  & 408 - 1500 & 62.0$^{+6.4}_{-5.6}$ \\
	    FR (mantle)  & 408 - 1500 & 33.0 $\pm$ 1.7 \Bstrut \\ 
	    \hline
	    \multicolumn{3}{c}{Reactor antineutrinos} \Tstrut\Bstrut \\
	    \hline
	    without 	& 408 - 1500  & 42.6 $\pm$ 0.7 \Tstrut \\
	    ``5\,MeV excess"          & 408 - 4000 & 97.6$^{+1.7}_{-1.6}$\\ [4pt]
		with & 408 - 1500  & 39.5 $\pm$ 0.7\\ 
		``5\,MeV excess"   & 408 - 4000 & 91.9$^{+1.6}_{-1.5}$ \Bstrut \\ 
		\hline
		\multicolumn{3}{c}{Atmospheric neutrinos} \Tstrut\Bstrut \\
		\hline
			          & 408 - 1500 & 2.2 $\pm$ 1.1 \Tstrut \\ 
			          & 408 - 4000 & 3.3 $\pm$ 1.6 \\ 
			          & 408 - 8000 & 9.2 $\pm$ 4.6 \Bstrut \\     
		\hline
		\multicolumn{3}{c}{1 TW Georeactor} \Tstrut\Bstrut \\
		\hline
		GR2: Earth's center  & 408 - 1500 & 3.6 $\pm$ 0.1 \Tstrut \\ 
			                 & 408 - 4000 & 8.9 $\pm$ 0.3\\ [4pt]
		GR1: CMB at 2900\,km & 408 - 1500 &  17.6  $\pm$ 0.5 \\ 
			                 & 408 - 4000 &  43.1$\pm$ 1.3\\ [4pt]
		GR3: CMB at 9842\,km & 408 - 1500 & 1.5 $\pm$ 0.04\\
			                 & 408 - 4000 & 3.7 $\pm$ 0.1 \Bstrut \\
		\hline
		\hline
	\end{tabular*}
\end{table}

        \subsection{Antineutrino events}
        \label{subsec:antinu_est_ev}
This Section summarizes the number of expected antineutrino events detected with the optimized selection cuts, assuming the expected antineutrino signals as given in Table~\ref{tab:antinu-signals-expected}. An overview is given in Table~\ref{tab:antinu-events-expected}.

        \subsection{Cosmogenic background}
        \label{subsec:cosmogenic_est}
 
The various cosmogenic backgrounds in Borexino which affect the geoneutrino analysis were explained in Sec.~\ref{subsec:cosmogenic}, while the analysis and technical evaluation of these backgrounds is explained in this Section.

    \paragraph{Hadronic background}
    The hadronic background, expected to be dominated by $^9$Li (Sec.~\ref{subsec:vetoes}), which remains after the detected muons out of the vetoed space and time is evaluated here. We first study the time and spatial distributions of the detected $^9$Li candidates with respect to the parent muon. After that, this background is evaluated for the different kinds of internal muons according to the respective vetoes applied after them, as previously described in Sec.~\ref{subsec:vetoes} and summarized in Fig.~\ref{fig:vetoes_scheme}. \\

	\begin{figure}
	\centering
	\subfigure[]{\includegraphics[width = 0.49\textwidth]{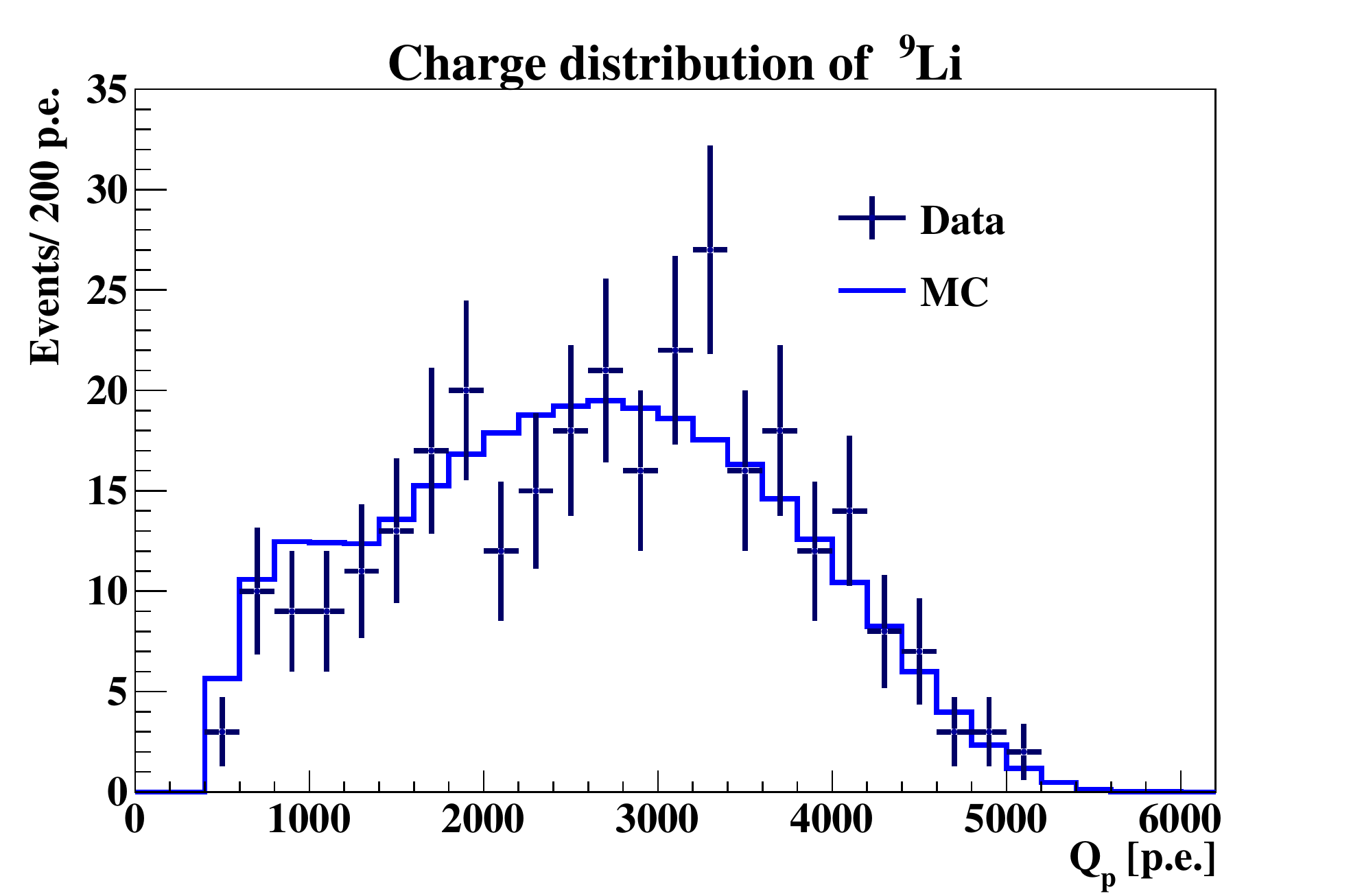}
	\label{fig:li9_charge}}
	\subfigure[]{\includegraphics[width =0.49\textwidth]{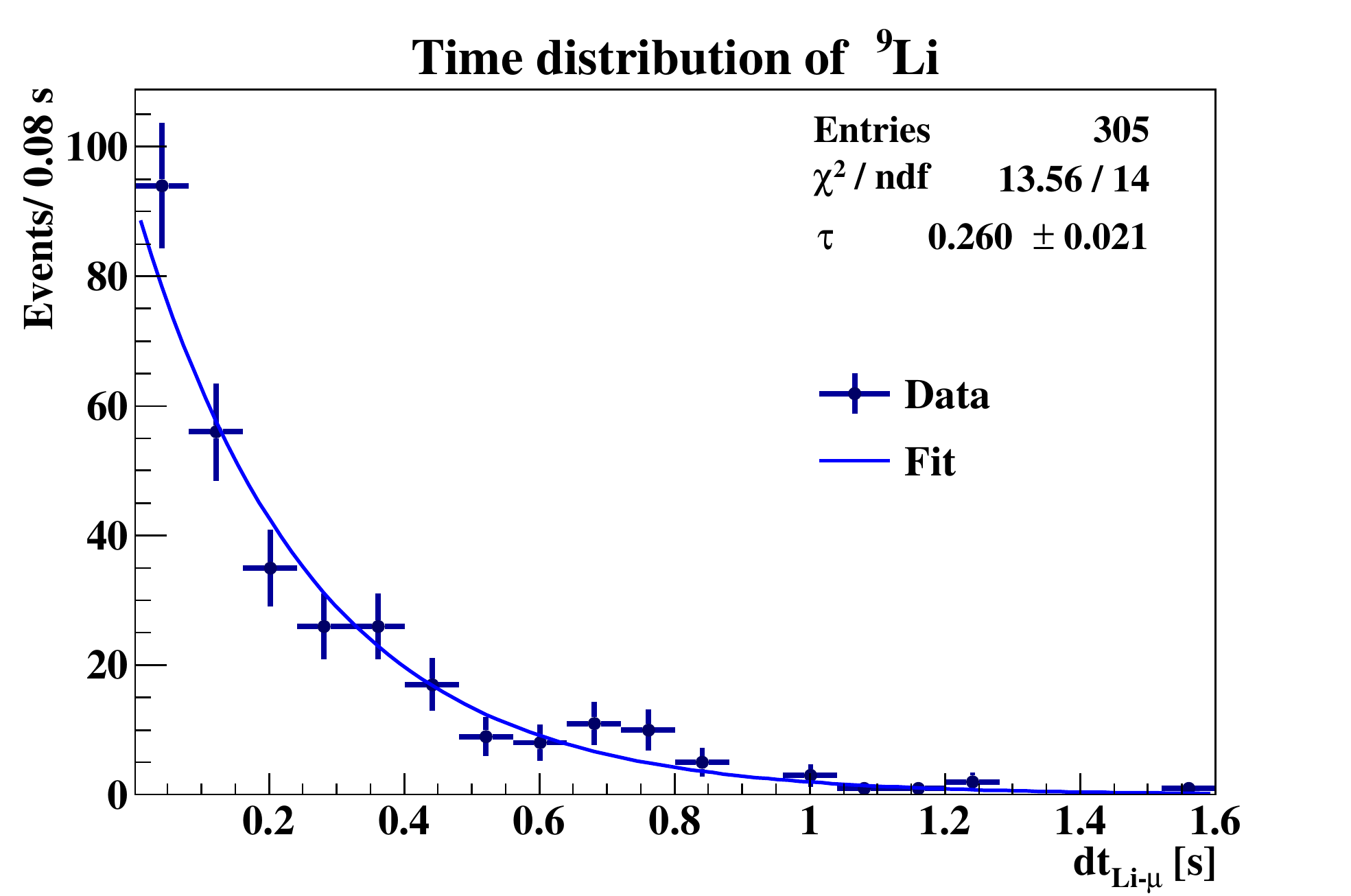}
	\label{fig:li9_dt}}
	\subfigure[]{\includegraphics[width = 0.49\textwidth]{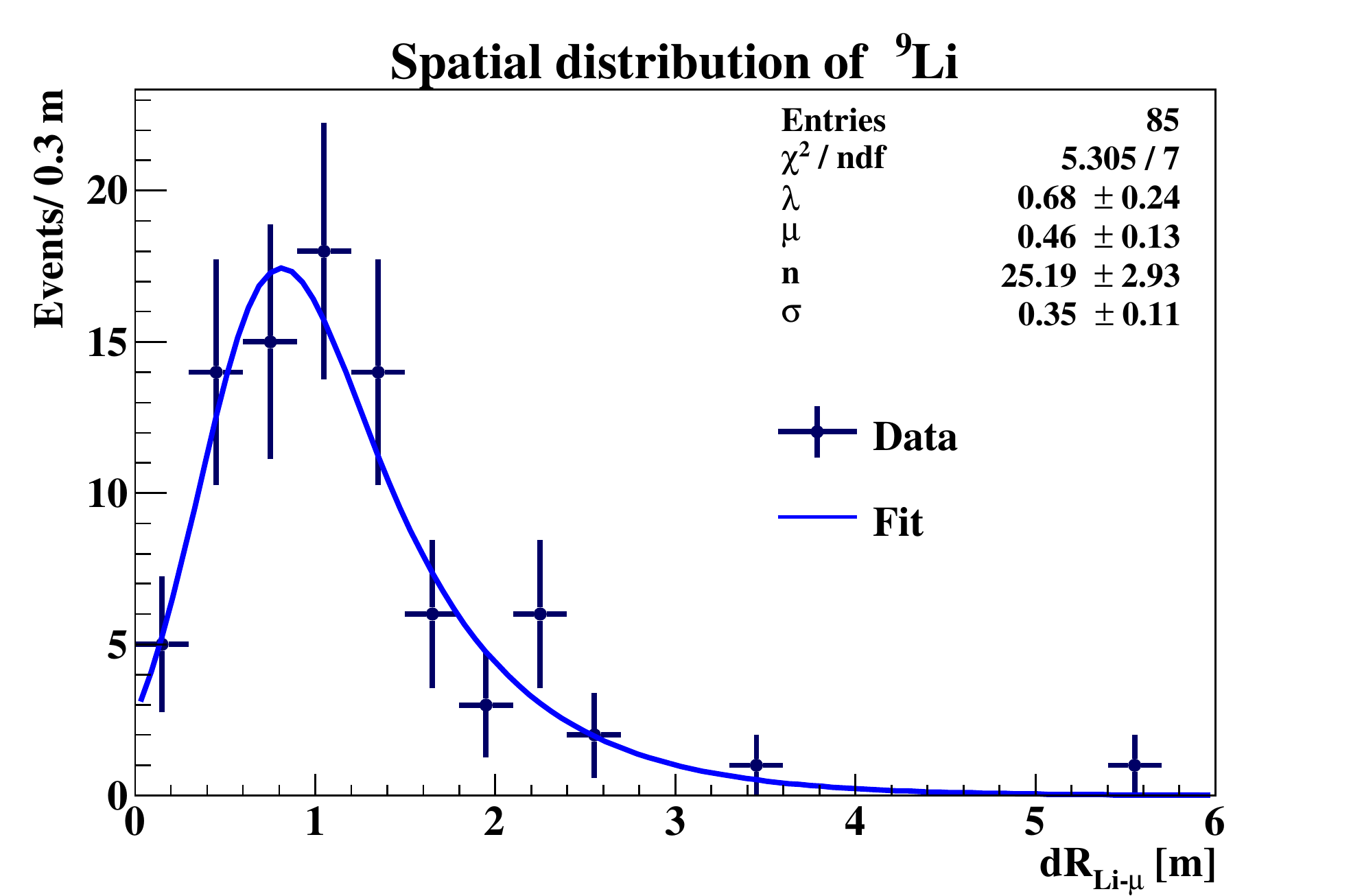}
	\label{fig:li9_dR_mu}}
	\caption{Distributions of the observed $^9$Li cosmogenic background, i.e. the IBD-like candidates after optimized selection cuts following muons after at least 2\,ms. (a) Comparison of the prompt $Q_{p}^{\mathrm{Li}}$ charge data spectra (points) with the MC spectrum  (solid line). (b) Distribution of $dt_{\mathrm{Li} - \mu}$, the time difference between the prompt and the preceding muon, fit with the exponential function. (c) Distribution of $dR_{\mathrm{Li} - \mu}$, the distance of the prompt from the well reconstructed muon track, fit with the function in Eq.~\ref{eq:dR-Li-mu}.}
	\end{figure}

    \noindent {\it Time distribution $dt_{\mathrm{Li} - \mu}$} \\
    
	 First, a search for IBD-like signals, passing the optimized selection cuts (Table~\ref{tab:sel_cuts}), is performed after the category of muons for which 1.6 or 2.0\,s veto is applied. We perform this search starting from 2\,ms after each muon, in order to remove cosmogenic neutrons. In total, we found 305 such IBD-like candidates, dominated by 282 candidates afer $(\mu + n)$ muons. The $Q_{p}^{\mathrm{Li}}$ charge energy spectrum of the prompts is compatible with the expected MC spectrum of $^{9}$Li, as it is shown in Fig.~\ref{fig:li9_charge}. The distribution of the time differences between the prompt and the preceding muon, $dt_{\mathrm{Li} - \mu}$, is shown in Fig.~\ref{fig:li9_dt}. As it can be seen, no events are observed after $dt_{\mathrm{Li} - \mu} > 1.6$\,s. The decay time $\tau$ is extracted by performing an exponential fit to the $dt_{\mathrm{Li} - \mu}$ distribution and is found to be (0.260 $\pm$ 0.021)\,s. This is compatible with $\tau_{^9 \mathrm{ Li}}$ = 0.257\,s decay time of $^{9}$Li.\\

    \noindent {\it Spatial distribution $dR_{\mathrm{Li} - \mu}$}\\
    
    The distance of the $^9$Li prompt from the muon track, $dR_{\mathrm{Li} - \mu}$, is studied for ($\mu$ + n) muons with reliably reconstructed tracks. It is shown for 85 candidates in Fig.~\ref{fig:li9_dR_mu}. This distribution is fit with the convolution of an exponential (with a characteristic length $\lambda$) with a Gaussian with parameters $\mu$ and $\sigma$, and a normalisation factor $n$:

    \begin{equation}
        \begin{split}
        \label{eq:dR-Li-mu}
	    f(dR_{\mathrm{Li} - \mu};\lambda,\sigma,\mu, n) = & \frac{n}{2\lambda} \\ 
    	\times&  \exp \bigg( \frac{2\mu + \sigma^{2}\lambda^{-1}-2dR_{\mathrm{Li} - \mu}}{2\lambda}\bigg) \\ 
	    \times&  \text{erfc} \bigg( \frac{\mu + \sigma^{2}\lambda^{-1}-dR_{\mathrm{Li} - \mu}}{\sqrt{2}\sigma}\bigg).
	    \end{split}
    \end{equation}
   The fit results in $\sigma$ = (0.35 $\pm$ 0.11)\,m, which represents well the combined position reconstruction of the muon track and the prompt, and in $\lambda$ = (0.68 $\pm$ 0.24)\,m. Considering these values and the fit function in Eq.~\ref{eq:dR-Li-mu}, 2.75\% of $^9$Li prompts would be reconstructed out of the cylinder with 3.0\,m radius around the muon track.
   Below, the hadronic background after different muon-veto categories (Fig.~\ref{fig:vetoes_scheme}) is evaluated, considering the $dt_{\mathrm{Li} - \mu}$ and $dR_{\mathrm{Li} - \mu}$ parametrisations described above.
   \begin{itemize}
   
   \item  {\it  1.6\,s and 2.0\,s vetoes of the whole detector}:
   
    For 8.1(7.1)\% of the total muons, we veto the whole detector for 1.6(2.0)\,s.
    In the time window [2\,ms, 1.6(2.0)\,s] after these muons, where IBD-like candidates were searched, falls 99.02(99.19)\% of the corresponding $^9$Li background. In this window, 282(23) candidates were found. Finally, in the time window after the time veto, the remaining background is $\exp(-1.6(2.0)/ \tau) = 0.21(0.05)\%$ of the total background. Adding the two components and including the statistical error and the error on $\tau$, the total amount of hadronic background in the antineutrino candidate sample is 0.18$^{+0.09}_{-0.06}$ events.

\item {\it 1.6\,s cylindrical veto}:

For 27.0\% of the muons with a reliably reconstructed track, only a cylinder with 3\,m radius is vetoed for 1.6\,s. In the [2\,ms, 1.6\,s] time window after these muons, 7 IBD-like events were observed in the whole detector. Thus, after 1.6\,s, we expect 0.015$^{+0.013}_{-0.011}$ events distributed in the whole detector. Of these, 2.75\% would be reconstructed out of the cylindrical veto, as mentioned before. This means, our expected background of this category can be conservatively set to (0.20 $\pm$ 0.08) events. By restricting the veto from the whole detector to only the cylindrical volume, the exposure increases by 1.47$\%$, corresponding to (2.2 $\pm$ 0.2) expected IBD events. We observe one additional candidate.

\item {\it 2\,ms veto of the whole detector}:

For 57.8\% of all muons which have a lower probability to produce detectable hadronic background (($\mu$ - n)$_{< 8000}$), we restrict the time veto of the whole detector to 2\,ms, to veto only the cosmogenic neutrons. Consequently, the exposure increases by 6.6\%, corresponding to (9.7 $\pm$ 0.8) expected IBD candidates. Seven candidates were observed, well within the expectation. However, this does not guarantee that we did not introduce additional $^9$Li background, that is estimated as follows.

Muons with less than 8000 hits produce less light because they pass mostly through the buffer region, where the neutrons are typically not detected or are below the threshold of this analysis. It is reasonable to assume that the production ratios for $^9$Li and cosmogenic neutrons are the same for $\mu_{<8000}$ and $\mu_{>8000}$ muon categories, since the muons have typically the same energies and the traversed media (LS and the buffer) have nearly the same density. It is also reasonable to assume, that for the $\mu_{<8000}$, the detection efficiency of the corresponding cosmogenic neutrons and neutrons from $^9$Li decays are the same. Thus, the equality of the ratios of the number of observed $^9$Li candidates (with decay neutron, $N_{\mathrm Li}$) and cosmogenic neutrons ($N_n$) should hold:

\begin{equation}
	\centering
     \frac{(N_\mathrm{Li} = 305)_{>8000}}{(N_n = 8.6 \times 10^5)_{>8000}} =  \frac{(N_{\mathrm{Li}})_{<8000}}{(N_n = 9181)_{<8000}}.
	\label{eq:Libgr}
	\end{equation}
From this equality, the expected number $(N_{{\mathrm Li}})_{<8000}$ is 3.2 events. This is the number of expected $^9$Li events produced by muons with less than 8000 hits, independent of whether this muon was followed by a neutron or not.
This is a conservative number for the background estimation due to the $(\mu - n)_{<8000}$ muons, after which we apply the reduced 2\,ms cut. Even if they represent about 99\% of all $\mu_{<8000}$ muons (Fig.~\ref{fig:vetoes_dechits}), $(\mu + n)_{<8000}$ muons can be expected to have a higher probability to produce observable $^9$Li than $(\mu - n)_{<8000}$ muons. However, we do not observe any $^9$Li candidate for $(\mu + n)_{<8000}$ muons. Thus to summarize, our expected $^9$Li background for $(\mu - n)_{<8000}$ muons is 3.2 $\pm$ 1.0 events, which includes larger systematic error.
\end{itemize}

\noindent In total, after summing all the contributions, the expected $^9$Li background within our golden IBD candidates is (3.6 $\pm$ 1.0) events.

            \paragraph{Untagged muons}

\noindent The (0.0013 $\pm$ 0.0005)$\%$ mutual inefficiency of the strict muon flags shown in Sec.~\ref{subsec:muon} corresponds to (195 $\pm$ 75) undetected muons in the entire data set. Following the discussion in Sec.~\ref{subsec:cosmogenic}, these muons could eventually cause background of three types:
\begin{itemize}
    \item  \textit{$\mu$ + $\mu$:} Considering the small amount of undetected muons in the entire data set, the probability that two undetected muons would fall in a 1260\,$\mu$s time window of the delayed coincidence is completely negligible.
    \item  \textit{$\mu$ + {\it n}:} The most dangerous are pairs of buffer muons (possibly fulfilling the $Q_p$ cut) followed by a single neutron (multiple neutrons are removed by the multiplicty cut). The probability that a ($\mu$ + {\it n}) pair falls within the IBD selection cuts, evaluated on the subset of MTB muons followed by the {\it TT128} trigger dedicated to neutron detection, is found to be (9.7 $\pm$ 0.003) $\times$ 10$^{-5}$. Hence, there will be (0.019 $\pm$ 0.007) events of this kind in the IBD sample due to the untagged muons. 
    \item \textit{Muon daughters:}  
     After $1.497 \times 10^{7}$ internal muons we have observed 305 IBD-like background events in a [2\,ms, 1.6\,s] time window, that covers 99.02\% of IBD-like candidates of the same type. Therefore, after (195 $\pm$ 75) undetected muons, we can 
     estimate to have (0.0040 $\pm$ 0.0015) IBD-like events created any time after these muons and falling within the selection cuts.
    \end{itemize}
    
    Summing all the three components, the estimated background originating from untagged muons is (0.023 $\pm$ 0.007) events in the IBD sample. We note that this is a very small number.

\paragraph{Fast neutrons}
 
As described in Sec.~\ref{subsec:cosmogenic}, undetected muons that pass the WT or the surrounding rocks, can produce fast neutrons that can give IBD-like signals. Fast neutrons from cosmic muons were simulated according to the energy spectrum  from~\cite{PhysRevD.73.053004}. We have found that the eventual signal from a scattered proton follows in nanosecond time scale after the neutron production, that is simultaneous with the muon signal. Considering the data structure detailed in Sec.~\ref{subsec:data}, this time range dictates the data selection cuts, as described below, in order to search for fast-neutron related IBD-like signals after the detected external muons pass the WT. Knowing the fraction of these muons creating IBD-like background, we can estimate the fast neutron background from the undetected muons that pass the WT. MC simulations are used to obtain an estimation of fast neutron background due to the muons passing through the surrounding rock and not the detector.
Both estimations are given below.

 {\it Water Tank muons:}
 
  In this search, the signal in the ID should correspond to a scattered proton, which is not tagged by the muon Inner Detector Flag (IDF). Without the proton signal in the ID, the external muon would be a {\it TT2 \& BTB4} event. The presence of the ID signal can, with lower than 100\% efficiency, turn the muon to be a {\it TT1 \& BTB4} event. Therefore, we search for two kinds of coincidences:
 \begin{itemize}
     \item The prompt signal is an internal muon {\it TT1 \& BTB4} that is not tagged by the IDF. The delayed signal is a neutron cluster found in the {\it TT128} gate which is opened immediately after the muon. 
     \item The prompt signal is an external muon {\it TT2 \& BTB4} that has a cluster inside the ID, and is not tagged by the IDF. The delayed signal follows within 2\,ms as a point-like {\it TT1 \& BTB0} event. 
 \end{itemize}
 This search was done with relaxed energy, $dt$, and $dR$ cuts, without any DFV or multiplicity cuts. This yielded 25 coincidences of the first kind and 12 coincidences of the second kind. However, only one coincidence satisfied all the geoneutrino selection cuts. The amount of these coincidences in the IBD data sample can be due to muons that go undetected by the OD. The average inefficiency of MTF with respect to the MCF and IDF flag is 0.27$\%$, as shown in Table~\ref{tab:mut_eff_tab}. This gives an upper limit of 0.013 IBD-like coincidences at 95\% C.L. due to the fast neutrons from undetected muons crossing the WT.

 {\it Surrounding rocks:}
   
In order to study the fast neutron background due to muons passing through the rocks surrounding the detector, we used the Borexino MC with the initial flux and energy spectrum of neutrons and their angular distributions taken from~\cite{PhysRevD.73.053004} for the specific case of LNGS. The total statistics of MC-generated neutrons corresponds to 3.3 times the exposure of this analysis. Fast neutrons with energies in diapason 1\,MeV - 3.5,GeV were simulated on the
surface of the Borexino outer water tank. Full simulation with tracking of each scintillation photon was done for the fast neutrons and other
particles penetrating inside the ID. Finally, we obtain only one IBD-like event for neutrons from
the rock passing the optimized IBD selection cuts, which corresponds to an upper limit of the corresponding background in our geoneutrino analysis of $<$1.43 events at 95 C.L.
 
  \subsection{Accidental coincidences}
    \label{subsec:acc_est}

In order to evaluate the amount of accidental coincidences in the antineutrino sample, coincidence events were searched for in the off-time interval $dt$ = [2\,s, 20\,s] and were then scaled to the 1270\,$\mu$s duration of the geoneutrino selection time window ($dt$ = [2.5\,$\mu$s, 12.5\,$\mu$s] + [20\,$\mu$s, 1280\,$\mu$s], Sec.~\ref{subsec:cuts}). In this scaling, a suppression factor due to the muon veto must be considered, as explained below.

\begin{figure}[t]
	\centering
	\subfigure[]{\includegraphics[width=0.49\textwidth]{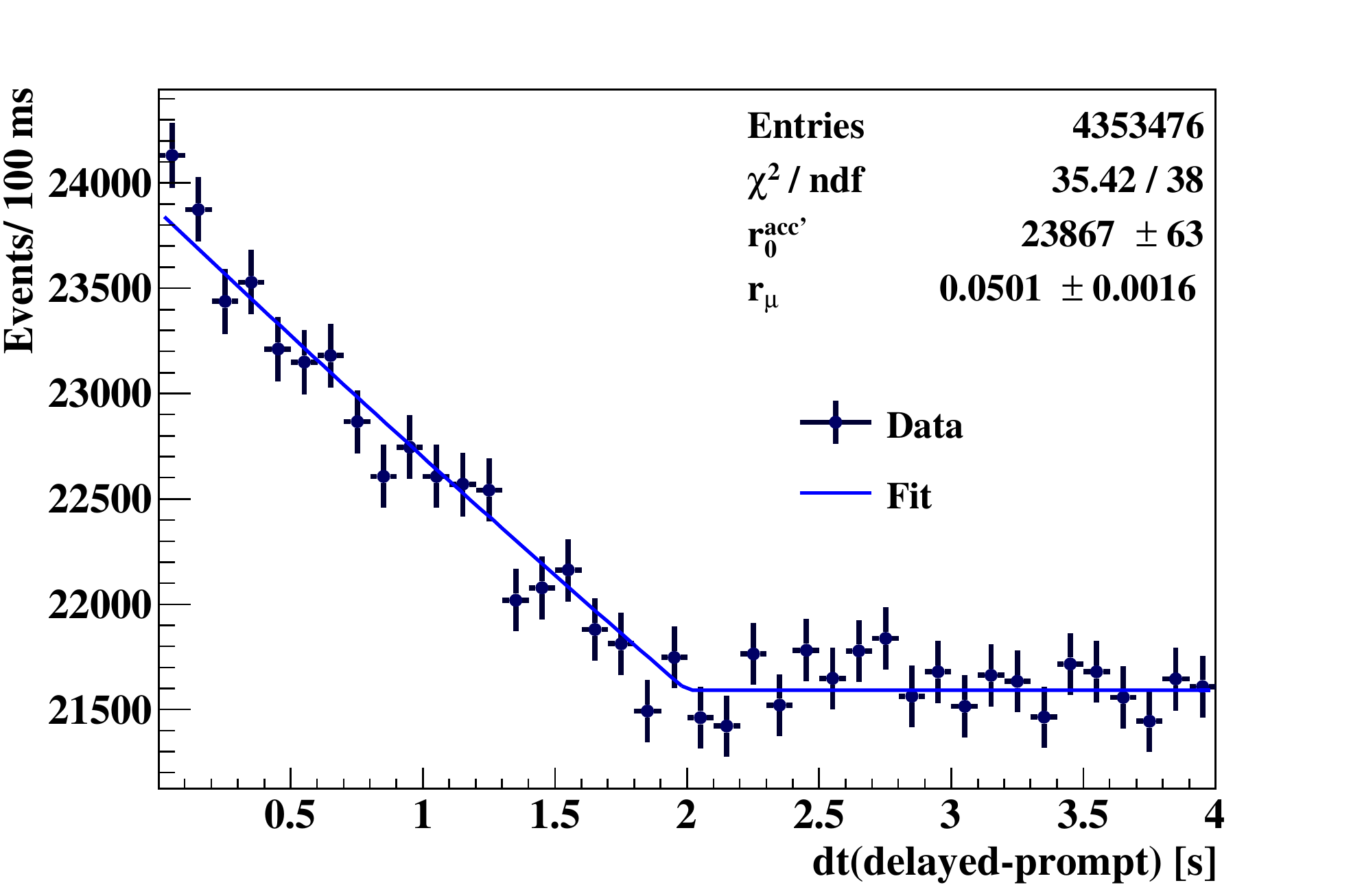}
	\label{fig:acc_2ms}}
	\subfigure[]{\includegraphics[width=0.49\textwidth]{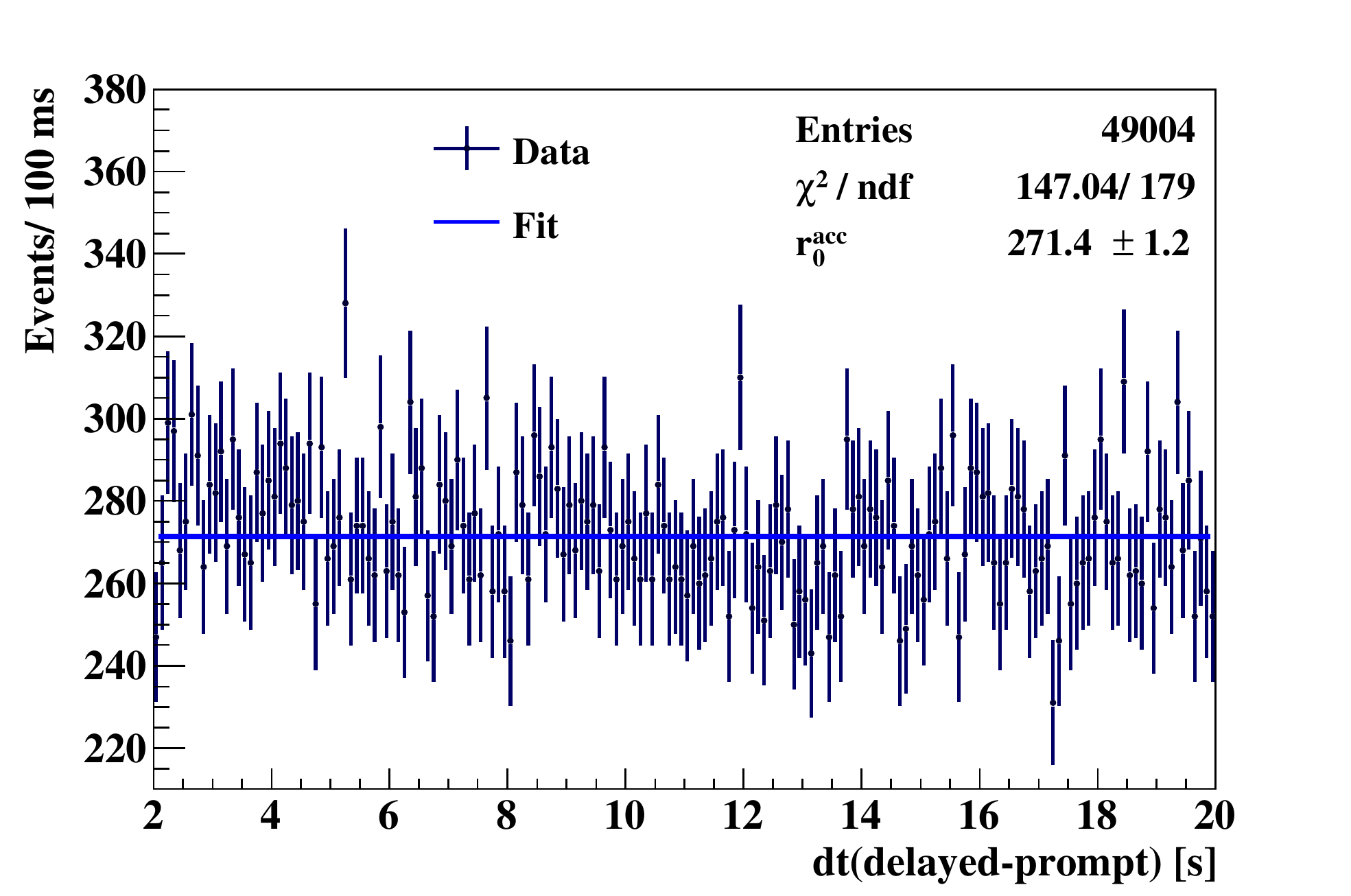}
   \label{fig:acc_2s}}
	\caption{Distribution of $dt$(delayed-prompt) for accidental coincidences (a) with relaxed selection cuts to show a decreasing trend until 2\,s and a constant trend after 2s and (b) with geoneutrino selection cuts in the time window [2\,s, 20\,s]. In both cases the search is performed by applying a 2\,s veto for all internal muons.}
\end{figure}

To evaluate the accidental background rate which is not biased by the cosmogenics, it is required not only for the prompt, but also for the delayed not to be preceded by a muon within 2\,s. This means, that once the prompt is accepted, there is no preceding muon within 2\,s before the prompt. After the prompt, as $dt$ between the prompt and a potential delayed increases, so does the probability that the delayed will be discarded due to the muon falling in between the prompt and the delayed. For time intervals longer than 2\,s, this probability becomes constant, because the muon veto is of 2\,s. This behavior is illustrated in Fig.~\ref{fig:acc_2ms}, which shows the time distribution between the prompt and delayed accidental signals in a time window $dt$ = [2\,ms, 4\,s]. One can see that until 2\,s, there is a decrease, while after 2\,s the distribution is flat. Note that this plot was constructed with relaxed selection cuts and serves only to demonstrate the suppression factor that depends only on the muon rate $r_{\mu}$ and the muon veto time. The fit function for this distribution in the interval $dt$ = [2\,ms, 2\,s]  is as follows:

\begin{equation}
    r_{\mathrm{acc'}} = r_0^{\mathrm{acc'}} \cdot \exp(-r_{\mu} \cdot dt), 
     \label{eq:dt1}
\end{equation}
where $r_0^{\mathrm{acc'}}$ would be the rate of accidental background with relaxed cuts and without the muon veto suppression factor. 
After 2\,s, the exponential suppression factor becomes constant and consequently the fit function for $dt > 2$\,s acquires a constant form:
\begin{equation}
    r_{\mathrm{acc'}} = r_0^{\mathrm{acc'}} \cdot \exp(-r_{\mu} \cdot 2\,\mathrm{s}).
     \label{eq:dt2}
\end{equation}
When fitting the spectrum with relaxed cuts, $r_{\mu}$ = (0.0501 $\pm$ 0.0016)\,s$^{-1}$ was obtained.  This is compatible with the measured rate of internal muons of (0.05311 $\pm$ 0.00001)\,s$^{-1}$. The validity of this behaviour has been also verified by a MC study.

The suppression factor $\exp(-r_{\mu} \cdot dt$) for the $dt$ of the real IBD selection is larger than 0.99993 and thus can be neglected. However, for the times $dt > 2$\,s, the suppression factor is 0.896 $\pm$ 0.003, conservatively considering also the difference between the $r_{\mu}$ resulting from the fit in Fig.~\ref{fig:acc_2ms} and just by measuring the rate of internal muons. 

In order to determine the rate of accidental coincidences $r_0^{\mathrm{acc}}$ for the geoneutrino measurement, the $dt$ = [2\,s, 20\,s] distribution of 49004 events selected with optimized IBD selection cuts was constructed, as shown in Fig.~\ref{fig:acc_2s}. This distribution is, as expected, flat and is fit with a function:
\begin{equation}
    r_{\mathrm{acc}} = r_0^{\mathrm{acc}} \cdot \exp(-r_{\mu} \cdot 2\,s).
     \label{eq:dt2-geo}
\end{equation}
The exponential suppression factor is set to 0.896 $\pm$ 0.003, the value discussed above, since the muon veto conditions are the same as in the accidental search with relaxed cuts. 

The resulting $r_0^{\mathrm{acc}}$ is $(3029.0 \pm 12.7)$\,s$^{-1}$. This means that the number of accidental coincidences among our IBD candidates can be estimated as $r_0^{\mathrm{acc}} \times 1270$\,$\mu$s, that is (3.846 $\pm$ 0.017) events.

\begin{figure} [t]
     \centering  
    \includegraphics[width = 0.49\textwidth]{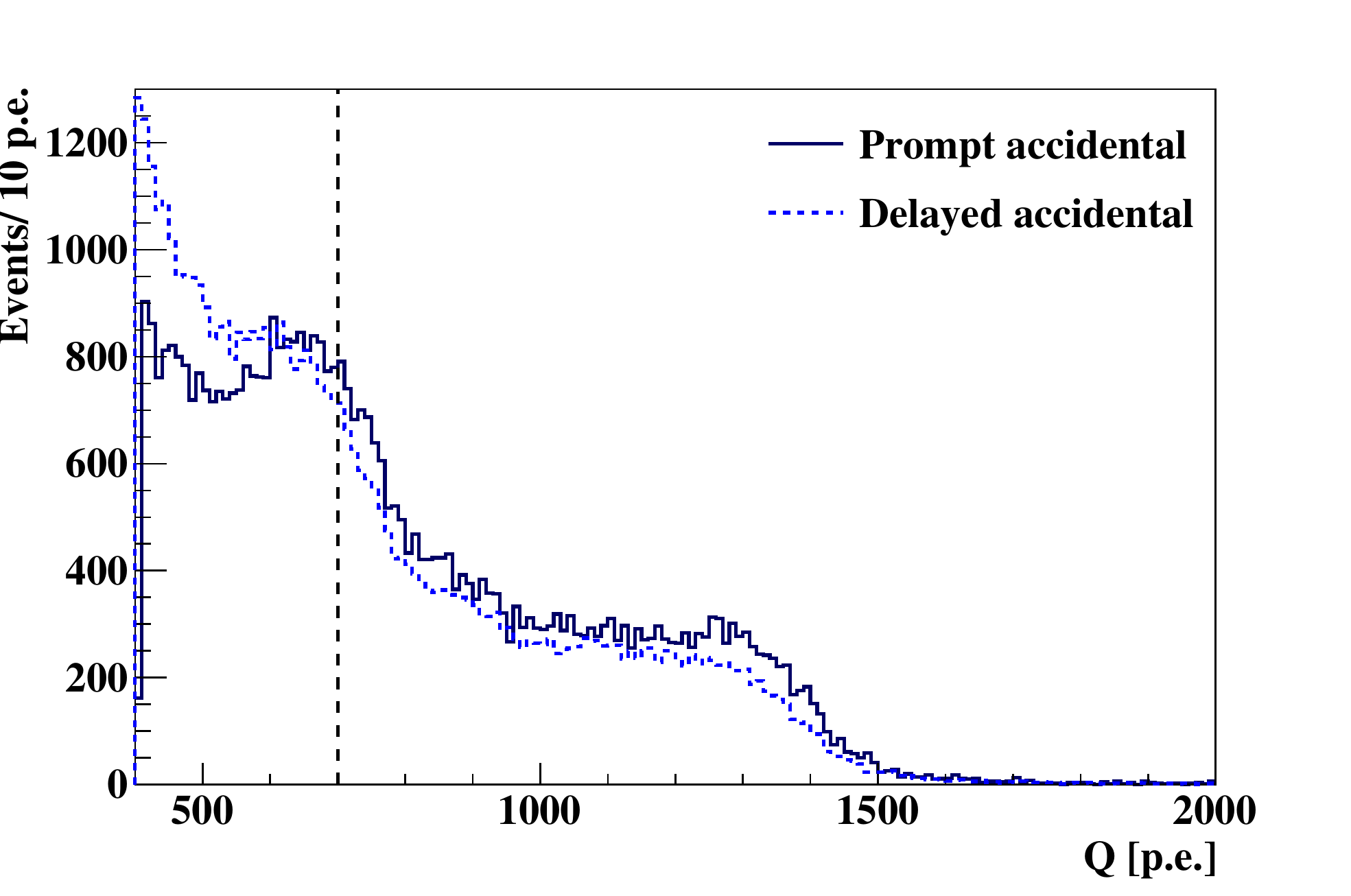} 
    \caption{The charge energy spectrum of the prompts (solid) of accidental coincidences selected in $dt$ = [2\,s, 20\,s] time window with the optimized IBD selection cuts. For the delayed signals (dotted), we show the spectrum with $Q^{\mathrm{min}}_{d}$ lowered to 400\,p.e. and scaled to the number of prompts in the solid-line spectrum. The dashed vertical line shows the chosen $Q^{\mathrm{min}}_{d}$ charge threshold of 700\,p.e.}
    \label{fig:acc_charge}
    \end{figure}

The $N_{pe}$ spectra of the prompt and delayed signals of the accidental coincidences, selected with optimized geoneutrino cuts in $dt$ = [2\,s, 20\,s] time window, are shown in Fig.~\ref{fig:acc_charge}.

  
        \subsection{($\alpha, n$) background}
        \label{subsec:alpha_n_est}
              
\begin{figure} [t]
     \centering  
     \includegraphics[width = 0.49\textwidth]{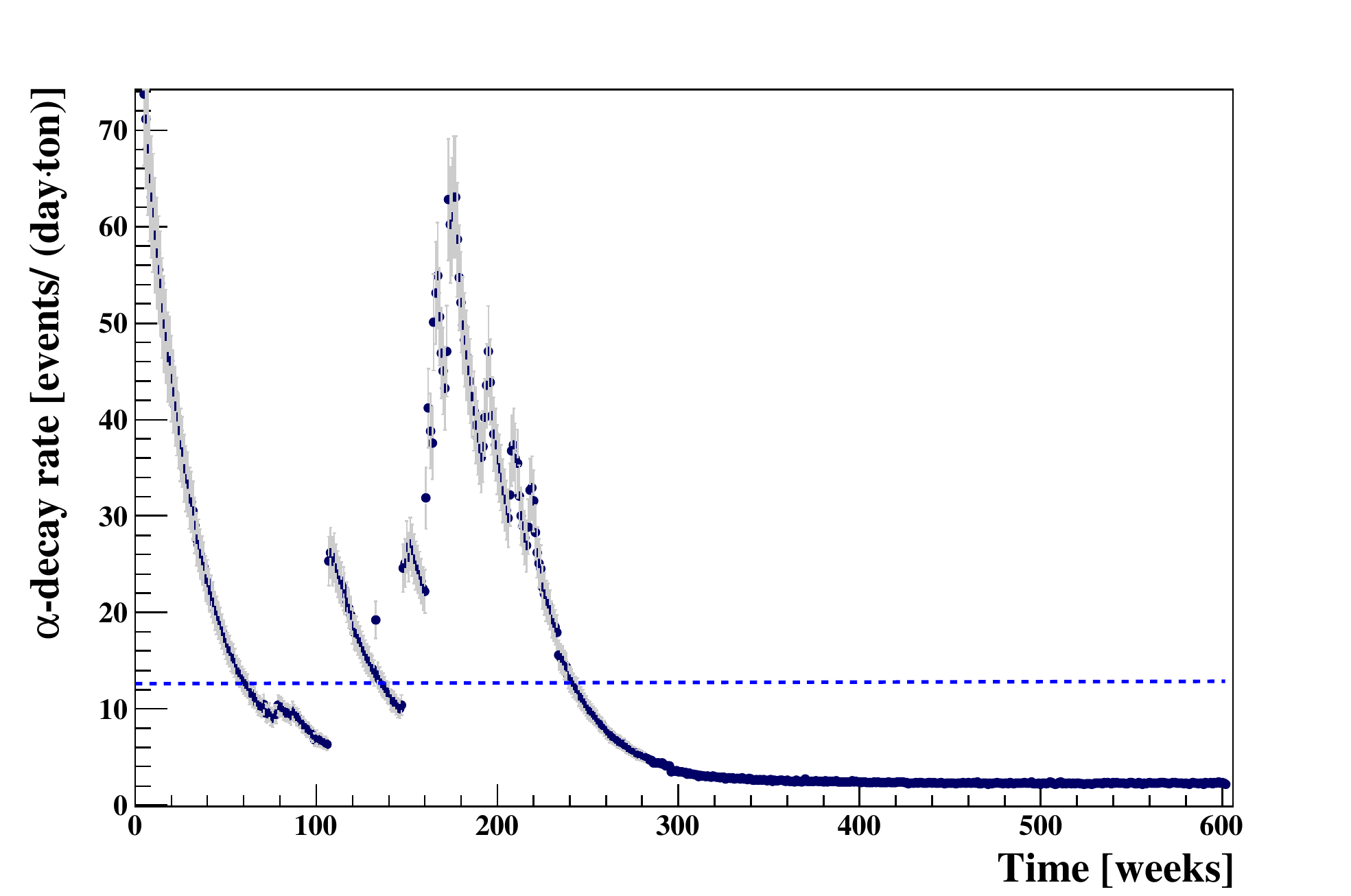}
      \caption{The evolution of the weekly $^{210}$Po $\alpha$-decay rates in the DFV of geoneutrino analysis, in the period from December 2007 to April 2019. The horizontal line shows the mean rate $\overline{R}_{\mathrm{DFV}}(^{210}\mathrm{Po})$ = (12.75 $\pm$ 0.08)\,events/(day$\cdot$ton).}
     \label{fig:210PovsTime}
    \end{figure}

	The ($\alpha, n$) background evaluation is done in three stages. First, the amount of $\alpha$ particles that could initiate this interaction is estimated. In Borexino, the only relevant isotope is $^{210}$Po that is found out of equilibrium with the rest of $^{238}$U chain~\cite{Bellini:2013lnn}. In the energy region of $^{210}$Po ($N_{pe}$ = 150 - 300\,p.e.), $\alpha$-like particles ($MLP < 0.3$) reconstructed in the DFV of the geoneutrino analysis are selected. The evolution of the weekly rates of such events for the whole analyzed period is shown in Fig.~\ref{fig:210PovsTime}. The mean rate of $\overline{R}_{\mathrm{DFV}}$($^{210}$Po) = (12.75 $\pm$ 0.08)\,events/(day$\cdot$ton) is used to evaluate the ($\alpha, n$) background from the $^{210}$Po contamination of the LS.
              
	In the second stage, the neutron yield, i.e. the probability that $^{210}$Po $\alpha$ would trigger an ($\alpha, n$) reaction in the LS, was calculated with the NeuCBOT program~\cite{Westerdale:NeuCBOT,Westerdale:2017kml,Westerdale:2018hck,Westerdale:2016}, which is based on the TALYS software for simulation of nuclear reactions~\cite{Talys:code,Talys:2008,Koning:2012zqy}. Only PC was considered as a target material. The contribution from PPO is negligible, as its relative mass fraction is small. According to a recent article~\cite{Mohr:2018alphaNBgr}, the analytical calculation of the ($\alpha, n$) cross section with TALYS provides a result similar to the experimental data for energies of the $^{210}$Po $\alpha$ particles.  This fact permits to apply as a relative uncertainty of our calculation the 15\% uncertainty of the experimental data~\citep{Mohr:2018alphaNBgr}. Taking this into account, the neutron yield $Y_n$ of the ($\alpha, n$) reaction in the PC is found to be $(1.45 \pm 0.22) \times 10^{-7}$ neutrons per a single $^{210}$Po decay. Note that the corresponding value used in the previous Borexino geoneutrino analysis, based on~\cite{McKee:2007bk}, was approximately three times smaller. Even if the $(\alpha, n)$ background is directly proportional to $Y_n$ (see Eq.~\ref{eq:alpha,n-BGR} below), this has a negligible impact on the final geoneutrino result, thanks to a high radio-purity of the Borexino scintillator.

	The final calculation of the number of IBD-like coincidences $N_{(\alpha,n)}$ triggered by the neutrons from the ($\alpha, n$) reaction over the whole analysis period can be computed using the following formula:

	\begin{equation}
		N_{(\alpha,n)} = \overline{R}_{\mathrm{DFV}}(^{210}\mathrm{Po})  \cdot \mathcal{E} \cdot Y_n \cdot\varepsilon_{(\alpha,n)},
		\label{eq:alpha,n-BGR}
	\end{equation}
    where $\mathcal{E}$ = (2145.8 $\pm$ 82.1)\,ton $\times$ yr (Sec.~\ref{subsec:exposure}) is the exposure and $\varepsilon_{(\alpha,n)} = 56\%$ is the probability of the ($\alpha, n$) interaction to produce an IBD-like signal passing all selection cuts, obtained with a full {\it G4Bx2} MC study.
    Based on this evaluation, the expected ($\alpha, n$) background due to the $^{210}$Po contamination of the LS is $(0.81 \pm 0.13)$\,events.
      
    Another potential source of background are ($\alpha, n$) reactions due to $^{210}$Po decays in the buffer. Based on a {\it G4Bx2} MC study, we have found that these interactions occurring in the outer buffer have negligible probability to create IBD-like background. However, for the interactions occurring in the inner buffer, this probability was estimated to be 0.23\% and the energy spectrum of prompts is very similar to the ($\alpha, n$) from the LS (middle panel of Fig.~\ref{fig:PDFs-backgrounds}). It is extremely difficult to determine the $^{210}$Po contamination of the buffer, since the $\alpha$ peak is completely quenched below the detection threshold. In 2009 we have estimated this contamination as $<$0.67\,mBq/kg~\cite{Bellini:2010geo} by employing the samples of buffer liquids in the center of the Counting Test Facility of Borexino~\cite{ALIMONTI1998411}, that not any more operational. This limit is several orders of magnitude above the contamination of the LS. DMP quencher, that is only present in the buffer, is considered to be the main source of the $^{210}$Po contamination in the buffer. In January 2010, the DMP concentration in the buffer was reduced to 2\,g/l (the original concentration was 5\,g/l), as discussed in Sec.~\ref{sec:det}. Since then, no further operations have been performed with the buffer and the $^{210}$Po contamination is expected only to decay ($\tau = 199.6$\,day) and to be suppressed in April 2019 by a factor $3.9\times10^{-8}$. In the present analysis, the estimated upper limit for this contamination is 0.14\,mBq/kg, which corresponds to an upper limit of 2.6 background events (from which only 0.3 events in the period from January 2010). We note however, that the original estimate of the $^{210}$Po rate in the buffer is very conservative, because of high risk of contamination of the samples during their handling. As it will be discussed in Sec.~\ref{subsec:golden_candidates}, the golden IBD candidates are evenly distributed in time and no excess close to the IV is observed.

        \subsection{($\gamma$, n) interactions and fission in PMTs}
        \label{subsec:gamma_n_est}
           
\begin{figure}[!t]
\includegraphics[width=0.49\textwidth]{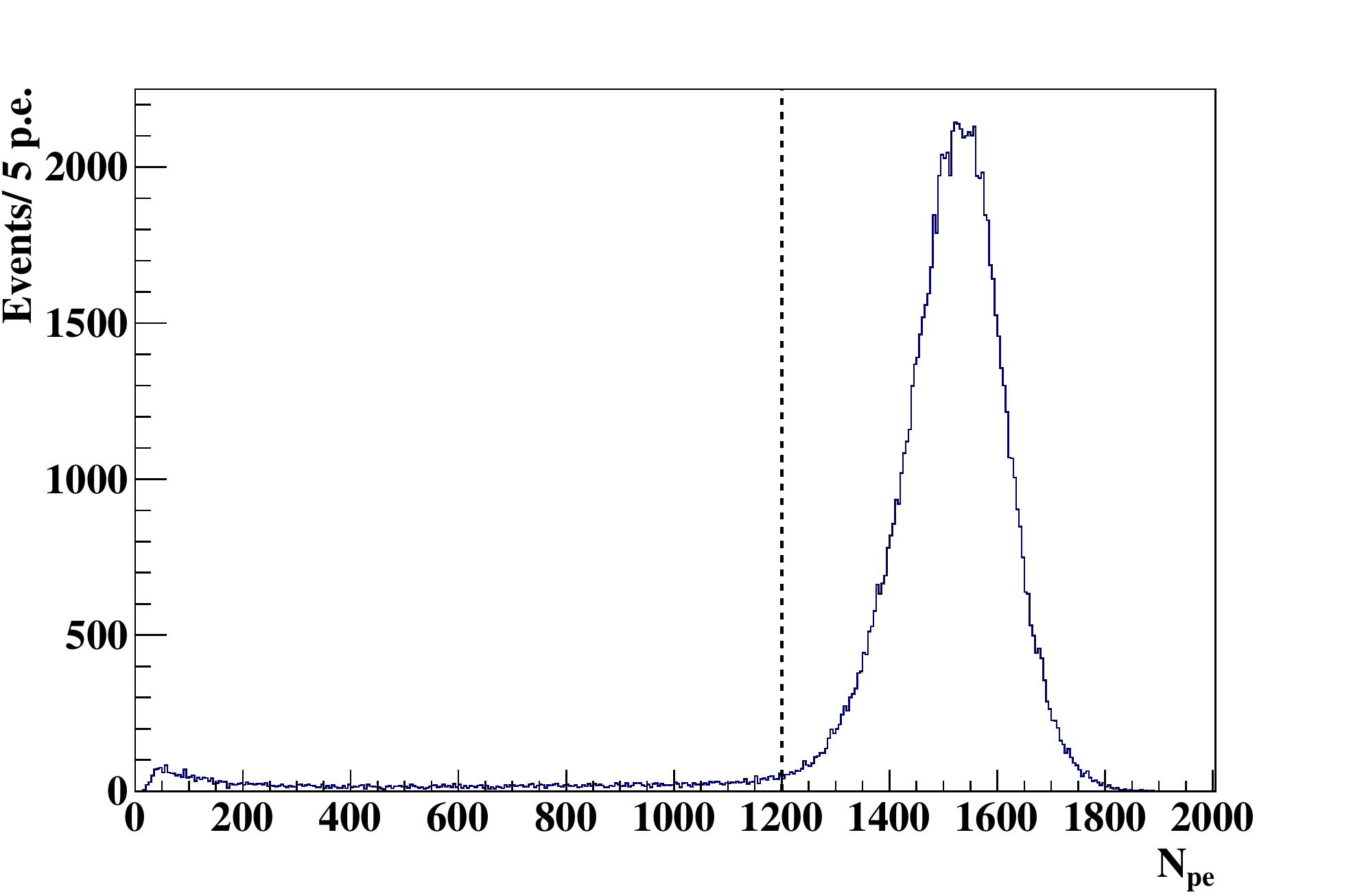}
\caption{The $N_{pe}$ distribution of 3\,MeV $\gamma$s generated in the DFV  and in the entire analyzed period of geoneutrino analysis. The vertical line indicates the 1200\,p.e. threshold used in the evaluation of the ($\gamma, n$) background.} 
\label{fig:3MeV}
\end{figure}

In order to obtain an upper limit to the possible background from ($\gamma$, n) reactions in the Borexino scintillator or in surrounding materials, we counted all the registered events with energies higher that 3\,MeV and we made the conservative assumption that they are only due to $\gamma$-ray interactions. Since the energy response of the detector is not uniform in space and time, an energy release of 3\,MeV does not correspond to a unique value of the registered charge $N_{pe}$. To consider this effect, 3\,MeV $\gamma$s have been generated with the {\it G4Bx2} MC code following the detector status during the whole analyzed period. According to Fig.~\ref{fig:3MeV}, a conservative charge threshold of 1200\,p.e. was chosen and a correction of 5.4\% for the inefficiency of the cut was then applied: 589,917 events were selected above 1200\,p.e., resulting in 623,571 hypothetical $\gamma$-rays after the correction.
Each of them can only interact  with the deuterons that meets along its path before being absorbed: an estimation for this background is then obtained by multiplying the numbers of gammas for the deuteron density, the interaction cross section, and the gamma's absorption length. According to the $\gamma$-ray attenuation coefficients calculated for the Borexino scintillator, the absorption length $\lambda$ for a 3\,MeV gamma is 29\,cm and the capture cross section on $^{2}$H is $\sigma_D$ = 1.6\,mb. Since the deuteron density is $\rho_D$ = 7.8~$\times$ 10$^{18}$ atoms/cm$^{3}$, the upper limit on the number of events $N_{(\gamma,n)}$ due to this background, taking into account the estimated detection efficiency $\varepsilon_{(\gamma,n)}$ = 50\%, is:

\begin{equation}
	N_{(\gamma,n)} < N_{\gamma} \cdot \rho_{D} \cdot \sigma_D \cdot 3\lambda \cdot \varepsilon_{(\gamma,n)} = 0.34 \, \mathrm{events}.
\label{gam_n}
\end{equation}
An attenuation length of 3$\lambda$ was chosen to obtain the 95\% C.L.
A possible contribution of neutron capture on $^{13}$C and $^{12}$C nuclei was also considered, but it was found to be more that a factor 10 smaller and therefore, neglected. 

In PMTs we have determined a $^{238}$U contamination of $(31 \pm 2)$\,ppb in the glass and $(60 \pm 4)$\,ppb in the dynodes. PMTs are located at about 6.85\,m from the center of the detector. To estimate the background induced by spontaneous fission, we consider that for each PMT the glass accounts for about 0.3\,kg and the dynodes for 0.05\,kg. In addition, we take into account the subtended solid angle by the IV and the neutron attenuation while propagating from the PMTs to the IV, which is of the order of 1\,m. The estimated number of neutrons reaching the scintillator is $(0.057\pm0.004)$ for the current exposure. The corresponding neutron-induced background will be negligibly small and we set for it a conservative upper limit of 0.057 events.

        \subsection{Radon background}
            \label{subsec:radon_est}
 \begin{figure} [t]
     \centering  
    \includegraphics[width = 0.49\textwidth]{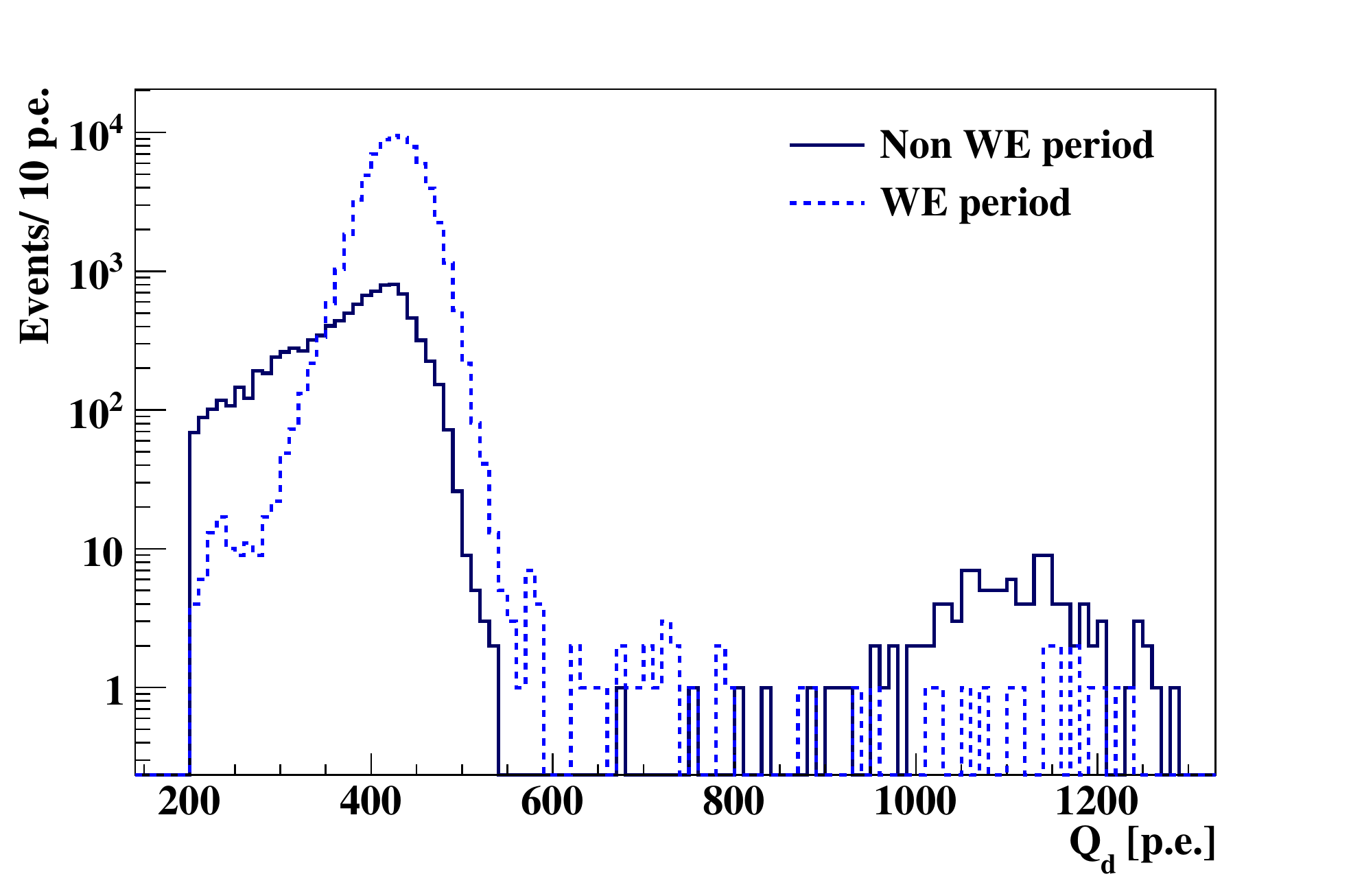}
        \caption{Charge distribution of delayed signals selected with the low-energy threshold of 200 p.e. and without any MLP cut for the WE period (dotted blue) and the rest of the data-taking (solid blue). The neutron capture peak of the IBDs can be seen at around 1100\,p.e. The peak at 400\,p.e. is the main $\alpha$ decay of $^{214}$Po due to the Radon events. The handful of events in between 600 and 860\,p.e. are due to the ($\alpha$ + $\gamma$) decay branch of $^{214}$Po, as shown in Table~\ref{tab:214Po}.}
        \label{fig:charge_we}
    \end{figure}

In Section~\ref{subsec:radon} we have discussed how the radon contamination of the LS can induce IBD-like background. Figure~\ref{fig:charge_we} demonstrates the increased Radon contamination during the WE period. A proper choice of the IBD selection cuts is extremely useful to reduce this kind of background and to safely include the WE period in the geoneutrino analysis.

\begin{figure}[h]
     \centering  
    \subfigure[]{\includegraphics[width = 0.49\textwidth]{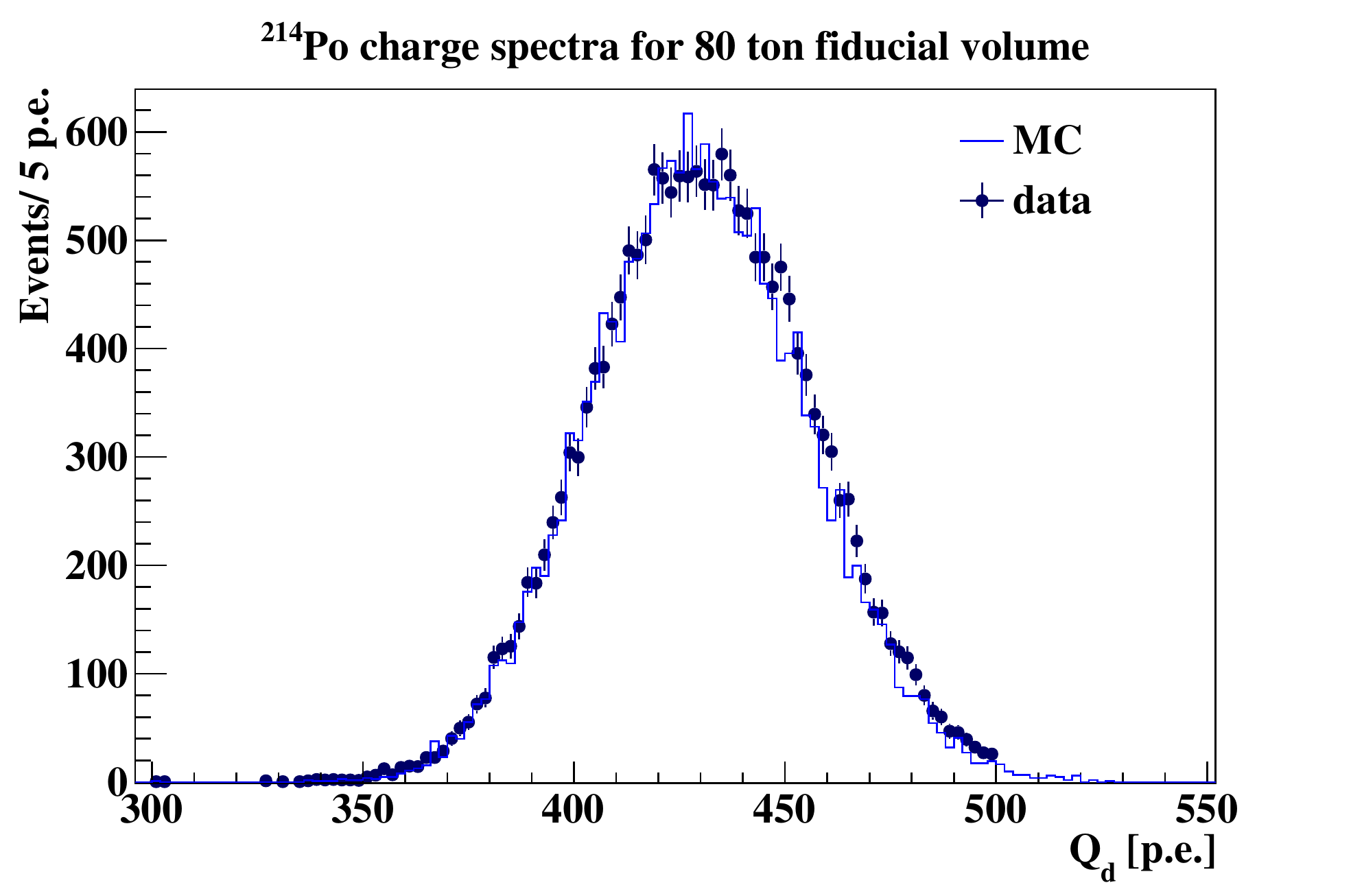}
    \label{fig:214Po_FV}}
    \subfigure[]{\includegraphics[width = 0.49\textwidth]{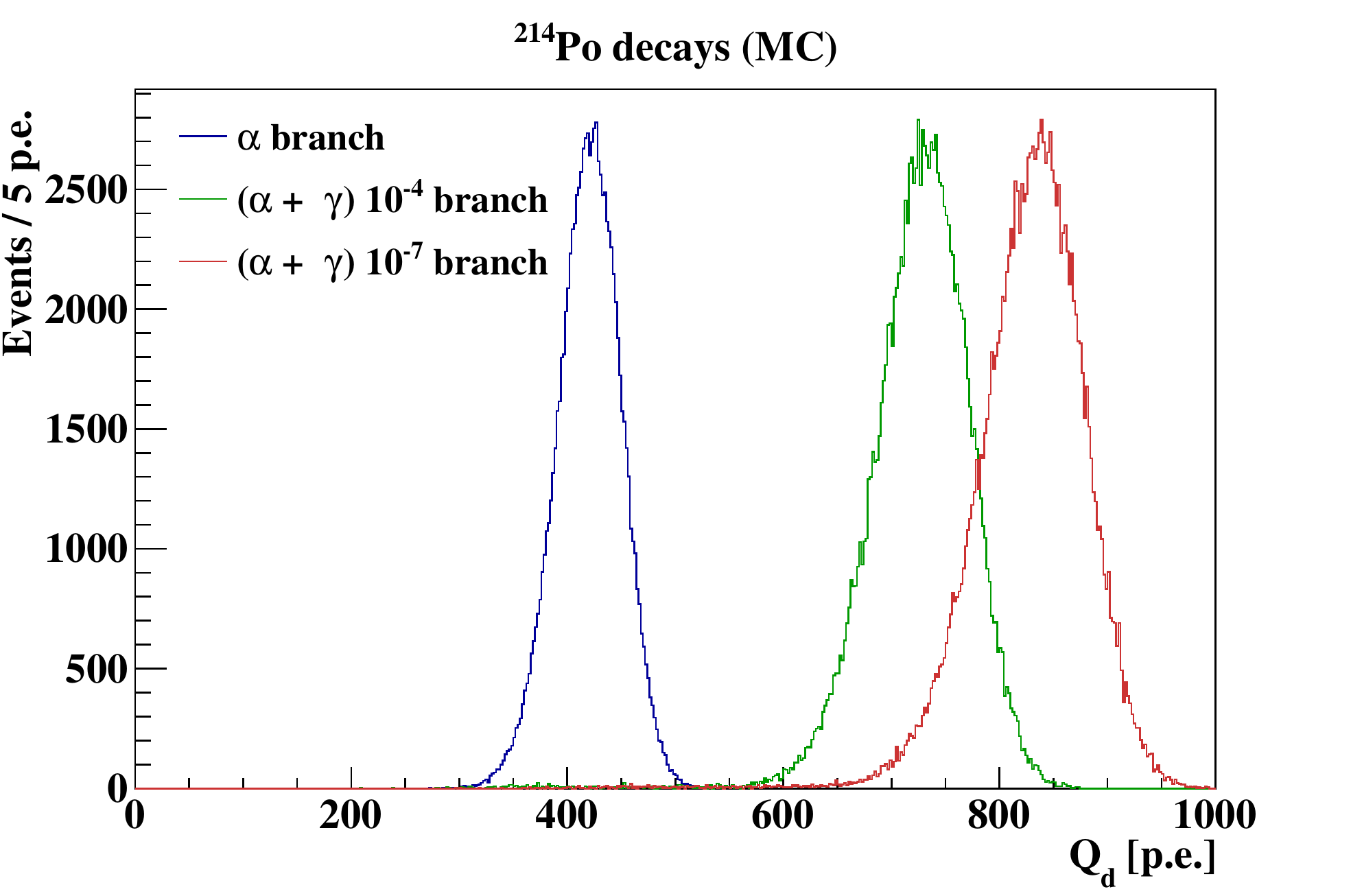}
    \label{fig:Radon_alphas}}
        \caption{Top: Comparison of the charge spectra of $\alpha$ decays from $^{214}$Po from data (circles with error bars) and MC (solid line) in the fiducial volume of $\sim$\,80 tons around the detector center chosen to tune the alpha particle quenching factor. Bottom: Charge distributions for the three $^{214}$Po decays (Table~\ref{tab:214Po}) obtained using MC with the $\alpha$ energy scale tuned on pure $\alpha$-decays (Fig.~\ref{fig:214Po_FV}). The main $\alpha$ decay (red line) and the two sub-dominant $(\alpha + \gamma)$ branches are shown: 10$^{-4}$ (green line) and 10$^{-7}$ (blue line) branch. The three spectra are normalised to have the same area.}
        \label{fig:214Po_tuned}
    \end{figure}

The energy scale of the $(\alpha + \gamma)$ decays of $^{214}$Po (Table~\ref{tab:214Po}) was evaluated. The energy scale in {\it G4Bx2}, including the overall light yield and the Birk's constant $kB$ important in the description of quenching (see Sec.~VII of~\cite{Bellini:2013lnn}), is tuned based on the calibration with $\gamma$ sources. Since the $kB$ is particle dependent, the energy scale for $\alpha$s must be further adjusted. For this purpose, the dominant pure $\alpha$ decay of $^{214}$Po was simulated and compared to the Radon events from the data (selected via the $^{214}$Bi$^{214}$Po delayed coincidence tag), as demonstrated in Fig.~\ref{fig:214Po_FV}. With both $\gamma$ and $\alpha$ energy scales fixed, the spectra for $(\alpha +\gamma)$ $^{214}$Po decays were simulated, as it is shown in Fig.~\ref{fig:Radon_alphas}. 
Since the overall radon statistics amounts to 1.1~$\times$~10$^5$ decays, the events due to the 10$^{-7}$ branch can be neglected, while $\sim$11 events are expected from the 10$^{-4}$ branch, when $^{214}$Po decays to $^{210}$Pb in the first excited state. In this case, the de-excitation gamma is emitted along with an $\alpha$-particle. In order to suppress these events, we keep the $Q^{\mathrm{min}}_{d}$ threshold fixed to 860 p.e. during the WE period, which effectively reduces this background by a factor of 10$^{3}$. Application of the pulse shape cut (MLP $>$ 0.8) on the delayed (Table~\ref{tab:sel_cuts}) further reduces the background by a factor of 5-6. During the analysis of non-WE period, the 10$^{-4}$ branch also becomes negligible, hence we can lower $Q^{\mathrm{min}}_{d}$ to 700\,p.e., safely above the $^{214}$Po $\alpha$-peak (Fig.~\ref{fig:Radon_alphas}). The total number of background events correlated with radon contamination is expected to be (0.003 $\pm$ 0.001), which is completely negligible.

        \subsection{$^{212}$Bi-$^{212}$Po background}
            \label{subsec:212BiPo_est}

\begin{figure}[t]
     \centering  
    \includegraphics[width = 0.49\textwidth]{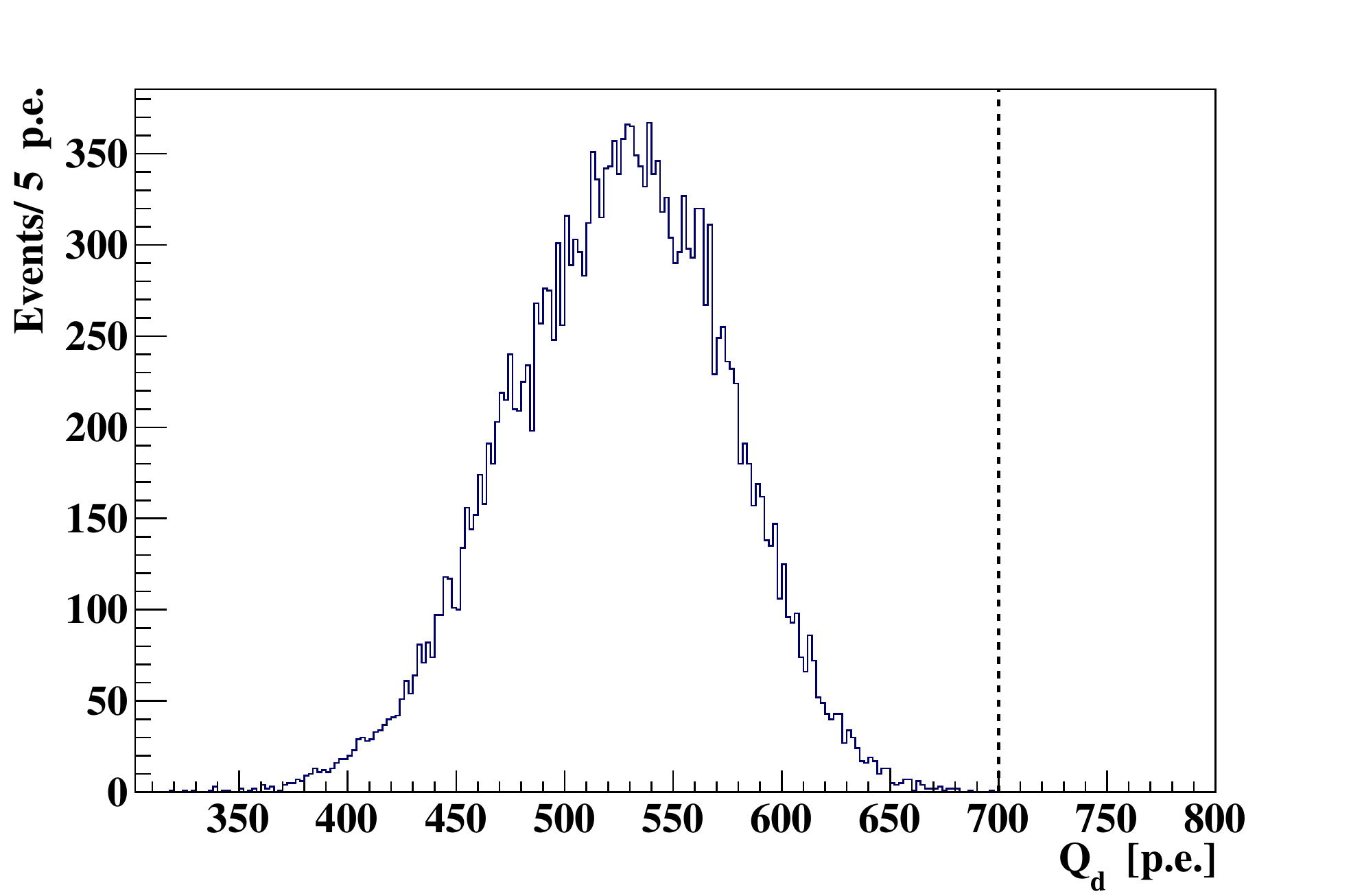}
        \caption{Charge spectrum of MC generated $^{212}$Po $\alpha$ peak. The vertical line at 700\,p.e. shows the optimized $Q^{\mathrm{min}}_{d}$ for IBD selection.}
        \label{fig:212BiPo}
    \end{figure}

A MC study proves that the cut on $Q^{\mathrm{min}}_{d} = 860$\,p.e., adopted to reject the radon contamination during the WE periods, is effective in removing the $^{212}$Bi-$^{212}$Po fast coincidences, as shown in Fig.~\ref{fig:212BiPo}. It can be seen that the end point of the $^{212}$Po $\alpha$ peak is around 700\, p.e. Therefore, a $Q^{\mathrm{min}}_{d}$ of 700\,p.e., combined with the MLP pulse shape cut on the delayed, makes the overall $^{212}$Bi-$^{212}$Po background fully negligible in geoneutrino analysis.
   
  \subsection{Summary of the estimated non-antineutrino background events}
    \label{subsec:bckg_sum_est}  
          
    Table~\ref{tab:backg_summary} summarizes the expected number of events from all non-antineutrino backgrounds passing the optimised selection cuts listed in Table~\ref{tab:sel_cuts}.

\begin{table} 
	\centering
	\caption{\label{tab:backg_summary} Summary of the expected number of events from non-antineutrino backgrounds in the antineutrino candidate sample (exposure $\mathcal{E}_p$ = (1.29 $\pm$ 0.05) $\times 10^{32}$\, protons $\times$ yr). The limits are 95\% C.L.} \vskip 2pt
	\begin{tabular*}{\columnwidth}{l @{\hskip 50pt} c}
		\hline
		\hline
		Background Type & Events \Tstrut\Bstrut  \\
		\hline
	    $^9$Li background& 3.6 $\pm$ 1.0 \Tstrut \\
		Untagged muons & 0.023 $\pm$ 0.007 \\
		Fast n's ($\mu$ in WT) & $<$0.013 \\
		Fast n's ($\mu$ in rock) & $<$1.43  \\
		Accidental coincidences & 3.846 $\pm$ 0.017 \\
		($\alpha$, n) in scintillator & 0.81 $\pm$ 0.13 \\ 
		($\alpha$, n) in buffer & $<$2.6  \\
		($\gamma$, n) & $<$0.34 \\
		Fission in PMTs & $<$0.057   \\
		$^{214}$Bi-$^{214}$Po & 0.003 $\pm$ 0.001 \Bstrut \\
		\hline 
		Total & 8.28 $\pm$ 1.01 \Tstrut\Bstrut \\			
		\hline
		\hline
	\end{tabular*}	
\end{table}

    \section{SENSITIVITY TO GEONEUTRINOS}
    \label{sec:sensitivity}
    
This Section describes the Borexino sensitivity to geoneutrinos and the MC based procedure with which it was evaluated. In Sec.~\ref{subsec:nutshell} the description of the basic ingredients of the analysis focuses on the spectral fit of the $Q_{\rm{p}}$ spectrum. Section~\ref{subsec:senstool} describes the sensitivity tool that performs such fits on 10,000 $Q_{\rm{p}}^{\rm {MC}}$ spectra, each corresponding to a MC generated pseudo-experiment. The expected precision for the Borexino geoneutrino measurement as well as its sensitivity to the mantle signal is discussed in Sec.~\ref{subsec:expsens}. This approach was also used in the optimization of the selection cuts, as mentioned in Sec.~\ref{sec:data_sel}. The systematic uncertainties given in Sec.~\ref{subsec:syst} are not included in the sensitivity studies and are only considered in the final results of Sec.~\ref{sec:results}. As it will be shown, the error on the geoneutrino measurement is largely dominated by the statistical error.

        \subsection{Geoneutrino analysis in a nutshell}
        \label{subsec:nutshell}

The geoneutrino signal is extracted from the spectral fit of the charges of the prompts of all selected IBD candidates. Since the number $N_{\mathrm {IBD}}$ of selected candidates is relatively small (in this analysis, $N_{\mathrm{IBD}} = 154$ candidates, see Sec.~\ref{subsec:golden_candidates}), an unbinned likelihood fit is used: 
\begin{equation}
  L = (\vec{\theta}; \vec{Q}_p) = \prod_{i=1}^{N_{\mathrm {IBD}}} L(\vec{\theta}; Q_p^i),
  \label{eq:Lkl}
\end{equation}
where $\vec{Q}_p$ is the vector of individual prompt charges $Q_p^i$, and index $i$ runs from 1 to $N_{\mathrm IBD}$. The symbol $\vec{\theta}$ indicates the set of the variables with respect to which the function is maximized, namely the number of events corresponding to individual spectral components with known shapes. In particular, we fit the number of geoneutrino and reactor antineutrino events as well as the number of events from several background components. The shapes of all spectral components are taken from the MC-constructed PDFs (see Figs.~\ref{fig:PDFs-Geo-Rea},~\ref{fig:PDFs-backgrounds}), with the exception of the accidental background, which can be measured with sufficient precision as shown in Fig.~\ref{fig:acc_charge} (prompt spectrum). Some of the spectral components are kept free (typically geoneutrinos and reactor antineutrinos), while others (typically other than reactor antineutrino backgrounds) are constrained using additional multiplicative Gaussian pullterms in the likelihood function of Eq.~\ref{eq:Lkl}. 

Naturally, the number of geoneutrinos is always kept free. One way of doing it is by having one free fit parameter for geoneutrinos, when we use the PDF in which the $^{232}$Th and $^{238}$U contributions are summed and weighted according to the chondritic mass ratio of 3.9, corresponding to $R_s$ signal ratio of 0.27 (Sec.~\ref{subsec:geo}). Alternatively, $^{232}$Th and $^{238}$U contributions can be fit as two independent contributions. Additional combinations are of course possible. For example, in the extraction of the geoneutrino signal from the mantle (Sec.~\ref{subsec:mantle}), we constrain the expected lithospheric contribution, while keeping the mantle contribution free.

The number of reactor antineutrino events is typically kept free. It is an important cross-check of our ability to measure electron antineutrinos, when we compare the unconstrained fit results (Sec.~\ref{subsubsec:UTh_fixed}) with the relatively-well known prediction of reactor antineutrino signal (Sec.~\ref{subsec:rea}). In addition, an eventual constraint on reactor antineutrino contribution does not significantly improve the precision of geoneutrinos, as verified and discussed below. The constrained reactor antineutrino signal is however used when extracting the limit on the hypothetical georeactor (Sec.~\ref{subsec:georeactor}), as it will be discussed in Sec.~\ref{subsec:georeactor-results}.

Typically, we include the following non-antineutrino backgrounds in the fit: cosmogenic $^9$Li, accidental coincidences, and ($\alpha$, n) interactions. Atmospheric neutrinos are included in the calculation of systematic uncertainties, as it will be described in Sec.~\ref{subsec:syst}. These background components are constrained in the fit, since independent analyses can yield the well constrained estimates of their rates, as they are summarized in Table~\ref{tab:backg_summary} for non-antineutrino backgrounds and in Table~\ref{tab:antinu-events-expected} given for atmospheric neutrinos. 

        \subsection{Sensitivity study}
        \label{subsec:senstool}
  
 A Monte Carlo approach was used in order to estimate the Borexino sensitivity to geoneutrinos, as well as to optimize the IBD selection cuts (Sec.~\ref{sec:data_sel}). This so-called {\it sensitivity study} can be divided in the following four steps:
%

\begin{figure*} [th]
  	\centering
   \includegraphics[width = \textwidth]{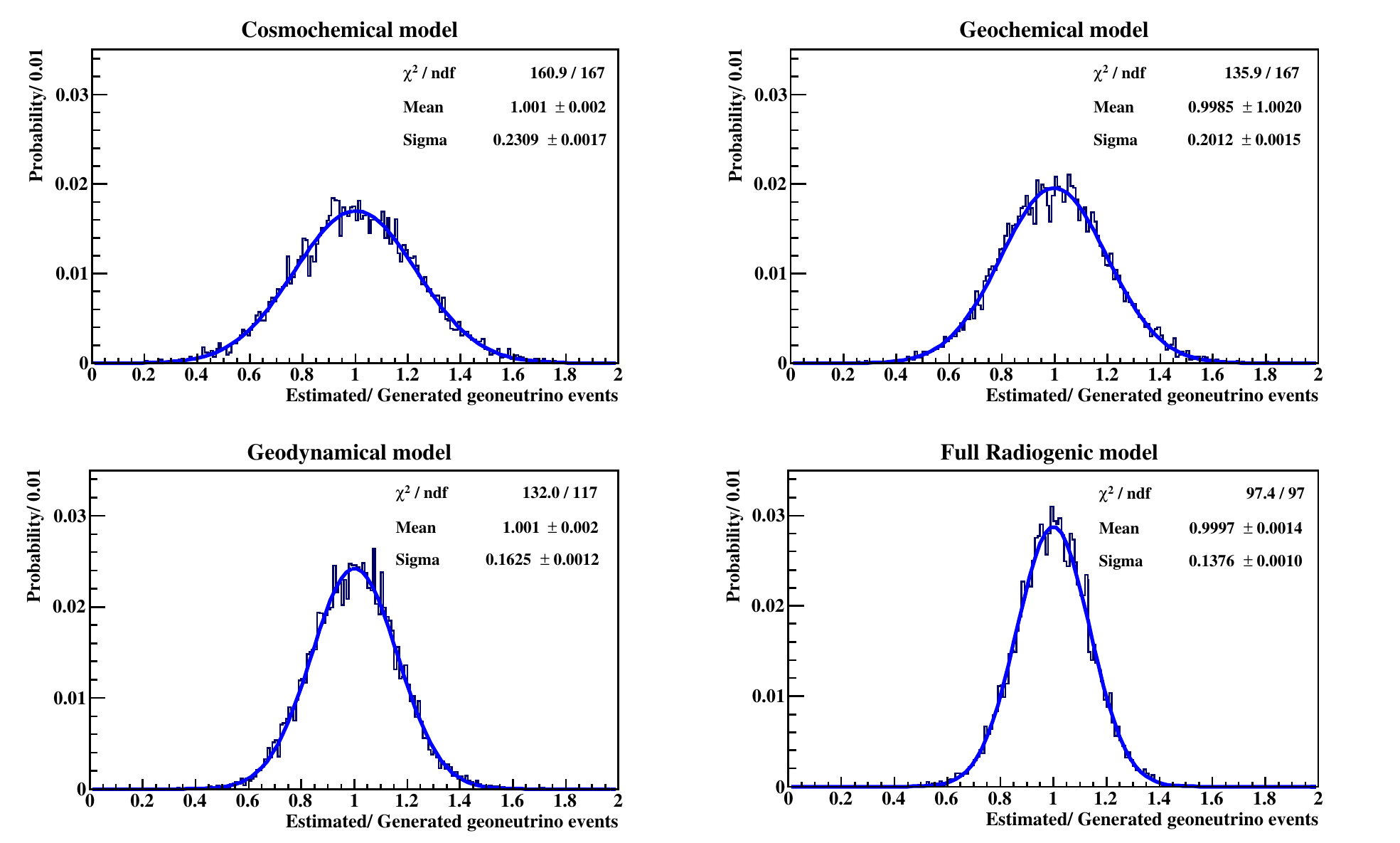}
  	\caption{Results of the sensitivity study for geoneutrinos considering the conditions of the presented analysis: the PDFs for the ratio {\it Estimated / Generated geoneutrino events} for the CC, GC, GD, and FR Earth models and correspoding fits. Each PDF is based on 10,000 generated spectra. The Gaussian fits are all centered at one, so no systematic bias is expected; their $\sigma$'s vary from (13.76 $\pm$ 0.10$)\%$ to (23.09 $\pm$ 0.17$)\%$, depending on the expected geoneutrino signal for different BSE models (Table~\ref{tab:antinu-events-expected}) and represent the expected statistical uncertainty of the measurement.}
  	\label{fig:optsel_sens}
  \end{figure*}
  
\begin{itemize}
        \item The arrays of charges of prompts for signal and backgrounds are generated from the PDFs including the detector response that were either created by the full {\it G4Bx2} MC code (Sec.~\ref{subsec:MCspectra}, Figs.~\ref{fig:PDFs-Geo-Rea} and \ref{fig:PDFs-backgrounds}) or measured, as for accidental background (Fig.~\ref{fig:acc_charge}, prompt spectrum). For each component, the number of generated charges is given by the expectations, as shown for antineutrino signals in Table~\ref{tab:antinu-events-expected} and for non-antineutrino backgrounds in Table~\ref{tab:backg_summary}. 
        \item{The generated spectra are fit in the same way as the data (Sec.~\ref{subsec:nutshell}), using in the fit the same PDFs that were used for the generation of these pseudo-experiments. This means, uncertainty due to the shape of the spectral components is not considered. This is justified by the fact, that Borexino's sensitivity to geoneutrinos is by far dominated by the statistical uncertainty.}
        \item{The procedure is repeated 10,000 times for each configuration. In each pseudo-experiment, the number of generated events for signal and all backgrounds is varied according to the statistical uncertainty.}
        \item{The distributions of ratios of the resulting fit value (estimated) over the MC-truth (generated) value in each individual fit are constructed for the parameters of interest. For example, such a distribution for the ratio of the number of geoneutrinos estimated from the fit over the number of generated geoneutrinos should be centered at one (when there is no systematic bias), while the width of this distribution corresponds to the expected statistical uncertainty of the measurement.}
\end{itemize}

      \subsection{Expected sensitivity}
      \label{subsec:expsens}
  
Using the sensitivity tool as explained in Sec.~\ref{subsec:senstool}, the expected statistical uncertainty of the Borexino geoneutrino measurement in the presented analysis varies from (13.76 $\pm$ 0.10$)\%$ to (23.09 $\pm$ 0.17$)\%$, depending on the expected signal for different geological models (Table~\ref{tab:antinu-events-expected}), as demonstrated in Fig.~\ref{fig:optsel_sens}. This study assumes the Th/U chondritic ratio to hold. In the previous 2015 Borexino geoneutrino analysis~\cite{Agostini:2015cba}, the statistical error was $\sim$26.2\%. 

\begin{figure} [t]
  	\centering
  	\vspace{1mm}
  	\includegraphics[width = 0.49\textwidth]{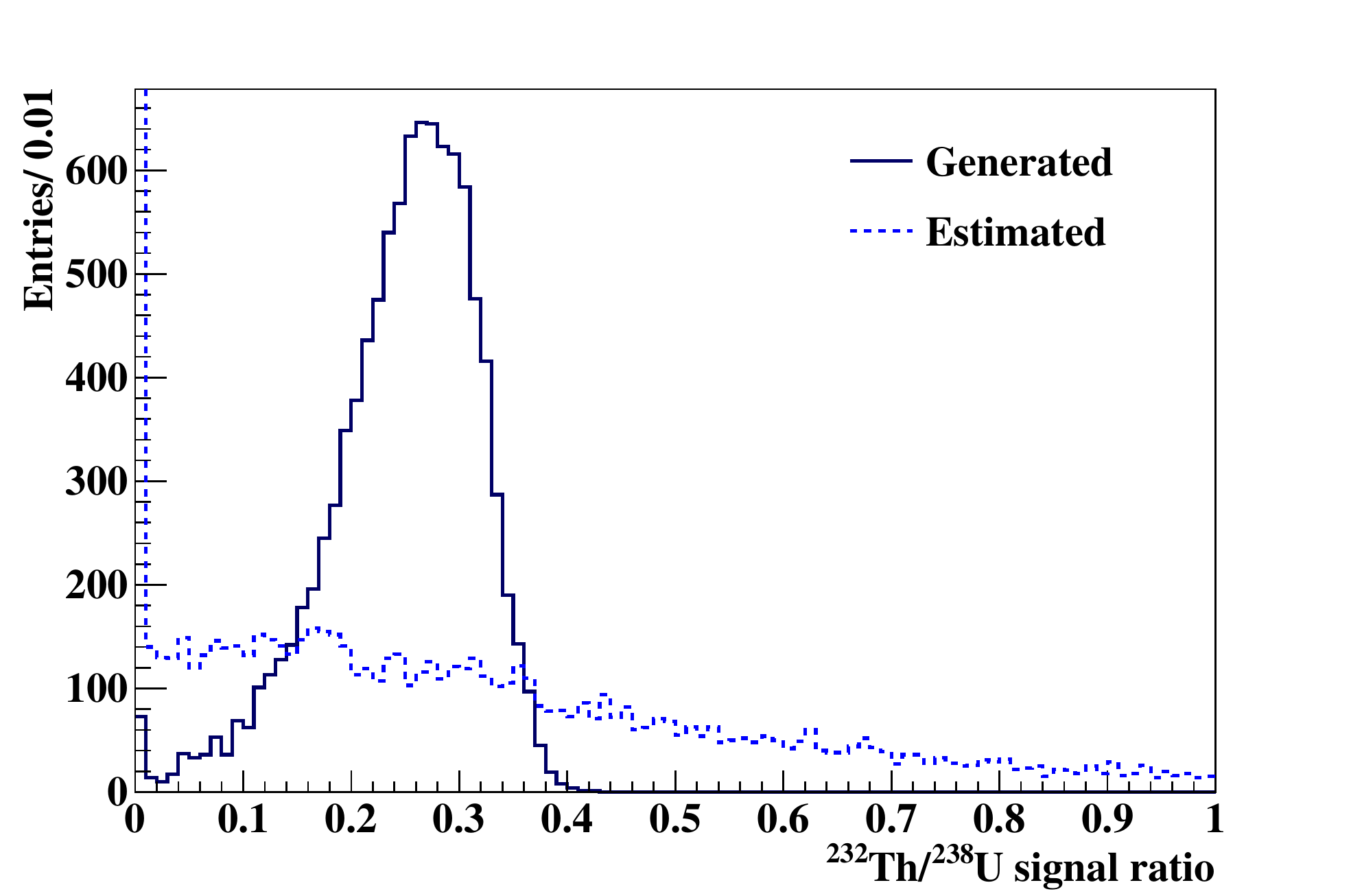}
  	\caption{Demonstration of no sensitivity of Borexino to measure $^{232}$Th/$^{238}$U geoneutrino signal ratio. Solid line shows the distribution of this ratio, assuming its chondritic value, for 10,000 generated pseudo-experiments. Distribution of this ratio, as obtained from the fit (dotted line), is nearly flat, with a clear peak at 0, due to the $^{232}$Th contribution railed to 0.}
  	\label{fig:UTh_sens}
  \end{figure}

The sensitivity of Borexino to measure the $^{232}$Th/$^{238}$U ratio was also studied. As it is shown in Fig.~\ref{fig:UTh_sens}, Borexino does not have any sensitivity to determine this ratio. Despite the input ratio assuming the chondritic value (considering the statistical fluctuations), the $^{232}$Th/$^{238}$U ratio resulting from the fit has nearly a flat distribution for the 10,000 pseudo-experiments. This will be also confirmed by large $^{232}$Th versus $^{238}$U contours shown in Fig.~\ref{fig:U_vs_Th_contour} for the fit of the data with free $^{238}$U and $^{232}$Th components.

The sensitivity of Borexino to measure the mantle signal was studied using the log-likelihood ratio method~\cite{Cowan2011} for the expectations according to four different geological models (CC, GC, GD, and FR, Table~\ref{tab:antinu-events-expected}). For each geological model, we have generated a set of 10,000 pseudo-experiments with the mantle geoneutrino component included. In addition, we have generated 1.2 million pseudo-experiments without the mantle contribution. In each data set, we have included the relatively-well known lithospheric contribution (Table~\ref{tab:S_litho}), as well as the reactor antineutrino ``without 5\,MeV excess" (Table~\ref{tab:antinu-events-expected}) and non-antineutrino backgrounds (Table~\ref{tab:backg_summary}).

Each pseudo-experiment from all five data sets (one without the mantle and four with mantle signal according to four geological models), are fit twice: with and without the mantle contribition. The best fit with the mantle contribution fixed to zero corresponds to the likelihood $L\{0\}$. The fit with the mantle component left free results in the likelihood $L\{\mu\}$. 
Obviously, for the data set without the mantle being generated, the two likelihoods tend to be the same. For the data sets with the mantle included, the $L\{0\}$ tends to be worse than $L\{\mu\}$: the bigger this difference, the better the sensitivity to observe the mantle signal. 

\begin{figure*}[h]
  	\centering
  	\includegraphics[width =  \textwidth]{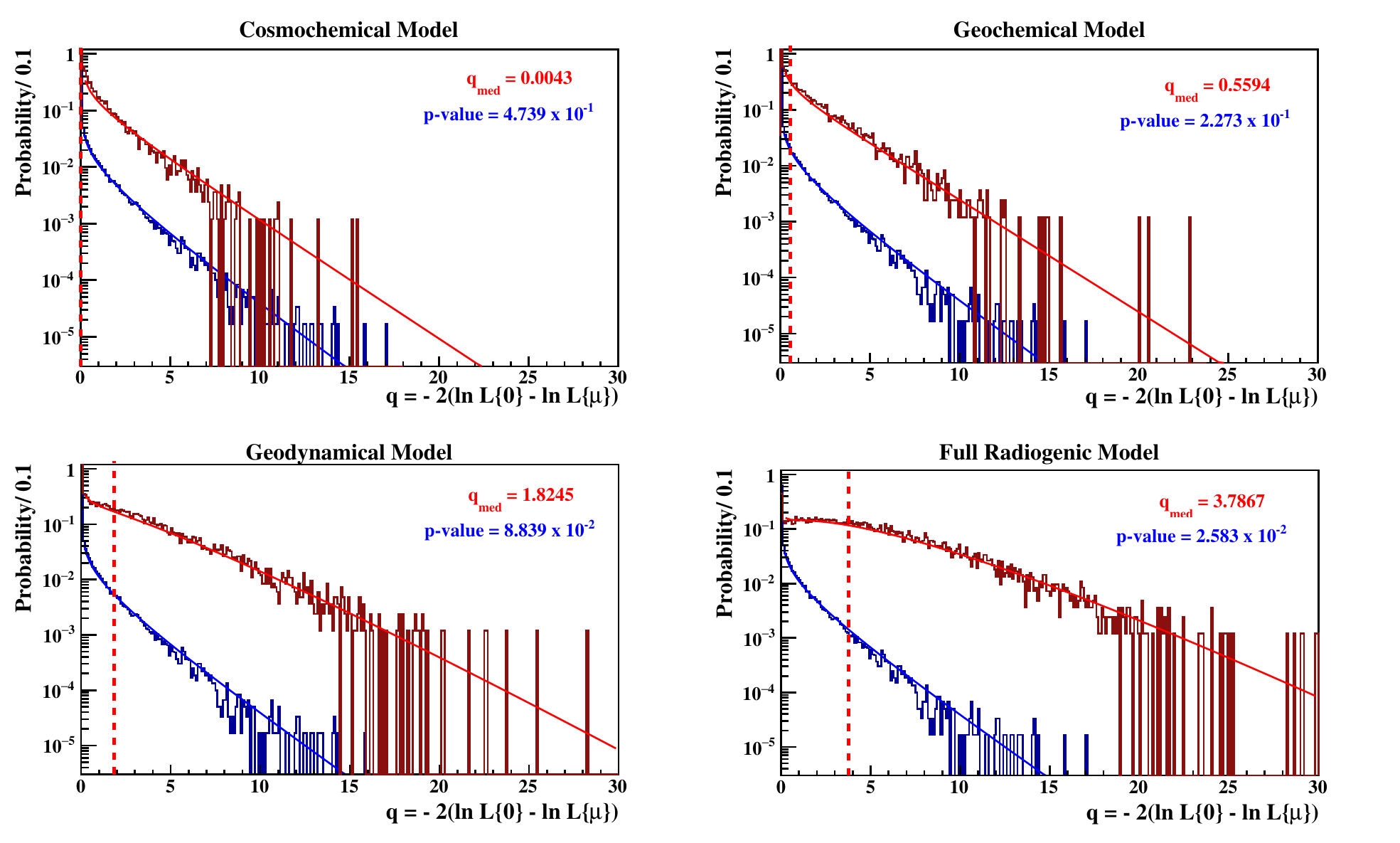}
  	\caption{Distributions of the test statistics $q  = -2(\ln L\{0\} - \ln L\{\mu\})$, where $L\{0\}$ and $L\{\mu\}$ are likelihoods of the best fits obatined with the mantle contribution fixed to zero and left free, respectively. Brown solid represents the $q$ values obtained from 10,000 pseudo-experiments with the generated mantle geoneutrino signal, based on the predictions of the different geological models (CC, GC, GD, and FR) and fit with $f(q|\mu)$ according to Eq.~\ref{eq:mantlesens}. The dark blue points show {\it q = q$_{0}$} test statistics obtained using 1.2M pseudo-experiments without any generated mantle signal and fit with $f(q|0)$ (Eq.~\ref{eq:mantlesens0}). The vertical dashed lines represent the medians $q_{med}$ of the $q$ distributions. The corresponding {\it p}-values are also shown. From the top left to the bottom right panels, the $q_{med}$ values are increasing because of increasing expected mantle signal (i.e. increasing predicted U and Th masses in the mantle).}
  	\label{fig:mantle_sens}
  \end{figure*}

We define the test statistics $q$ ($q \ge 0$):
\begin{equation}
   q  = -2\bigg(\ln L\{0\} - \ln L\{\mu\}\bigg),
   \label{eq:q}
\end{equation}
that we call $q_0$ for the data set without the mantle generated. The $q_0$ and the four $q$ distributions for different geological models are shown in Fig.~\ref{fig:mantle_sens}. The $q_0$ corresponds to the theoretical $f(q|0)$ distribution:
\begin{equation}
 f(q|0) = \frac{1}{2}\delta(q) + \frac{1}{2\sqrt{2\pi q}}\exp\Bigg(-\frac{1}{2}q\Bigg).
   \label{eq:mantlesens0}
\end{equation}
The four $q$ distributions we fit with the $f(q | \mu)$
\begin{equation}
 f(q|\mu) = \Bigg(1 - \Phi\bigg(\frac{\mu}{\sigma}\bigg)\Bigg) \delta(q) + \frac{1}{2\sqrt{2\pi q}}\exp\Bigg({-\frac{1}{2}\bigg(\sqrt{q} - \frac{\mu}{\sigma}\bigg)^{2}}\Bigg),
   \label{eq:mantlesens}
\end{equation}
where $\Phi$ stands for a cumulative Gaussian distribution with mean $\mu$ and standard deviation $\sigma$. For high statistical significance, $\mu/\sigma$ is very large and $\Phi(\frac{\mu}{\sigma}) \rightarrow 1$.
In Figure~\ref{fig:mantle_sens} we show also $q_{med} = (\mu/\sigma)^{2}$, the median value of $f(q|\mu)$, for the four different geological models. We express the Borexino sensitivity to measure the mantle geoneutrino signal, according to these four geological models, in terms of the {\it p}-value, which is given by:
\begin{equation}
p = \int_{q_{med}}^{\infty} f(q|0).
\label{eq:p-value}
\end{equation}
The differences in $q_{med}$ values shown in Fig.~\ref{fig:mantle_sens} for the 4 geological models, that correspond to different $p$-values and different sensitivity to observe the mantle signal, are to be ascribed only to the differences in the central values of the expected signals (Table~
\ref{tab:antinu-events-expected}), which in turn come from the different central values of U and Th masses associated to the different models (Table~\ref{tab:S_mantle}).

The $q_{obs}$ from the data fit should be used to obtain the final statistical significance of the mantle signal, which will be described in Sec.~\ref{subsec:mantle}. 

    \section{RESULTS}
    \label{sec:results}
   
This Section describes the results of our analysis. In Sec.~\ref{subsec:golden_candidates} the final IBD candidates selected with the optimized selection cuts are presented. In Sec.~\ref{subsec:ngeo} the analysis, and in particular the spectral fit with the $^{238}$U/$^{232}$Th ratio fixed to the chondritic value (Sec.~\ref{subsubsec:UTh_fixed}) or left free (Sec.~\ref{subsubsec:UTh_free}), is described. The systematic uncertainties are discussed in Sec.~\ref{subsec:syst}. A summary of the geoneutrino signal as measured at the LNGS is given in Sec.~\ref{subsec:geo_lngs}. Considering the expected signal from the bulk lithosphere (Table~\ref{tab:S_litho}), we estimate the geoneutrino signal from the mantle in Sec.~\ref{subsec:mantle}. The consequences with regard to the Earth radiogenic heat are then presented in Sec.~\ref{subsec:radiogenic}. Finally, in Sec.~\ref{subsec:georeactor-results} the constraints on the power of a hypothetical georeactor (Sec.~\ref{subsec:georeactor}) are set.

        \subsection{Golden candidates}
        \label{subsec:golden_candidates}
    
    \begin{figure*} [t]
     \centering  
     \subfigure[]{\includegraphics[width = 0.49\textwidth]{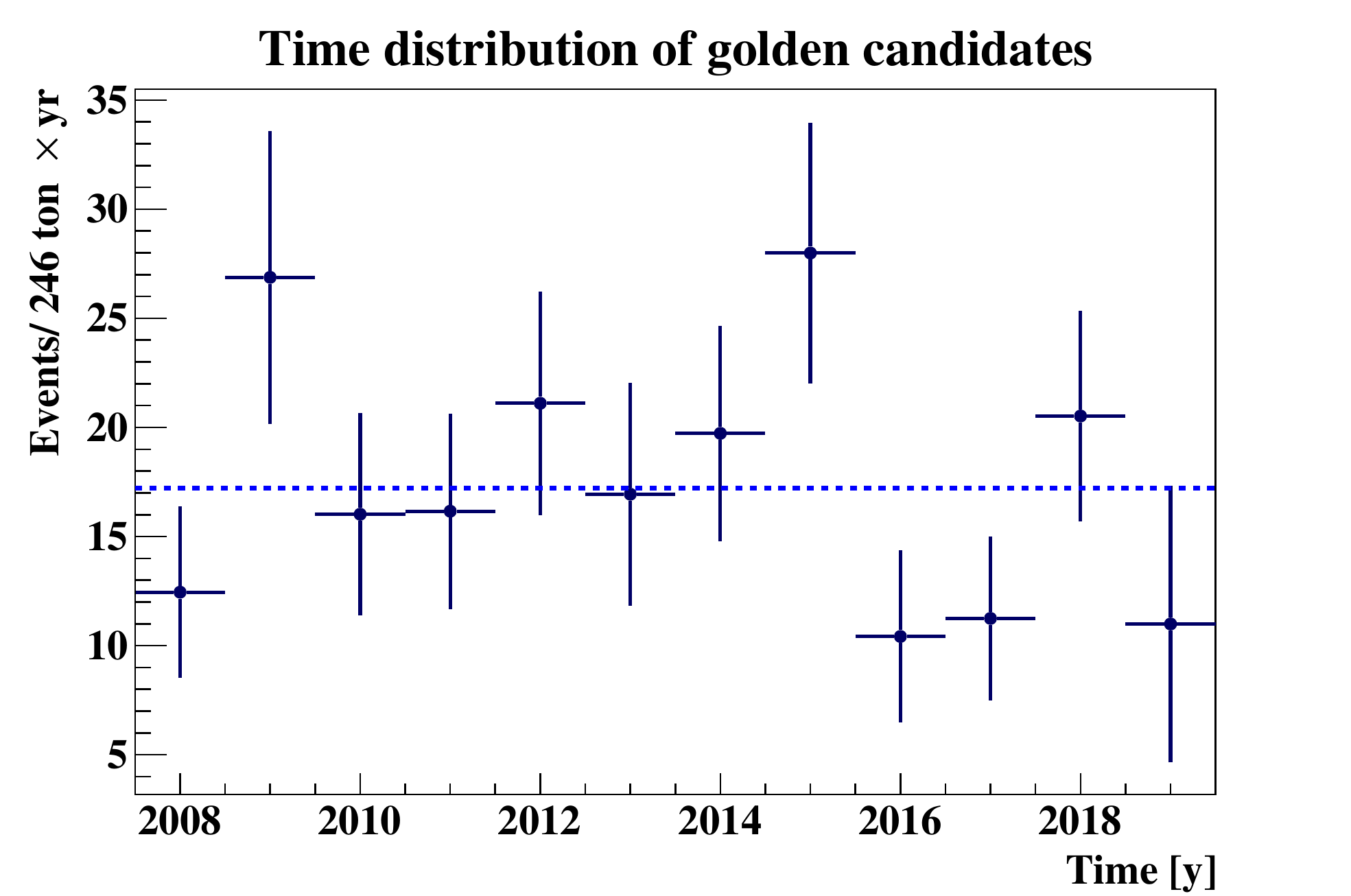}
      \label{fig:time_dis}}
        \subfigure[]{\includegraphics[width = 0.49\textwidth]{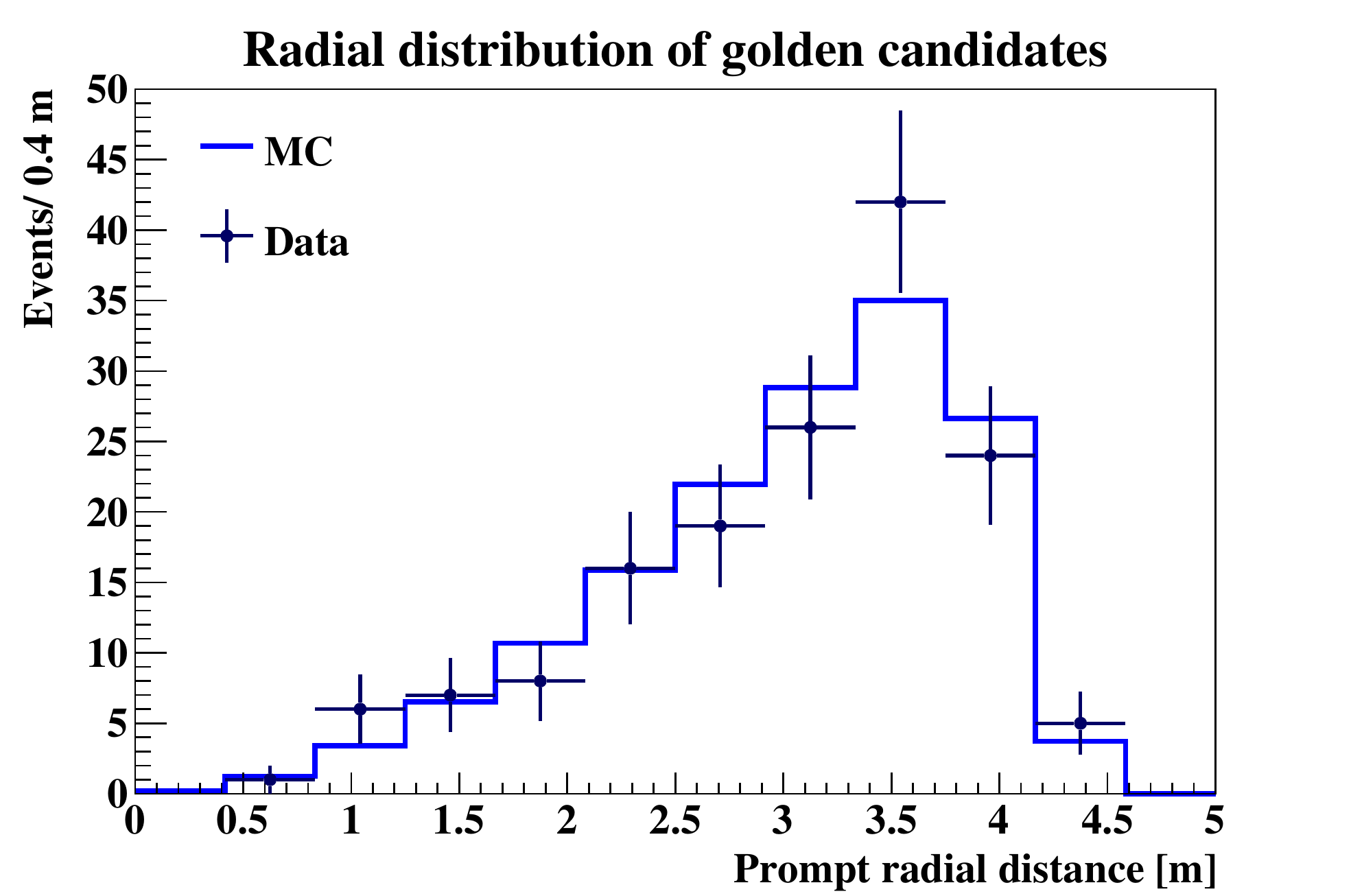} 
    \label{fig:rad_dis}}
   \subfigure[]{\includegraphics[width = 0.49\textwidth]{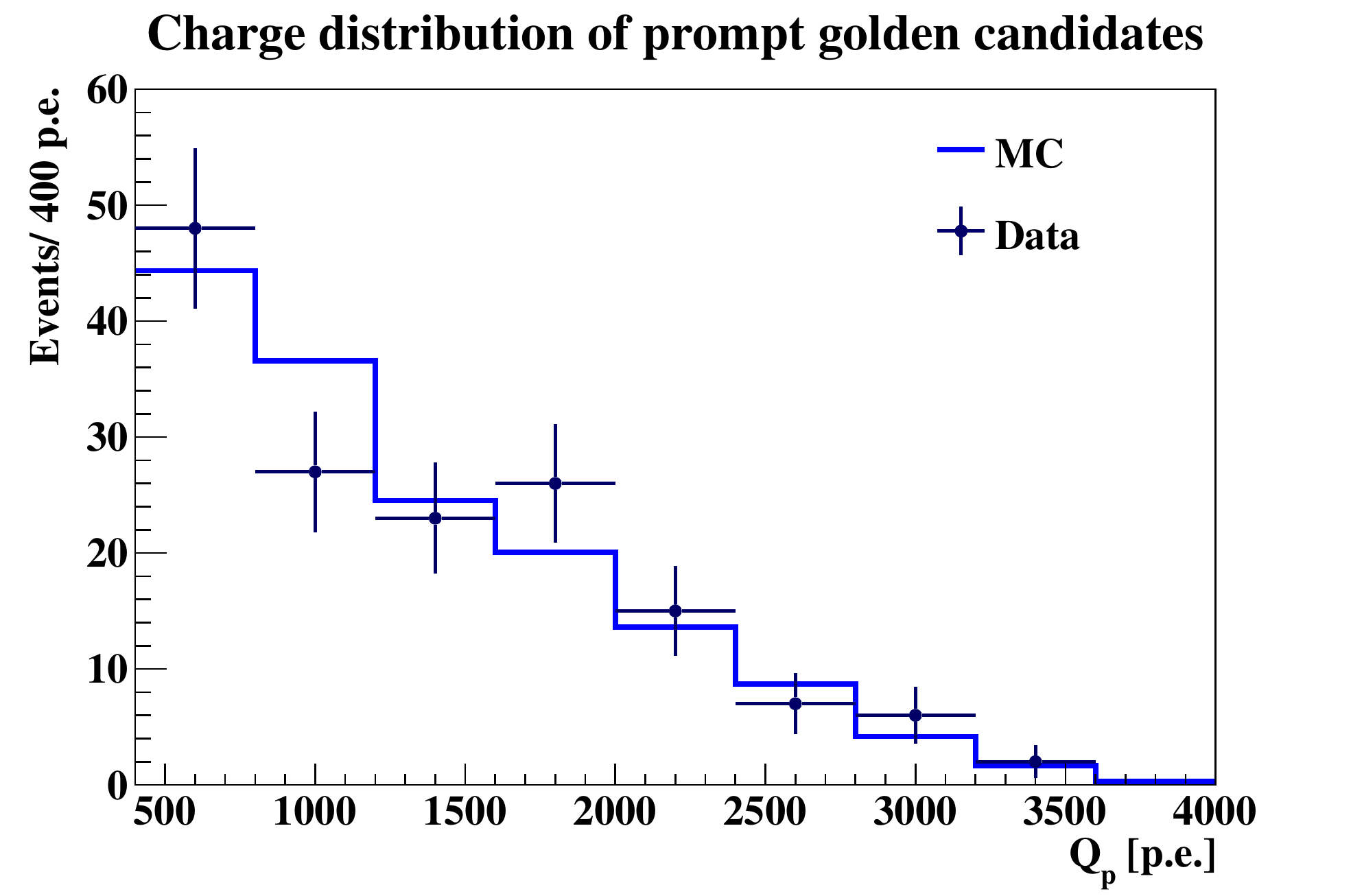}
  \label{fig:Qp_golden}}
   \subfigure[]{\includegraphics[width = 0.49\textwidth]{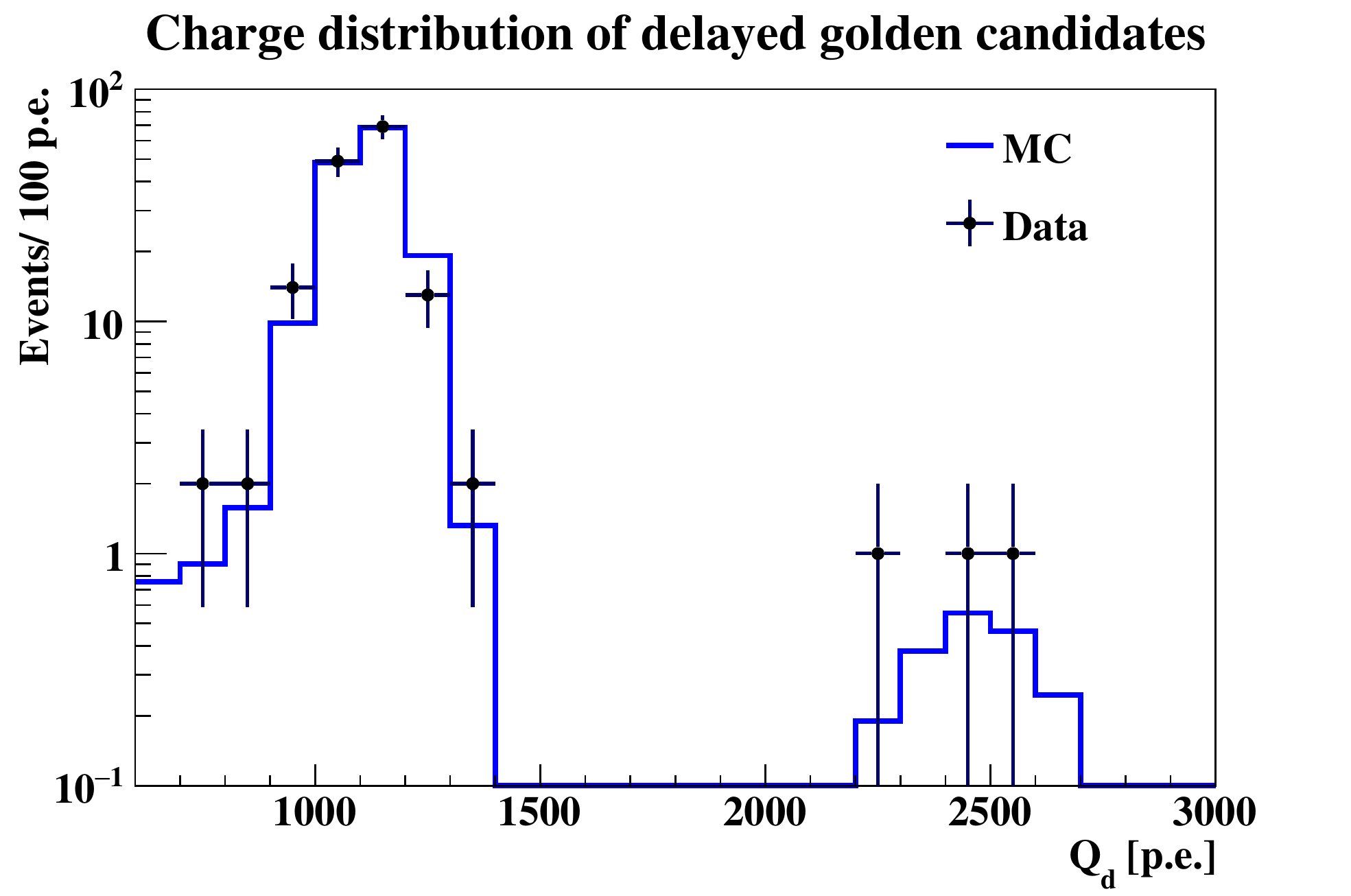}
  \label{fig:Qd_golden}}
    \caption{Distributions for 154 golden IBD candidates (black data points). MC distributions (blue solid lines) are all normalized to the same number of events. (a) Observed IBD-rate (per average FV and one year) as a function of time in one year bins (December 2007 is included in 2008 data point). The dashed line represents the average IBD rate in the entire data set. 
    (b) Radial distribution of the prompt signals compared to the MC expectation. (c) Charge distribution of the prompts compared to the MC, assuming the geoneutrino and reactor antineutrino events follow the expectations as in Table~\ref{tab:antinu-events-expected}. (d) Charge distribution of the delayed compared to MC. The two peaks due to the captures on proton and on $^{12}$C are clearly visible.}
    \label{time_rad_golden}
    \end{figure*}

    In the period between December 9, 2007 and April 28, 2019, corresponding to 3262.74\,days of data acquisition, $N_{\mathrm{IBD}} = 154$ golden IBD candidates were observed to pass the data selection cuts described in Sec.~\ref{sec:data_sel}. The events are evenly distributed in time (Fig.~\ref{time_rad_golden}~\subref{fig:time_dis}) and radially in the FV (Fig.~\ref{time_rad_golden}~\subref{fig:rad_dis}). The charge distributions of the prompt and delayed signals are also compatible with the expectations, as shown in Figs.~\ref{fig:Qp_golden} and~\ref{fig:Qd_golden}. 
        
\begin{figure*} [t]
     \centering  
  \subfigure[]{\includegraphics[width = 0.5\textwidth]{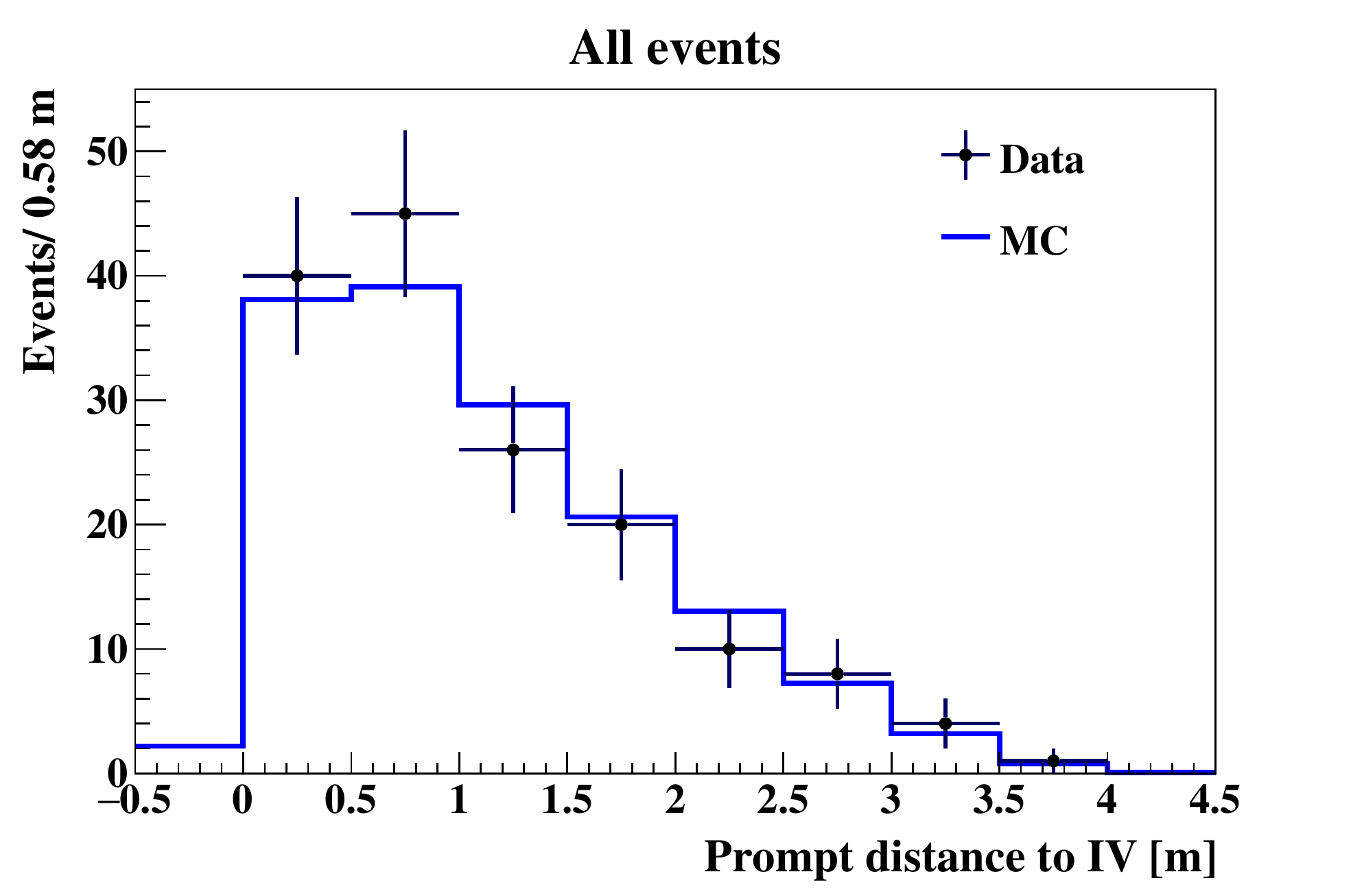}
  \label{fig:dIV_all}}
    \subfigure[]{\includegraphics[width = 0.49\textwidth]{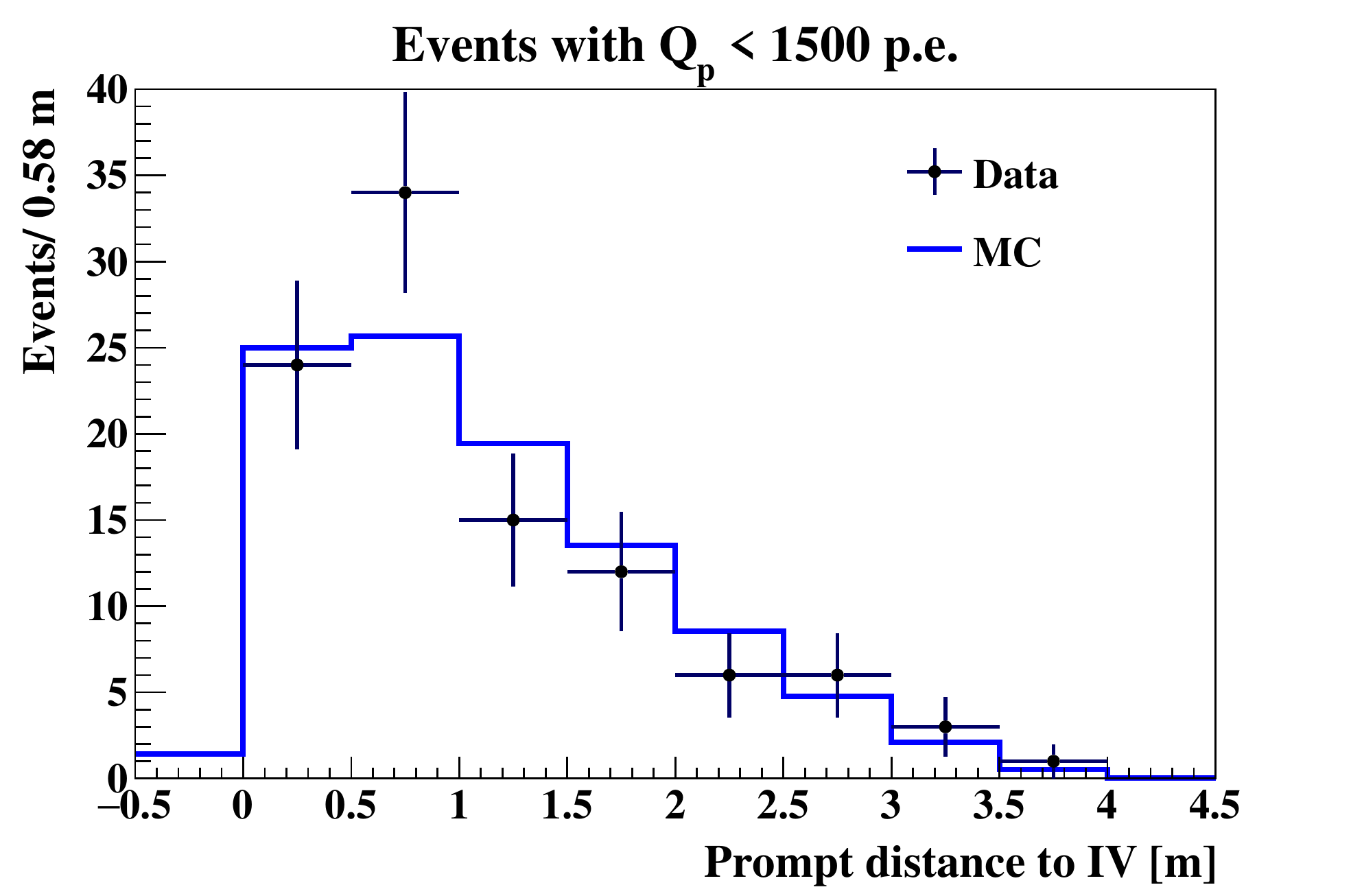}
    \label{fig:dIV_l33}}
   \subfigure[]{\includegraphics[width = 0.49\textwidth]{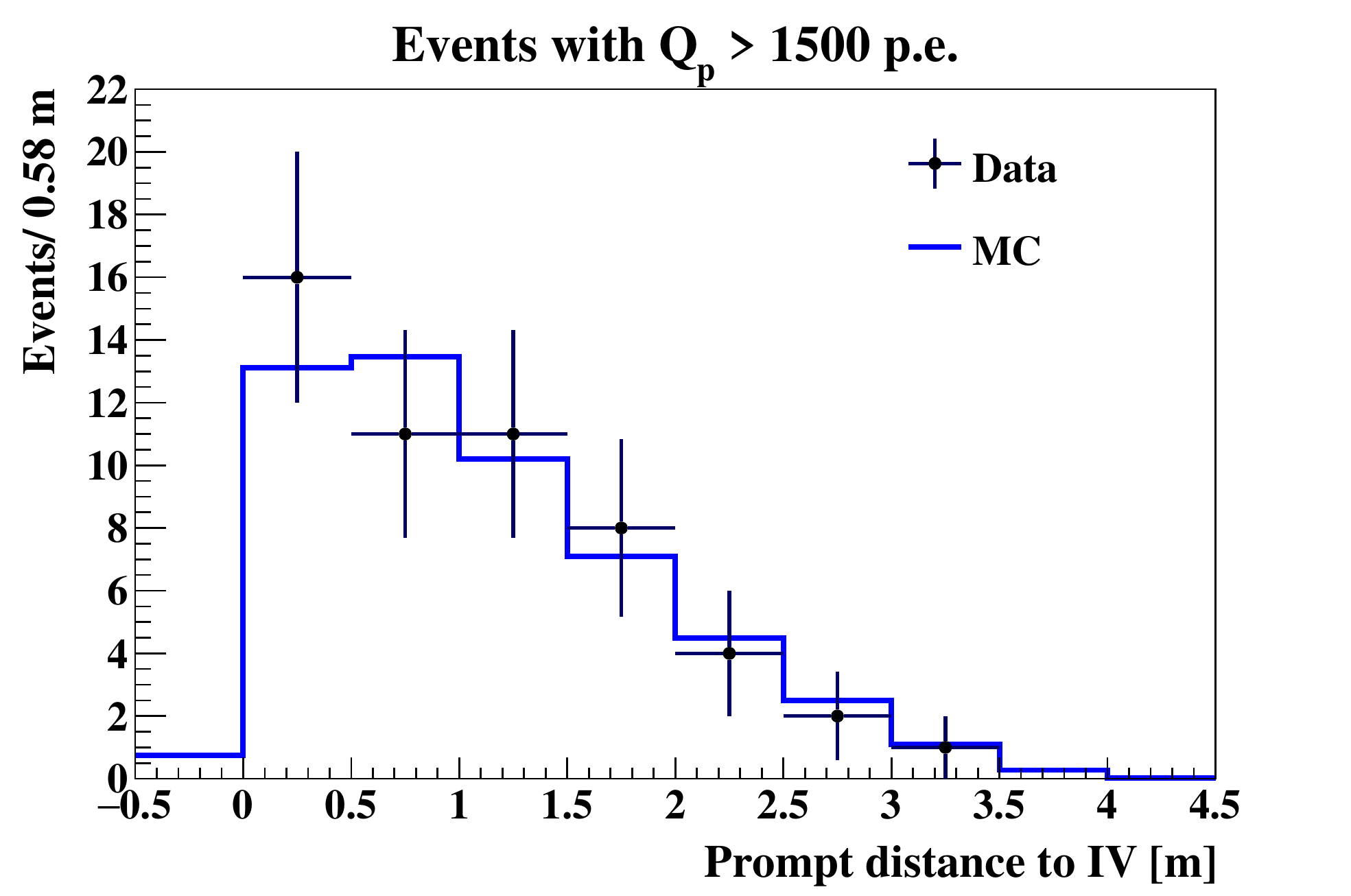} 
  \label{fig:dIV_g33}}
     \caption{(a) Prompt's distance to the IV for the 154 golden candidates (black data points) compared to MC (solid blue line) scaled to the same number of events. (b) and (c) show the same distribution but split in two energy windows: below and above the end-point of the geoneutrinos at 1500\,p.e.}
     \label{fig:dIV}
   \end{figure*}

    The distance to the IV of the prompt signal was also studied. This test would be particularly sensitive to a potential background originated from the IV itself or from the buffer: in the radial distribution of Fig.~\ref{time_rad_golden}~\subref{fig:rad_dis}, due to the changing IV shape, a small excess of this origin could have been smeared. In fact, in a deformed IV, the points characterised by the same distance from the IV (and thus a potential source of background) can correspond to different radii. As it is shown in Fig.~\ref{fig:dIV}, this test was done for all candidates, as well as separately for the geoneutrino energy window (below 1500 p.e.) and above. No excess was observed.

        \subsection{Analysis}
        \label{subsec:ngeo}
        \begin{figure*}
     \centering  
    \subfigure[]{\includegraphics[width = 0.49\textwidth]{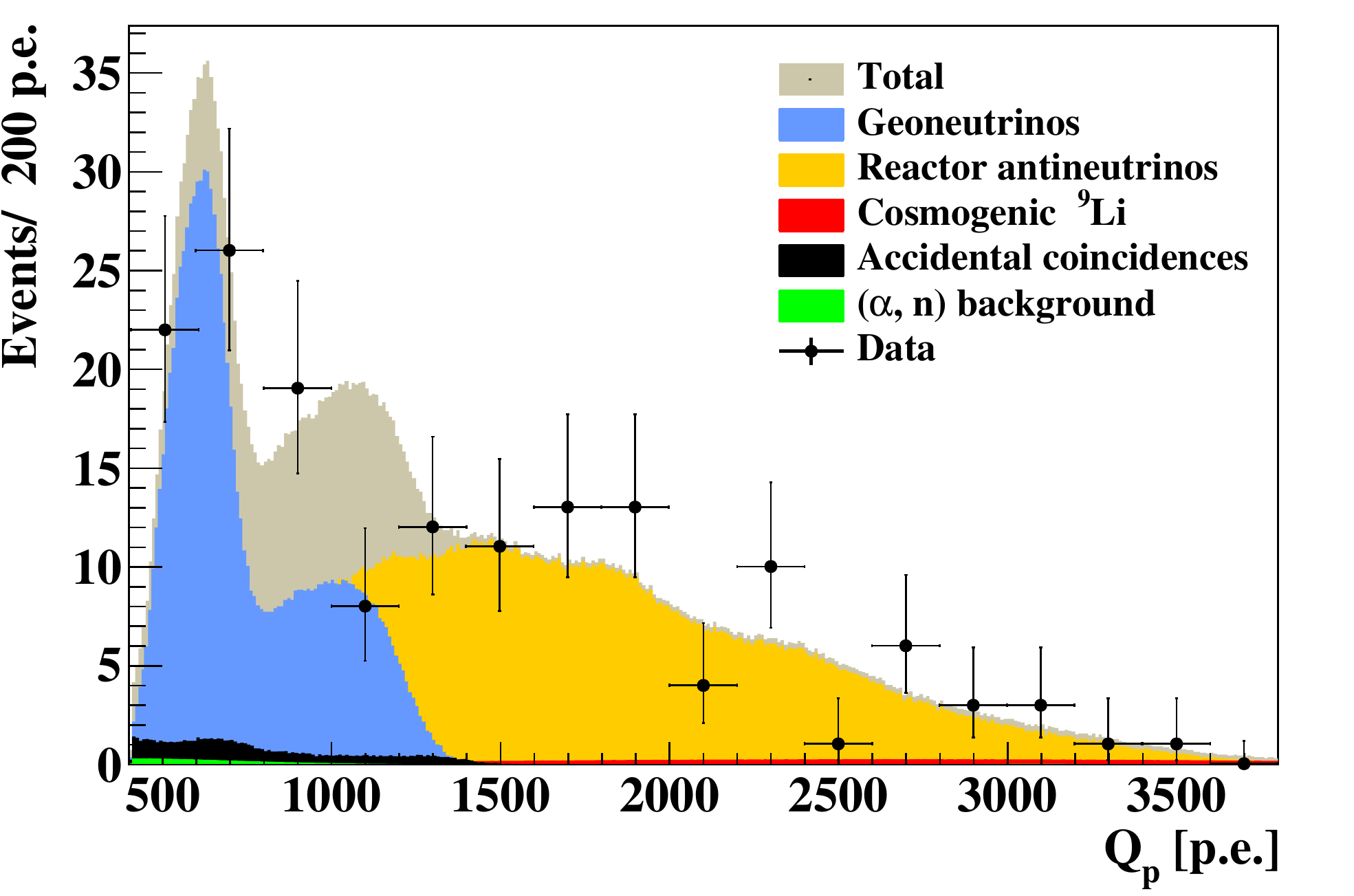}      
    \label{fig:fit_UTh-fixed}}
    \subfigure[]{\includegraphics[width = 0.49\textwidth]{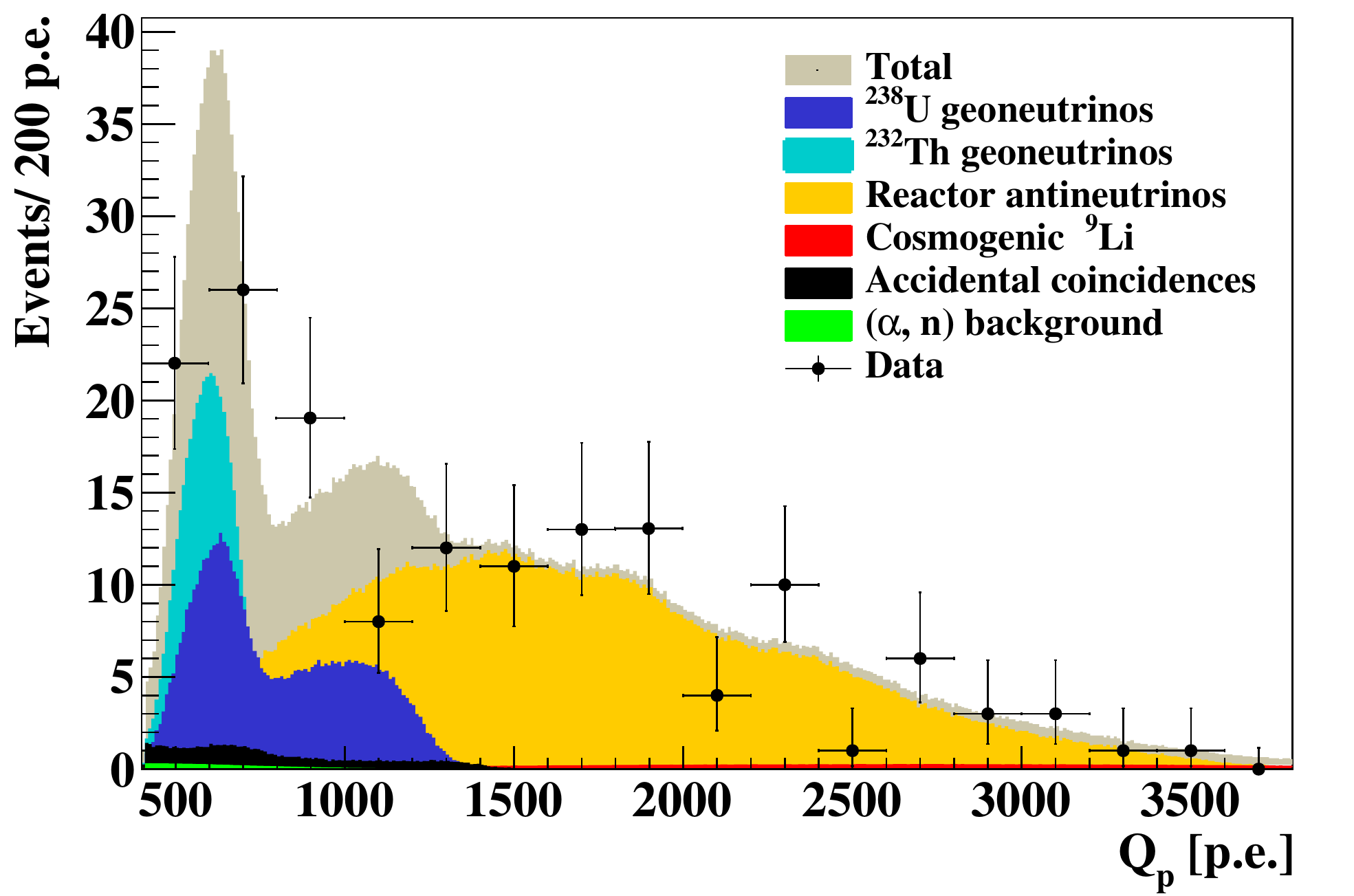}
    \label{fig:fit_UTh-free}}
    \subfigure[]{\includegraphics[width = 0.49\textwidth]{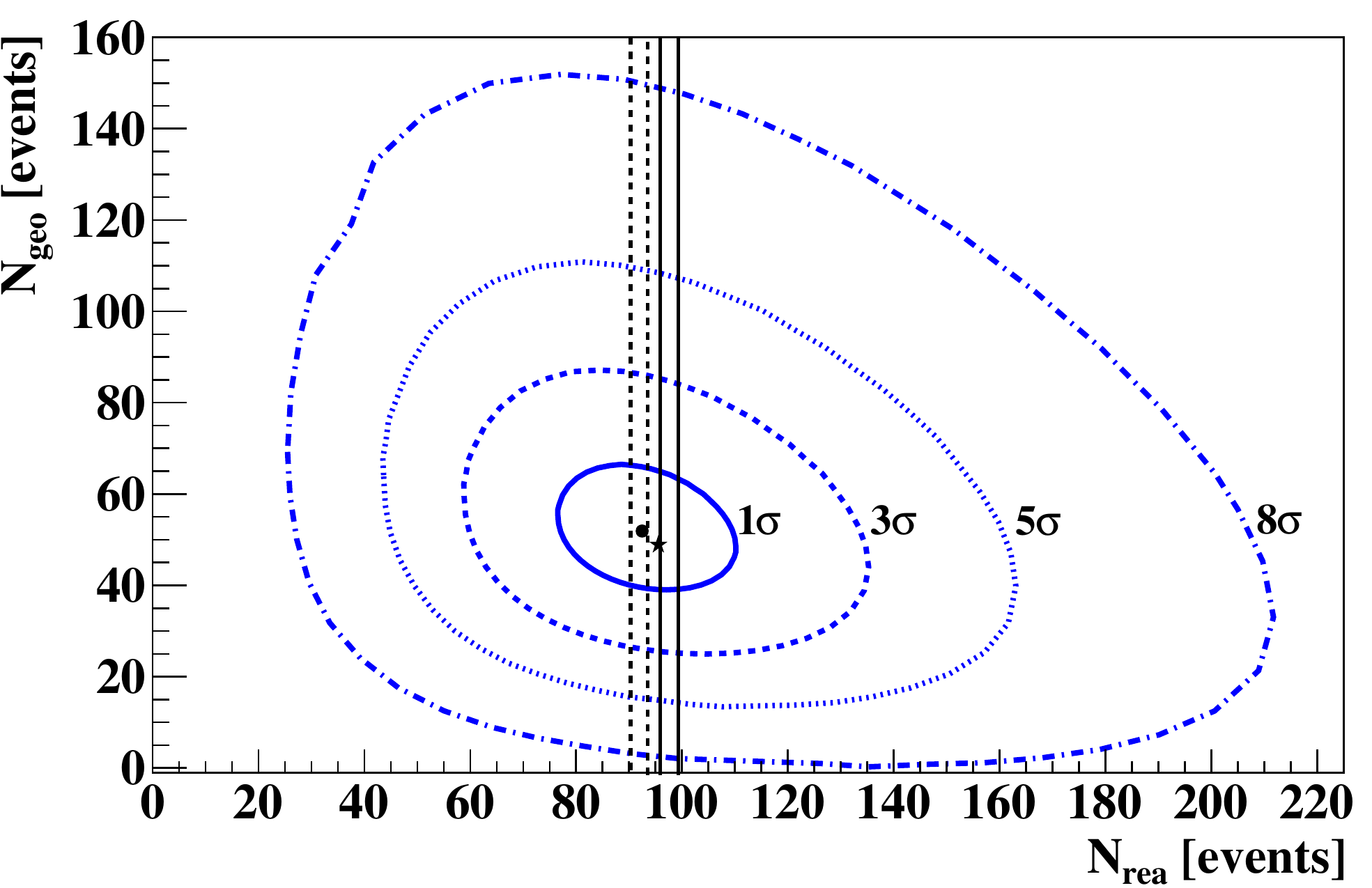}    \label{fig:geo_vs_rea_contour}}
    \subfigure[]{\includegraphics[width = 0.49\textwidth]{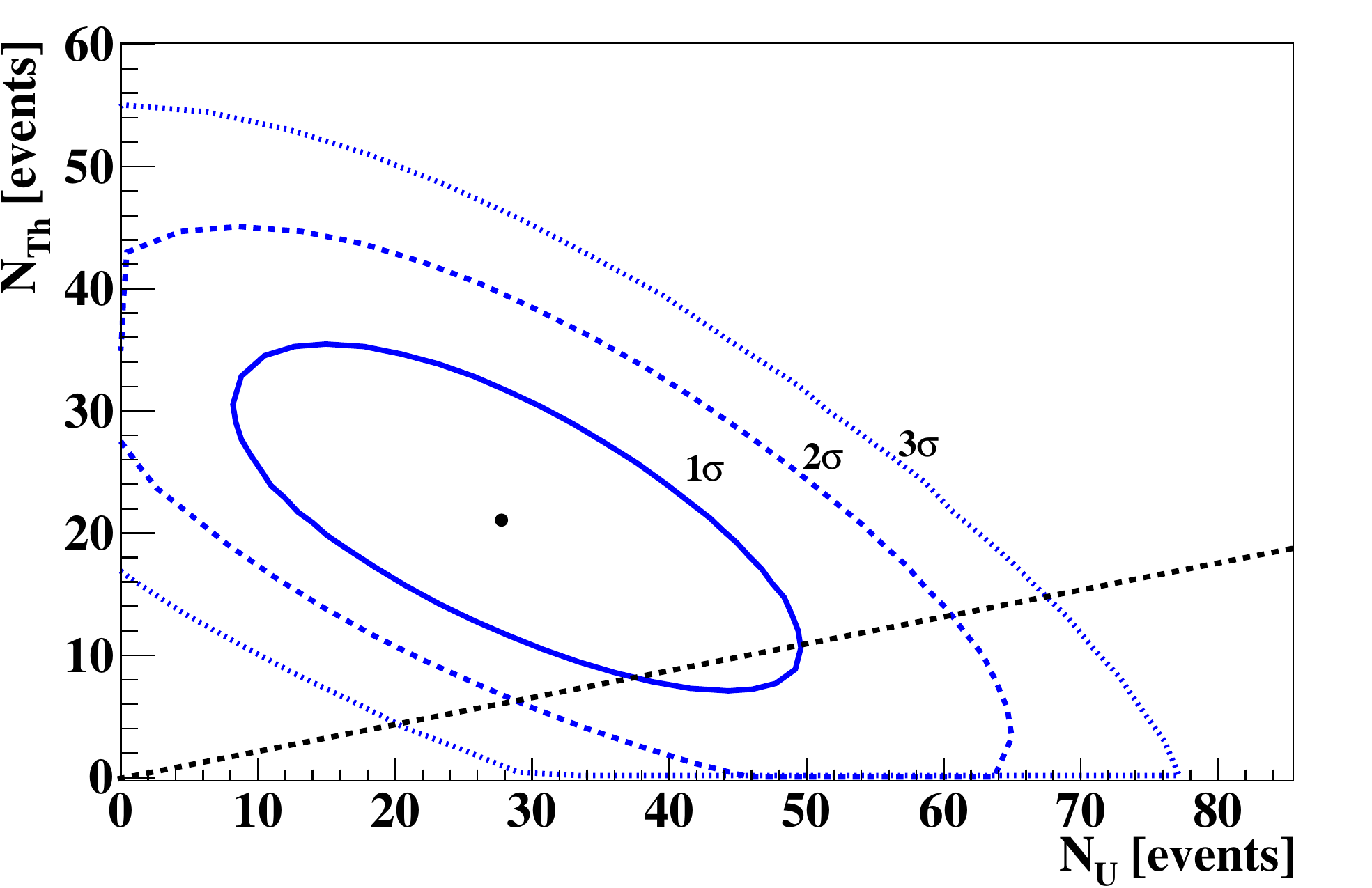} 
    \label{fig:U_vs_Th_contour}}
     \caption{Results of the analysis of 154 golden IBD candidates. (a) Spectral fit of the data (black points with Poissonian errors) assuming the chondritic Th/U ratio. The total fit function containing all signal and background components is shown in brownish-grey. Geoneutrinos (blue) and reactor antineutrinos (yellow) were kept as free fit parameters. Other non-antineutrino backgrounds were constrained in the fit. (b) Similar fit as in (a) but with $^{238}$U (dark blue) and $^{232}$Th (cyan) contributions as free and independent fit components. (c) The best fit point (black dot) and the contours for the 2D coverage of 68, 99.7, (100 - $5.7 \times 10^{-5})$\%, and (100 - $1.2 \times 10^{-13})$\%, (corresponding to 1, 3, 5, and 8$\sigma$, respectively), for N$_{\mathrm{geo}}$ versus N$_{\mathrm{rea}}$ assuming Th/U chondritic ratio. The vertical lines mark the 1$\sigma$ bands of the expected reactor antineutrino signal (solid - without ``5\,MeV excess", dashed - with ``5\,MeV excess"). For comparison, the star shows the best fit performed assuming the $^{238}$U and $^{232}$Th contributions as free and independent fit components. (d) The best fit (black dot) and the 68, 95.5, and 99.7\% coverage contours (corresponding to 1$\sigma$, 2 $\sigma$, and 3$\sigma$ contours) N$_{\mathrm{Th}}$ versus N$_{\mathrm{U}}$. The dashed line represents the chondritic Th/U ratio.}
    \label{fig:fits_contour}
    \end{figure*} 
       
  An unbinned likelihood fit, as described in Sec.~\ref{subsec:nutshell}, was performed with the prompt charge of the 154 golden candidates shown in Sec.~\ref{subsec:golden_candidates}. The three major non-antineutrino backgrounds, namely, the cosmogenic $^{9}$Li background, the ($\alpha$, n) background from the scintillator, and accidental coincidences were included in the fit using the PDFs shown in Fig.~\ref{fig:PDFs-backgrounds} and Fig.~\ref{fig:acc_2s}, respectively. These components were constrained according to values in Table~\ref{tab:backg_summary} with Gaussian pull terms. Reactor antineutrinos were unconstrained in the fit, using the PDF as in Fig.~\ref{fig:PDF_rea}.
  The differences in the shape of the reactor antineutrino spectra ``with 5\,MeV excess" and ``without 5\,MeV excess" (bottom right in Fig.~\ref{fig:PDF_bump}) are included in the systematic uncertainty calculation (Sec.~\ref{subsec:syst}).
 Obviously, geoneutrinos were also kept unconstrained. The fit was performed in two different ways with respect to the relative ratio of the $^{232}$Th and $^{238}$U contributions, as detailed in the next two sub-sections.
  
 The presented fit results are obtained following the recommendations given under the Statistics Chapter of~\cite{PhysRevD.98.030001} for cases, when there are 
 physical boundaries on the possible parameter values. In our case, all the fit parameters must have non-negative values. For the main parameters resulting from the fit, the profiles of the likelihood $L$ (Eq.~\ref{eq:Lkl}) are provided, and in addition to the best fit values, the mean, median, as well as the 68\% and 99.7\% coverage intervals for non-negative parameter values, are provided in the summary Table~\ref{tab:summary_results}.
 
  \subsubsection{Th/U fixed to chondritic ratio}
  \label{subsubsec:UTh_fixed}
    
  The fit was performed assuming the Th/U chondritic ratio and using the corresponding PDF shown in the top-left of Fig.~\ref{fig:PDF_geo}. The resulting spectral fit is shown in Fig.~\ref{fig:fit_UTh-fixed} and the numerical results are summarized in Table~\ref{tab:summary_results}. The likelihood profile for the number of geoneutrinos $N_{\mathrm {geo}}$ (Fig.~\ref{fig:lkl_geo}), yields the best fit value $N_{\mathrm {geo}}^{\mathrm{best}}$ = 51.9, the median value $N_{\mathrm {geo}}^{\mathrm{med}}$ = 52.6, and the 68\% coverage interval $I_{N \mathrm {geo}}^{68\mathrm{stat}}$ = [44.0 - 62.0]\,events. The likelihood profile for the number of reactor antineutrinos is shown in Fig.~\ref{fig:lkl_rea}: $N_{\mathrm {rea}}^{\mathrm{best}}$ = 92.5\,events was obatined with the median value $N_{\mathrm {rea}}^{\mathrm{med}}$ = 93.4 and the 68\% coverage interval $I_{N \mathrm {rea}}^{68\mathrm{stat}}$ = [82.6 - 104.7]\,events. This is compatible with the reactor antineutrino expectation of (97.6 $\pm$ 1.7)\,events (without ``5\,MeV excess") as well as (91.9 $\pm$ 1.6) events (with ``5\,MeV excess"), given in Table~\ref{tab:antinu-events-expected}. Thus, from the total of 154 golden IBD candidates, the number of detected antineutrinos (geo + reactor) is $N_{\mathrm {antinu}}^{\mathrm{best}}$ = 144.4\,events. This leaves the number of background events compatible with the expectation (Table~\ref{tab:backg_summary}). The contour plot for N$_{\mathrm{geo}}$ versus N$_{\mathrm{rea}}$ is shown in Fig.~\ref{fig:geo_vs_rea_contour}. The fit was also performed by constraining the expected number of reactor antineutrino events to (97.6 $\pm$ 1.7)\,events (Table~\ref{tab:antinu-events-expected}). The result (the best fit value $N_{\mathrm {geo}}^{\mathrm{best}}$ = 51.3, the median $N_{\mathrm {geo}}^{\mathrm{med}}$ = 52.0, and the 68\% coverage interval $I_{N \mathrm {geo}}^{68\mathrm{stat}}$ = [43.6 - 61.1]\,events) is nearly unchanged with respect to that obtained when leaving the reactor antineutrino contribution free. The best fit value is shifted by about 1.5\% and the error is only marginally reduced. This fit stability is due to the fact that above the geoneutrino energy window there is almost no non-antineutrino background, and thus the data above the geoneutrino end point well constrain the reactor antineutrino contribution also in the geoneutrino window. The fact that without any constraint on N$_{\mathrm {rea}}$ the fit returns a value compatible with expectation is an important confirmation of the Borexino ability to measure electron antineutrinos.
  %

  \subsubsection{Th and U as free fit parameters}
  \label{subsubsec:UTh_free}
    
\begin{figure*}[t]
     \centering  
   \subfigure[]{\includegraphics[width = 0.46\textwidth]{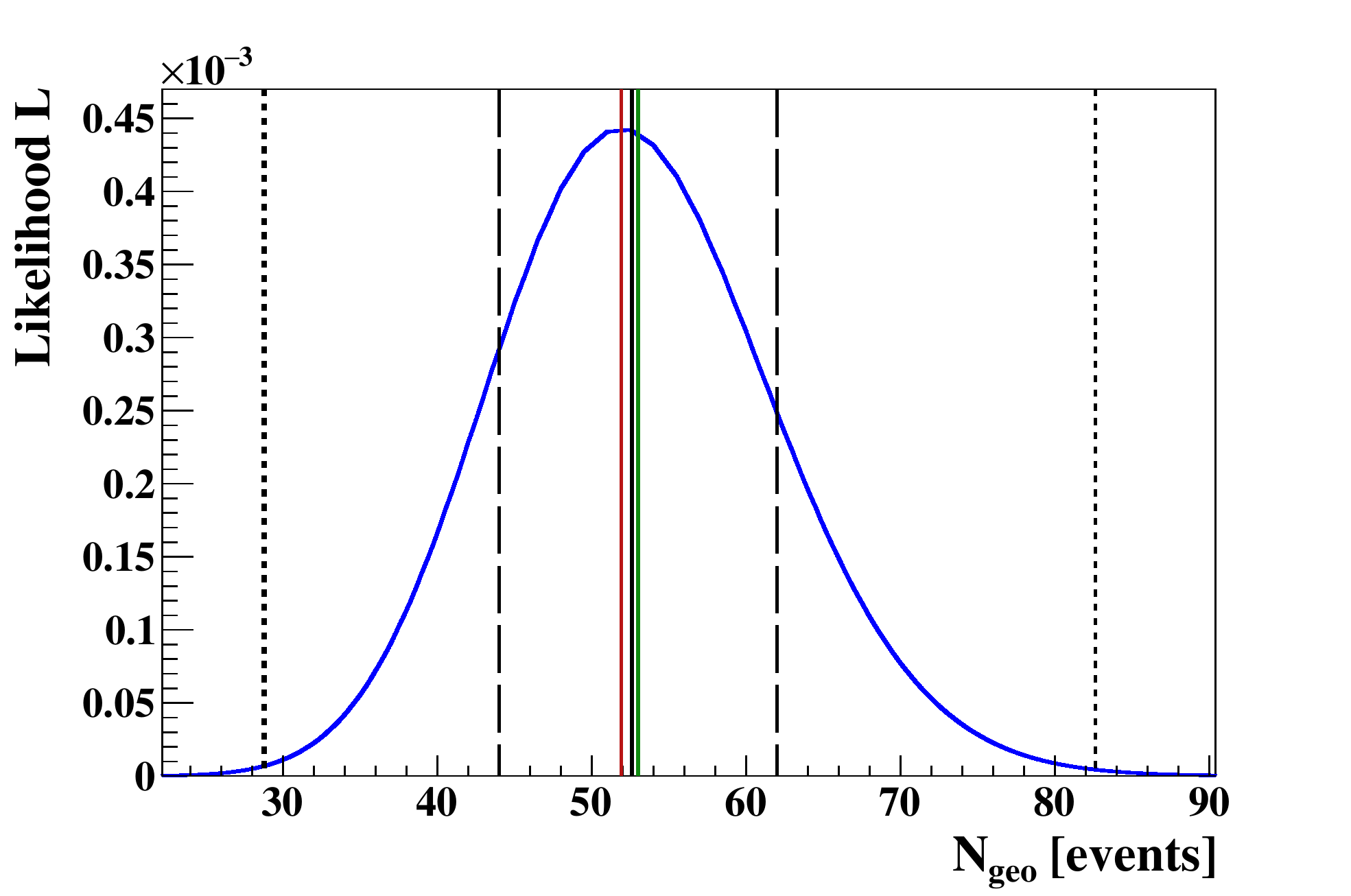}      
    \label{fig:lkl_geo}}
  \subfigure[]{\includegraphics[width = 0.46\textwidth]{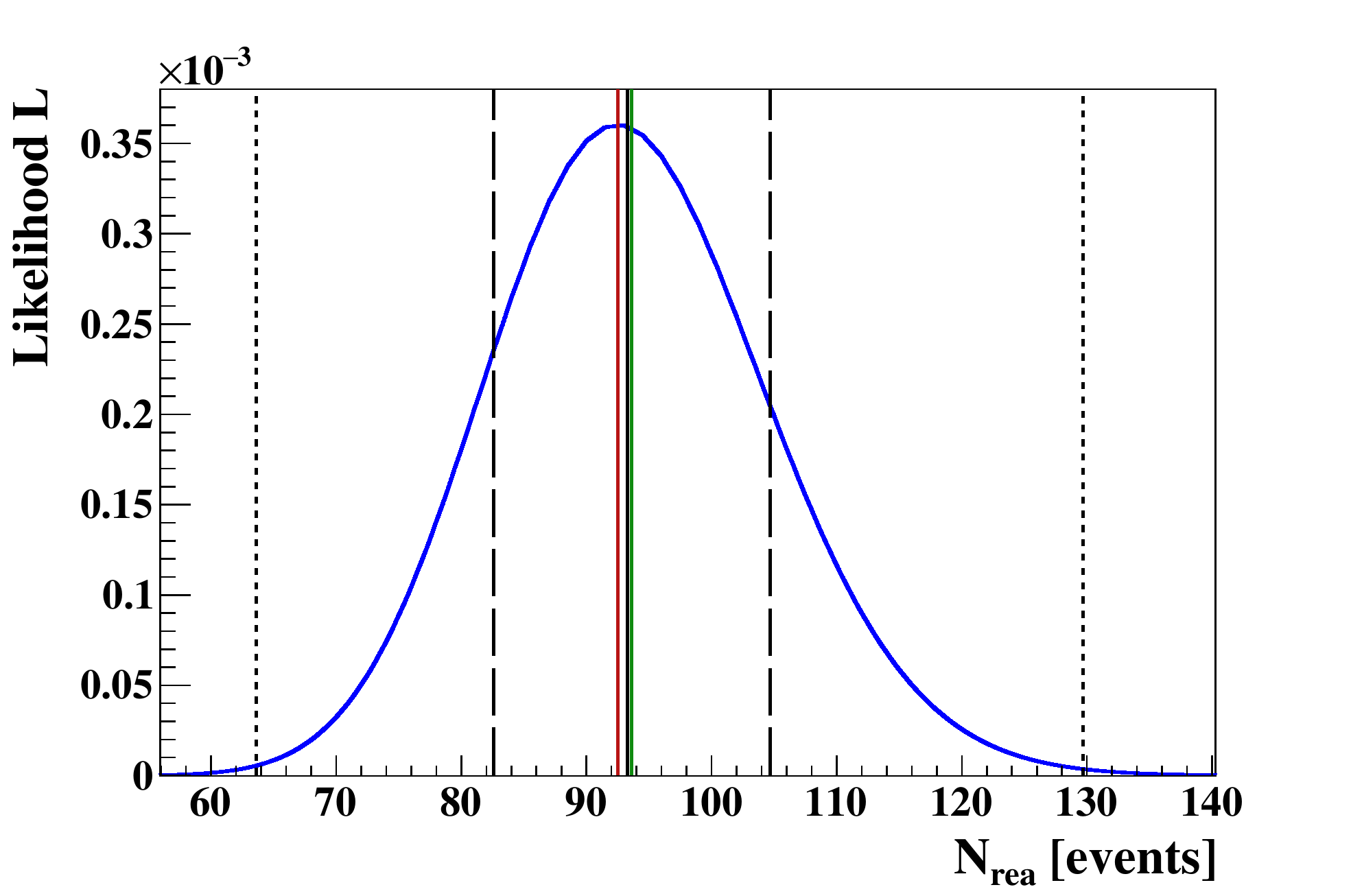}
  \label{fig:lkl_rea}}
   \subfigure[]{\includegraphics[width = 0.46\textwidth]{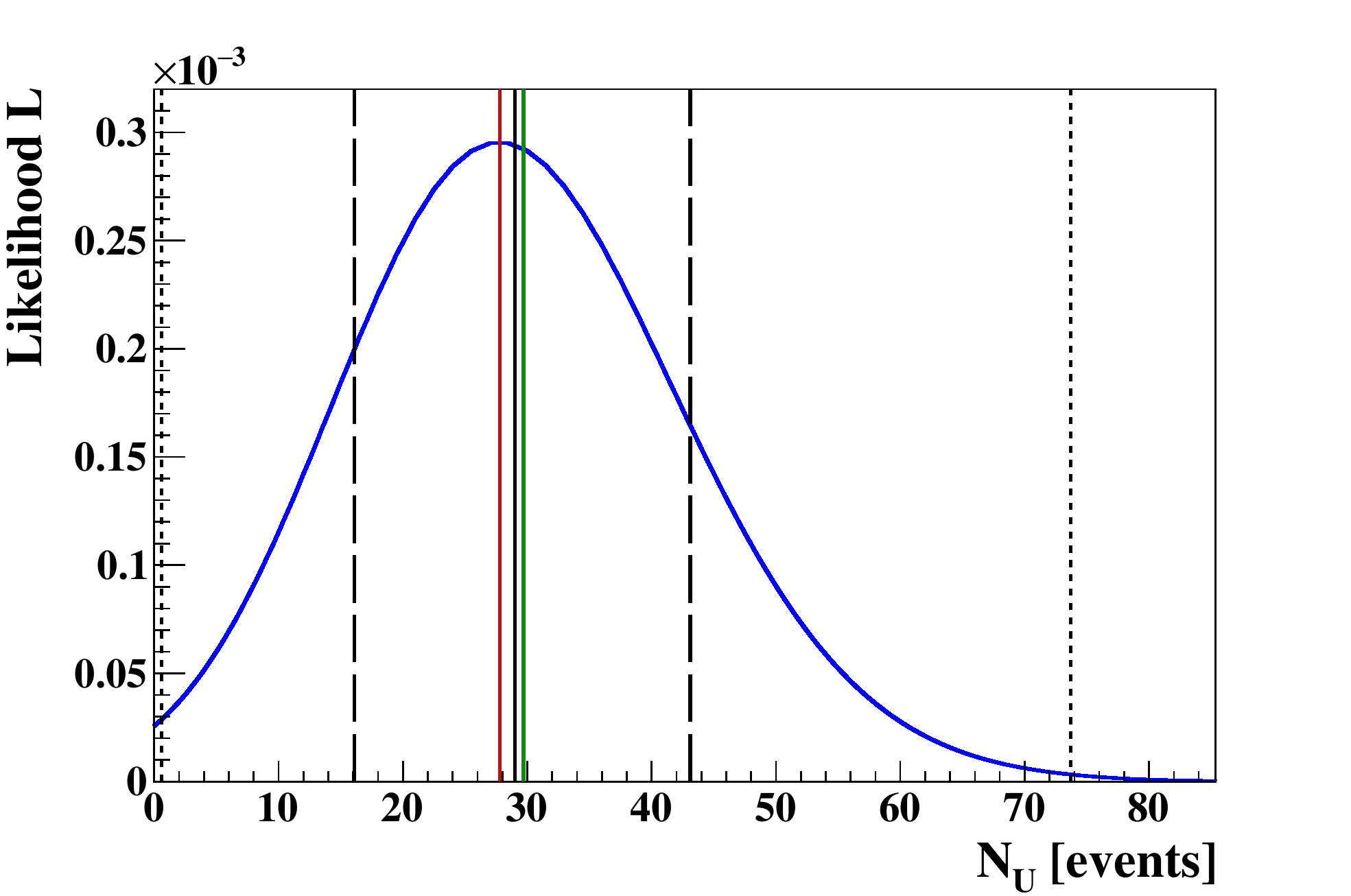}
   \label{fig:lkl_u}}
    \subfigure[]{\includegraphics[width = 0.46\textwidth]{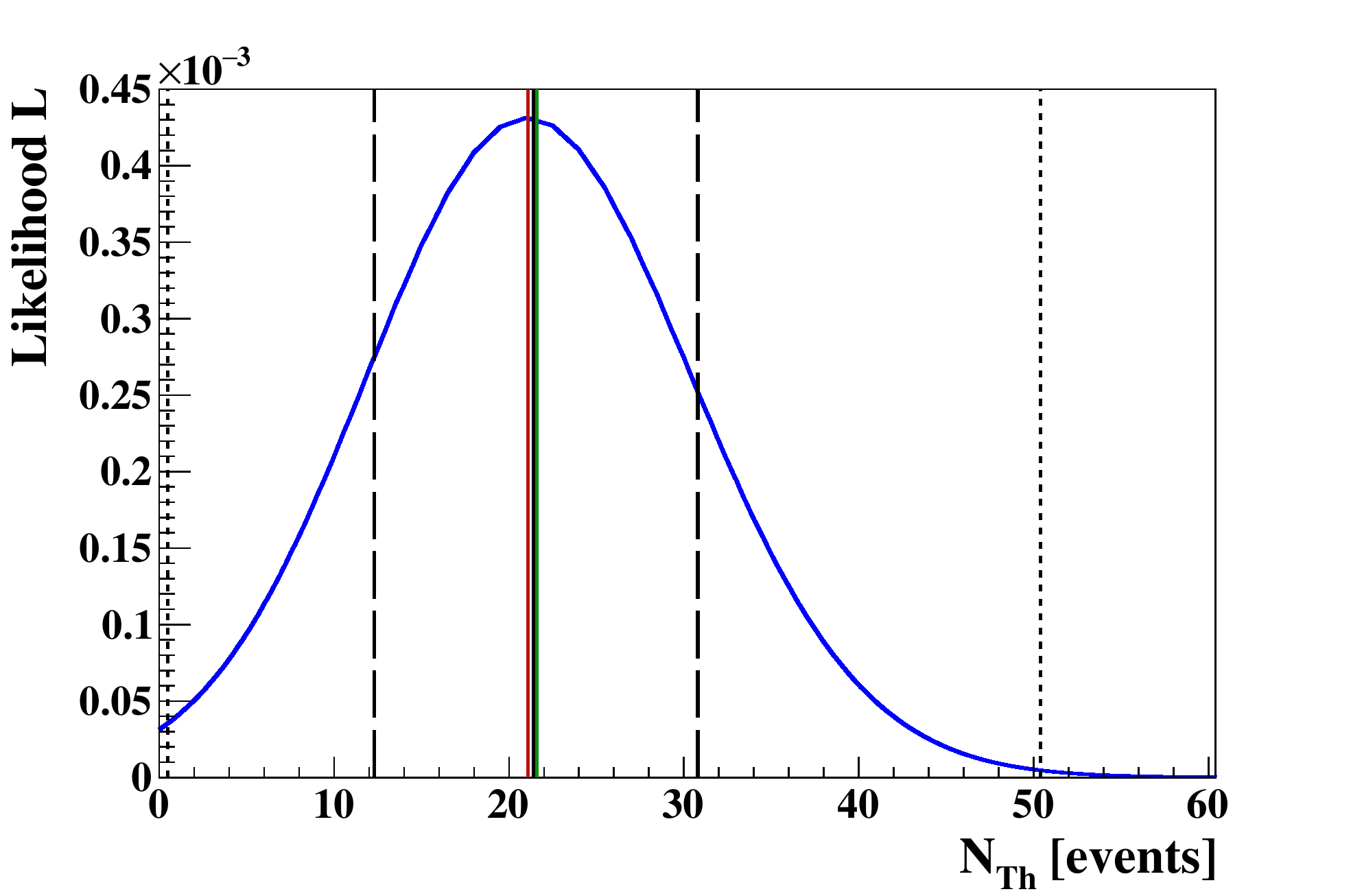}
 \label{fig:lkl_th}}
        \caption{The likelihood profiles for the number of geoneutrino events $N_{\mathrm{geo}}$ (a) and reactor antineutrino events $N_{\mathrm{rea}}$ (b) obtained from the fit assuming the chondritic Th/U ratio. The lower row shows the likelihood profiles for the number of $^{238}$U (c) and $^{232}$Th (d) events obtained from the fit assuming the $^{238}$U and $^{232}$Th contributions as the two independent free fit components. In each plot, the vertical solid red line indicates the best fit, while the vertical solid black and green lines indicate the median and mean values of the distributions, respectively. The vertical dashed/dotted lines show the 68\%/99.7\% confidence intervals of the distributions, corresponding to the signal values given in Table~\ref{tab:summary_results}.}
        \label{fig:lkl}
    \end{figure*}

 The second type of fit was performed by treating $^{238}$U and $^{232}$Th contributions as free and independent fit components. The corresponding MC PDFs from Fig.~\ref{fig:PDF_UTh} were used. The spectral fit is shown in Fig.~\ref{fig:fit_UTh-free} and the numerical results are summarized in Table~\ref{tab:summary_results}. The likelihood profiles for the number of $^{238}$U and $^{232}$Th geoneutrinos are shown in Figs.~\ref{fig:lkl_u} and \ref{fig:lkl_th}, respectively. The fit yielded $N_{\mathrm {U}}^{\mathrm{best}}$ = 27.8, $N_{\mathrm {U}}^{\mathrm{med}}$ = 29.0, and the 68\% coverage interval $I_{N \mathrm {U}}^{68\mathrm{stat}}$ = [16.1 - 43.1]\,events for the Uranium contribution and $N_{\mathrm {Th}}^{\mathrm{best}}$ = 21.1, $N_{\mathrm {Th}}^{\mathrm{med}}$ = 21.4, and the 68\% coverage interval $I_{N \mathrm {Th}}^{68\mathrm{stat}}$ = [12.2 - 30.8]\,events for the Th contribution. The best fit leads to 48.9 geoneutrinos in total, which is fully compatible with 51.9 geoneutrino events obtained in the case when Th/U ratio was fixed to the chondritic value. The only difference is significantly larger error in case of the fit with free U and Th contributions. For reactor antineutrinos, $N_{\mathrm {rea}}^{\mathrm{best}}$ = 95.8 and $I_{N \mathrm {rea}}^{68\mathrm{stat}}$ = [85.2 - 109.0]\,events were obtained, which is also compatible with the expectation. The total number of detected antineutrinos (geo + reactor) is $N_{\mathrm {antinu}}^{\mathrm{best}}$ = 144.7\,events. The contour plot for N$_{\mathrm{geo}}$ versus N$_{\mathrm{rea}}$ is shown in Fig.~\ref{fig:geo_vs_rea_contour}.

 The contour plot for N$_{\mathrm{U}}$ versus N$_{\mathrm{Th}}$ is shown in Fig.~\ref{fig:U_vs_Th_contour}. The results obtained after constraining the expected N$_{\mathrm{rea}}$ were
 again fully compatible with the results obtained when leaving the reactor antineutrino component free and without any significant reduction on error.
 
        \subsection{Systematic uncertainties}
        \label{subsec:syst}
         
 This section discusses the different sources of systematic uncertainty in the geoneutrino and reactor antineutrino measurement. They are detailed below and summarized in Table~\ref{tab:summary_sys}. 
  
\begin{table}[h]
	\centering
	\caption{\label{tab:summary_sys} Summary of the different sources of systematic uncertainty in the geoneutrino and reactor antineutrino measurement. Different contributions are summed up as uncorrelated.} \vskip 2pt
	\begin{tabular*}{\columnwidth}{l @{\hskip 25pt} c @{\hskip 25pt} c}
		\hline \hline
		Source & Geo &  Reactor \Tstrut \\
               & Error &  Error \\
               &  [\%] &  [\%]  \Bstrut \\
		\hline
		Atmospheric neutrinos & {\Large $^{+0.00}_{-0.38}$} & {\Large $^{+0.00}_{-3.90}$} \TstrutLarge \\ [8pt]
		Shape of reactor spectrum &  {\Large  $^{+0.00}_{-0.57}$} & {\Large $^{+0.04}_{-0.00}$} \\ [8pt]
		Vessel shape & {\Large $^{+3.46}_{-0.00}$} & {\Large $^{+3.25}_{-0.00}$} \\ [8pt]
		Efficiency & 1.5 & 1.5 \\ [8pt]
		Position reconstruction & 3.6  & 3.6 \Bstrut \\
		\hline
		Total &  {\Large $^{+5.2}_{-4.0}$} & {\Large $^{+5.1}_{-5.5}$} \TstrutLarge\BstrutLarge \\
		\hline
		\hline
	\end{tabular*}
\end{table}
 
   \paragraph{Atmospheric neutrinos}
    
Atmospheric neutrinos as the source of background were discussed in Sec.~\ref{subsec:atm}, while the expected number of IBD-like events passing the geoneutrino selection cuts in different energy regions was given in Table~\ref{tab:antinu-events-expected}. The uncertainty of this prediction is large, estimated to be 50\%. In addition, there is an indication of some over-estimation of this background, since above the end-point of the reactor spectrum, where we would expect (3.3 $\pm$ 1.6)\,atmospheric events, no IBD candidates are observed. In the estimation of the systematic uncertainty due to atmospheric neutrinos, two fits were preformed which were similar to that shown in Fig.~\ref{fig:fit_UTh-fixed}, but with additional contribution due to atmospheric neutrinos.
These are represented by the PDF shown in the top part of Fig.~\ref{fig:PDFs-backgrounds}. The fit was performed in two energy ranges and the number of events from atmospheric neutrino background was constrained according to the values in Table~\ref{tab:antinu-events-expected}. First, the fit was performed for data up to the end point of the reactor antineutrino spectrum at 4000\,p.e., which is the interval containing 63\% of atmospheric neutrino background. The resulting number of atmospheric neutrino events is (4.6 $\pm$ 3.2) and is compatible with the expectation of (6.7 $\pm$ 3.4). The geoneutrino signal is almost unchanged, while $N_{\mathrm{rea}}$ decreased to (89.0 $\pm$ 11.3)\,events. Second, we have performed the fit up to the end-point of the atmospheric-neutrino background passing our IBD selection criteria (7500\,p.e.). In this case, due to the fact that no IBD candidates are observed above the reactor antineutrino energy window, the resulting number of atmospheric neutrino background events is very low and with a large error, (1.2 $\pm$ 4.1)\,events. Fortunately, the resulting $N_{\mathrm{geo}}$ and $N_{\mathrm{rea}}$ are nearly unchanged. To summarize, we estimate the respective systematic uncertainty on geoneutrinos as $^{+0.00}_{-0.38}$\% and on reactor antineutrinos as $^{+0.00}_{-3.90}$\%.
    
\paragraph{Shape of the reactor spectrum}
  
  The likelihood fit, as described in Sec.~\ref{subsec:ngeo} was performed using the MC PDF of reactor antineutrinos without any ``5\,MeV excess", based on the flux prediction of~\cite{mueller2011improved}, as discussed in Sec.~\ref{subsec:rea}. In order to study the changes that might arise due to the observed ``5\,MeV excess", the fit was also performed using the corresponding MC PDF as shown in Fig.~\ref{fig:PDFs-Geo-Rea}, based on the measured Daya Bay spectrum~\cite{an2016measurement}. Since there is no constraint on $N_{\mathrm{rea}}$ and the two spectral shapes are relatively similar, the change in $N_{\mathrm{rea}}$ is very small: we observe an increase of 0.05\,events. In case of $N_{\mathrm{geo}}$, we observe a decrease of 0.3\,events.
    
\paragraph{Inner Vessel shape reconstruction}
  
  We consider a conservative 5\,cm error on the IV position (Sec.~\ref{subsec:IV}). This means that the function defining our DFV ($d_{\mathrm {IV}} = 10$\,cm) inward from the IV is inside the scintillator with high probability. This implies that the systematic uncertainty on the FV defintion due to the IV shape reconstruction is negligible. However, there is a systematic uncertainty due to the selection of the IBD candidates using the DFV cut, which was evaluated by smearing the distance-to-IV of each IBD candidate with a Gaussian function with $\sigma$ = 5\,cm. Consequently, the DFV cut was applied on the smeared distances and the spectral fit was performed on newly selected candidates. This procedure was repeated 50 times. The distributions of the differences between the resulting $N_{\mathrm{geo}}$ and $N_{\mathrm{rea}}$ values with respect to the default fit have positive offsets, which were then conservatively taken as the systematic uncertainty due to the IV shape reconstruction. We estimate the respective systematic uncertainty on geoneutrinos as $^{+3.46}_{-0.00}$\% and on reactor antineutrinos as $^{+3.25}_{-0.00}$\%.

\paragraph{MC efficiency}
    
The major source of uncertainty for the MC efficiency arises from the event losses close to the IV edges, especially near the south pole because of the combined effect of a large number of broken PMTs and the IV deformation. The trigger efficiency for the 2.2\,MeV gamma from $^{241}$Am-$^{9}$Be calibration source compared to MC simulations for different source positions was studied. The uncertainty in the efficiency was then set to a conservative limit of 1.5\%. 

\paragraph{Position reconstruction}
    
The position of events in Borexino is calculated using the photon arrival times. Since the events are selected inside the DFV based on the reconstructed position, the uncertainty in the position reconstruction of events affects the error on the fiducial volume, and thus, on the resulting exposure. This uncertainty is obtained using the calibration campaign performed in 2009~\cite{Back:2012awa}. Data from the $^{222}$Rn and $^{241}$Am-$^{9}$Be sources placed at 182 and 29 positions in the scintillator, respectively, was used for this. The reconstructed position of the source was compared to the nominal source position measured by the CCD camera inside the detector. The uncertainty in position reconstruction for the geoneutrino analysis was calculated using the shift in the positions for the $^{241}$Am-$^{9}$Be source. The maximal resulting uncertainty in the position was observed to be 5\,cm. Considering the nominal spherical radius of our FV of 4.15\,m, this gives an uncertainty of 3.6\% in the fiducial volume and consequently, in the corresponding exposure.

\subsection{Geoneutrino signal at LNGS}
\label{subsec:geo_lngs}

\begin{figure}[t]
\centering  
\includegraphics[width = 0.46\textwidth]{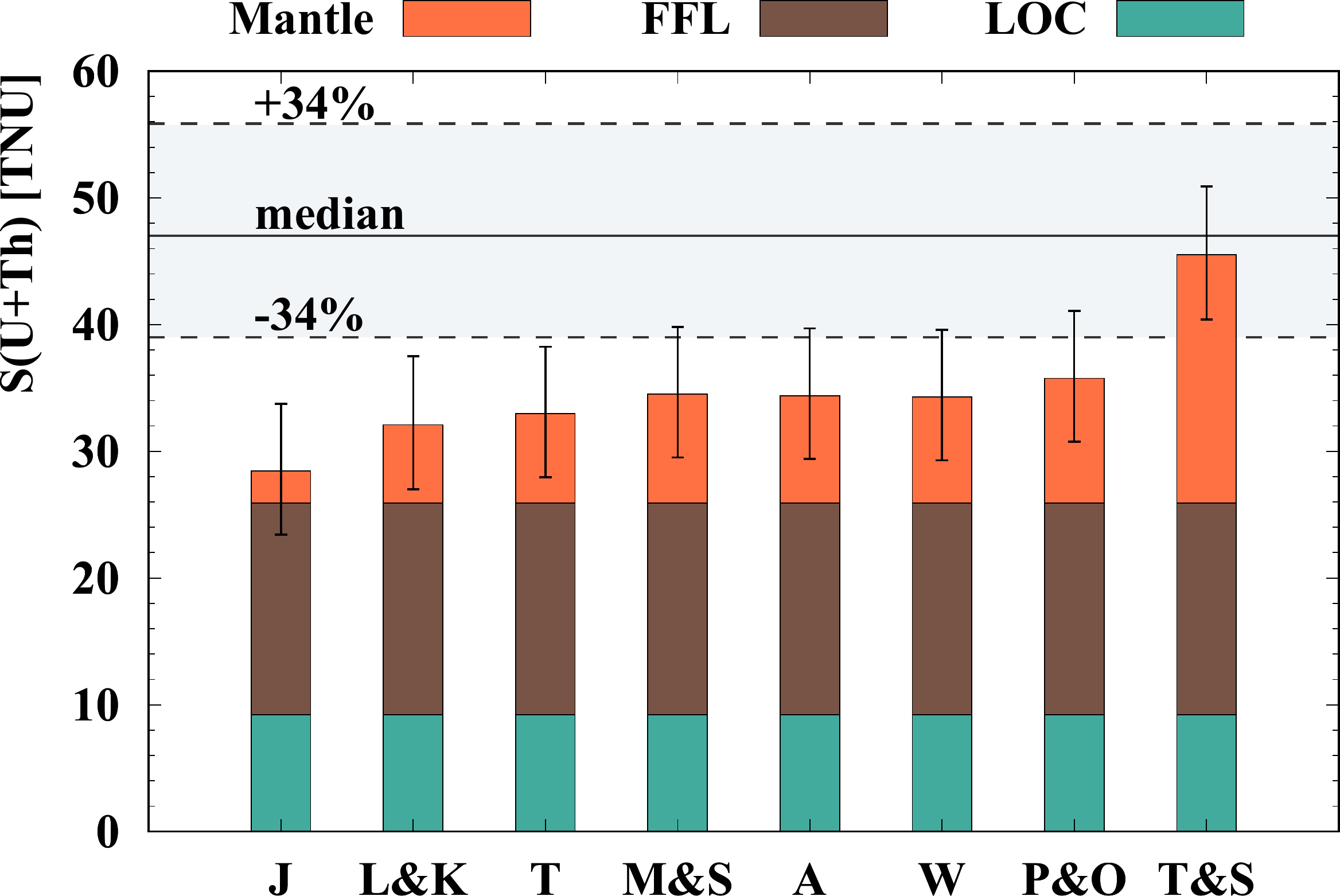}
\vspace{1mm}
\caption{Comparison of the expected geoneutrino signal $S_{\mathrm{geo}}$(U+Th) at LNGS (calculated according to different BSE models, see Sec.~\ref{subsec:geo}) with the Borexino measurement. For each model, the LOC and FFL contributions are the same (Table~\ref{tab:S_litho}), while the mantle signal is obtained considering an {\it intermediate scenario} (Fig.~\ref{fig:MantleScenarios}(b)). The error bars represent the 1$\sigma$ uncertainties of the total signal $S$(U+Th). The horizontal solid back line represents the geoneutrino signal $S_{\mathrm {geo}}^{\mathrm{med}}$, while the grey band the $I_{S \mathrm {geo}}^{68\mathrm{full}}$ interval as measured by Borexino.}
\label{fig:Sgeo_Bx_vs_model} 
\end{figure}

\begin{figure}[t]
\centering  
\vspace{-4mm}
\includegraphics[width = 0.5\textwidth]{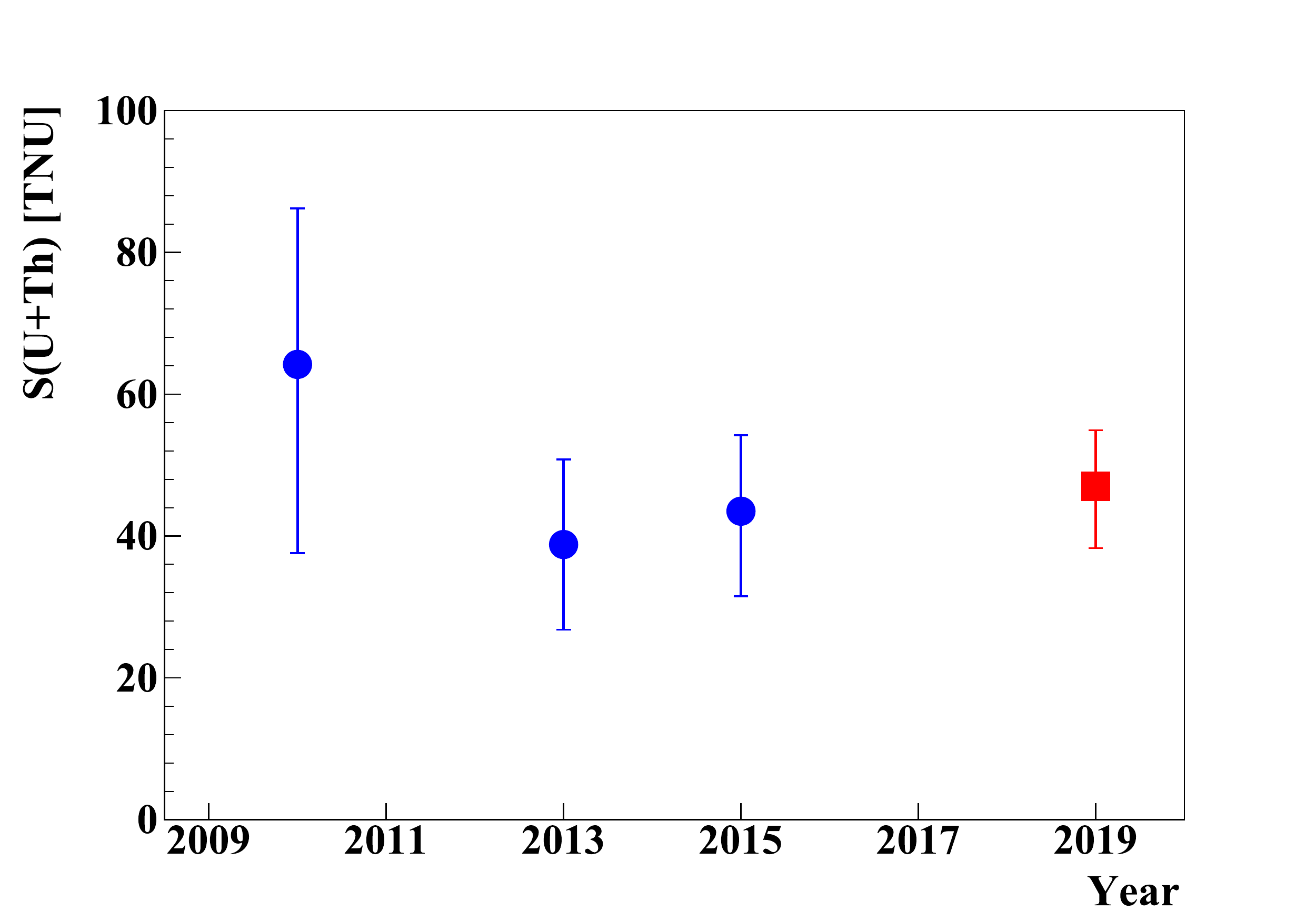}
\caption{Comparison of the geoneutrino signal $S_{\mathrm{geo}}$(U+Th) at LNGS as measured by Borexino. Blue circles indicate the results from 2010~\cite{Bellini:2010geo}, 2013~\cite{Bellini:2013geo}, and 2015~\cite{Agostini:2015cba}, while the red square demonstrates the current analysis.}
\label{fig:geo_signal_LNGS} 
\end{figure}

This Section details the conversion of the number of geoneutrino events $N_{\mathrm{geo}}$, resulting from the spectral fits described in Sec.~\ref{subsec:ngeo}, to the geoneutrino signal $S_{\mathrm{geo}}$ expressed in TNU, the unit introduced in Sec.~\ref{subsec:geo}:
\begin{equation}
S_{\mathrm{geo}} [\mathrm{TNU}] = \frac {N_{\mathrm{geo}}} { \varepsilon_{\mathrm{geo}} \cdot \frac{\mathcal{E}_p}{10^{32}}} = \frac {N_{\mathrm{geo}}} { \frac{\mathcal{E'}_p}{10^{32}}},
\label{eq:TNU}
\end{equation}  
where the detection efficiency $\varepsilon_{\mathrm{geo}}$ = 0.8698 $\pm$ 0.0150 (Table~\ref{tab:eff}) and the exposure $\mathcal{E}_p$ = (1.29 $\pm$ 0.05)$\times 10^{32}$\, protons $\times$ yr (Sec.~\ref{subsec:data}). We obtain $S_{\mathrm {geo}}^{\mathrm{best}}$ = 46.3\,TNU, the median value $S_{\mathrm {geo}}^{\mathrm{med}}$ = 47.0\,TNU, and including the systematic uncertainties from Table~\ref{tab:summary_sys}, the 68\% coverage interval $I_{S \mathrm {geo}}^{68\mathrm{full}}$ = (38.9 - 55.6)\,TNU. This results in a final precision of our measurement of $^{+18.3}_{-17.2}$\% with respect to $S_{\mathrm {geo}}^{\mathrm{med}}$. The comparison of the result, obtained assuming the chondritic Th/U mass ratio of 3.9, with the expected geoneutrino signal considering different geological models (Sec.~\ref{subsec:geo}) is shown in Fig.~\ref{fig:Sgeo_Bx_vs_model}.
Figure~\ref{fig:geo_signal_LNGS}
shows the time evolution of the Borexino measurements of the geoneutrino signal $S_{\mathrm{geo}}$(U+Th) at LNGS from 2010 up to the current result.
Table~\ref{tab:summary_results} summarizes the signals, expressed in TNU, for geoneutrinos and reactor antineutrinos obtained with the two fits, assuming Th/U chondritic ratio and keeping U and Th contributions as free fit parameters, as described in Sec.~\ref{subsec:ngeo}. It was shown in Sec.~\ref{subsec:expsens} that Borexino does not have any sensitivity to measure the Th/U ratio with the current exposure. Therefore, the ratio obtained from the fit when U and Th are free parameters is not discussed.

\subsection{Extraction of mantle signal}
\label{subsec:mantle}

  \begin{figure}[t]
     \centering  
    \subfigure[]{\includegraphics[width = 0.49\textwidth]{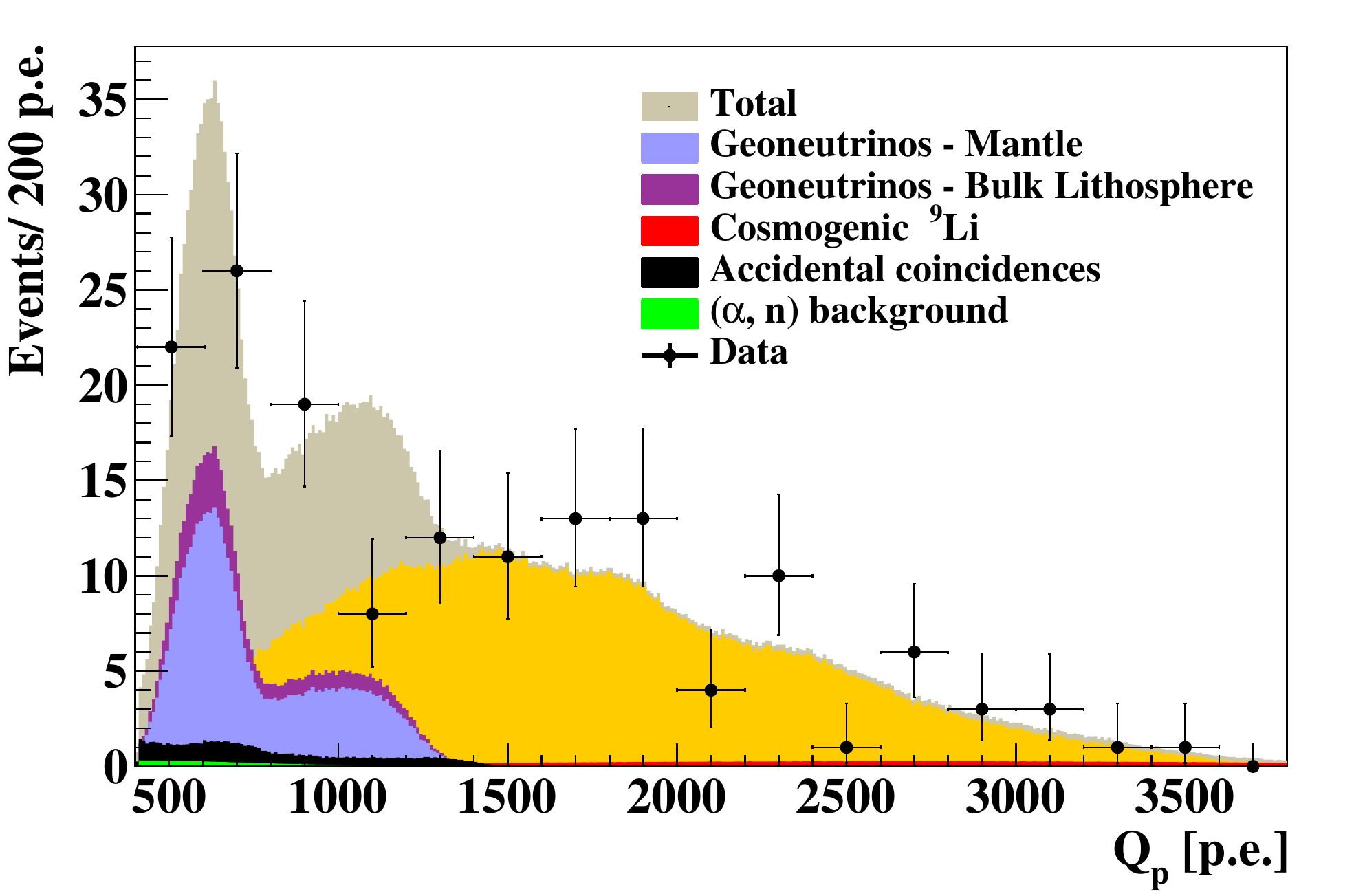}       
   \label{fig:fit_mantle}}
    \subfigure[]{\includegraphics[width = 0.49\textwidth]{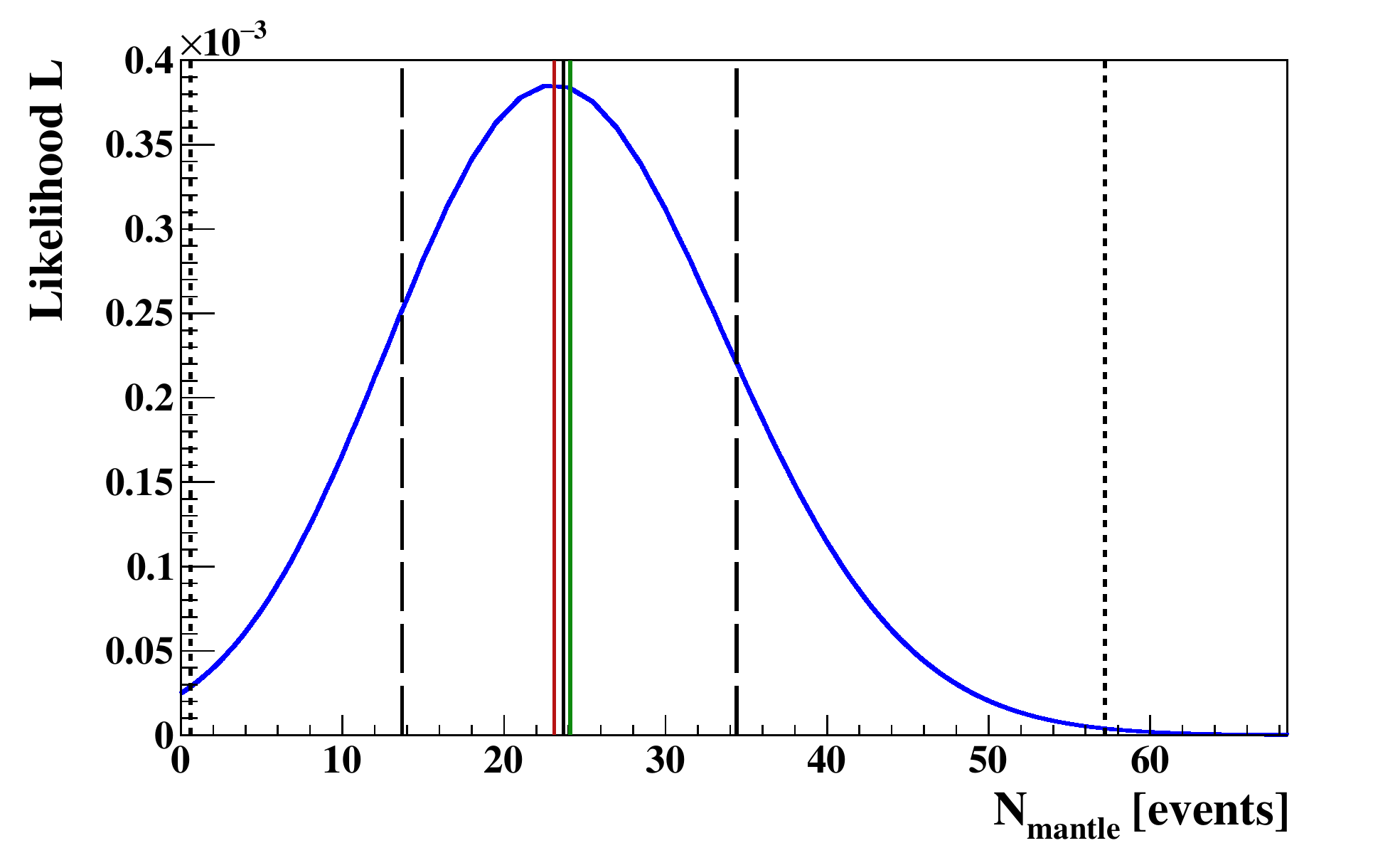}
   \label{fig:lkl_mantle}}
        \caption{(a) Spectral fit to extract the mantle signal after constraining the contribution of the bulk lithosphere. The grey shaded area shows the summed PDFs of all the signal and background components. (b) The likelihood profile for $N_{\mathrm {mantle}}$, the number of mantle geoneutrino events. The vertical solid red line indicates the best fit, while the vertical solid black and green lines indicate the median and mean values of the distributions, respectively. The vertical dashed/dotted lines represent the 68\%/99.7\%\, confidence intervals of the distribution.}
    \end{figure}
    
  \begin{figure} [t]
  	\centering
  	\includegraphics[width = 0.49\textwidth]{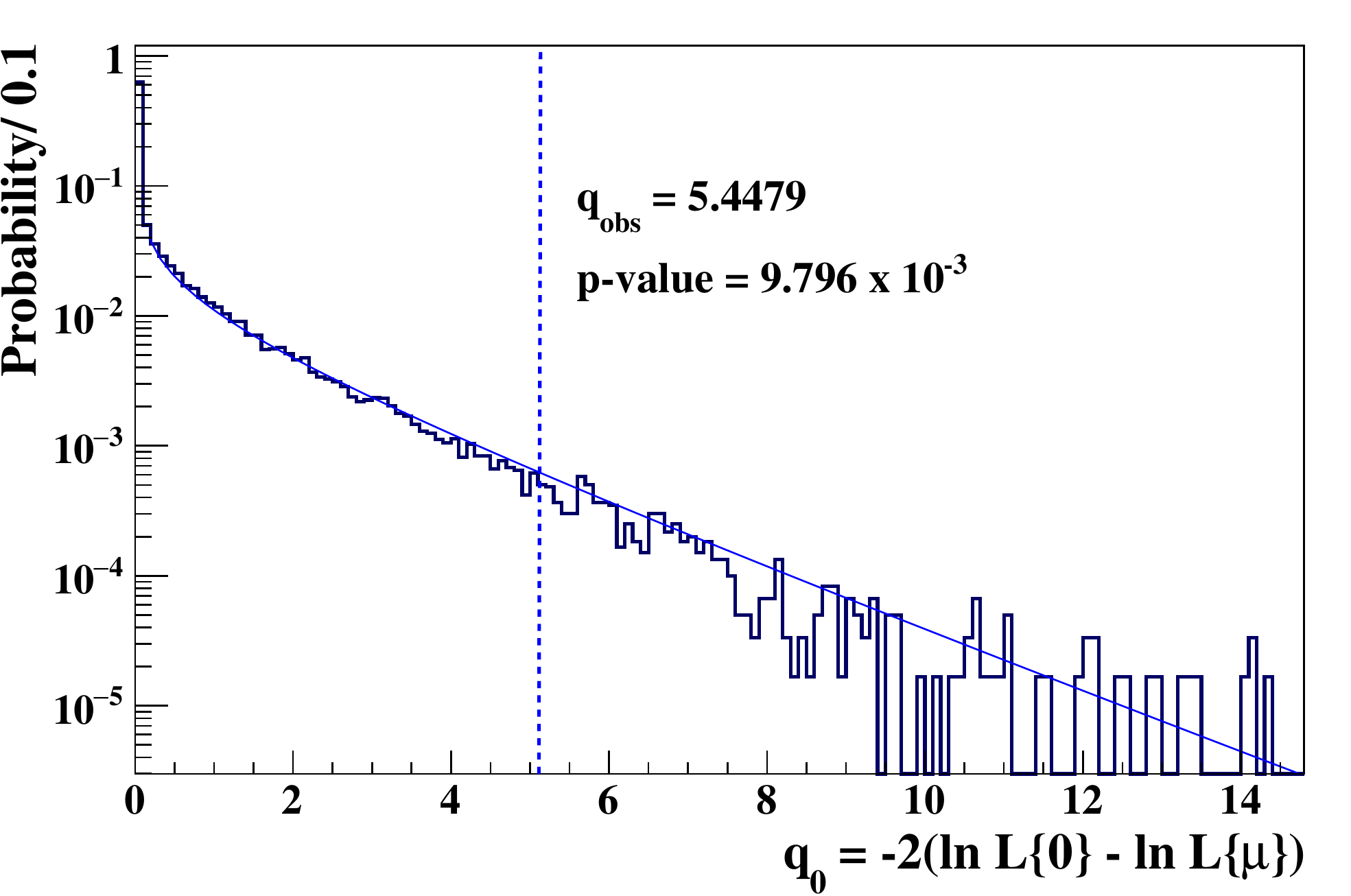}
  	\caption{The $q_{0}$ distribution for 1.2 million pseudo-experiments without any generated mantle signal fitted with $f(q|0)$ (Eq.~\ref{eq:mantlesens0}). The vertical dashed line represents $q_{\mathrm{obs}}$ obtained from the data. The indicated $p$-value, calculated following Eq.~\ref{eq:p-value} by setting $q_{\mathrm{med}}$ = $q_{\mathrm{obs}}$, represents the statistical significance of the Borexino observation of the mantle signal.}
  	\label{fig:mantle_obs}
  \end{figure}
  
  The mantle signal was extracted from the spectral fit by constraining the contribution from the bulk lithosphere according to the expectation discussed in Sec.~\ref{subsec:geo} and given in Table~\ref{tab:antinu-events-expected} as 28.8$^{+5.5}_{-4.6}$\,events. The corresponding MC PDF was constructed from the PDFs of $^{232}$Th and $^{238}$U geoneutrinos shown in Fig.~\ref{fig:PDFs-Geo-Rea}. They were scaled with the lithospheric Th/U signal ratio equal to 0.29 (Table~\ref{tab:S_litho}). The MC PDF used for the mantle was also constructed from the $^{232}$Th and $^{238}$U PDFs, but the applied Th/U signal ratio was 0.26, the value discussed in Sec.~\ref{subsec:geo}. The mantle signal, as well as the reactor antineutrino contribution were free in the fit. The best fit is shown in Fig.~\ref{fig:fit_mantle}. It resulted in a mantle signal of $N_{\mathrm {mantle}}^{\mathrm{best}}$ = 23.1\,events, with the median value $N_{\mathrm {mantle}}^{\mathrm{med}}$ = 23.7\,events, and the 68\% coverage interval $I_{N \mathrm {mantle}}^{68\mathrm{stat}}$ = (13.7 - 34.4)\,events. The likelihood profile of the mantle signal is shown in Fig.~\ref{fig:lkl_mantle}. After considering the systematic uncertainties, the final mantle signal can be given as $S_{\mathrm {mantle}}^{\mathrm{best}}$ = 20.6\,TNU, with the median value $S_{\mathrm {mantle}}^{\mathrm{med}}$ = 21.2\,TNU, and the 68\% coverage interval $I_{S \mathrm {mantle}}^{68\mathrm{full}}$ = [12.2 - 30.8]\,TNU, as shown also in Table~\ref{tab:summary_results}.

The statistical significance of the mantle signal was studied using MC pseudo-experiments with and without a generated mantle signal as described in Sec~\ref{subsec:expsens}. The $q_{\mathrm{obs}}$ obtained from the spectral fit is 5.4479, and it is compared with the theoretical function $f(q|0)$, described in Sec.~\ref{subsec:expsens}, Eq.~\ref{eq:mantlesens0}, as shown in Fig.~\ref{fig:mantle_obs}. The corresponding {\it p}-value is 9.796 $\times$ 10$^{-3}$. Therefore, in conclusion the null-hypothesis of the mantle signal can be rejected with 99.0\%\,C.L. (corresponding to 2.3$\sigma$ significance). 
The Borexino mantle signal can be compared with calculations according to a wide spectrum of BSE models (Table~\ref{tab:S_mantle}). The Borexino measurement constrains at 90(95)\% C.L. a mantle composition with $a_{\mathrm{mantle}}$(U) $>$ 13(9)\,ppb and $a_{\mathrm{mantle}}$(Th) $>$ 48(34)\,ppb assuming for the mantle homogeneous distribution of U and Th and a Th/U mass ratio of 3.7.

          \subsection{Estimated radiogenic heat}
        \label{subsec:radiogenic}
 \begin{figure}[t]
\centering  
\includegraphics[width = 0.48\textwidth]{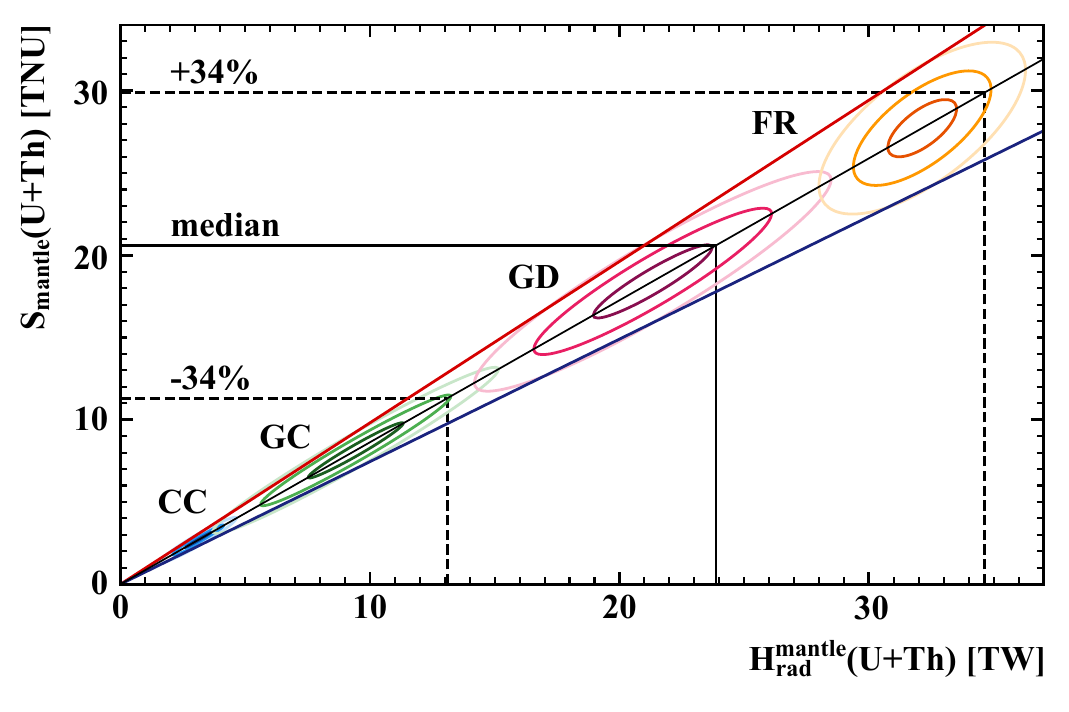}
\caption{Mantle geoneutrino signal expected in Borexino as a function of U and Th mantle radiogenic heat: the area between the red and blue lines denotes the full range allowed between a homogeneous mantle (high scenario - Fig.~\ref{fig:MantleScenarios}c) and a unique rich layer just above the CMB (low scenario - Fig.~\ref{fig:MantleScenarios}a). The slope of the central inclined black line ($\beta_{\mathrm{centr}}$ = 0.86\,TNU/TW) is the average of the slopes of the blue and red lines. The blue, green, red, and yellow ellipses are calculated with the following U and Th mantle radiogenic power $H_{\mathrm{rad}}^{\mathrm {mantle}}$(U+Th) (with 1$\sigma$ error) according to different BSE models: CC model $(3.1 \pm 0.5)$\,TW, GC model $(9.5 \pm 1.9)$\,TW, GD model $(21.3 \pm 2.4)$\,TW, and FR model $(32.2 \pm 1.4)$\,TW. For each model darker to lighter shades of respective colours represent 1, 2, and 3$\sigma$ contours. The black horizontal lines represent the mantle signal measured by Borexino: the median mantle signal (solid line) and the 68\% coverage interval (dashed lines).}
\label{fig:SUTh_vs_Heat} 
\end{figure}

The global HPEs' masses in the Earth are estimated by matching geophysical, geochemical, and cosmochemical arguments. Direct samplings of the accessible lithosphere constrain the radiogenic heat of $H_{\mathrm{rad}}^{\mathrm{LSp}}$(U+Th+K) = 8.1$^{+1.9}_{-1.4}$\,TW (Table~\ref{tab:MH_litho}), corresponding to $\sim$17\% of the total terrestrial heat power $H_{\mathrm{tot}}$ = $(47\pm 2)$\,TW. The radiogenic heat from the unexplored mantle could embrace a wide range of $H_{\mathrm{rad}}^{\mathrm{mantle}}$(U+Th+K) = (1.2 – 39.8)\,TW (Table~\ref{tab:S_mantle}), where the highest values are obtained for a Fully Radiogenic Earth model.

The total amount of HPEs, as well as their distribution in the deep Earth, affect the geoneutrino flux. We will express the dependence of the expected mantle geoneutrino signal $S_{\mathrm{mantle}}$(U+Th) on the mantle radiogenic power $H_{\mathrm{rad}}^{\mathrm {mantle}}$(U+Th). The unequivocal relation between the radiogenic power and the HPEs' masses can be expressed via the constant U and Th specific heats $h$(U) = 98.5\,$\mu$W/kg and $h$(Th) = 26.3\,$\mu$W/kg~\cite{Dye:2011mc}:
 \begin{eqnarray}
 \label{eq:H_on_M}
H_{\mathrm{rad}}^{\mathrm{mantle}}& &(\mathrm{U + Th}) = \\
\nonumber
&=& h(\mathrm{U}) \cdot M_{\mathrm{mantle}}(\mathrm{U}) + h(\mathrm{Th}) \cdot M_{\mathrm{mantle}}(\mathrm{Th})\\
\nonumber
&=& [h(\mathrm{U}) + 3.7 \cdot h(\mathrm{Th})] \cdot M_{\mathrm{mantle}}(\mathrm{U}),
 \end{eqnarray}
where $M_{\mathrm{mantle}}$(U) is the U mass in the mantle (Table~\ref{tab:S_mantle}). In the last passage as well as in all calculations below, we assume the mantle Th/U mass ratio of 3.7. With this assumption, for a given detector site the ratio:
\begin{equation}
\beta = S_{\mathrm{mantle}} \mathrm{(U+Th)} / H_{\mathrm{rad}}^{\mathrm{mantle}}(\mathrm{U + Th})
\label{eq:beta}
\end{equation}
depends only on U and Th distribution in the mantle. For Borexino, the calculated $\beta$ ranges between $\beta_{\mathrm{low}} = 0.75$\,TNU/TW and $\beta_{\mathrm{high}} = 0.98$\,TNU/TW, obtained assuming the HPEs placed in an unique HPEs-rich layer just above the CMB (i.e. low scenario, Fig.~\ref{fig:MantleScenarios}a)) and homogeneously distributed in the mantle (high scenario, Fig.~\ref{fig:MantleScenarios}c), respectively. Considering then Eq.~\ref{eq:H_on_M}, the linear relation between the mantle signal $S_{\mathrm{mantle}}$(U+Th) and radiogenic power $H_{\mathrm{rad}}^{\mathrm {mantle}}$(U+Th) can be expressed:
\begin{eqnarray}
\label{eq:S_vs_H}
\nonumber
S_{\mathrm {mantle}} & & (\mathrm {U+Th}) \\
\nonumber
&=& \beta \cdot \left [ h(\mathrm{U})+3.7 \cdot h(\mathrm{Th})
\right ] \cdot M_{\mathrm{mantle}}(\mathrm{U}) \\
&=& \beta \cdot H_{\mathrm{rad}}^{\mathrm{mantle}}(\mathrm{U + Th})
\end{eqnarray}
is reported in Fig.~\ref{fig:SUTh_vs_Heat}, where the slope of central line (i.e. $\beta_{\mathrm{centr}}$ = 0.86\,TNU/TW, black line) is the average of $\beta_{\mathrm{low}}$ (blue line) and $\beta_{\mathrm{high}}$ (red line). The area between the two extreme lines denotes the region allowed by all possible U and Th distributions in the mantle, assuming that the abundances in this reservoir are radial, non-decreasing function of the depth and in a fixed ratio $M_{\mathrm{mantle}}$(Th)/$M_{\mathrm{mantle}}$(U) = 3.7. The maximal and minimal excursions of mantle geoneutrino signal is taken as a proxy for the 3$\sigma$ error range.

Since the radiogenic heat power of the lithosphere is independent from the BSE model, the discrimination capability of Borexino geoneutrino measurement among the different BSE models can be studied in the space $S_{\mathrm{mantle}}$ (U + Th) vs $H_{\mathrm{rad}}^{\mathrm{mantle}}$ (U + Th). In Figure~\ref{fig:SUTh_vs_Heat}, the solid black horizontal line represents the Borexino measurement, the median $S^{\mathrm {med}}_{\mathrm {mantle}}$, which falls within prediction of the Geodynamical model (GD). The 68\% coverage interval $I_{S \mathrm {mantle}}^{68\mathrm{full}}$, also represented in Fig.~\ref{fig:SUTh_vs_Heat} by horizontal black dashed lines, covers the area of prediction of the GD and the Fully Radiogenic (FR) models. We are least compatible with the Cosmochemical model (CC), which central value agrees with our measurement at 2.4$\sigma$ level.

\begin{figure}[t]
\centering  
\includegraphics[width = 0.46\textwidth]{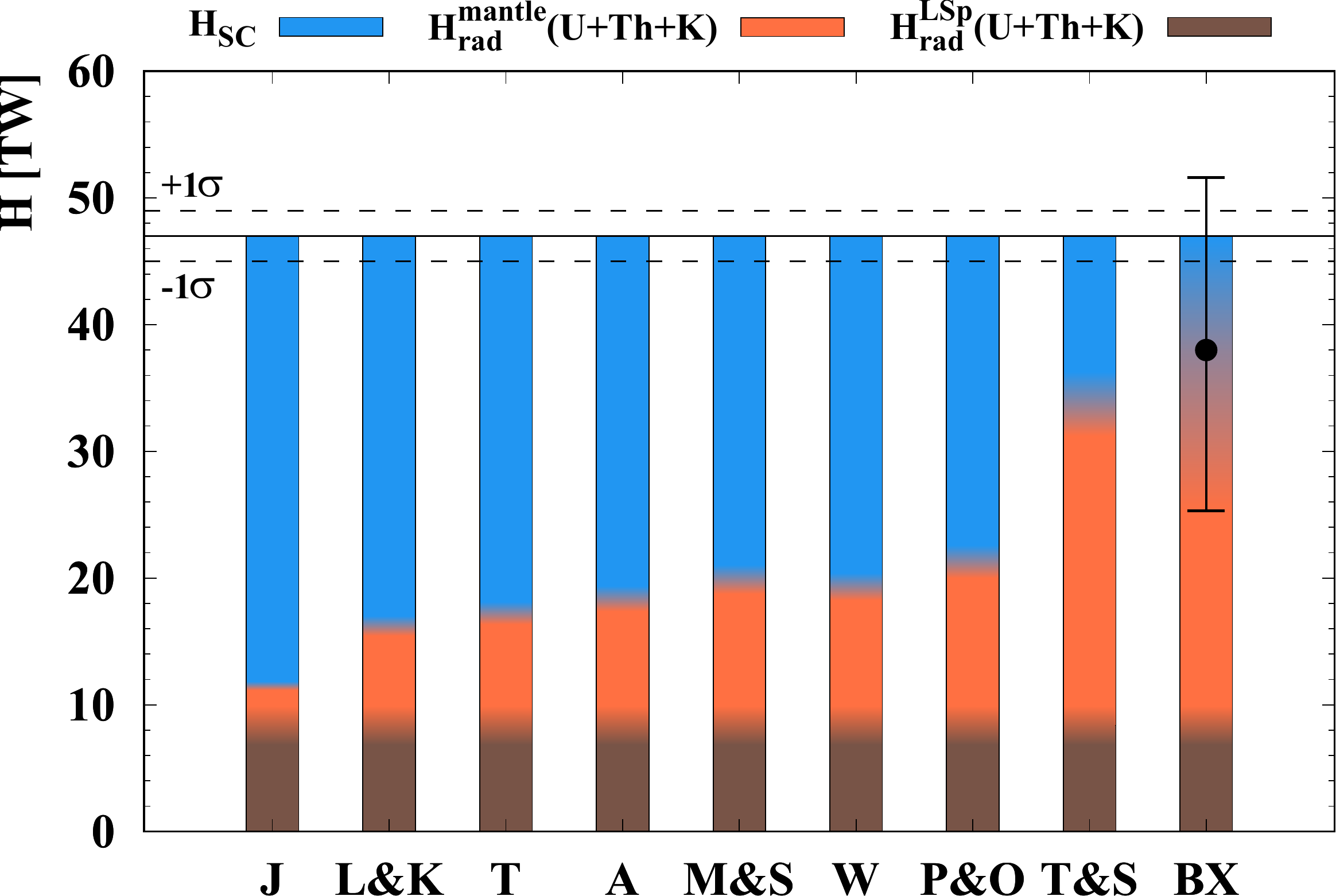}
\vspace{4.2mm}
\caption{Decomposition of the Earth's total surface heat flux $H_{\mathrm{tot}}$ = $(47\pm2)$\,TW (horizontal black lines) into its three major contributions - lithospheric (brown) and mantle (orange) radiogenic heat $H_{\mathrm{rad}}^{\mathrm{LSp}}$ and $H_{\mathrm{rad}}^{\mathrm{mantle}}$, respectively, and secular cooling $H_{\mathrm{SC}}$ (blue). 
 The labels on the $x$ axis identify different BSE models (Table~\ref{tab:S_mantle}), while the last bar labeled BX represents the Borexino measurement. The lithospheric contribution $H_{\mathrm{rad}}^{\mathrm{LSp}}$ = $8.1^{+1.9}_{-1.4}$\,TW (Table~\ref{tab:MH_litho}) is the same for all bars. The amount of HPEs predicted by BSE models determines the mantle radiogenic heat (Table~\ref{tab:S_mantle}), while for Borexino the value of 30.0$^{+13.5}_{-12.7}$\,TW is inferred from the extracted mantle signal. The difference between $H_{\mathrm{tot}}$ and the respective total radiogenic heat is assigned to the heat from secular cooling of the Earth.}
\label{fig:H_vs_BSE} 
\end{figure}

The mantle signal measured by Borexino can be converted to the corresponding radiogenic heat by inverting the Eq.~\ref{eq:S_vs_H}. Since the experimental error on the mantle signal is much larger than the systematic variability associated to the U and Th distribution in the mantle, the radiogenic power from U and Th in the mantle $H_{\mathrm{rad}}^{\mathrm{mantle}}$(U + Th)
inferred from the Borexino signal $S_{\mathrm{mantle}}$(U+Th)
can be obtained with:
\begin{eqnarray}
     \label{eq:StoH}
     \nonumber
H_{\mathrm{rad}}^{\mathrm{mantle}} (\mathrm{U + Th}) &=& (1/ \beta_{\mathrm{centr}}) \cdot S_{\mathrm{mantle}} (\mathrm{U + Th}) \\
&=& 1.16 \cdot S_{\mathrm{mantle}} (\mathrm{U + Th}).
\end{eqnarray}
Adopting $S^{\mathrm{med}}_{\mathrm{mantle}}$(U+Th) = 21.2\,TNU together with the 68\% C.L. interval including both statistical and systematic errors (Table~\ref{tab:summary_results}), we obtain:
\begin{equation}
     \begin{split}
     H_{\mathrm{rad}}^{\mathrm {mantle-med}}(\mathrm{U + Th}) &= 24.6\,\text{TW}\\
     I_{H ^{\mathrm {mantle}}_{\mathrm{rad}} }^{68\mathrm{full}} (\mathrm{U + Th}) & = 14.2 - 35.7\,\text{TW}.
     \end{split}
     \label{eq:H_BX_UTh}
\end{equation}
Summing the radiogenic power of U and Th in the lithosphere $H^{\mathrm{LSp}}_{\mathrm{rad}}$(U+Th) = 6.9$^{+1.6}_{-1.2}$\,TW, the Earth's radiogenic power from U and Th is $H_{\mathrm{rad}}$(U+Th) = 31.7$^{+14.4}_{-9.2}$\,TW.

Assuming the contribution from $^{\mathrm{40}}$K to be 18\% of the total mantle radiogenic heat (Sec.~\ref{sec:geo}), the total radiogenic mantle signal can be expressed as $H^{\mathrm {mantle}}_{\mathrm {rad}}(\mathrm{U + Th + K})$ = 30.0$^{+13.5}_{-12.7}$\,TW, where we have expressed the 1$\sigma$ errors with respect to the median.
If we further add the lithospheric contribution $H_{\mathrm{rad}}^{\mathrm{LSp}}$(U+Th+K) = 8.1$^{+1.9}_{-1.4}$\,TW, we get the 68\% coverage interval for the Earth's radiogenice heat $H_{\mathrm {rad}}$(U + Th + K) = 38.2$^{+13.6}_{-12.7}$\,TW, as shown in Fig.~\ref{fig:H_vs_BSE}.

\begin{figure}[t]
\centering  
\includegraphics[width = 0.47\textwidth]{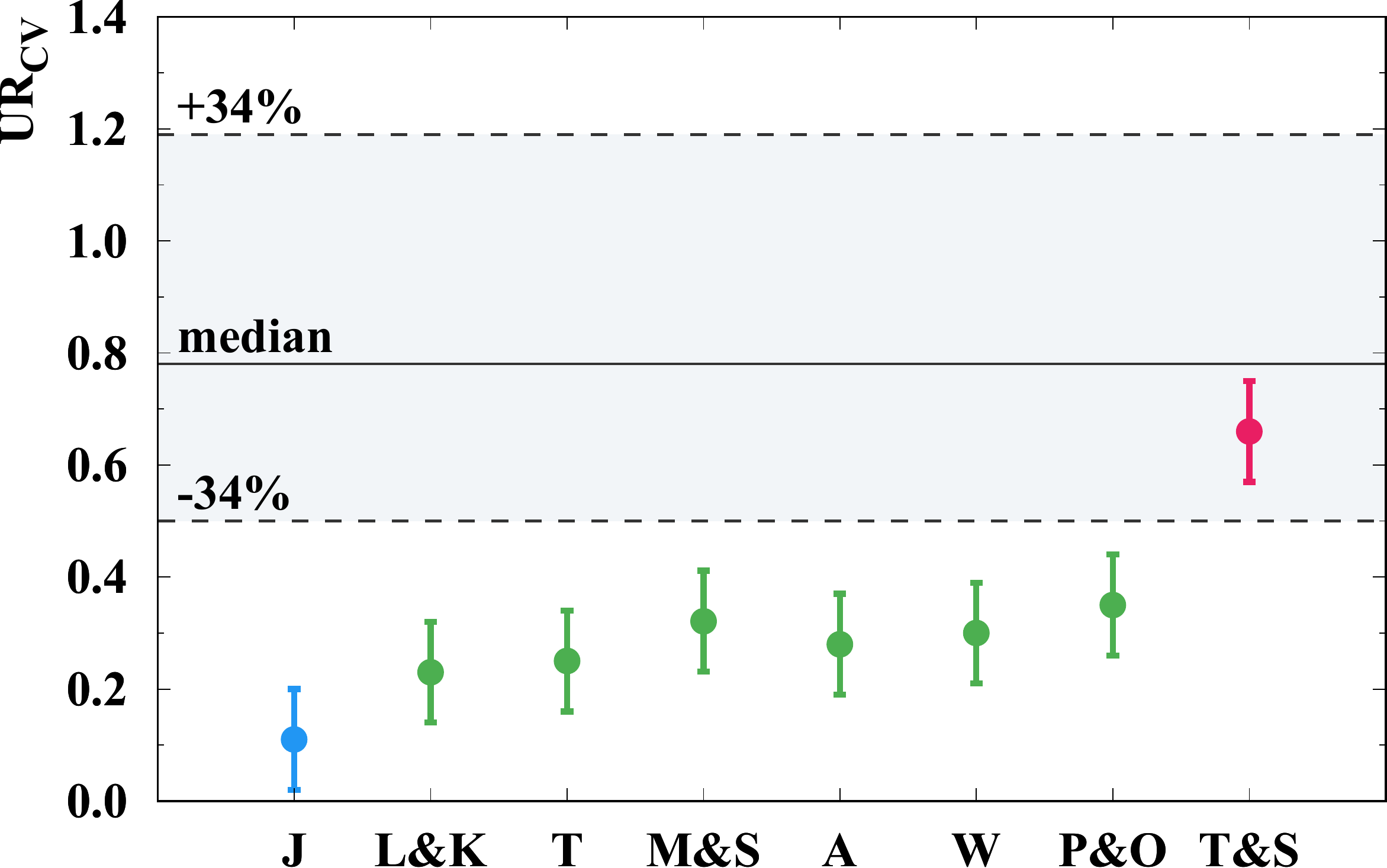}
\vspace{0.5mm}
\caption{Comparison of Borexino constraints (horizontal band) with predictions of the BSE models (points with $\pm$3$\sigma$ error bars, Table~\ref{tab:S_mantle}) for the convective Urey ratio $UR_{\mathrm{CV}}$ (Eq.~\ref{eq:URCV}), assuming the total heat flux $H_{\mathrm{tot}}$ = $(47\pm 2)$\,TW and the radiogenic heat of the continental crust $H_{\mathrm{rad}}^{\mathrm{CC}}$= $6.8 ^{\mathrm{+1.4}}_{\mathrm{-1.1}}$\,TW (Table~\ref{tab:MH_litho}). The blue, green, and red colours represent different BSE models (CC, GC, and GD; Table~\ref{tab:S_mantle}, respectively).}
\label{fig:UR_vs_BSE} 
\end{figure}

 The experimental error on the Earth's radiogenic heat power estimated by Borexino is comparable with the spread of power predictions derived from the eight BSE models reported in Table~\ref{tab:BSE}. This comparison is represented in Fig.~\ref{fig:H_vs_BSE}. Among these, a preference is found for models with relatively high radiogenic power, which correspond to a cool initial environment at early Earth's formation stages and small values of the current heat coming from the secular cooling. However, no model can be excluded at 3$\sigma$ level.

 The total radiogenic heat estimated by Borexino can be used to extract the convective Urey ratio according to Eq.~\ref{eq:URCV}. The resulting value of $UR_{\mathrm{CV}}$ = 0.78$^{+0.41}_{-0.28}$ is compared to the $UR_{\mathrm{CV}}$ predicted by different BSE models in Fig.~\ref{fig:UR_vs_BSE}. The Borexino geoneutrino measurement constrains at 90(95)\% C.L. a mantle radiogenic heat power to be $H_{\mathrm{rad}}^{\mathrm{mantle}}$(U+Th) $>$ 10(7)\,TW and $H_{\mathrm{rad}}^{\mathrm{mantle}}$(U+Th+K) $>$ 12.2(8.6)\,TW and the convective Urey ratio $UR_{CV}$ $>$ 0.13(0.04).

        \subsection{Testing the georeactor hypothesis}
        \label{subsec:georeactor-results}

\begin{figure}[t]
        \centering
        \vspace{-4mm}
        \includegraphics[width = 0.50 \textwidth]{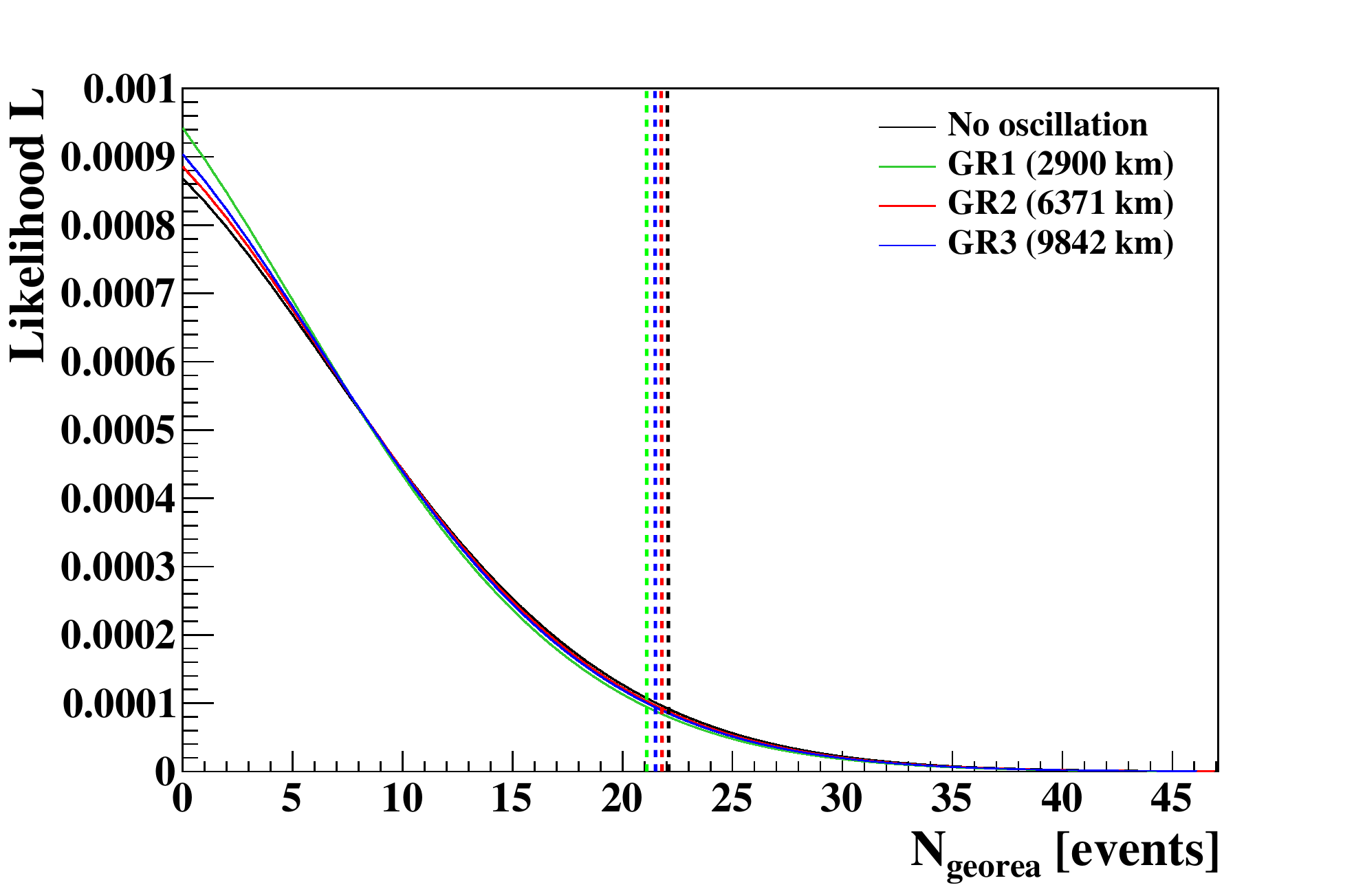}
        \caption{Likelihood profiles for the number of georeactor events $N_{\mathrm {georea}}$ obtained in the spectral fits with the constrained number of reactor antineutrino events. The four profiles represent the fits that differ in the shape of the PDF used for the georeactor contribution (Fig.~\ref{fig:PDFs-georeactor}), corresponding to different source positions in the deep Earth (GR1, GR2, GR3, Fig.~\ref{fig:GeoReactorPosition}) and to a no-oscillation hypothesis. The vertical dashed lines indicate the upper limit at 95\%\, C.L. for $N_{\mathrm {georea}}$ as obtained in each of the four fits. In setting the upper limit on the power of the georeactor, that depends on the assumed location of the georeactor, we use conservatively the highest limit on $N_{\mathrm {georea}}$ equal to 21.7\,events.}
        \label{fig:chi_georea}
    \end{figure}

    The georeactor hypothesis described in Sec.~\ref{subsec:georeactor} was tested by performing the spectral fit after constraining the expected number of reactor antineutrino events (Table~\ref{tab:antinu-events-expected}) to 97.6 $\pm$ 1.7 (stat) $\pm$ 5.2 (syst). The geoneutrino (Th/U fixed to chondritic mass ratio of 3.9) and georeactor contributions were left free in the fit. For each georeactor location (Fig.~\ref{fig:GeoReactorPosition}), we have used the respective georeactor PDF as in Fig.~\ref{fig:PDFs-georeactor}. However, their shapes are practically identical and Borexino does not have any sensitivity to distinguish them. The different likelihood profiles obtained using different georeactor PDFs are very similar (including the PDF constructed assuming no neutrino oscillations), as shown in Fig.~\ref{fig:chi_georea}. The vertical lines represent the 95\% C.L. limits for the number of georeactor events $N_{\mathrm {georea}}$ obtained in fits with different georeactor PDFs. In setting the upper limit on the power of the georeactor, that depends on the assumed location of the georeactor, we use conservatively the highest limit on $N_{\mathrm {georea}}$ equal to 21.7\,events. The latter is transformed to the signal $S_{\mathrm{georea}}$ of 18.7\,TNU, as the 95\% C.L. upper limit on the signal coming from a hypothetical georeactor. 

    Considering the values from Table~\ref{tab:antinu-signals-expected}, that is, the predicted georeactor signal expressed in TNU for a 1\,TW georeactor in different locations, these upper limits on the georeactor power are set: 2.4\,TW for the location in the Earth center (GR2) and 0.5\,TW and 5.7\,TW for the georeactor placed at the CMB at 2900\,km (GR1) and 9842\,km (GR3), respectively. Therefore, we exclude the existence of a georeactor with a power greater than 0.5/2.4/5.7\,TW at 95\% C.L., assuming its location at 2900/6371/9842\,km distance from the detector. 
    
 \begin{table*}[h]
	\centering
	\caption{\label{tab:summary_results} Summary of the number of geoneutrino and reactor antineutrino events and the corresponding signals in TNU as well as the fluxes obtained from this work. The systematic uncertainties (Table~\ref{tab:summary_sys}) are given for the median values.} \vskip 2pt
	\begin{tabular*}{\textwidth}{l @{\hskip 24pt} c @{\hskip 18pt} c @{\hskip 18pt} c @{\hskip 18pt} c @{\hskip 18pt} c @{\hskip 18pt} c @{\hskip 18pt} c @{\hskip 18pt} c @{\hskip 18pt} c @{\hskip 24pt} c}
		\hline
		\hline
		Criteria & Best fit & Median & Mean & \thead{{\normalsize 68\%\,C.L.} \\ {\normalsize stat.}} & \thead{{\normalsize 99.7\%\,C.L.} \\ {\normalsize stat.}} & \thead{{\normalsize Sys.} \\ {\normalsize error}} & \thead{{\normalsize 68\%\,C.L.} \\ {\normalsize stat. \& sys.}}  \Tstrut\Bstrut \\
		\hline
		 \multicolumn{8}{c}{$R_s$ = $S$(Th)/$S$(U) = 0.27}  \\ [2pt]
		\hline
	 	$N_{\mathrm{geo}}$ [events] & 51.9 & 52.6 & 53.0 & 44.0 - 62.0 & 28.8 - 82.6 & {\Large $^{+2.7}_{-2.1}$} & 43.6 - 62.2 \TstrutLarge \\ [7pt]
    		$N_{\mathrm{rea}}$ [events] & 92.5 & 93.4 & 93.6 & 82.6 - 104.7 &63.6 - 129.7 &  {\Large $^{+4.8}_{-5.1}$} & 81.6 - 105.8 \\ [7pt]
    		$N_{\mathrm{U}}$ [events] & 40.5 & 41.1 & 41.4 & 34.4 - 48.4 & 22.5 - 64.5 &  {\Large $^{+2.1}_{-1.6}$} & 34.0 - 48.6 \\ [7pt]
    		$N_{\mathrm{Th}}$ [events] & 11.4 & 11.5 & 11.6 & 9.7 - 13.6 & 6.3 - 18.1 & {\Large  $^{+0.6}_{-0.5}$} & 9.6 - 13.7\\ [12pt]
    		$S_{\mathrm{geo}}$ [TNU] & 46.3 & 47.0 & 47.3 & 39.3 - 55.4  & 25.7 - 73.8  & {\Large $^{+2.4}_{-1.9}$} & 38.9 - 55.6  \\ [7pt]
    		$S_{\mathrm{rea}}$ [TNU] & 79.7 & 80.5 & 80.7 & 71.2 - 90.3 & 54.8 - 111.8 & {\Large $^{+4.1}_{-4.4}$} & 70.3 - 91.2 \\ [7pt]
    		$S_{\mathrm{U}}$ [TNU] & 35.9 & 36.3 & 36.6 & 30.4 - 42.8 & 19.9 - 57.1 & {\Large $^{+1.9}_{-1.5}$} & 30.1 - 43.0 \\ [7pt]
   		$S_{\mathrm{Th}}$ [TNU] & 10.4 & 10.5 & 10.6 & 8.7 - 12.4 & 5.7 - 16.5 & {\Large $^{+0.6}_{-0.4}$} & 8.8 - 12.6 \\ [12pt]
    		$\phi_{\mathrm{U}}$ [10$^{6}$ cm$^{-2}$s$^{-1}$] & 2.8 & 2.8 & 2.9 & 2.4 - 3.4 & 1.6 - 4.5 & {\Large $^{+0.2}_{-0.1}$}  & 2.4 - 3.4  \\ [7pt]
    		$\phi_{\mathrm{Th}}$ [10$^{6}$ cm$^{-2}$s$^{-1}$] & 2.5 & 2.6 & 2.6  & 2.2 - 3.0  & 1.4 - 4.1 & {\Large $^{+0.2}_{-0.1}$}  & 2.1 - 3.1 \BstrutLarge \\
    		\hline
    	    \multicolumn{8}{c}{$R_s$ (lithosphere) = $S$(Th)/$S$(U) = 0.29; $R_s$ (mantle) = $S$(Th)/$S$(U) = 0.26} \\ [2pt]
    	    \hline
    		$N_{\mathrm{mantle}}$ [events] & 23.1 & 23.7 & 24.1 & 13.7 - 34.4 & 0.6 - 57.2 &  {\Large $^{+1.2}_{-1.0}$} & 13.6 - 34.4 \TstrutLarge \\ [7pt]	
    		$S_{\mathrm{mantle}}$ [TNU] & 20.6 & 21.2 & 21.5 & 12.2 - 30.7 & 0.5 - 51.1 & {\Large $^{+1.1}_{-0.9}$} & 12.2 - 30.8 \BstrutLarge \\ 
    		\hline 
    		\multicolumn{8}{c}{$S$(Th) and $S$(U) independent}  \\ [2pt]
		\hline
	 	$N_{\mathrm{geo}}$ [events] & 48.9 & 50.4 & 51.3 & 28.4 - 73.9 & 1.1 - 124.1 & {\Large $^{+2.6}_{-2.0}$} & 28.2 - 74.0 \TstrutLarge \\ [7pt]
    		$N_{\mathrm{rea}}$ [events] &  95.8 & 96.7 & 97.1 & 85.2 - 109.0 &65.1 - 136.1 & {\Large $^{+4.9}_{-5.3}$}  &  84.2 - 110.1 \\ [7pt]
    		$N_{\mathrm{U}}$ [events] &  27.8 & 29.0 & 29.7 & 16.1 - 43.1 & 0.6 - 73.7& {\Large $^{+1.5}_{-1.2}$} & 16.0 - 43.2  \\ [7pt]
    		$N_{\mathrm{Th}}$ [events] &  21.1 & 21.4 & 21.6 & 12.3 - 30.8 & 0.5 - 50.4 & {\Large $^{+1.1}_{-0.9}$} & 12.2 - 30.8 \\ [12pt]
    		$S_{\mathrm{geo}}$ [TNU] & 43.7 & 45.0 & 45.8 & 25.4 - 66.0 & 1.0 - 110.8 & {\Large $^{+2.3}_{-1.8}$} & 25.2 - 66.1\\ [7pt]
   		$S_{\mathrm{rea}}$ [TNU] & 82.6 & 83.4 & 83.7 & 73.5 - 94.0 & 56.1 - 117.3 & {\Large $^{+4.2}_{-4.6}$}  & 72.6 - 94.9 \\ [7pt]
    		$S_{\mathrm{U}}$ [TNU] & 24.6 & 25.7 & 26.3 & 14.3 - 38.1 & 0.5 - 65.2 & {\Large $^{+1.3}_{-1.0}$} & 14.2 - 38.2 \\ [7pt]
   		$S_{\mathrm{Th}}$ [TNU] & 19.2  & 19.5 & 19.6 & 11.2 - 28.0 & 0.5 - 45.8 & {\Large $^{+1.0}_{-0.8}$} & 11.1 - 28.0 \\ [12pt]
		$\phi_{\mathrm{U}}$ [10$^{6}$ cm$^{-2}$s$^{-1}$] &  1.9 & 2.0 & 2.1 & 1.1 - 3.0 & 0.04 - 5.1 & {\Large $^{+0.1}_{-0.1}$} & 1.1 - 3.0 \\ [7pt]
		$\phi_{\mathrm{Th}}$ [10$^{6}$ cm$^{-2}$s$^{-1}$] & 4.7 & 4.8 & 4.9 & 2.8 - 6.9 & 0.1 - 11.3 &  {\Large $^{+0.3}_{-0.2}$} & 2.7 - 6.9 \BstrutLarge \\ 
		\hline
		\hline
	\end{tabular*}
\end{table*}         

    \section{CONCLUSIONS}
    \label{sec:conclusion}

    Borexino is 280-ton liquid scintillator neutrino detector located at Laboratori Nazionali del Gran Sasso (LNGS) in Italy, and has been acquiring data since 2007. It has proven to be a successful neutrino observatory, which went well beyond its original proposal to observe $^7$Be solar neutrinos. In addition to solar neutrino measurements, Borexino has proven to be able to detect also antineutrinos. Radiopurity of the detector, its calibration, stable performance, its relatively large distance to nuclear reactors, as well as depth of the LNGS laboratory to guarantee smallness of cosmogenic background, are the main building blocks for a geoneutrino measurement with systematic uncertainty below 5\%.
   
    The focus of the paper is to provide the scientific community with a comprehensive study that combines the expertise of neutrino physicists and geoscientists. The paper provides an in-depth motivation and description of geoneutrino measurement, as well as the geological interpretations of the result. It presents in detail the analysis of 3262.74\,days of Borexino data taken between December 2007 and April 2019 and provides, with some assumptions, a measurement of the Uranium and Thorium content of the Earth's mantle and its radiogenic heat.

    Borexino detects geoneutrinos from $^{238}$U and $^{232}$Th through inverse beta decay, in which electron flavour antineutrinos with energies above 1.8\,MeV interact with free protons of the LS. The detection efficiency for optimized data selection cuts is $(87.0 \pm 1.5)$\%. This interaction is the only channel presently available for detection of MeV-scale electron antineutrinos. Optimized data selection including an enlarged fiducial volume and a sophisticated cosmogenic veto resulted in an exposure of (1.29 $\pm$ 0.05) $\times$ 10$^{32}$ protons $\times$ year. This represents an increase by a factor of two over the previous Borexino analysis reported in 2015.
   
    The paper documents improved techniques in the in-depth analysis of the Borexino data, and provides future experiments with a description of the substantial effort required to extract geoneutrino signals. We have underlined the importance of muon detection (in particular special categories of muon events that become crucial in low-rate measurements), as well as the $\alpha$/$\beta$ pulse shape discrimination techniques. The optimization of data selection cuts, chosen to maximize Borexino's sensitivity to measure geoneutrinos, has been described. All kinds of background types considered important for geoneutrino measurement have also been discussed, including approaches of their estimation either through theoretical calculation and Monte Carlo simulation, or by analysis of independent data. Borexino ability to measure electron antineutrinos is calibrated via reactor antineutrino background, that is not constrained in geoneutrino analysis and has been found to be in agreement with the expectations. By observing $52.6 ^{+9.4}_{-8.6}\,({\rm stat}) ^{+2.7}_{-2.1}\,({\rm sys})$ geoneutrinos (68\% interval) from $^{238}$U and $^{232}$Th, a geoneutrino signal of $47.0^{+8.4}_{-7.7}\,({\rm stat)}^{+2.4}_{-1.9}\,({\rm sys})$\,TNU has been obtained. The total precision of $^{+18.3}_{-17.2}$\% is found to be in agreement with the expected sensitivity. This result assumes a Th/U mass ratio of 3.9, as found in chondritic CI meteorites, and is compatible with result when contributions from $^{238}$U and $^{232}$Th were both fit as free parameters.

   Importance of the knowledge of abundances and distributions of U and Th in the Earth, and in particular around the detector, for both the signal prediction as well as interpretation of results, have been discussed. The measured geoneutrino signal is found to be in agreement with the predictions of different geological models with a preference for those predicting the highest concentrations of heat producing elements. The hypothesis of observing a null mantle signal has been excluded at 99\% C.L. when exploiting detailed knowledge of the local crust near the LNGS. The latter is characterized by the presence of thick, U and Th depleted sediments. We note that geophysical and geochemical observations constrain the Th/U mass ratio for the bulk lihtosphere to a value of 4.3. Maintaining the global chondritic ratio of 3.9 for the bulk Earth, the inferred Th/U mass ratio for the mantle is 3.7. Assuming the latter value, we have observed mantle signal of $21.2 ^{+9.5}_{-9.0}\,({\rm stat}) ^{+1.1}_{-0.9}\,({\rm sys})$\,TNU.
   
   Considering different scenarios about the U and Th distribution in the mantle, the measured mantle geoneutrino signal has been converted to radiogenic heat from U and Th in the mantle of $24.6 ^{+11.1}_{-10.4}$\,TW (68\% interval). Assuming the contribution of 18\% from $^{40}$K in the mantle and adding the relatively-well known lithospheric radiogenic heat of  $8.1^{+1.9}_{-1.4}$\,TW, Borexino has estimated the total radiogenic heat of the Earth to be $38.2 ^{+13.6}_{-12.7}$\,TW. The latter is found to be compatible with different geological predictions. However, there is a $\sim$2.4$\sigma$ tension with Earth models predicting the lowest concentration of heat-producing elements. The total radiogenic heat estimated by Borexino can be used to extract a convective Urey ratio of 0.78$^{+0.41}_{-0.28}$. In conclusion, Borexino geoneutrino measurement has constrained at 90\%\,C.L. the mantle composition to $a_{\mathrm{mantle}}$(U) $>$ 13\,ppb and $a_{\mathrm{mantle}}$(Th) $>$ 48\,ppb, the mantle radiogenic heat power to $H_{\mathrm{rad}}^{\mathrm{mantle}}$(U+Th) $>$ 10\,TW and $H_{\mathrm{rad}}^{\mathrm{mantle}}$(U+Th+K) $>$ 12.2\,TW, as well as the convective Urey ratio to $UR_{CV}$ $>$ 0.13.

   With the application of a constraint on the number of expected reactor antineutrino events, Borexino has placed an upper limit on the number of events from  a hypothetical georeactor inside the Earth. Assuming the georeactor located at the center of the Earth, its existence with a power greater than 2.4\,TW has been excluded at 95\% C.L.
   
    In conclusion, Borexino confirms the feasibility of geoneutrino measurements as well as the validity of different geological models predicting the U and Th abundances in the Earth. This is an enormous success of both neutrino physics and geosciences. However, in spite of some preference of Borexino results for the models predicting high U and Th abundances, additional and more precise measurements are needed in order to extract firm geological results. The next generation of large volume liquid scintillator detectors has a strong potential to provide fundamental information about our planet.

\section*{APPENDIX - LIST OF ACRONYMS}
    \label{sec:acronyms}
    
\noindent 
$\alpha$ - alpha particle\\
A - BSE model Anderson, 2007~\cite{RN372}\\
$\beta$ - beta particle\\
BDT - Boosted Decision Tree\\
BSE - Bulk Silicate Earth \\
BTB - Borexino Trigger Board\\
BTB4 - the same as MTB flag, see below\\
CC - continental crust \\
CC model - Cosmochemical Bulk Silicate Earth model\\
C.L. - confidence level \\
CLM - continental lithospheric margin\\
CMB - core-mantle boundary \\
CT - Central Tile\\
DAQ - data acquisition\\
DFV - Dynamical Fiducial Volume \\
DM - depleted mantle\\
DMP - dimethylphthalate (DMP, C$_6$H$_4$(COOCH$_3$)$_2$)\\
$e^{-}$ or $\beta^{-}$ - electron\\
$e^{+}$ or $\beta^{+}$ - positron\\
EM - enriched mantle\\
$E_{\mathrm  {p}}$ - energy of the prompt IBD candidate\\
$E_{\mathrm  {d}}$ - energy of the delayed IBD candidate\\
FADC - Flash Analog-to-Digital Converter\\
FEB - front end board\\
FFL - far field lithosphere\\
FR model - Fully Radiogenic Bulk Silicate Earth model\\
FWFD - Fast Wave Form Digitizer\\
$\gamma$ - gamma ray\\
G - Gatti parameter\\
{\it G4Bx2} - Geant4 based Borexino Monte Carlo code\\
GC model - Geochemical Bulk Silicate Earth model\\
GD model - Geodynamical Bulk Silicate Earth model\\
$GR_{1}$, $GR_{2}$, $GR_{3}$ - 3 studied positions of georeactor inside the Earth\\
$H_{\mathrm {rad}}$ - Earth's radiogenic heat \\
$H^{\mathrm {CC}}_{\mathrm {rad}}$ - Earth's continental crust radiogenic heat \\
$H^{\mathrm {mantle}}_{\mathrm {rad}}$ - Earth's mantle radiogenic heat \\
$H^{\mathrm {LSp}}_{\mathrm {rad}}$ - Earth's lithosphere radiogenic heat \\
$H_{\mathrm {SC}}$ - Earth's heat from the secular cooling\\
$H_{\mathrm{tot}}$ - integrated total surface heat flux of the Earth\\
HPEs - heat producing elements \\
HSc - high scenario of the mantle signal prediction\\
IBD - Inverse Beta Decay\\
ID - Inner Detector \\
IDF - Inner Detector Flag\\
IV - Inner Vessel\\
ISc - intermediate scenario of the mantle signal prediction\\
J - BSE model Javoy et al., 2010~\cite{RN367}\\
LF - load factor of nuclear power plants\\
L \& K - BSE model Lyubetskaya \& Korenaga, 2007~\cite{RN747}\\
LNGS - Laboratori Nazionali del Gran Sasso\\
LOC - local crust\\
LS - liquid scintillator\\
LSc - low scenario of the mantle signal prediction\\
LSp - lithosphere\\
$\mu$ - muon\\
MC - Monte Carlo\\
MLP - Multi-Layer Perceptron\\
M \& S - BSE model McDonough \& Sun, 1995~\cite{RN1380}\\
MTB - Muon Trigger Board\\
MTF - Muon Trigger Flag\\
m w.e. - meter water equivalent\\
$\nu$ - neutrino\\
$\bar{\nu}$ - antineutrino\\
$\bar{\nu}_{e}$ - electron flavour antineutrino\\
$n$ - neutron\\
$N_h$ - number of detected hits\\
$N_{P}$ - number of triggered PMTs\\
$N_{pe}$ - number of detected photoelectrons\\
OC - oceanic crust\\
OD - Outer Detector\\
OV - Outer Vessel\\
$p$ - proton\\
$P_{ee}$ - survival probability of electron flavour neutrino\\
PC - Pseudocumene liquid scintillator, C$_6$H$_3$(CH$_3$)$_3$, 1,2,4-trimethylbenzene\\
PDF - probability distribution function\\
p.e. - photoelectron(s)\\
PID - particle identification\\
PM - Primitive Mantle\\
PMNS - Pontecorvo–Maki–Nakagawa–Sakata mixing matrix\\
PMTs - photo-multiplier Tubes\\
PPO - fluorescent dye, C$_{15}$H$_{11}$NO, 2,5-diphenyloxazole\\
P \& O - BSE model Palme and O'Neil, 2003~\cite{RN400} \\
$Q_{\mathrm  {p}}$ - charge of the prompt IBD candidate\\
$Q_{\mathrm  {d}}$ - charge of the delayed IBD candidate\\
RR - Rest of the Region\\
SSS - Stanless Steel Sphere\\
SVM - Support Vector Machine\\
T - BSE model Taylor, 1980~\cite{RN1378}\\
TMVA - Toolkit for Multivariate Data Analysis\\
TNU - Terrestrial Neutrino Unit\\
T \& S - BSE model Turcotte \& Schubert, 2002~\cite{RN368} \\
$UR_{\mathrm{CV}}$ - convective Urey ratio\\
W - BSE model Wang et al., 2018~\cite{RN1319} \\
WE - water extraction procedure of LS-purification\\
WT - Water Tank\\

 \addcontentsline{toc}{section}{APPENDIX - LIST OF ACRONYMS} 

    \section*{ACKNOWLEDGEMENTS}
    \label{sec:ackn}
The Borexino program is made possible by funding from Istituto Nazionale di Fisica Nucleare (INFN) (Italy), National Science Foundation (NSF) (USA), Deutsche Forschungsgemeinschaft (DFG) and Helmholtz-Gemeinschaft (HGF) (Germany), Russian Foundation for Basic Research (RFBR) (Grants No. 16-29-13014ofi-m, No. 17-02-00305A, and No. 19-02-00097A) and Russian Science Foundation (RSF) (Grant No. 17-12-01009) (Russia), and Narodowe Centrum Nauki (NCN) (Grant No. UMO 2017/26/M/ST2/00915) (Poland).
\bibliographystyle{apsrev4-1}
\bibliography{main}

\begin{thebibliography}{136}%
\makeatletter
\providecommand \@ifxundefined [1]{%
 \@ifx{#1\undefined}
}%
\providecommand \@ifnum [1]{%
 \ifnum #1\expandafter \@firstoftwo
 \else \expandafter \@secondoftwo
 \fi
}%
\providecommand \@ifx [1]{%
 \ifx #1\expandafter \@firstoftwo
 \else \expandafter \@secondoftwo
 \fi
}%
\providecommand \natexlab [1]{#1}%
\providecommand \enquote  [1]{``#1''}%
\providecommand \bibnamefont  [1]{#1}%
\providecommand \bibfnamefont [1]{#1}%
\providecommand \citenamefont [1]{#1}%
\providecommand \href@noop [0]{\@secondoftwo}%
\providecommand \href [0]{\begingroup \@sanitize@url \@href}%
\providecommand \@href[1]{\@@startlink{#1}\@@href}%
\providecommand \@@href[1]{\endgroup#1\@@endlink}%
\providecommand \@sanitize@url [0]{\catcode `\\12\catcode `\$12\catcode
  `\&12\catcode `\#12\catcode `\^12\catcode `\_12\catcode `\%12\relax}%
\providecommand \@@startlink[1]{}%
\providecommand \@@endlink[0]{}%
\providecommand \url  [0]{\begingroup\@sanitize@url \@url }%
\providecommand \@url [1]{\endgroup\@href {#1}{\urlprefix }}%
\providecommand \urlprefix  [0]{URL }%
\providecommand \Eprint [0]{\href }%
\providecommand \doibase [0]{http://dx.doi.org/}%
\providecommand \selectlanguage [0]{\@gobble}%
\providecommand \bibinfo  [0]{\@secondoftwo}%
\providecommand \bibfield  [0]{\@secondoftwo}%
\providecommand \translation [1]{[#1]}%
\providecommand \BibitemOpen [0]{}%
\providecommand \bibitemStop [0]{}%
\providecommand \bibitemNoStop [0]{.\EOS\space}%
\providecommand \EOS [0]{\spacefactor3000\relax}%
\providecommand \BibitemShut  [1]{\csname bibitem#1\endcsname}%
\let\auto@bib@innerbib\@empty
\bibitem [{\citenamefont {Winter}(2016)}]{Winter:2015zwx}%
  \BibitemOpen
  \bibfield  {author} {\bibinfo {author} {\bibfnamefont {W.}~\bibnamefont
  {Winter}},\ }\href {\doibase 10.1016/j.nuclphysb.2016.03.033} {\bibfield
  {journal} {\bibinfo  {journal} {Nucl. Phys.}\ }\textbf {\bibinfo {volume}
  {B908}},\ \bibinfo {pages} {250} (\bibinfo {year} {2016})},\ \Eprint
  {http://arxiv.org/abs/1511.05154} {arXiv:1511.05154 [hep-ph]} \BibitemShut
  {NoStop}%
\bibitem [{\citenamefont {Donini}\ \emph {et~al.}(2019)\citenamefont {Donini},
  \citenamefont {Palomares-Ruiz},\ and\ \citenamefont
  {Salvado}}]{Donini:2018tsg}%
  \BibitemOpen
  \bibfield  {author} {\bibinfo {author} {\bibfnamefont {A.}~\bibnamefont
  {Donini}}, \bibinfo {author} {\bibfnamefont {S.}~\bibnamefont
  {Palomares-Ruiz}}, \ and\ \bibinfo {author} {\bibfnamefont {J.}~\bibnamefont
  {Salvado}},\ }\href {\doibase 10.1038/s41567-018-0319-1} {\bibfield
  {journal} {\bibinfo  {journal} {Nature Phys.}\ }\textbf {\bibinfo {volume}
  {15}},\ \bibinfo {pages} {37} (\bibinfo {year} {2019})},\ \Eprint
  {http://arxiv.org/abs/1803.05901} {arXiv:1803.05901 [hep-ph]} \BibitemShut
  {NoStop}%
\bibitem [{\citenamefont {Bourret}\ \emph {et~al.}(2017)\citenamefont
  {Bourret}, \citenamefont {Coelho},\ and\ \citenamefont
  {Van~Elewyck}}]{Bourret:2017tkw}%
  \BibitemOpen
  \bibfield  {author} {\bibinfo {author} {\bibfnamefont {S.}~\bibnamefont
  {Bourret}}, \bibinfo {author} {\bibfnamefont {J.~A.~B.}\ \bibnamefont
  {Coelho}}, \ and\ \bibinfo {author} {\bibfnamefont {V.}~\bibnamefont
  {Van~Elewyck}} (\bibinfo {collaboration} {KM3NeT}),\ }\bibfield  {booktitle}
  {\emph {\bibinfo {booktitle} {{Proceedings, 27th International Conference on
  Neutrino Physics and Astrophysics (Neutrino 2016): London, United Kingdom,
  July 4-9, 2016}}},\ }\href {\doibase 10.1088/1742-6596/888/1/012114}
  {\bibfield  {journal} {\bibinfo  {journal} {J. Phys. Conf. Ser.}\ }\textbf
  {\bibinfo {volume} {888}},\ \bibinfo {pages} {012114} (\bibinfo {year}
  {2017})},\ \Eprint {http://arxiv.org/abs/1702.03723} {arXiv:1702.03723
  [physics.ins-det]} \BibitemShut {NoStop}%
\bibitem [{IAE()}]{IAEA}%
  \BibitemOpen
  \href@noop {} {\enquote {\bibinfo {title} {{IAEA} nuclear data section},}\
  }\bibinfo {note}
  {\href{https://www-nds.iaea.org/relnsd/NdsEnsdf/QueryForm.html}{IAEA
  database}}\BibitemShut {NoStop}%
\bibitem [{\citenamefont {Fiorentini}\ \emph {et~al.}(2007)\citenamefont
  {Fiorentini}, \citenamefont {Lissia},\ and\ \citenamefont
  {Mantovani}}]{Fiorentini:2007te}%
  \BibitemOpen
  \bibfield  {author} {\bibinfo {author} {\bibfnamefont {G.}~\bibnamefont
  {Fiorentini}}, \bibinfo {author} {\bibfnamefont {M.}~\bibnamefont {Lissia}},
  \ and\ \bibinfo {author} {\bibfnamefont {F.}~\bibnamefont {Mantovani}},\
  }\href {\doibase 10.1016/j.physrep.2007.09.001} {\bibfield  {journal}
  {\bibinfo  {journal} {Phys. Rept.}\ }\textbf {\bibinfo {volume} {453}},\
  \bibinfo {pages} {117} (\bibinfo {year} {2007})},\ \Eprint
  {http://arxiv.org/abs/0707.3203} {arXiv:0707.3203 [physics.geo-ph]}
  \BibitemShut {NoStop}%
\bibitem [{\citenamefont {Marx}\ and\ \citenamefont
  {Menyh\'ard}(1960)}]{Marx1960}%
  \BibitemOpen
  \bibfield  {author} {\bibinfo {author} {\bibfnamefont {G.}~\bibnamefont
  {Marx}}\ and\ \bibinfo {author} {\bibfnamefont {N.}~\bibnamefont
  {Menyh\'ard}},\ }\href@noop {} {\bibfield  {journal} {\bibinfo  {journal}
  {Mitteilungen der Sternw\"{a}rte Budapest}\ }\textbf {\bibinfo {volume}
  {44}},\ \bibinfo {pages} {1} (\bibinfo {year} {1960})}\BibitemShut {NoStop}%
\bibitem [{\citenamefont {Eder}(1966)}]{EDER1966657}%
  \BibitemOpen
  \bibfield  {author} {\bibinfo {author} {\bibfnamefont {G.}~\bibnamefont
  {Eder}},\ }\href {\doibase 10.1016/0029-5582(66)90903-5} {\bibfield
  {journal} {\bibinfo  {journal} {Nuclear Physics}\ }\textbf {\bibinfo {volume}
  {78}},\ \bibinfo {pages} {657} (\bibinfo {year} {1966})}\BibitemShut
  {NoStop}%
\bibitem [{\citenamefont {Marx}(1969)}]{Marx1969}%
  \BibitemOpen
  \bibfield  {author} {\bibinfo {author} {\bibfnamefont {G.}~\bibnamefont
  {Marx}},\ }\href {\doibase 10.1007/BF01698889} {\bibfield  {journal}
  {\bibinfo  {journal} {Czechoslovak Journal of Physics B}\ }\textbf {\bibinfo
  {volume} {19}},\ \bibinfo {pages} {1471} (\bibinfo {year}
  {1969})}\BibitemShut {NoStop}%
\bibitem [{\citenamefont {Krauss}\ \emph {et~al.}(1984)\citenamefont {Krauss},
  \citenamefont {Glashow},\ and\ \citenamefont
  {Schramm}}]{KraussGlashowSchramm}%
  \BibitemOpen
  \bibfield  {author} {\bibinfo {author} {\bibfnamefont {L.~M.}\ \bibnamefont
  {Krauss}}, \bibinfo {author} {\bibfnamefont {S.~L.}\ \bibnamefont {Glashow}},
  \ and\ \bibinfo {author} {\bibfnamefont {D.~N.}\ \bibnamefont {Schramm}},\
  }\href {\doibase 10.1038/310191a0} {\bibfield  {journal} {\bibinfo  {journal}
  {Nature}\ }\textbf {\bibinfo {volume} {310}},\ \bibinfo {pages} {191}
  (\bibinfo {year} {1984})}\BibitemShut {NoStop}%
\bibitem [{\citenamefont {Rothschild}\ \emph {et~al.}(1998)\citenamefont
  {Rothschild}, \citenamefont {Chen},\ and\ \citenamefont
  {Calaprice}}]{Rothschild:1997dd}%
  \BibitemOpen
  \bibfield  {author} {\bibinfo {author} {\bibfnamefont {C.~G.}\ \bibnamefont
  {Rothschild}}, \bibinfo {author} {\bibfnamefont {M.~C.}\ \bibnamefont
  {Chen}}, \ and\ \bibinfo {author} {\bibfnamefont {F.~P.}\ \bibnamefont
  {Calaprice}},\ }\href {\doibase 10.1029/98GL50667} {\bibfield  {journal}
  {\bibinfo  {journal} {Geophys. Res. Lett.}\ }\textbf {\bibinfo {volume}
  {25}},\ \bibinfo {pages} {1083} (\bibinfo {year} {1998})},\ \Eprint
  {http://arxiv.org/abs/nucl-ex/9710001} {arXiv:nucl-ex/9710001 [nucl-ex]}
  \BibitemShut {NoStop}%
\bibitem [{\citenamefont {Raghavan}\ \emph {et~al.}(1998)\citenamefont
  {Raghavan}, \citenamefont {Sch{\"o}nert}, \citenamefont {Enomoto},
  \citenamefont {Shirai}, \citenamefont {Suekane},\ and\ \citenamefont
  {Suzuki}}]{Raghavan:1997gw}%
  \BibitemOpen
  \bibfield  {author} {\bibinfo {author} {\bibfnamefont {R.~S.}\ \bibnamefont
  {Raghavan}}, \bibinfo {author} {\bibfnamefont {S.}~\bibnamefont
  {Sch{\"o}nert}}, \bibinfo {author} {\bibfnamefont {S.}~\bibnamefont
  {Enomoto}}, \bibinfo {author} {\bibfnamefont {J.}~\bibnamefont {Shirai}},
  \bibinfo {author} {\bibfnamefont {F.}~\bibnamefont {Suekane}}, \ and\
  \bibinfo {author} {\bibfnamefont {A.}~\bibnamefont {Suzuki}},\ }\href
  {\doibase 10.1103/PhysRevLett.80.635} {\bibfield  {journal} {\bibinfo
  {journal} {Phys. Rev. Lett.}\ }\textbf {\bibinfo {volume} {80}},\ \bibinfo
  {pages} {635} (\bibinfo {year} {1998})}\BibitemShut {NoStop}%
\bibitem [{\citenamefont {Araki}\ \emph {et~al.}(2005)\citenamefont {Araki}
  \emph {et~al.}}]{Araki:2005qa}%
  \BibitemOpen
  \bibfield  {author} {\bibinfo {author} {\bibfnamefont {T.}~\bibnamefont
  {Araki}} \emph {et~al.},\ }\href {\doibase 10.1038/nature03980} {\bibfield
  {journal} {\bibinfo  {journal} {Nature}\ }\textbf {\bibinfo {volume} {436}},\
  \bibinfo {pages} {499} (\bibinfo {year} {2005})}\BibitemShut {NoStop}%
\bibitem [{\citenamefont {Abe}\ \emph {et~al.}(2008)\citenamefont {Abe} \emph
  {et~al.}}]{Abe:2008aa}%
  \BibitemOpen
  \bibfield  {author} {\bibinfo {author} {\bibfnamefont {S.}~\bibnamefont
  {Abe}} \emph {et~al.} (\bibinfo {collaboration} {KamLAND}),\ }\href {\doibase
  10.1103/PhysRevLett.100.221803} {\bibfield  {journal} {\bibinfo  {journal}
  {Phys. Rev. Lett.}\ }\textbf {\bibinfo {volume} {100}},\ \bibinfo {pages}
  {221803} (\bibinfo {year} {2008})},\ \Eprint {http://arxiv.org/abs/0801.4589}
  {arXiv:0801.4589 [hep-ex]} \BibitemShut {NoStop}%
\bibitem [{\citenamefont {Gando}\ \emph {et~al.}(2011)\citenamefont {Gando}
  \emph {et~al.}}]{Gando:1900zz}%
  \BibitemOpen
  \bibfield  {author} {\bibinfo {author} {\bibfnamefont {A.}~\bibnamefont
  {Gando}} \emph {et~al.} (\bibinfo {collaboration} {KamLAND}),\ }\href
  {\doibase 10.1038/ngeo1205} {\bibfield  {journal} {\bibinfo  {journal}
  {Nature Geo.}\ }\textbf {\bibinfo {volume} {4}},\ \bibinfo {pages} {647}
  (\bibinfo {year} {2011})}\BibitemShut {NoStop}%
\bibitem [{\citenamefont {Gando}\ \emph {et~al.}(2013)\citenamefont {Gando}
  \emph {et~al.}}]{Gando:2013nba}%
  \BibitemOpen
  \bibfield  {author} {\bibinfo {author} {\bibfnamefont {A.}~\bibnamefont
  {Gando}} \emph {et~al.} (\bibinfo {collaboration} {KamLAND}),\ }\href
  {\doibase 10.1103/PhysRevD.88.033001} {\bibfield  {journal} {\bibinfo
  {journal} {Phys. Rev.}\ }\textbf {\bibinfo {volume} {D88}},\ \bibinfo {pages}
  {033001} (\bibinfo {year} {2013})},\ \Eprint {http://arxiv.org/abs/1303.4667}
  {arXiv:1303.4667 [hep-ex]} \BibitemShut {NoStop}%
\bibitem [{\citenamefont {Bellini}\ \emph {et~al.}(2010)\citenamefont {Bellini}
  \emph {et~al.}}]{Bellini:2010geo}%
  \BibitemOpen
  \bibfield  {author} {\bibinfo {author} {\bibfnamefont {G.}~\bibnamefont
  {Bellini}} \emph {et~al.} (\bibinfo {collaboration} {Borexino}),\ }\href
  {\doibase 10.1016/j.physletb.2010.03.051} {\bibfield  {journal} {\bibinfo
  {journal} {Physics Letters}\ }\textbf {\bibinfo {volume} {B687}},\ \bibinfo
  {pages} {299} (\bibinfo {year} {2010})},\ \Eprint
  {http://arxiv.org/abs/1003.0284} {arXiv:1003.0284 [hep-ex]} \BibitemShut
  {NoStop}%
\bibitem [{\citenamefont {Bellini}\ \emph
  {et~al.}(2013{\natexlab{a}})\citenamefont {Bellini} \emph
  {et~al.}}]{Bellini:2013geo}%
  \BibitemOpen
  \bibfield  {author} {\bibinfo {author} {\bibfnamefont {G.}~\bibnamefont
  {Bellini}} \emph {et~al.} (\bibinfo {collaboration} {Borexino}),\ }\href
  {\doibase 10.1016/j.physletb.2013.04.030} {\bibfield  {journal} {\bibinfo
  {journal} {Physice Letters.}\ }\textbf {\bibinfo {volume} {B722}},\ \bibinfo
  {pages} {295} (\bibinfo {year} {2013}{\natexlab{a}})},\ \Eprint
  {http://arxiv.org/abs/1506.04610} {arXiv:1506.04610 [hep-ex]} \BibitemShut
  {NoStop}%
\bibitem [{\citenamefont {Agostini}\ \emph {et~al.}(2015)\citenamefont
  {Agostini} \emph {et~al.}}]{Agostini:2015cba}%
  \BibitemOpen
  \bibfield  {author} {\bibinfo {author} {\bibfnamefont {M.}~\bibnamefont
  {Agostini}} \emph {et~al.} (\bibinfo {collaboration} {Borexino}),\ }\href
  {\doibase 10.1103/PhysRevD.92.031101} {\bibfield  {journal} {\bibinfo
  {journal} {Phys. Rev.}\ }\textbf {\bibinfo {volume} {D92}},\ \bibinfo {pages}
  {031101} (\bibinfo {year} {2015})},\ \Eprint
  {http://arxiv.org/abs/1506.04610} {arXiv:1506.04610 [hep-ex]} \BibitemShut
  {NoStop}%
\bibitem [{\citenamefont {Vinyoles}\ \emph {et~al.}(2017)\citenamefont
  {Vinyoles}, \citenamefont {Serenelli}, \citenamefont {Villante},
  \citenamefont {Basu}, \citenamefont {Bergström}, \citenamefont
  {Gonzalez-Garcia}, \citenamefont {Maltoni}, \citenamefont {Pe\~{n}a Garay},\
  and\ \citenamefont {Song}}]{Vinyoles2016}%
  \BibitemOpen
  \bibfield  {author} {\bibinfo {author} {\bibfnamefont {N.}~\bibnamefont
  {Vinyoles}}, \bibinfo {author} {\bibfnamefont {A.~M.}\ \bibnamefont
  {Serenelli}}, \bibinfo {author} {\bibfnamefont {F.~L.}\ \bibnamefont
  {Villante}}, \bibinfo {author} {\bibfnamefont {S.}~\bibnamefont {Basu}},
  \bibinfo {author} {\bibfnamefont {J.}~\bibnamefont {Bergström}}, \bibinfo
  {author} {\bibfnamefont {M.~C.}\ \bibnamefont {Gonzalez-Garcia}}, \bibinfo
  {author} {\bibfnamefont {M.}~\bibnamefont {Maltoni}}, \bibinfo {author}
  {\bibfnamefont {C.}~\bibnamefont {Pe\~{n}a Garay}}, \ and\ \bibinfo {author}
  {\bibfnamefont {N.}~\bibnamefont {Song}},\ }\href {\doibase
  10.3847/1538-4357/835/2/202} {\bibfield  {journal} {\bibinfo  {journal}
  {Astrophys. J.}\ }\textbf {\bibinfo {volume} {835}},\ \bibinfo {pages} {202}
  (\bibinfo {year} {2017})},\ \Eprint {http://arxiv.org/abs/1611.09867}
  {arXiv:1611.09867 [astro-ph.SR]} \BibitemShut {NoStop}%
\bibitem [{\citenamefont {Bellini}\ \emph
  {et~al.}(2014{\natexlab{a}})\citenamefont {Bellini} \emph
  {et~al.}}]{Bellini:2013lnn}%
  \BibitemOpen
  \bibfield  {author} {\bibinfo {author} {\bibfnamefont {G.}~\bibnamefont
  {Bellini}} \emph {et~al.} (\bibinfo {collaboration} {Borexino}),\ }\href
  {\doibase 10.1103/PhysRevD.89.112007} {\bibfield  {journal} {\bibinfo
  {journal} {Phys. Rev.}\ }\textbf {\bibinfo {volume} {D89}},\ \bibinfo {pages}
  {112007} (\bibinfo {year} {2014}{\natexlab{a}})},\ \Eprint
  {http://arxiv.org/abs/1303.2571} {arXiv:1303.2571 [hep-ex]} \BibitemShut
  {NoStop}%
\bibitem [{\citenamefont {Bellini}\ \emph
  {et~al.}(2014{\natexlab{b}})\citenamefont {Bellini} \emph
  {et~al.}}]{Bellini:2014uqa}%
  \BibitemOpen
  \bibfield  {author} {\bibinfo {author} {\bibfnamefont {G.}~\bibnamefont
  {Bellini}} \emph {et~al.} (\bibinfo {collaboration} {Borexino}),\ }\href
  {\doibase 10.1038/nature13702} {\bibfield  {journal} {\bibinfo  {journal}
  {Nature}\ }\textbf {\bibinfo {volume} {512}},\ \bibinfo {pages} {383}
  (\bibinfo {year} {2014}{\natexlab{b}})}\BibitemShut {NoStop}%
\bibitem [{\citenamefont {Agostini}\ \emph
  {et~al.}(2018{\natexlab{a}})\citenamefont {Agostini} \emph
  {et~al.}}]{Agostini:2018uly}%
  \BibitemOpen
  \bibfield  {author} {\bibinfo {author} {\bibfnamefont {M.}~\bibnamefont
  {Agostini}} \emph {et~al.} (\bibinfo {collaboration} {Borexino}),\ }\href
  {\doibase 10.1038/s41586-018-0624-y} {\bibfield  {journal} {\bibinfo
  {journal} {Nature}\ }\textbf {\bibinfo {volume} {562}},\ \bibinfo {pages}
  {505} (\bibinfo {year} {2018}{\natexlab{a}})}\BibitemShut {NoStop}%
\bibitem [{\citenamefont {Chen}(2006)}]{Chen:2005zza}%
  \BibitemOpen
  \bibfield  {author} {\bibinfo {author} {\bibfnamefont {M.~C.}\ \bibnamefont
  {Chen}},\ }\bibfield  {booktitle} {\emph {\bibinfo {booktitle} {{Neutrino
  geophysics. Proceedings, Conference, Neutrino Sciences 2005, Honolulu, USA,
  December 14-16, 2005}}},\ }\href {\doibase 10.1007/s11038-006-9116-4}
  {\bibfield  {journal} {\bibinfo  {journal} {Earth Moon Planets}\ }\textbf
  {\bibinfo {volume} {99}},\ \bibinfo {pages} {221} (\bibinfo {year}
  {2006})}\BibitemShut {NoStop}%
\bibitem [{\citenamefont {An}\ \emph {et~al.}(2016{\natexlab{a}})\citenamefont
  {An} \emph {et~al.}}]{An:2015jdp}%
  \BibitemOpen
  \bibfield  {author} {\bibinfo {author} {\bibfnamefont {F.}~\bibnamefont {An}}
  \emph {et~al.} (\bibinfo {collaboration} {JUNO}),\ }\href {\doibase
  10.1088/0954-3899/43/3/030401} {\bibfield  {journal} {\bibinfo  {journal} {J.
  Phys.}\ }\textbf {\bibinfo {volume} {G43}},\ \bibinfo {pages} {030401}
  (\bibinfo {year} {2016}{\natexlab{a}})},\ \Eprint
  {http://arxiv.org/abs/1507.05613} {arXiv:1507.05613 [physics.ins-det]}
  \BibitemShut {NoStop}%
\bibitem [{\citenamefont {Wan}\ \emph {et~al.}(2017)\citenamefont {Wan},
  \citenamefont {Hussain}, \citenamefont {Wang},\ and\ \citenamefont
  {Chen}}]{Wan:2016nhe}%
  \BibitemOpen
  \bibfield  {author} {\bibinfo {author} {\bibfnamefont {L.}~\bibnamefont
  {Wan}}, \bibinfo {author} {\bibfnamefont {G.}~\bibnamefont {Hussain}},
  \bibinfo {author} {\bibfnamefont {Z.}~\bibnamefont {Wang}}, \ and\ \bibinfo
  {author} {\bibfnamefont {S.}~\bibnamefont {Chen}},\ }\href {\doibase
  10.1103/PhysRevD.95.053001} {\bibfield  {journal} {\bibinfo  {journal} {Phys.
  Rev.}\ }\textbf {\bibinfo {volume} {D95}},\ \bibinfo {pages} {053001}
  (\bibinfo {year} {2017})},\ \Eprint {http://arxiv.org/abs/1612.00133}
  {arXiv:1612.00133 [hep-ex]} \BibitemShut {NoStop}%
\bibitem [{\citenamefont {Dye}(2012)}]{Dye:2011mc}%
  \BibitemOpen
  \bibfield  {author} {\bibinfo {author} {\bibfnamefont {S.}~\bibnamefont
  {Dye}},\ }\href {\doibase 10.1029/2012RG000400} {\bibfield  {journal}
  {\bibinfo  {journal} {Rev. Geophys.}\ }\textbf {\bibinfo {volume} {50}},\
  \bibinfo {pages} {RG3007} (\bibinfo {year} {2012})},\ \Eprint
  {http://arxiv.org/abs/1111.6099} {arXiv:1111.6099 [nucl-ex]} \BibitemShut
  {NoStop}%
\bibitem [{\citenamefont {Agostini}\ \emph
  {et~al.}(2018{\natexlab{b}})\citenamefont {Agostini} \emph
  {et~al.}}]{Agostini:2017aaa}%
  \BibitemOpen
  \bibfield  {author} {\bibinfo {author} {\bibfnamefont {M.}~\bibnamefont
  {Agostini}} \emph {et~al.} (\bibinfo {collaboration} {Borexino}),\ }\href
  {\doibase 10.1016/j.astropartphys.2017.10.003} {\bibfield  {journal}
  {\bibinfo  {journal} {Astropart. Phys.}\ }\textbf {\bibinfo {volume} {97}},\
  \bibinfo {pages} {136} (\bibinfo {year} {2018}{\natexlab{b}})},\ \Eprint
  {http://arxiv.org/abs/1704.02291} {arXiv:1704.02291 [physics.ins-det]}
  \BibitemShut {NoStop}%
\bibitem [{\citenamefont {Gaschnig}\ \emph {et~al.}(2016)\citenamefont
  {Gaschnig}, \citenamefont {Rudnick}, \citenamefont {McDonough}, \citenamefont
  {Kaufman}, \citenamefont {Valley}, \citenamefont {Hu}, \citenamefont {Gao},\
  and\ \citenamefont {Beck}}]{RN1100}%
  \BibitemOpen
  \bibfield  {author} {\bibinfo {author} {\bibfnamefont {R.~M.}\ \bibnamefont
  {Gaschnig}}, \bibinfo {author} {\bibfnamefont {R.~L.}\ \bibnamefont
  {Rudnick}}, \bibinfo {author} {\bibfnamefont {W.~F.}\ \bibnamefont
  {McDonough}}, \bibinfo {author} {\bibfnamefont {A.~J.}\ \bibnamefont
  {Kaufman}}, \bibinfo {author} {\bibfnamefont {J.}~\bibnamefont {Valley}},
  \bibinfo {author} {\bibfnamefont {Z.}~\bibnamefont {Hu}}, \bibinfo {author}
  {\bibfnamefont {S.}~\bibnamefont {Gao}}, \ and\ \bibinfo {author}
  {\bibfnamefont {M.~L.}\ \bibnamefont {Beck}},\ }\href {\doibase
  10.1016/j.gca.2016.03.020} {\bibfield  {journal} {\bibinfo  {journal}
  {Geochimica et Cosmochimica Acta}\ }\textbf {\bibinfo {volume} {186}},\
  \bibinfo {pages} {316} (\bibinfo {year} {2016})}\BibitemShut {NoStop}%
\bibitem [{\citenamefont {{National Research Council}}\ \emph
  {et~al.}(2008)\citenamefont {{National Research Council}} \emph
  {et~al.}}]{national2008origin}%
  \BibitemOpen
  \bibfield  {author} {\bibinfo {author} {\bibnamefont {{National Research
  Council}}} \emph {et~al.},\ }\href {\doibase 10.17226/12161} {\emph {\bibinfo
  {title} {Origin and evolution of Earth: Research questions for a changing
  planet}}}\ (\bibinfo  {publisher} {The National Academies Press},\ \bibinfo
  {year} {2008})\BibitemShut {NoStop}%
\bibitem [{\citenamefont {McDonough}(2014)}]{RN1398}%
  \BibitemOpen
  \bibfield  {author} {\bibinfo {author} {\bibfnamefont {W.~F.}\ \bibnamefont
  {McDonough}},\ }in\ \href {\doibase 10.1016/B978-0-08-095975-7.00215-1}
  {\emph {\bibinfo {booktitle} {Treatise on Geochemistry}}},\ Vol.~\bibinfo
  {volume} {3},\ \bibinfo {editor} {edited by\ \bibinfo {editor} {\bibfnamefont
  {H.~D.}\ \bibnamefont {Holland}}\ and\ \bibinfo {editor} {\bibfnamefont
  {K.~K.}\ \bibnamefont {Turekian}}}\ (\bibinfo  {publisher} {Elsevier},\
  \bibinfo {year} {2014})\ \bibinfo {edition} {2nd}\ ed.,\ pp.\ \bibinfo
  {pages} {559--577}\BibitemShut {NoStop}%
\bibitem [{\citenamefont {Rubie}\ \emph {et~al.}(2011)\citenamefont {Rubie},
  \citenamefont {Frost}, \citenamefont {Mann}, \citenamefont {Asahara},
  \citenamefont {Nimmo}, \citenamefont {Tsuno}, \citenamefont {Kegler},
  \citenamefont {Holzheid},\ and\ \citenamefont {Palme}}]{RN607}%
  \BibitemOpen
  \bibfield  {author} {\bibinfo {author} {\bibfnamefont {D.~C.}\ \bibnamefont
  {Rubie}}, \bibinfo {author} {\bibfnamefont {D.~J.}\ \bibnamefont {Frost}},
  \bibinfo {author} {\bibfnamefont {U.}~\bibnamefont {Mann}}, \bibinfo {author}
  {\bibfnamefont {Y.}~\bibnamefont {Asahara}}, \bibinfo {author} {\bibfnamefont
  {F.}~\bibnamefont {Nimmo}}, \bibinfo {author} {\bibfnamefont
  {K.}~\bibnamefont {Tsuno}}, \bibinfo {author} {\bibfnamefont
  {P.}~\bibnamefont {Kegler}}, \bibinfo {author} {\bibfnamefont
  {A.}~\bibnamefont {Holzheid}}, \ and\ \bibinfo {author} {\bibfnamefont
  {H.}~\bibnamefont {Palme}},\ }\href {\doibase 10.1016/j.epsl.2010.11.030}
  {\bibfield  {journal} {\bibinfo  {journal} {Earth and Planetary Science
  Letters}\ }\textbf {\bibinfo {volume} {301}},\ \bibinfo {pages} {31}
  (\bibinfo {year} {2011})}\BibitemShut {NoStop}%
\bibitem [{\citenamefont {Bellini}\ \emph
  {et~al.}(2013{\natexlab{b}})\citenamefont {Bellini}, \citenamefont {Ianni},
  \citenamefont {Ludhova}, \citenamefont {Mantovani},\ and\ \citenamefont
  {McDonough}}]{BELLINI20131}%
  \BibitemOpen
  \bibfield  {author} {\bibinfo {author} {\bibfnamefont {G.}~\bibnamefont
  {Bellini}}, \bibinfo {author} {\bibfnamefont {A.}~\bibnamefont {Ianni}},
  \bibinfo {author} {\bibfnamefont {L.}~\bibnamefont {Ludhova}}, \bibinfo
  {author} {\bibfnamefont {F.}~\bibnamefont {Mantovani}}, \ and\ \bibinfo
  {author} {\bibfnamefont {W.~F.}\ \bibnamefont {McDonough}},\ }\href {\doibase
  10.1016/j.ppnp.2013.07.001} {\bibfield  {journal} {\bibinfo  {journal} {Prog.
  Part. Nucl. Phys.}\ }\textbf {\bibinfo {volume} {73}},\ \bibinfo {pages} {1}
  (\bibinfo {year} {2013}{\natexlab{b}})},\ \Eprint
  {http://arxiv.org/abs/1310.3732} {arXiv:1310.3732 [physics.geo-ph]}
  \BibitemShut {NoStop}%
\bibitem [{\citenamefont {Deuss}\ and\ \citenamefont
  {Woodhouse}(2001)}]{RN1391}%
  \BibitemOpen
  \bibfield  {author} {\bibinfo {author} {\bibfnamefont {A.}~\bibnamefont
  {Deuss}}\ and\ \bibinfo {author} {\bibfnamefont {J.}~\bibnamefont
  {Woodhouse}},\ }\href {\doibase 10.1126/science.1063524} {\bibfield
  {journal} {\bibinfo  {journal} {Science}\ }\textbf {\bibinfo {volume}
  {294}},\ \bibinfo {pages} {354} (\bibinfo {year} {2001})}\BibitemShut
  {NoStop}%
\bibitem [{\citenamefont {Javoy}\ \emph {et~al.}(2010)\citenamefont {Javoy}
  \emph {et~al.}}]{RN367}%
  \BibitemOpen
  \bibfield  {author} {\bibinfo {author} {\bibfnamefont {M.}~\bibnamefont
  {Javoy}} \emph {et~al.},\ }\href {\doibase 10.1016/j.epsl.2010.02.033}
  {\bibfield  {journal} {\bibinfo  {journal} {Earth and Planetary Science
  Letters}\ }\textbf {\bibinfo {volume} {293}},\ \bibinfo {pages} {259}
  (\bibinfo {year} {2010})}\BibitemShut {NoStop}%
\bibitem [{\citenamefont {Huang}\ \emph {et~al.}(2013)\citenamefont {Huang},
  \citenamefont {Chubakov}, \citenamefont {Mantovani}, \citenamefont
  {Rudnick},\ and\ \citenamefont {McDonough}}]{Huang2013}%
  \BibitemOpen
  \bibfield  {author} {\bibinfo {author} {\bibfnamefont {Y.}~\bibnamefont
  {Huang}}, \bibinfo {author} {\bibfnamefont {V.}~\bibnamefont {Chubakov}},
  \bibinfo {author} {\bibfnamefont {F.}~\bibnamefont {Mantovani}}, \bibinfo
  {author} {\bibfnamefont {R.~L.}\ \bibnamefont {Rudnick}}, \ and\ \bibinfo
  {author} {\bibfnamefont {W.~F.}\ \bibnamefont {McDonough}},\ }\href {\doibase
  10.1002/ggge.20129} {\bibfield  {journal} {\bibinfo  {journal} {Geochemistry,
  Geophysics, Geosystems}\ }\textbf {\bibinfo {volume} {14}},\ \bibinfo {pages}
  {2023} (\bibinfo {year} {2013})}\BibitemShut {NoStop}%
\bibitem [{\citenamefont {Simmons}\ \emph {et~al.}(2010)\citenamefont
  {Simmons}, \citenamefont {Forte}, \citenamefont {Boschi},\ and\ \citenamefont
  {Grand}}]{RN1394}%
  \BibitemOpen
  \bibfield  {author} {\bibinfo {author} {\bibfnamefont {N.~A.}\ \bibnamefont
  {Simmons}}, \bibinfo {author} {\bibfnamefont {A.~M.}\ \bibnamefont {Forte}},
  \bibinfo {author} {\bibfnamefont {L.}~\bibnamefont {Boschi}}, \ and\ \bibinfo
  {author} {\bibfnamefont {S.~P.}\ \bibnamefont {Grand}},\ }\href {\doibase
  10.1029/2010jb007631} {\bibfield  {journal} {\bibinfo  {journal} {Journal of
  Geophysical Research}\ }\textbf {\bibinfo {volume} {115}},\ \bibinfo {pages}
  {B12310} (\bibinfo {year} {2010})}\BibitemShut {NoStop}%
\bibitem [{\citenamefont {Hofmeister}\ and\ \citenamefont
  {Criss}(2005)}]{RN1384}%
  \BibitemOpen
  \bibfield  {author} {\bibinfo {author} {\bibfnamefont {A.~M.}\ \bibnamefont
  {Hofmeister}}\ and\ \bibinfo {author} {\bibfnamefont {R.~E.}\ \bibnamefont
  {Criss}},\ }\href {\doibase 10.1016/j.tecto.2004.09.006} {\bibfield
  {journal} {\bibinfo  {journal} {Tectonophysics}\ }\textbf {\bibinfo {volume}
  {395}},\ \bibinfo {pages} {159} (\bibinfo {year} {2005})}\BibitemShut
  {NoStop}%
\bibitem [{\citenamefont {Williams}\ and\ \citenamefont
  {Von~Herzen}(1974)}]{RN1386}%
  \BibitemOpen
  \bibfield  {author} {\bibinfo {author} {\bibfnamefont {D.~L.}\ \bibnamefont
  {Williams}}\ and\ \bibinfo {author} {\bibfnamefont {R.~P.}\ \bibnamefont
  {Von~Herzen}},\ }\href {\doibase
  10.1130/0091-7613(1974)2<327:Hlften>2.0.Co;2} {\bibfield  {journal} {\bibinfo
   {journal} {Geology}\ }\textbf {\bibinfo {volume} {2}},\ \bibinfo {pages}
  {327} (\bibinfo {year} {1974})}\BibitemShut {NoStop}%
\bibitem [{\citenamefont {Davies}(1980)}]{RN1385}%
  \BibitemOpen
  \bibfield  {author} {\bibinfo {author} {\bibfnamefont {G.~F.}\ \bibnamefont
  {Davies}},\ }\href {\doibase 10.1029/RG018i003p00718} {\bibfield  {journal}
  {\bibinfo  {journal} {Reviews of Geophysics}\ }\textbf {\bibinfo {volume}
  {18}} (\bibinfo {year} {1980}),\ 10.1029/RG018i003p00718}\BibitemShut
  {NoStop}%
\bibitem [{\citenamefont {Sclater}\ \emph {et~al.}(1980)\citenamefont
  {Sclater}, \citenamefont {Jaupart},\ and\ \citenamefont {Galson}}]{RN1381}%
  \BibitemOpen
  \bibfield  {author} {\bibinfo {author} {\bibfnamefont {J.~G.}\ \bibnamefont
  {Sclater}}, \bibinfo {author} {\bibfnamefont {C.}~\bibnamefont {Jaupart}}, \
  and\ \bibinfo {author} {\bibfnamefont {D.}~\bibnamefont {Galson}},\ }\href
  {\doibase 10.1029/RG018i001p00269} {\bibfield  {journal} {\bibinfo  {journal}
  {Reviews of Geophysics}\ }\textbf {\bibinfo {volume} {18}} (\bibinfo {year}
  {1980}),\ 10.1029/RG018i001p00269}\BibitemShut {NoStop}%
\bibitem [{\citenamefont {Pollack}\ \emph {et~al.}(1993)\citenamefont
  {Pollack}, \citenamefont {Hurter},\ and\ \citenamefont {Johnson}}]{RN1382}%
  \BibitemOpen
  \bibfield  {author} {\bibinfo {author} {\bibfnamefont {H.~N.}\ \bibnamefont
  {Pollack}}, \bibinfo {author} {\bibfnamefont {S.~J.}\ \bibnamefont {Hurter}},
  \ and\ \bibinfo {author} {\bibfnamefont {J.~R.}\ \bibnamefont {Johnson}},\
  }\href {\doibase 10.1029/93rg01249} {\bibfield  {journal} {\bibinfo
  {journal} {Reviews of Geophysics}\ }\textbf {\bibinfo {volume} {31}},\
  \bibinfo {pages} {267} (\bibinfo {year} {1993})}\BibitemShut {NoStop}%
\bibitem [{\citenamefont {Jaupart}\ and\ \citenamefont
  {Mareschal}(2007)}]{RN1387}%
  \BibitemOpen
  \bibfield  {author} {\bibinfo {author} {\bibfnamefont {C.}~\bibnamefont
  {Jaupart}}\ and\ \bibinfo {author} {\bibfnamefont {J.~C.}\ \bibnamefont
  {Mareschal}},\ }in\ \href {\doibase 10.1016/C2009-1-28330-4} {\emph {\bibinfo
  {booktitle} {Treatise on Geophysics}}},\ Vol.~\bibinfo {volume} {6},\
  \bibinfo {editor} {edited by\ \bibinfo {editor} {\bibfnamefont
  {G.}~\bibnamefont {Schubert}}}\ (\bibinfo  {publisher} {Elsevier B.V.},\
  \bibinfo {year} {2007})\ pp.\ \bibinfo {pages} {217--251}\BibitemShut
  {NoStop}%
\bibitem [{\citenamefont {Davies}\ and\ \citenamefont {Davies}(2010)}]{RN593}%
  \BibitemOpen
  \bibfield  {author} {\bibinfo {author} {\bibfnamefont {J.~H.}\ \bibnamefont
  {Davies}}\ and\ \bibinfo {author} {\bibfnamefont {D.~R.}\ \bibnamefont
  {Davies}},\ }\href {\doibase 10.5194/se-1-5-2010} {\bibfield  {journal}
  {\bibinfo  {journal} {Solid Earth}\ }\textbf {\bibinfo {volume} {1}},\
  \bibinfo {pages} {5} (\bibinfo {year} {2010})}\BibitemShut {NoStop}%
\bibitem [{\citenamefont {Korenaga}(2008)}]{RN404}%
  \BibitemOpen
  \bibfield  {author} {\bibinfo {author} {\bibfnamefont {J.}~\bibnamefont
  {Korenaga}},\ }\href {\doibase 10.1029/2007rg000241} {\bibfield  {journal}
  {\bibinfo  {journal} {Reviews of Geophysics}\ }\textbf {\bibinfo {volume}
  {46}} (\bibinfo {year} {2008}),\ 10.1029/2007rg000241}\BibitemShut {NoStop}%
\bibitem [{\citenamefont {Jaupart}\ \emph {et~al.}(2015)\citenamefont
  {Jaupart}, \citenamefont {Labrosse}, \citenamefont {Lucazeau},\ and\
  \citenamefont {Mareschal}}]{RN1395}%
  \BibitemOpen
  \bibfield  {author} {\bibinfo {author} {\bibfnamefont {C.}~\bibnamefont
  {Jaupart}}, \bibinfo {author} {\bibfnamefont {S.}~\bibnamefont {Labrosse}},
  \bibinfo {author} {\bibfnamefont {F.}~\bibnamefont {Lucazeau}}, \ and\
  \bibinfo {author} {\bibfnamefont {J.~C.}\ \bibnamefont {Mareschal}},\ }in\
  \href {\doibase 10.1016/b978-0-444-53802-4.00126-3} {\emph {\bibinfo
  {booktitle} {Treatise on Geophysics}}},\ Vol.~\bibinfo {volume} {7},\
  \bibinfo {editor} {edited by\ \bibinfo {editor} {\bibfnamefont
  {G.}~\bibnamefont {Schubert}}}\ (\bibinfo  {publisher} {Elsevier B.V.},\
  \bibinfo {year} {2015})\ pp.\ \bibinfo {pages} {223--270}\BibitemShut
  {NoStop}%
\bibitem [{\citenamefont {Crosby}\ \emph {et~al.}(2006)\citenamefont {Crosby},
  \citenamefont {McKenzie},\ and\ \citenamefont {Sclater}}]{RN1397}%
  \BibitemOpen
  \bibfield  {author} {\bibinfo {author} {\bibfnamefont {A.~G.}\ \bibnamefont
  {Crosby}}, \bibinfo {author} {\bibfnamefont {D.}~\bibnamefont {McKenzie}}, \
  and\ \bibinfo {author} {\bibfnamefont {J.~G.}\ \bibnamefont {Sclater}},\
  }\href {\doibase 10.1111/j.1365-246X.2006.03015.x} {\bibfield  {journal}
  {\bibinfo  {journal} {Geophysical Journal International}\ }\textbf {\bibinfo
  {volume} {166}},\ \bibinfo {pages} {553} (\bibinfo {year}
  {2006})}\BibitemShut {NoStop}%
\bibitem [{\citenamefont {Lyubetskaya}\ and\ \citenamefont
  {Korenaga}(2007)}]{RN747}%
  \BibitemOpen
  \bibfield  {author} {\bibinfo {author} {\bibfnamefont {T.}~\bibnamefont
  {Lyubetskaya}}\ and\ \bibinfo {author} {\bibfnamefont {J.}~\bibnamefont
  {Korenaga}},\ }\href {\doibase 10.1029/2005JB004224} {\bibfield  {journal}
  {\bibinfo  {journal} {Journal of Geophysical Research: Solid Earth}\ }\textbf
  {\bibinfo {volume} {112}},\ \bibinfo {pages} {B03212} (\bibinfo {year}
  {2007})}\BibitemShut {NoStop}%
\bibitem [{\citenamefont {Taylor}(1980)}]{RN1378}%
  \BibitemOpen
  \bibfield  {author} {\bibinfo {author} {\bibfnamefont {S.}~\bibnamefont
  {Taylor}},\ }\href@noop {} {\bibfield  {journal} {\bibinfo  {journal} {Proc.
  Lunar Planet. Sci. Conf.}\ }\textbf {\bibinfo {volume} {11}},\ \bibinfo
  {pages} {333} (\bibinfo {year} {1980})}\BibitemShut {NoStop}%
\bibitem [{\citenamefont {McDonough}\ and\ \citenamefont {Sun}(1995)}]{RN1380}%
  \BibitemOpen
  \bibfield  {author} {\bibinfo {author} {\bibfnamefont {W.~F.}\ \bibnamefont
  {McDonough}}\ and\ \bibinfo {author} {\bibfnamefont {S.}~\bibnamefont
  {Sun}},\ }\href {\doibase 10.1016/0009-2541(94)00140-4} {\bibfield  {journal}
  {\bibinfo  {journal} {Chemical Geology}\ }\textbf {\bibinfo {volume} {120}},\
  \bibinfo {pages} {223} (\bibinfo {year} {1995})}\BibitemShut {NoStop}%
\bibitem [{\citenamefont {Anderson}(2007)}]{RN372}%
  \BibitemOpen
  \bibfield  {author} {\bibinfo {author} {\bibfnamefont {D.~L.}\ \bibnamefont
  {Anderson}},\ }\href {\doibase 10.1017/CBO9781139167291} {\emph {\bibinfo
  {title} {New theory of the Earth}}},\ \bibinfo {edition} {2nd}\ ed.\
  (\bibinfo  {publisher} {Cambridge University Press},\ \bibinfo {address} {New
  York, USA},\ \bibinfo {year} {2007})\BibitemShut {NoStop}%
\bibitem [{\citenamefont {Wang}\ \emph {et~al.}(2018)\citenamefont {Wang},
  \citenamefont {Lineweaver},\ and\ \citenamefont {Ireland}}]{RN1319}%
  \BibitemOpen
  \bibfield  {author} {\bibinfo {author} {\bibfnamefont {H.~S.}\ \bibnamefont
  {Wang}}, \bibinfo {author} {\bibfnamefont {C.~H.}\ \bibnamefont
  {Lineweaver}}, \ and\ \bibinfo {author} {\bibfnamefont {T.~R.}\ \bibnamefont
  {Ireland}},\ }\href {\doibase 10.1016/j.icarus.2017.08.024} {\bibfield
  {journal} {\bibinfo  {journal} {Icarus}\ }\textbf {\bibinfo {volume} {299}},\
  \bibinfo {pages} {460} (\bibinfo {year} {2018})}\BibitemShut {NoStop}%
\bibitem [{\citenamefont {Palme}\ and\ \citenamefont {O'Neill}(2003)}]{RN400}%
  \BibitemOpen
  \bibfield  {author} {\bibinfo {author} {\bibfnamefont {H.}~\bibnamefont
  {Palme}}\ and\ \bibinfo {author} {\bibfnamefont {H.}~\bibnamefont
  {O'Neill}},\ }in\ \href {\doibase 10.1016/B0-08-043751-6/02177-0} {\emph
  {\bibinfo {booktitle} {Treatise on Geochemistry}}},\ Vol.~\bibinfo {volume}
  {2},\ \bibinfo {editor} {edited by\ \bibinfo {editor} {\bibfnamefont {H.~D.}\
  \bibnamefont {Holland}}\ and\ \bibinfo {editor} {\bibfnamefont {K.~K.}\
  \bibnamefont {Turekian}}}\ (\bibinfo  {publisher} {Elsevier},\ \bibinfo
  {year} {2003})\ pp.\ \bibinfo {pages} {1--38}\BibitemShut {NoStop}%
\bibitem [{\citenamefont {Turcotte}\ and\ \citenamefont
  {Schubert}(2002)}]{RN368}%
  \BibitemOpen
  \bibfield  {author} {\bibinfo {author} {\bibfnamefont {D.~L.}\ \bibnamefont
  {Turcotte}}\ and\ \bibinfo {author} {\bibfnamefont {G.}~\bibnamefont
  {Schubert}},\ }\href@noop {} {\emph {\bibinfo {title} {Geodynamics,
  applications of continuum physics to geological problems}}},\ \bibinfo
  {edition} {2nd}\ ed.\ (\bibinfo  {publisher} {Cambridge University Press},\
  \bibinfo {year} {2002})\BibitemShut {NoStop}%
\bibitem [{\citenamefont {\v{S}r\'{a}mek}\ \emph {et~al.}(2013)\citenamefont
  {\v{S}r\'{a}mek}, \citenamefont {McDonough}, \citenamefont {Kite},
  \citenamefont {Leki\'{c}}, \citenamefont {Dye},\ and\ \citenamefont
  {Zhong}}]{RN630}%
  \BibitemOpen
  \bibfield  {author} {\bibinfo {author} {\bibfnamefont {O.}~\bibnamefont
  {\v{S}r\'{a}mek}}, \bibinfo {author} {\bibfnamefont {W.~F.}\ \bibnamefont
  {McDonough}}, \bibinfo {author} {\bibfnamefont {E.~S.}\ \bibnamefont {Kite}},
  \bibinfo {author} {\bibfnamefont {V.}~\bibnamefont {Leki\'{c}}}, \bibinfo
  {author} {\bibfnamefont {S.~T.}\ \bibnamefont {Dye}}, \ and\ \bibinfo
  {author} {\bibfnamefont {S.}~\bibnamefont {Zhong}},\ }\href {\doibase
  10.1016/j.epsl.2012.11.001} {\bibfield  {journal} {\bibinfo  {journal} {Earth
  and Planetary Science Letters}\ }\textbf {\bibinfo {volume} {361}},\ \bibinfo
  {pages} {356} (\bibinfo {year} {2013})}\BibitemShut {NoStop}%
\bibitem [{\citenamefont {Mantovani}\ \emph {et~al.}(2004)\citenamefont
  {Mantovani}, \citenamefont {Carmignani}, \citenamefont {Fiorentini},\ and\
  \citenamefont {Lissia}}]{RN356}%
  \BibitemOpen
  \bibfield  {author} {\bibinfo {author} {\bibfnamefont {F.}~\bibnamefont
  {Mantovani}}, \bibinfo {author} {\bibfnamefont {L.}~\bibnamefont
  {Carmignani}}, \bibinfo {author} {\bibfnamefont {G.}~\bibnamefont
  {Fiorentini}}, \ and\ \bibinfo {author} {\bibfnamefont {M.}~\bibnamefont
  {Lissia}},\ }\href {\doibase 10.1103/PhysRevD.69.013001} {\bibfield
  {journal} {\bibinfo  {journal} {Phys. Rev.}\ }\textbf {\bibinfo {volume}
  {D69}},\ \bibinfo {pages} {013001} (\bibinfo {year} {2004})},\ \Eprint
  {http://arxiv.org/abs/hep-ph/0309013} {arXiv:hep-ph/0309013 [hep-ph]}
  \BibitemShut {NoStop}%
\bibitem [{\citenamefont {Arevalo}\ \emph {et~al.}(2009)\citenamefont
  {Arevalo}, \citenamefont {McDonough},\ and\ \citenamefont {Luong}}]{RN361}%
  \BibitemOpen
  \bibfield  {author} {\bibinfo {author} {\bibfnamefont {R.}~\bibnamefont
  {Arevalo}}, \bibinfo {author} {\bibfnamefont {W.~F.}\ \bibnamefont
  {McDonough}}, \ and\ \bibinfo {author} {\bibfnamefont {M.}~\bibnamefont
  {Luong}},\ }\href {\doibase 10.1016/j.epsl.2008.12.023} {\bibfield  {journal}
  {\bibinfo  {journal} {Earth and Planetary Science Letters}\ }\textbf
  {\bibinfo {volume} {278}},\ \bibinfo {pages} {361} (\bibinfo {year}
  {2009})}\BibitemShut {NoStop}%
\bibitem [{\citenamefont {Bouvier}\ and\ \citenamefont {Boyet}(2016)}]{RN1396}%
  \BibitemOpen
  \bibfield  {author} {\bibinfo {author} {\bibfnamefont {A.}~\bibnamefont
  {Bouvier}}\ and\ \bibinfo {author} {\bibfnamefont {M.}~\bibnamefont
  {Boyet}},\ }\href {\doibase 10.1038/nature19351} {\bibfield  {journal}
  {\bibinfo  {journal} {Nature}\ }\textbf {\bibinfo {volume} {537}},\ \bibinfo
  {pages} {399} (\bibinfo {year} {2016})}\BibitemShut {NoStop}%
\bibitem [{\citenamefont {Wohlers}\ and\ \citenamefont {Wood}(2017)}]{RN1399}%
  \BibitemOpen
  \bibfield  {author} {\bibinfo {author} {\bibfnamefont {A.}~\bibnamefont
  {Wohlers}}\ and\ \bibinfo {author} {\bibfnamefont {B.~J.}\ \bibnamefont
  {Wood}},\ }\href {\doibase 10.1016/j.gca.2017.01.050} {\bibfield  {journal}
  {\bibinfo  {journal} {Geochimica et Cosmochimica Acta}\ }\textbf {\bibinfo
  {volume} {205}},\ \bibinfo {pages} {226} (\bibinfo {year}
  {2017})}\BibitemShut {NoStop}%
\bibitem [{\citenamefont {Blanchard}\ \emph {et~al.}(2017)\citenamefont
  {Blanchard}, \citenamefont {Siebert}, \citenamefont {Borensztajn},\ and\
  \citenamefont {Badro}}]{RN1400}%
  \BibitemOpen
  \bibfield  {author} {\bibinfo {author} {\bibfnamefont {I.}~\bibnamefont
  {Blanchard}}, \bibinfo {author} {\bibfnamefont {J.}~\bibnamefont {Siebert}},
  \bibinfo {author} {\bibfnamefont {S.}~\bibnamefont {Borensztajn}}, \ and\
  \bibinfo {author} {\bibfnamefont {J.}~\bibnamefont {Badro}},\ }\href
  {\doibase 10.7185/geochemlet.1737} {\bibfield  {journal} {\bibinfo  {journal}
  {Geochemical Perspectives Letters}\ }\textbf {\bibinfo {volume} {5}},\
  \bibinfo {pages} {1} (\bibinfo {year} {2017})}\BibitemShut {NoStop}%
\bibitem [{\citenamefont {Chidester}\ \emph {et~al.}(2017)\citenamefont
  {Chidester}, \citenamefont {Rahman}, \citenamefont {Righter},\ and\
  \citenamefont {Campbell}}]{RN1401}%
  \BibitemOpen
  \bibfield  {author} {\bibinfo {author} {\bibfnamefont {B.~A.}\ \bibnamefont
  {Chidester}}, \bibinfo {author} {\bibfnamefont {Z.}~\bibnamefont {Rahman}},
  \bibinfo {author} {\bibfnamefont {K.}~\bibnamefont {Righter}}, \ and\
  \bibinfo {author} {\bibfnamefont {A.~J.}\ \bibnamefont {Campbell}},\ }\href
  {\doibase 10.1016/j.gca.2016.11.035} {\bibfield  {journal} {\bibinfo
  {journal} {Geochimica et Cosmochimica Acta}\ }\textbf {\bibinfo {volume}
  {199}},\ \bibinfo {pages} {1} (\bibinfo {year} {2017})}\BibitemShut {NoStop}%
\bibitem [{\citenamefont {Blichert-Toft}\ \emph {et~al.}(2010)\citenamefont
  {Blichert-Toft}, \citenamefont {Zanda}, \citenamefont {Ebel},\ and\
  \citenamefont {Albar\'ede}}]{RN1377}%
  \BibitemOpen
  \bibfield  {author} {\bibinfo {author} {\bibfnamefont {J.}~\bibnamefont
  {Blichert-Toft}}, \bibinfo {author} {\bibfnamefont {B.}~\bibnamefont
  {Zanda}}, \bibinfo {author} {\bibfnamefont {D.~S.}\ \bibnamefont {Ebel}}, \
  and\ \bibinfo {author} {\bibfnamefont {F.}~\bibnamefont {Albar\'ede}},\
  }\href {\doibase 10.1016/j.epsl.2010.10.001} {\bibfield  {journal} {\bibinfo
  {journal} {Earth and Planetary Science Letters}\ }\textbf {\bibinfo {volume}
  {300}},\ \bibinfo {pages} {152} (\bibinfo {year} {2010})}\BibitemShut
  {NoStop}%
\bibitem [{\citenamefont {Wipperfurth}\ \emph {et~al.}(2018)\citenamefont
  {Wipperfurth}, \citenamefont {Guo}, \citenamefont {Šrámek},\ and\
  \citenamefont {McDonough}}]{RN1328}%
  \BibitemOpen
  \bibfield  {author} {\bibinfo {author} {\bibfnamefont {S.~A.}\ \bibnamefont
  {Wipperfurth}}, \bibinfo {author} {\bibfnamefont {M.}~\bibnamefont {Guo}},
  \bibinfo {author} {\bibfnamefont {O.}~\bibnamefont {Šrámek}}, \ and\
  \bibinfo {author} {\bibfnamefont {W.~F.}\ \bibnamefont {McDonough}},\ }\href
  {\doibase 10.1016/j.epsl.2018.06.029} {\bibfield  {journal} {\bibinfo
  {journal} {Earth and Planetary Science Letters}\ }\textbf {\bibinfo {volume}
  {498}},\ \bibinfo {pages} {196} (\bibinfo {year} {2018})}\BibitemShut
  {NoStop}%
\bibitem [{\citenamefont {Strati}\ \emph {et~al.}(2017)\citenamefont {Strati},
  \citenamefont {Wipperfurth}, \citenamefont {Baldoncini}, \citenamefont
  {McDonough},\ and\ \citenamefont {Mantovani}}]{RN1290}%
  \BibitemOpen
  \bibfield  {author} {\bibinfo {author} {\bibfnamefont {V.}~\bibnamefont
  {Strati}}, \bibinfo {author} {\bibfnamefont {S.~A.}\ \bibnamefont
  {Wipperfurth}}, \bibinfo {author} {\bibfnamefont {M.}~\bibnamefont
  {Baldoncini}}, \bibinfo {author} {\bibfnamefont {W.~F.}\ \bibnamefont
  {McDonough}}, \ and\ \bibinfo {author} {\bibfnamefont {F.}~\bibnamefont
  {Mantovani}},\ }\href {\doibase 10.1002/2017gc007067} {\bibfield  {journal}
  {\bibinfo  {journal} {Geochemistry, Geophysics, Geosystems}\ }\textbf
  {\bibinfo {volume} {18}},\ \bibinfo {pages} {4326} (\bibinfo {year}
  {2017})}\BibitemShut {NoStop}%
\bibitem [{\citenamefont {Huang}\ \emph {et~al.}(2014)\citenamefont {Huang},
  \citenamefont {Strati}, \citenamefont {Mantovani}, \citenamefont {Shirey},\
  and\ \citenamefont {McDonough}}]{RN872}%
  \BibitemOpen
  \bibfield  {author} {\bibinfo {author} {\bibfnamefont {Y.}~\bibnamefont
  {Huang}}, \bibinfo {author} {\bibfnamefont {V.}~\bibnamefont {Strati}},
  \bibinfo {author} {\bibfnamefont {F.}~\bibnamefont {Mantovani}}, \bibinfo
  {author} {\bibfnamefont {S.~B.}\ \bibnamefont {Shirey}}, \ and\ \bibinfo
  {author} {\bibfnamefont {W.~F.}\ \bibnamefont {McDonough}},\ }\href {\doibase
  10.1002/2014gc005397} {\bibfield  {journal} {\bibinfo  {journal}
  {Geochemistry, Geophysics, Geosystems}\ }\textbf {\bibinfo {volume} {15}},\
  \bibinfo {pages} {3925} (\bibinfo {year} {2014})}\BibitemShut {NoStop}%
\bibitem [{\citenamefont {Coltorti}\ \emph {et~al.}(2011)\citenamefont
  {Coltorti} \emph {et~al.}}]{coltorti}%
  \BibitemOpen
  \bibfield  {author} {\bibinfo {author} {\bibfnamefont {M.}~\bibnamefont
  {Coltorti}} \emph {et~al.},\ }\href {\doibase 10.1016/j.gca.2011.01.024}
  {\bibfield  {journal} {\bibinfo  {journal} {Geochim. Cosmochim. Acta}\
  }\textbf {\bibinfo {volume} {75}},\ \bibinfo {pages} {2271} (\bibinfo {year}
  {2011})},\ \Eprint {http://arxiv.org/abs/1102.1335} {arXiv:1102.1335
  [astro-ph.EP]} \BibitemShut {NoStop}%
\bibitem [{\citenamefont {Herndon}(1993)}]{herndon1993feasibility}%
  \BibitemOpen
  \bibfield  {author} {\bibinfo {author} {\bibfnamefont {J.~M.}\ \bibnamefont
  {Herndon}},\ }\href {\doibase 10.5636/jgg.45.423} {\bibfield  {journal}
  {\bibinfo  {journal} {Journal of geomagnetism and geoelectricity}\ }\textbf
  {\bibinfo {volume} {45}},\ \bibinfo {pages} {423} (\bibinfo {year}
  {1993})}\BibitemShut {NoStop}%
\bibitem [{\citenamefont {Herndon}(1996)}]{herndon1996substructure}%
  \BibitemOpen
  \bibfield  {author} {\bibinfo {author} {\bibfnamefont {J.~M.}\ \bibnamefont
  {Herndon}},\ }\href {\doibase 10.1073/pnas.93.2.646} {\bibfield  {journal}
  {\bibinfo  {journal} {Proceedings of the National Academy of Sciences}\
  }\textbf {\bibinfo {volume} {93}},\ \bibinfo {pages} {646} (\bibinfo {year}
  {1996})}\BibitemShut {NoStop}%
\bibitem [{\citenamefont {Rusov}\ \emph {et~al.}(2007)\citenamefont {Rusov}
  \emph {et~al.}}]{rusov2007geoantineutrino}%
  \BibitemOpen
  \bibfield  {author} {\bibinfo {author} {\bibfnamefont {V.~D.}\ \bibnamefont
  {Rusov}} \emph {et~al.},\ }\href {\doibase 10.1029/2005JB004212} {\bibfield
  {journal} {\bibinfo  {journal} {Journal of Geophysical Research: Solid
  Earth}\ }\textbf {\bibinfo {volume} {112}},\ \bibinfo {pages} {203} (\bibinfo
  {year} {2007})}\BibitemShut {NoStop}%
\bibitem [{\citenamefont {Meijer}\ and\ \citenamefont
  {Westrenen}(2008)}]{de2008feasibility}%
  \BibitemOpen
  \bibfield  {author} {\bibinfo {author} {\bibfnamefont {R.}~\bibnamefont
  {Meijer}}\ and\ \bibinfo {author} {\bibfnamefont {W.}~\bibnamefont
  {Westrenen}},\ }\href {https://hdl.handle.net/10520/EJC96784} {\bibfield
  {journal} {\bibinfo  {journal} {South African Journal of Science}\ }\textbf
  {\bibinfo {volume} {104}},\ \bibinfo {pages} {111} (\bibinfo {year}
  {2008})}\BibitemShut {NoStop}%
\bibitem [{\citenamefont {Wang}\ and\ \citenamefont {Wen}(2004)}]{wang}%
  \BibitemOpen
  \bibfield  {author} {\bibinfo {author} {\bibfnamefont {Y.}~\bibnamefont
  {Wang}}\ and\ \bibinfo {author} {\bibfnamefont {L.}~\bibnamefont {Wen}},\
  }\href {\doibase 10.1029/2003JB002674} {\bibfield  {journal} {\bibinfo
  {journal} {Journal of Geophysical Research}\ }\textbf {\bibinfo {volume}
  {109}},\ \bibinfo {pages} {B10305} (\bibinfo {year} {2004})}\BibitemShut
  {NoStop}%
\bibitem [{\citenamefont {Szczerbinska}\ \emph {et~al.}(2011)\citenamefont
  {Szczerbinska}, \citenamefont {Alyssa},\ and\ \citenamefont
  {Dongming}}]{Szczerbinska2011}%
  \BibitemOpen
  \bibfield  {author} {\bibinfo {author} {\bibfnamefont {B.}~\bibnamefont
  {Szczerbinska}}, \bibinfo {author} {\bibfnamefont {D.}~\bibnamefont
  {Alyssa}}, \ and\ \bibinfo {author} {\bibfnamefont {M.}~\bibnamefont
  {Dongming}},\ }\bibfield  {booktitle} {\emph {\bibinfo {booktitle}
  {{Proceedings of the South Dakota Academy of Science}}},\ }\href@noop {}
  {\bibfield  {journal} {\bibinfo  {journal} {PoS}\ }\textbf {\bibinfo {volume}
  {90}},\ \bibinfo {pages} {13} (\bibinfo {year} {2011})}\BibitemShut {NoStop}%
\bibitem [{\citenamefont {Cabrera}\ \emph {et~al.}(2019)\citenamefont {Cabrera}
  \emph {et~al.}}]{OpaqueDetector}%
  \BibitemOpen
  \bibfield  {author} {\bibinfo {author} {\bibfnamefont {A.}~\bibnamefont
  {Cabrera}} \emph {et~al.},\ }\href@noop {} {\  (\bibinfo {year} {2019})},\
  \Eprint {http://arxiv.org/abs/1908.02859} {arXiv:1908.02859
  [physics.ins-det]} \BibitemShut {NoStop}%
\bibitem [{\citenamefont {Goettel}(1976)}]{Goettel1976}%
  \BibitemOpen
  \bibfield  {author} {\bibinfo {author} {\bibfnamefont {K.~A.}\ \bibnamefont
  {Goettel}},\ }\href {\doibase 10.1007/BF01454192} {\bibfield  {journal}
  {\bibinfo  {journal} {Geophysical surveys}\ }\textbf {\bibinfo {volume}
  {2}},\ \bibinfo {pages} {369} (\bibinfo {year} {1976})}\BibitemShut {NoStop}%
\bibitem [{\citenamefont {Alimonti}\ \emph {et~al.}(2009)\citenamefont
  {Alimonti} \emph {et~al.}}]{Alimonti:2008gc}%
  \BibitemOpen
  \bibfield  {author} {\bibinfo {author} {\bibfnamefont {G.}~\bibnamefont
  {Alimonti}} \emph {et~al.} (\bibinfo {collaboration} {Borexino}),\ }\href
  {\doibase 10.1016/j.nima.2008.11.076} {\bibfield  {journal} {\bibinfo
  {journal} {Nucl. Instrum. Meth.}\ }\textbf {\bibinfo {volume} {A600}},\
  \bibinfo {pages} {568} (\bibinfo {year} {2009})},\ \Eprint
  {http://arxiv.org/abs/0806.2400} {arXiv:0806.2400 [physics.ins-det]}
  \BibitemShut {NoStop}%
\bibitem [{\citenamefont {Agostini}\ \emph {et~al.}(2019)\citenamefont
  {Agostini} \emph {et~al.}}]{Agostini:2018fnx}%
  \BibitemOpen
  \bibfield  {author} {\bibinfo {author} {\bibfnamefont {M.}~\bibnamefont
  {Agostini}} \emph {et~al.} (\bibinfo {collaboration} {Borexino}),\ }\href
  {\doibase 10.1088/1475-7516/2019/02/046} {\bibfield  {journal} {\bibinfo
  {journal} {JCAP}\ }\textbf {\bibinfo {volume} {1902}},\ \bibinfo {pages}
  {046} (\bibinfo {year} {2019})},\ \Eprint {http://arxiv.org/abs/1808.04207}
  {arXiv:1808.04207 [hep-ex]} \BibitemShut {NoStop}%
\bibitem [{\citenamefont {Back}\ \emph {et~al.}(2012)\citenamefont {Back} \emph
  {et~al.}}]{Back:2012awa}%
  \BibitemOpen
  \bibfield  {author} {\bibinfo {author} {\bibfnamefont {H.}~\bibnamefont
  {Back}} \emph {et~al.} (\bibinfo {collaboration} {Borexino}),\ }\href
  {\doibase 10.1088/1748-0221/7/10/P10018} {\bibfield  {journal} {\bibinfo
  {journal} {JINST}\ }\textbf {\bibinfo {volume} {7}},\ \bibinfo {pages}
  {P10018} (\bibinfo {year} {2012})},\ \Eprint {http://arxiv.org/abs/1207.4816}
  {arXiv:1207.4816 [physics.ins-det]} \BibitemShut {NoStop}%
\bibitem [{\citenamefont {Bellini}\ \emph {et~al.}(2011)\citenamefont {Bellini}
  \emph {et~al.}}]{Bellini:2011yd}%
  \BibitemOpen
  \bibfield  {author} {\bibinfo {author} {\bibfnamefont {G.}~\bibnamefont
  {Bellini}} \emph {et~al.} (\bibinfo {collaboration} {Borexino}),\ }\href
  {\doibase 10.1088/1748-0221/6/05/P05005} {\bibfield  {journal} {\bibinfo
  {journal} {JINST}\ }\textbf {\bibinfo {volume} {6}},\ \bibinfo {pages}
  {P05005} (\bibinfo {year} {2011})},\ \Eprint {http://arxiv.org/abs/1101.3101}
  {arXiv:1101.3101 [physics.ins-det]} \BibitemShut {NoStop}%
\bibitem [{\citenamefont {Voss}\ \emph {et~al.}(2007)\citenamefont {Voss},
  \citenamefont {Hocker}, \citenamefont {Stelzer},\ and\ \citenamefont
  {Tegenfeldt}}]{Voss:2007jxm}%
  \BibitemOpen
  \bibfield  {author} {\bibinfo {author} {\bibfnamefont {H.}~\bibnamefont
  {Voss}}, \bibinfo {author} {\bibfnamefont {A.}~\bibnamefont {Hocker}},
  \bibinfo {author} {\bibfnamefont {J.}~\bibnamefont {Stelzer}}, \ and\
  \bibinfo {author} {\bibfnamefont {F.}~\bibnamefont {Tegenfeldt}},\ }\bibfield
   {booktitle} {\emph {\bibinfo {booktitle} {{Proceedings, 11th International
  Workshop on Advanced computing and analysis techniques in physics research
  (ACAT 2007): Amsterdam, Netherlands, April 23-27, 2007}}},\ }\href {\doibase
  10.22323/1.050.0040} {\bibfield  {journal} {\bibinfo  {journal} {PoS}\
  }\textbf {\bibinfo {volume} {ACAT}},\ \bibinfo {pages} {040} (\bibinfo {year}
  {2007})}\BibitemShut {NoStop}%
\bibitem [{\citenamefont {Cortes}\ and\ \citenamefont
  {Vapnik}(1995)}]{Cortes:1995}%
  \BibitemOpen
  \bibfield  {author} {\bibinfo {author} {\bibfnamefont {C.}~\bibnamefont
  {Cortes}}\ and\ \bibinfo {author} {\bibfnamefont {V.}~\bibnamefont
  {Vapnik}},\ }\href {\doibase 10.1007/BF00994018} {\bibfield  {journal}
  {\bibinfo  {journal} {Machine Learning}\ }\textbf {\bibinfo {volume} {20}},\
  \bibinfo {pages} {273} (\bibinfo {year} {1995})}\BibitemShut {NoStop}%
\bibitem [{\citenamefont {Freund}\ and\ \citenamefont
  {Schapire}(1997)}]{Freund:1997xna}%
  \BibitemOpen
  \bibfield  {author} {\bibinfo {author} {\bibfnamefont {Y.}~\bibnamefont
  {Freund}}\ and\ \bibinfo {author} {\bibfnamefont {R.~E.}\ \bibnamefont
  {Schapire}},\ }\href {\doibase 10.1006/jcss.1997.1504} {\bibfield  {journal}
  {\bibinfo  {journal} {J. Comput. Syst. Sci.}\ }\textbf {\bibinfo {volume}
  {55}},\ \bibinfo {pages} {119} (\bibinfo {year} {1997})}\BibitemShut
  {NoStop}%
\bibitem [{\citenamefont {Lukyanchenko}(2017)}]{Lukyanchenko:2017}%
  \BibitemOpen
  \bibfield  {author} {\bibinfo {author} {\bibfnamefont {G.}~\bibnamefont
  {Lukyanchenko}},\ }\emph {\bibinfo {title} {The DAQ system based on Fast Wave
  Form Digitizers in the Borexino detector for the registration of neutrino
  radiation from astrophysical sources}},\ \href@noop {} {Ph.D. thesis},\
  \bibinfo  {school} {National Research Center Kurchatov Institute} (\bibinfo
  {year} {2017}),\ \bibinfo {note} {(in Russian)
  \href{http://www.nrcki.ru/files/pdf/1490777684.pdf}{online
  version}}\BibitemShut {NoStop}%
\bibitem [{\citenamefont {Back}\ \emph {et~al.}(2008)\citenamefont {Back} \emph
  {et~al.}}]{Back:2008}%
  \BibitemOpen
  \bibfield  {author} {\bibinfo {author} {\bibfnamefont {H.}~\bibnamefont
  {Back}} \emph {et~al.} (\bibinfo {collaboration} {Borexino}),\ }\href
  {\doibase 10.1016/j.nima.2007.09.036} {\bibfield  {journal} {\bibinfo
  {journal} {Nucl. Instrum. Meth.}\ }\textbf {\bibinfo {volume} {A584}},\
  \bibinfo {pages} {98} (\bibinfo {year} {2008})},\ \Eprint
  {http://arxiv.org/abs/0705.0239} {arXiv:0705.0239 [physics.ins-det]}
  \BibitemShut {NoStop}%
\bibitem [{\citenamefont {Agostini}\ \emph {et~al.}(2017)\citenamefont
  {Agostini} \emph {et~al.}}]{Agostini:Jun2017}%
  \BibitemOpen
  \bibfield  {author} {\bibinfo {author} {\bibfnamefont {M.}~\bibnamefont
  {Agostini}} \emph {et~al.} (\bibinfo {collaboration} {Borexino}),\ }\href
  {\doibase 10.1016/j.astropartphys.2017.04.004} {\bibfield  {journal}
  {\bibinfo  {journal} {Astroparticle Physics}\ }\textbf {\bibinfo {volume}
  {92}},\ \bibinfo {pages} {21} (\bibinfo {year} {2017})},\ \Eprint
  {http://arxiv.org/abs/1701.07970} {arXiv:1701.07970 [hep-ex]} \BibitemShut
  {NoStop}%
\bibitem [{\citenamefont {Strumia}\ and\ \citenamefont
  {Vissani}(2003)}]{strumia2003precise}%
  \BibitemOpen
  \bibfield  {author} {\bibinfo {author} {\bibfnamefont {A.}~\bibnamefont
  {Strumia}}\ and\ \bibinfo {author} {\bibfnamefont {F.}~\bibnamefont
  {Vissani}},\ }\href {\doibase 10.1016/S0370-2693(03)00616-6} {\bibfield
  {journal} {\bibinfo  {journal} {Phys. Lett.}\ }\textbf {\bibinfo {volume}
  {B564}},\ \bibinfo {pages} {42} (\bibinfo {year} {2003})},\ \Eprint
  {http://arxiv.org/abs/astro-ph/0302055} {arXiv:astro-ph/0302055 [astro-ph]}
  \BibitemShut {NoStop}%
\bibitem [{\citenamefont {Enomoto}()}]{Enomoto}%
  \BibitemOpen
  \bibfield  {author} {\bibinfo {author} {\bibfnamefont {S.}~\bibnamefont
  {Enomoto}},\ }\href@noop {} {\enquote {\bibinfo {title} {Geoneutrino spectrum
  and luminosity},}\ }\bibinfo {note}
  {{\href{https://www.awa.tohoku.ac.jp/~sanshiro/research/geoneutrino/spectrum/}{online
  version}}}\BibitemShut {NoStop}%
\bibitem [{\citenamefont {Capozzi}\ \emph {et~al.}(2014)\citenamefont
  {Capozzi}, \citenamefont {Lisi},\ and\ \citenamefont
  {Marrone}}]{capozzi2014neutrino}%
  \BibitemOpen
  \bibfield  {author} {\bibinfo {author} {\bibfnamefont {F.}~\bibnamefont
  {Capozzi}}, \bibinfo {author} {\bibfnamefont {E.}~\bibnamefont {Lisi}}, \
  and\ \bibinfo {author} {\bibfnamefont {A.}~\bibnamefont {Marrone}},\ }\href
  {\doibase 10.1103/PhysRevD.89.013001} {\bibfield  {journal} {\bibinfo
  {journal} {Phys. Rev.}\ }\textbf {\bibinfo {volume} {D89}},\ \bibinfo {pages}
  {013001} (\bibinfo {year} {2014})},\ \Eprint {http://arxiv.org/abs/1309.1638}
  {arXiv:1309.1638 [hep-ph]} \BibitemShut {NoStop}%
\bibitem [{\citenamefont {Esteban}\ \emph {et~al.}(2018)\citenamefont {Esteban}
  \emph {et~al.}}]{NUFIT2018}%
  \BibitemOpen
  \bibfield  {author} {\bibinfo {author} {\bibfnamefont {I.}~\bibnamefont
  {Esteban}} \emph {et~al.},\ }\href@noop {} {\enquote {\bibinfo {title} {Nufit
  3.2},}\ } (\bibinfo {year} {2018}),\ \bibinfo {note}
  {\href{http://www.nu-fit.org/?q=node/166}{NuFIT website}}\BibitemShut
  {NoStop}%
\bibitem [{\citenamefont {Baldoncini}\ \emph {et~al.}(2015)\citenamefont
  {Baldoncini}, \citenamefont {Callegari}, \citenamefont {Fiorentini},
  \citenamefont {Mantovani}, \citenamefont {Ricci}, \citenamefont {Strati},\
  and\ \citenamefont {Xhixha}}]{baldoncini2015reference}%
  \BibitemOpen
  \bibfield  {author} {\bibinfo {author} {\bibfnamefont {M.}~\bibnamefont
  {Baldoncini}}, \bibinfo {author} {\bibfnamefont {I.}~\bibnamefont
  {Callegari}}, \bibinfo {author} {\bibfnamefont {G.}~\bibnamefont
  {Fiorentini}}, \bibinfo {author} {\bibfnamefont {F.}~\bibnamefont
  {Mantovani}}, \bibinfo {author} {\bibfnamefont {B.}~\bibnamefont {Ricci}},
  \bibinfo {author} {\bibfnamefont {V.}~\bibnamefont {Strati}}, \ and\ \bibinfo
  {author} {\bibfnamefont {G.}~\bibnamefont {Xhixha}},\ }\href {\doibase
  10.1103/PhysRevD.91.065002} {\bibfield  {journal} {\bibinfo  {journal} {Phys.
  Rev.}\ }\textbf {\bibinfo {volume} {D91}},\ \bibinfo {pages} {065002}
  (\bibinfo {year} {2015})},\ \Eprint {http://arxiv.org/abs/1411.6475}
  {arXiv:1411.6475 [physics.ins-det]} \BibitemShut {NoStop}%
\bibitem [{\citenamefont {Dziewonski}\ and\ \citenamefont
  {Anderson}(1981)}]{Dziewonski:1981xy}%
  \BibitemOpen
  \bibfield  {author} {\bibinfo {author} {\bibfnamefont {A.~M.}\ \bibnamefont
  {Dziewonski}}\ and\ \bibinfo {author} {\bibfnamefont {D.~L.}\ \bibnamefont
  {Anderson}},\ }\href {\doibase 10.1016/0031-9201(81)90046-7} {\bibfield
  {journal} {\bibinfo  {journal} {Phys. Earth Planet. Interiors}\ }\textbf
  {\bibinfo {volume} {25}},\ \bibinfo {pages} {297} (\bibinfo {year}
  {1981})}\BibitemShut {NoStop}%
\bibitem [{\citenamefont {Fiorentini}\ \emph {et~al.}(2010)\citenamefont
  {Fiorentini}, \citenamefont {Ianni}, \citenamefont {Korga}, \citenamefont
  {Lissia}, \citenamefont {Mantovani}, \citenamefont {Miramonti}, \citenamefont
  {Oberauer}, \citenamefont {Obolensky}, \citenamefont {Smirnov},\ and\
  \citenamefont {Suvorov}}]{RN885}%
  \BibitemOpen
  \bibfield  {author} {\bibinfo {author} {\bibfnamefont {G.}~\bibnamefont
  {Fiorentini}}, \bibinfo {author} {\bibfnamefont {A.}~\bibnamefont {Ianni}},
  \bibinfo {author} {\bibfnamefont {G.}~\bibnamefont {Korga}}, \bibinfo
  {author} {\bibfnamefont {M.}~\bibnamefont {Lissia}}, \bibinfo {author}
  {\bibfnamefont {F.}~\bibnamefont {Mantovani}}, \bibinfo {author}
  {\bibfnamefont {L.}~\bibnamefont {Miramonti}}, \bibinfo {author}
  {\bibfnamefont {L.}~\bibnamefont {Oberauer}}, \bibinfo {author}
  {\bibfnamefont {M.}~\bibnamefont {Obolensky}}, \bibinfo {author}
  {\bibfnamefont {O.}~\bibnamefont {Smirnov}}, \ and\ \bibinfo {author}
  {\bibfnamefont {Y.}~\bibnamefont {Suvorov}},\ }\href {\doibase
  10.1103/PhysRevC.81.034602} {\bibfield  {journal} {\bibinfo  {journal} {Phys.
  Rev.}\ }\textbf {\bibinfo {volume} {C81}},\ \bibinfo {pages} {034602}
  (\bibinfo {year} {2010})},\ \Eprint {http://arxiv.org/abs/0908.3433}
  {arXiv:0908.3433 [nucl-ex]} \BibitemShut {NoStop}%
\bibitem [{\citenamefont {Finetti}(2005)}]{RN572}%
  \BibitemOpen
  \bibinfo {editor} {\bibfnamefont {I.}~\bibnamefont {Finetti}},\ ed.,\
  \href@noop {} {\emph {\bibinfo {title} {CROP Project: Deep Seismic
  Exploration of the Central Mediterranean and Italy}}},\ Vol.~\bibinfo
  {volume} {1}\ (\bibinfo  {publisher} {Elsevier Science},\ \bibinfo {year}
  {2005})\BibitemShut {NoStop}%
\bibitem [{\citenamefont {Plank}(2014)}]{RN392}%
  \BibitemOpen
  \bibfield  {author} {\bibinfo {author} {\bibfnamefont {T.}~\bibnamefont
  {Plank}},\ }in\ \href {\doibase 10.1016/B978-0-08-095975-7.00319-3} {\emph
  {\bibinfo {booktitle} {Treatise of Geochemistry}}},\ Vol.~\bibinfo {volume}
  {4},\ \bibinfo {editor} {edited by\ \bibinfo {editor} {\bibfnamefont {H.~D.}\
  \bibnamefont {Holland}}\ and\ \bibinfo {editor} {\bibfnamefont {K.~K.}\
  \bibnamefont {Turekian}}}\ (\bibinfo  {publisher} {Elsevier},\ \bibinfo
  {year} {2014})\ \bibinfo {edition} {2nd}\ ed.,\ pp.\ \bibinfo {pages}
  {607--629}\BibitemShut {NoStop}%
\bibitem [{\citenamefont {Bassin}\ \emph {et~al.}(2000)\citenamefont {Bassin},
  \citenamefont {Laske},\ and\ \citenamefont {Masters}}]{bassin2000current}%
  \BibitemOpen
  \bibfield  {author} {\bibinfo {author} {\bibfnamefont {C.}~\bibnamefont
  {Bassin}}, \bibinfo {author} {\bibfnamefont {G.}~\bibnamefont {Laske}}, \
  and\ \bibinfo {author} {\bibfnamefont {G.}~\bibnamefont {Masters}},\
  }\href@noop {} {\bibfield  {journal} {\bibinfo  {journal} {Eos}\ }\textbf
  {\bibinfo {volume} {81}},\ \bibinfo {pages} {F897} (\bibinfo {year}
  {2000})}\BibitemShut {NoStop}%
\bibitem [{\citenamefont {Laske}\ \emph {et~al.}(2001)\citenamefont {Laske},
  \citenamefont {Masters},\ and\ \citenamefont {Reif}}]{RN896}%
  \BibitemOpen
  \bibfield  {author} {\bibinfo {author} {\bibfnamefont {G.}~\bibnamefont
  {Laske}}, \bibinfo {author} {\bibfnamefont {G.}~\bibnamefont {Masters}}, \
  and\ \bibinfo {author} {\bibfnamefont {C.}~\bibnamefont {Reif}},\ }\href@noop
  {} {\enquote {\bibinfo {title} {Crust 2.0. {A} new global crustal model at $2
  \times 2$ degrees},}\ } (\bibinfo {year} {2001}),\ \bibinfo {note}
  {\href{http://igppweb.ucsd.edu/~gabi/crust2.html}{CRUST 2.0
  website}}\BibitemShut {NoStop}%
\bibitem [{\citenamefont {Shapiro}\ and\ \citenamefont
  {Ritzwoller}(2002)}]{RN345}%
  \BibitemOpen
  \bibfield  {author} {\bibinfo {author} {\bibfnamefont {N.}~\bibnamefont
  {Shapiro}}\ and\ \bibinfo {author} {\bibfnamefont {M.}~\bibnamefont
  {Ritzwoller}},\ }\href {\doibase 10.1046/j.1365-246X.2002.01742.x} {\bibfield
   {journal} {\bibinfo  {journal} {Geophysical Journal International}\ }\textbf
  {\bibinfo {volume} {151}},\ \bibinfo {pages} {88} (\bibinfo {year}
  {2002})}\BibitemShut {NoStop}%
\bibitem [{\citenamefont {Reguzzoni}\ and\ \citenamefont
  {Sampietro}(2015)}]{RN1096}%
  \BibitemOpen
  \bibfield  {author} {\bibinfo {author} {\bibfnamefont {M.}~\bibnamefont
  {Reguzzoni}}\ and\ \bibinfo {author} {\bibfnamefont {D.}~\bibnamefont
  {Sampietro}},\ }\href {\doibase 10.1016/j.jag.2014.04.002} {\bibfield
  {journal} {\bibinfo  {journal} {International Journal of Applied Earth
  Observation and Geoinformation}\ }\textbf {\bibinfo {volume} {35}},\ \bibinfo
  {pages} {31} (\bibinfo {year} {2015})}\BibitemShut {NoStop}%
\bibitem [{\citenamefont {Laske}\ and\ \citenamefont {Masters}(1997)}]{RN895}%
  \BibitemOpen
  \bibfield  {author} {\bibinfo {author} {\bibfnamefont {G.}~\bibnamefont
  {Laske}}\ and\ \bibinfo {author} {\bibfnamefont {T.}~\bibnamefont
  {Masters}},\ }\href@noop {} {\bibfield  {journal} {\bibinfo  {journal} {EOS
  Trans. AGU}\ }\textbf {\bibinfo {volume} {78}},\ \bibinfo {pages} {F483}
  (\bibinfo {year} {1997})}\BibitemShut {NoStop}%
\bibitem [{\citenamefont {{World Nuclear Association
  Database}}(2015)}]{coresposition}%
  \BibitemOpen
  \bibfield  {author} {\bibinfo {author} {\bibnamefont {{World Nuclear
  Association Database}}},\ }\href@noop {} {\enquote {\bibinfo {title} {Reactor
  database},}\ } (\bibinfo {year} {2015}),\ \bibinfo {note}
  {\href{http://world‐nuclear.org/NuclearDatabase/Default.aspx?id=27232}{Database
  website}}\BibitemShut {NoStop}%
\bibitem [{\citenamefont {Ma}\ \emph {et~al.}(2013)\citenamefont {Ma},
  \citenamefont {Zhong}, \citenamefont {Wang}, \citenamefont {Chen},\ and\
  \citenamefont {Cao}}]{ma2013improved}%
  \BibitemOpen
  \bibfield  {author} {\bibinfo {author} {\bibfnamefont {X.~B.}\ \bibnamefont
  {Ma}}, \bibinfo {author} {\bibfnamefont {W.~L.}\ \bibnamefont {Zhong}},
  \bibinfo {author} {\bibfnamefont {L.~Z.}\ \bibnamefont {Wang}}, \bibinfo
  {author} {\bibfnamefont {Y.~X.}\ \bibnamefont {Chen}}, \ and\ \bibinfo
  {author} {\bibfnamefont {J.}~\bibnamefont {Cao}},\ }\href {\doibase
  10.1103/PhysRevC.88.014605} {\bibfield  {journal} {\bibinfo  {journal} {Phys.
  Rev.}\ }\textbf {\bibinfo {volume} {C88}},\ \bibinfo {pages} {014605}
  (\bibinfo {year} {2013})},\ \Eprint {http://arxiv.org/abs/1212.6625}
  {arXiv:1212.6625 [nucl-ex]} \BibitemShut {NoStop}%
\bibitem [{\citenamefont {IAEA}(2019)}]{iaeaLF}%
  \BibitemOpen
  \bibfield  {author} {\bibinfo {author} {\bibnamefont {IAEA}},\ }\href@noop {}
  {\enquote {\bibinfo {title} {Power reactor information system ({PRIS})},}\ }
  (\bibinfo {year} {2019}),\ \bibinfo {note} {\href{www.iaea.org/pris}{PRIS
  webpage}}\BibitemShut {NoStop}%
\bibitem [{\citenamefont {An}\ \emph {et~al.}(2016{\natexlab{b}})\citenamefont
  {An} \emph {et~al.}}]{an2016measurement}%
  \BibitemOpen
  \bibfield  {author} {\bibinfo {author} {\bibfnamefont {F.~P.}\ \bibnamefont
  {An}} \emph {et~al.} (\bibinfo {collaboration} {Daya Bay}),\ }\href {\doibase
  10.1103/PhysRevLett.116.061801, 10.1103/PhysRevLett.118.099902} {\bibfield
  {journal} {\bibinfo  {journal} {Phys. Rev. Lett.}\ }\textbf {\bibinfo
  {volume} {116}},\ \bibinfo {pages} {061801} (\bibinfo {year}
  {2016}{\natexlab{b}})},\ \bibinfo {note} {[Erratum: Phys. Rev.
  Lett.118,no.9,099902(2017)]},\ \Eprint {http://arxiv.org/abs/1508.04233}
  {arXiv:1508.04233 [hep-ex]} \BibitemShut {NoStop}%
\bibitem [{\citenamefont {Abe}\ \emph {et~al.}(2014)\citenamefont {Abe} \emph
  {et~al.}}]{abe2014improved}%
  \BibitemOpen
  \bibfield  {author} {\bibinfo {author} {\bibfnamefont {Y.}~\bibnamefont
  {Abe}} \emph {et~al.} (\bibinfo {collaboration} {Double Chooz}),\ }\href
  {\doibase 10.1007/JHEP02(2015)074, 10.1007/JHEP10(2014)086} {\bibfield
  {journal} {\bibinfo  {journal} {JHEP}\ }\textbf {\bibinfo {volume} {10}},\
  \bibinfo {pages} {086} (\bibinfo {year} {2014})},\ \bibinfo {note} {[Erratum:
  JHEP02,074(2015)]},\ \Eprint {http://arxiv.org/abs/1406.7763}
  {arXiv:1406.7763 [hep-ex]} \BibitemShut {NoStop}%
\bibitem [{\citenamefont {Pac}(2018)}]{reno2018nufact}%
  \BibitemOpen
  \bibfield  {author} {\bibinfo {author} {\bibfnamefont {M.~Y.}\ \bibnamefont
  {Pac}} (\bibinfo {collaboration} {RENO}),\ }\bibfield  {booktitle} {\emph
  {\bibinfo {booktitle} {{Proceedings, 2017 International Workshop on Neutrinos
  from Accelerators (NuFact17): Uppsala University Main Building, Uppsala,
  Sweden, September 25-30, 2017}}},\ }\href {\doibase 10.22323/1.295.0038}
  {\bibfield  {journal} {\bibinfo  {journal} {PoS}\ }\textbf {\bibinfo {volume}
  {NuFact2017}},\ \bibinfo {pages} {038} (\bibinfo {year} {2018})},\ \Eprint
  {http://arxiv.org/abs/1801.04049} {arXiv:1801.04049 [hep-ex]} \BibitemShut
  {NoStop}%
\bibitem [{\citenamefont {Siyeon}(2018)}]{neos2017}%
  \BibitemOpen
  \bibfield  {author} {\bibinfo {author} {\bibfnamefont {K.}~\bibnamefont
  {Siyeon}} (\bibinfo {collaboration} {NEOS}),\ }\bibfield  {booktitle} {\emph
  {\bibinfo {booktitle} {{The Fluorescence detector Array of Single-pixel
  Telescopes: Contributions to the 35th International Cosmic Ray Conference
  (ICRC 2017)}}},\ }\href {\doibase 10.22323/1.301.1024} {\bibfield  {journal}
  {\bibinfo  {journal} {PoS}\ }\textbf {\bibinfo {volume} {ICRC2017}},\
  \bibinfo {pages} {1024} (\bibinfo {year} {2018})}\BibitemShut {NoStop}%
\bibitem [{\citenamefont {Mueller}\ \emph {et~al.}(2011)\citenamefont {Mueller}
  \emph {et~al.}}]{mueller2011improved}%
  \BibitemOpen
  \bibfield  {author} {\bibinfo {author} {\bibfnamefont {T.~A.}\ \bibnamefont
  {Mueller}} \emph {et~al.},\ }\href {\doibase 10.1103/PhysRevC.83.054615}
  {\bibfield  {journal} {\bibinfo  {journal} {Phys. Rev.}\ }\textbf {\bibinfo
  {volume} {C83}},\ \bibinfo {pages} {054615} (\bibinfo {year} {2011})},\
  \Eprint {http://arxiv.org/abs/1101.2663} {arXiv:1101.2663 [hep-ex]}
  \BibitemShut {NoStop}%
\bibitem [{\citenamefont {An}\ \emph {et~al.}(2017)\citenamefont {An} \emph
  {et~al.}}]{PhysRevLett.118.251801}%
  \BibitemOpen
  \bibfield  {author} {\bibinfo {author} {\bibfnamefont {F.~P.}\ \bibnamefont
  {An}} \emph {et~al.} (\bibinfo {collaboration} {Daya Bay}),\ }\href {\doibase
  10.1103/PhysRevLett.118.251801} {\bibfield  {journal} {\bibinfo  {journal}
  {Phys. Rev. Lett.}\ }\textbf {\bibinfo {volume} {118}},\ \bibinfo {pages}
  {251801} (\bibinfo {year} {2017})},\ \Eprint
  {http://arxiv.org/abs/1704.01082} {arXiv:1704.01082 [hep-ex]} \BibitemShut
  {NoStop}%
\bibitem [{\citenamefont {Mention}\ \emph {et~al.}(2017)\citenamefont
  {Mention}, \citenamefont {Vivier}, \citenamefont {Gaffiot}, \citenamefont
  {Lasserre}, \citenamefont {Letourneau},\ and\ \citenamefont
  {Materna}}]{MENTION2017307}%
  \BibitemOpen
  \bibfield  {author} {\bibinfo {author} {\bibfnamefont {G.}~\bibnamefont
  {Mention}}, \bibinfo {author} {\bibfnamefont {M.}~\bibnamefont {Vivier}},
  \bibinfo {author} {\bibfnamefont {J.}~\bibnamefont {Gaffiot}}, \bibinfo
  {author} {\bibfnamefont {T.}~\bibnamefont {Lasserre}}, \bibinfo {author}
  {\bibfnamefont {A.}~\bibnamefont {Letourneau}}, \ and\ \bibinfo {author}
  {\bibfnamefont {T.}~\bibnamefont {Materna}},\ }\href {\doibase
  10.1016/j.physletb.2017.08.035} {\bibfield  {journal} {\bibinfo  {journal}
  {Phys. Lett.}\ }\textbf {\bibinfo {volume} {B773}},\ \bibinfo {pages} {307}
  (\bibinfo {year} {2017})},\ \Eprint {http://arxiv.org/abs/1705.09434}
  {arXiv:1705.09434 [hep-ex]} \BibitemShut {NoStop}%
\bibitem [{\citenamefont {Ejiri}\ \emph {et~al.}(2019)\citenamefont {Ejiri},
  \citenamefont {Suhonen},\ and\ \citenamefont {Zuber}}]{Ejiri2019}%
  \BibitemOpen
  \bibfield  {author} {\bibinfo {author} {\bibfnamefont {H.}~\bibnamefont
  {Ejiri}}, \bibinfo {author} {\bibfnamefont {J.}~\bibnamefont {Suhonen}}, \
  and\ \bibinfo {author} {\bibfnamefont {K.}~\bibnamefont {Zuber}},\ }\href
  {\doibase 10.1016/j.physrep.2018.12.001} {\bibfield  {journal} {\bibinfo
  {journal} {Phys. Rept.}\ }\textbf {\bibinfo {volume} {797}},\ \bibinfo
  {pages} {1} (\bibinfo {year} {2019})}\BibitemShut {NoStop}%
\bibitem [{\citenamefont {Honda}\ \emph {et~al.}(2015)\citenamefont {Honda},
  \citenamefont {Athar}, \citenamefont {Kajita}, \citenamefont {Kasahara},\
  and\ \citenamefont {Midorikawa}}]{Honda:2015fha}%
  \BibitemOpen
  \bibfield  {author} {\bibinfo {author} {\bibfnamefont {M.}~\bibnamefont
  {Honda}}, \bibinfo {author} {\bibfnamefont {M.~S.}\ \bibnamefont {Athar}},
  \bibinfo {author} {\bibfnamefont {T.}~\bibnamefont {Kajita}}, \bibinfo
  {author} {\bibfnamefont {K.}~\bibnamefont {Kasahara}}, \ and\ \bibinfo
  {author} {\bibfnamefont {S.}~\bibnamefont {Midorikawa}},\ }\href {\doibase
  10.1103/PhysRevD.92.023004} {\bibfield  {journal} {\bibinfo  {journal} {Phys.
  Rev.}\ }\textbf {\bibinfo {volume} {D92}},\ \bibinfo {pages} {023004}
  (\bibinfo {year} {2015})},\ \Eprint {http://arxiv.org/abs/1502.03916}
  {arXiv:1502.03916 [astro-ph.HE]} \BibitemShut {NoStop}%
\bibitem [{\citenamefont {Battistoni}\ \emph {et~al.}(2005)\citenamefont
  {Battistoni}, \citenamefont {Ferrari}, \citenamefont {Montaruli},\ and\
  \citenamefont {Sala}}]{Battistoni:2005pd}%
  \BibitemOpen
  \bibfield  {author} {\bibinfo {author} {\bibfnamefont {G.}~\bibnamefont
  {Battistoni}}, \bibinfo {author} {\bibfnamefont {A.}~\bibnamefont {Ferrari}},
  \bibinfo {author} {\bibfnamefont {T.}~\bibnamefont {Montaruli}}, \ and\
  \bibinfo {author} {\bibfnamefont {P.~R.}\ \bibnamefont {Sala}},\ }\href
  {\doibase 10.1016/j.astropartphys.2005.03.006} {\bibfield  {journal}
  {\bibinfo  {journal} {Astropart. Phys.}\ }\textbf {\bibinfo {volume} {23}},\
  \bibinfo {pages} {526} (\bibinfo {year} {2005})}\BibitemShut {NoStop}%
\bibitem [{\citenamefont {Wendell}(2012)}]{bib:ProbPP}%
  \BibitemOpen
  \bibfield  {author} {\bibinfo {author} {\bibfnamefont {R.}~\bibnamefont
  {Wendell}},\ }\href@noop {} {\enquote {\bibinfo {title} {Prob3++ software for
  computing three flavor neutrino oscillation probabilities},}\ }\bibinfo
  {howpublished} {\href{http://www.phy.duke.edu/~raw22/public/Prob3++}{Prob3++
  website}} (\bibinfo {year} {2012})\BibitemShut {NoStop}%
\bibitem [{\citenamefont {Andreopoulos}\ \emph {et~al.}(2010)\citenamefont
  {Andreopoulos} \emph {et~al.}}]{Andreopoulos:2009rq}%
  \BibitemOpen
  \bibfield  {author} {\bibinfo {author} {\bibfnamefont {C.}~\bibnamefont
  {Andreopoulos}} \emph {et~al.},\ }\href {\doibase 10.1016/j.nima.2009.12.009}
  {\bibfield  {journal} {\bibinfo  {journal} {Nucl. Instrum. Meth.}\ }\textbf
  {\bibinfo {volume} {A614}},\ \bibinfo {pages} {87} (\bibinfo {year}
  {2010})},\ \Eprint {http://arxiv.org/abs/0905.2517} {arXiv:0905.2517
  [hep-ph]} \BibitemShut {NoStop}%
\bibitem [{\citenamefont {Farley}\ and\ \citenamefont
  {Neroda}(1998)}]{farley1998noble}%
  \BibitemOpen
  \bibfield  {author} {\bibinfo {author} {\bibfnamefont {K.}~\bibnamefont
  {Farley}}\ and\ \bibinfo {author} {\bibfnamefont {E.}~\bibnamefont
  {Neroda}},\ }\href {\doibase 10.1146/annurev.earth.26.1.189} {\bibfield
  {journal} {\bibinfo  {journal} {Annual Review of Earth and Planetary
  Sciences}\ }\textbf {\bibinfo {volume} {26}},\ \bibinfo {pages} {189}
  (\bibinfo {year} {1998})}\BibitemShut {NoStop}%
\bibitem [{\citenamefont {Dye}(2009)}]{dye2009neutrino}%
  \BibitemOpen
  \bibfield  {author} {\bibinfo {author} {\bibfnamefont {S.~T.}\ \bibnamefont
  {Dye}},\ }\href {\doibase 10.1016/j.physletb.2009.07.010} {\bibfield
  {journal} {\bibinfo  {journal} {Phys. Lett.}\ }\textbf {\bibinfo {volume}
  {B679}},\ \bibinfo {pages} {15} (\bibinfo {year} {2009})},\ \Eprint
  {http://arxiv.org/abs/0905.0523} {arXiv:0905.0523 [nucl-ex]} \BibitemShut
  {NoStop}%
\bibitem [{\citenamefont {Rusov}\ \emph {et~al.}(2013)\citenamefont {Rusov}
  \emph {et~al.}}]{rusov2010kamland}%
  \BibitemOpen
  \bibfield  {author} {\bibinfo {author} {\bibfnamefont {V.~D.}\ \bibnamefont
  {Rusov}} \emph {et~al.},\ }\href {\doibase 10.4236/jmp.2013.44075} {\bibfield
   {journal} {\bibinfo  {journal} {J. Mod. Phys.}\ }\textbf {\bibinfo {volume}
  {4}},\ \bibinfo {pages} {528} (\bibinfo {year} {2013})},\ \Eprint
  {http://arxiv.org/abs/1011.3568} {arXiv:1011.3568 [astro-ph.EP]} \BibitemShut
  {NoStop}%
\bibitem [{\citenamefont {Herndon}\ and\ \citenamefont
  {Edgerley}(2005)}]{herndon2005background}%
  \BibitemOpen
  \bibfield  {author} {\bibinfo {author} {\bibfnamefont {J.~M.}\ \bibnamefont
  {Herndon}}\ and\ \bibinfo {author} {\bibfnamefont {D.~A.}\ \bibnamefont
  {Edgerley}},\ }\href@noop {} {\bibfield  {journal} {\bibinfo  {journal}
  {Submitted to: Proc. Roy. Soc. Lond.}\ } (\bibinfo {year} {2005})},\ \Eprint
  {http://arxiv.org/abs/hep-ph/0501216} {arXiv:hep-ph/0501216 [hep-ph]}
  \BibitemShut {NoStop}%
\bibitem [{\citenamefont {Bellini}\ \emph
  {et~al.}(2013{\natexlab{c}})\citenamefont {Bellini} \emph
  {et~al.}}]{Bellini:2013cosmo}%
  \BibitemOpen
  \bibfield  {author} {\bibinfo {author} {\bibfnamefont {G.}~\bibnamefont
  {Bellini}} \emph {et~al.} (\bibinfo {collaboration} {Borexino}),\ }\href
  {\doibase 10.1088/1475-7516/2013/08/049} {\bibfield  {journal} {\bibinfo
  {journal} {Journal of Cosmology and Astroparticle Physics}\ }\textbf
  {\bibinfo {volume} {2013}},\ \bibinfo {pages} {49} (\bibinfo {year}
  {2013}{\natexlab{c}})},\ \Eprint {http://arxiv.org/abs/1304.7381}
  {arXiv:1304.7381 [physics.ins-det]} \BibitemShut {NoStop}%
\bibitem [{\citenamefont {Kelley}\ \emph {et~al.}(2017)\citenamefont {Kelley},
  \citenamefont {Purcell},\ and\ \citenamefont {Sheu}}]{lightnuclei}%
  \BibitemOpen
  \bibfield  {author} {\bibinfo {author} {\bibfnamefont {J.~H.}\ \bibnamefont
  {Kelley}}, \bibinfo {author} {\bibfnamefont {J.~E.}\ \bibnamefont {Purcell}},
  \ and\ \bibinfo {author} {\bibfnamefont {C.~G.}\ \bibnamefont {Sheu}},\
  }\href {\doibase 10.1016/j.nuclphysa.2017.07.015} {\bibfield  {journal}
  {\bibinfo  {journal} {Nucl. Phys.}\ }\textbf {\bibinfo {volume} {A968}},\
  \bibinfo {pages} {71} (\bibinfo {year} {2017})}\BibitemShut {NoStop}%
\bibitem [{\citenamefont {de~Kerret}\ \emph {et~al.}(2018)\citenamefont
  {de~Kerret} \emph {et~al.}}]{deKerret:2018fqd}%
  \BibitemOpen
  \bibfield  {author} {\bibinfo {author} {\bibfnamefont {H.}~\bibnamefont
  {de~Kerret}} \emph {et~al.} (\bibinfo {collaboration} {Double Chooz}),\
  }\href {\doibase 10.1007/JHEP11(2018)053} {\bibfield  {journal} {\bibinfo
  {journal} {JHEP}\ }\textbf {\bibinfo {volume} {11}},\ \bibinfo {pages} {053}
  (\bibinfo {year} {2018})},\ \Eprint {http://arxiv.org/abs/1802.08048}
  {arXiv:1802.08048 [hep-ex]} \BibitemShut {NoStop}%
\bibitem [{\citenamefont {Mohr}(2018)}]{Mohr:2018alphaNBgr}%
  \BibitemOpen
  \bibfield  {author} {\bibinfo {author} {\bibfnamefont {P.}~\bibnamefont
  {Mohr}},\ }\href {\doibase 10.1103/PhysRevC.97.064613} {\bibfield  {journal}
  {\bibinfo  {journal} {Phys. Rev. C}\ }\textbf {\bibinfo {volume} {97}},\
  \bibinfo {pages} {064613} (\bibinfo {year} {2018})},\ \Eprint
  {http://arxiv.org/abs/1806.02722} {arXiv:1806.02722 [nucl-ex]} \BibitemShut
  {NoStop}%
\bibitem [{\citenamefont {Harissopulos}\ \emph {et~al.}(2005)\citenamefont
  {Harissopulos}, \citenamefont {Becker}, \citenamefont {Hammer}, \citenamefont
  {Lagoyannis}, \citenamefont {Rolfs},\ and\ \citenamefont
  {Strieder}}]{Harissopulos:2005alphaNBgr}%
  \BibitemOpen
  \bibfield  {author} {\bibinfo {author} {\bibfnamefont {S.}~\bibnamefont
  {Harissopulos}}, \bibinfo {author} {\bibfnamefont {H.~W.}\ \bibnamefont
  {Becker}}, \bibinfo {author} {\bibfnamefont {J.~W.}\ \bibnamefont {Hammer}},
  \bibinfo {author} {\bibfnamefont {A.}~\bibnamefont {Lagoyannis}}, \bibinfo
  {author} {\bibfnamefont {C.}~\bibnamefont {Rolfs}}, \ and\ \bibinfo {author}
  {\bibfnamefont {F.}~\bibnamefont {Strieder}},\ }\href {\doibase
  10.1103/PhysRevC.72.062801} {\bibfield  {journal} {\bibinfo  {journal} {Phys.
  Rev. C}\ }\textbf {\bibinfo {volume} {72}},\ \bibinfo {pages} {062801}
  (\bibinfo {year} {2005})},\ \Eprint {http://arxiv.org/abs/nucl-ex/0509014}
  {arXiv:nucl-ex/0509014 [nucl-ex]} \BibitemShut {NoStop}%
\bibitem [{\citenamefont {Koning}\ and\ \citenamefont
  {Rochman}(2012)}]{Koning:2012zqy}%
  \BibitemOpen
  \bibfield  {author} {\bibinfo {author} {\bibfnamefont {A.~J.}\ \bibnamefont
  {Koning}}\ and\ \bibinfo {author} {\bibfnamefont {D.}~\bibnamefont
  {Rochman}},\ }\href {\doibase 10.1016/j.nds.2012.11.002} {\bibfield
  {journal} {\bibinfo  {journal} {Nucl. Data Sheets}\ }\textbf {\bibinfo
  {volume} {113}},\ \bibinfo {pages} {2841} (\bibinfo {year}
  {2012})}\BibitemShut {NoStop}%
\bibitem [{\citenamefont {Bellini}\ \emph
  {et~al.}(2013{\natexlab{d}})\citenamefont {Bellini} \emph
  {et~al.}}]{Bellini2013}%
  \BibitemOpen
  \bibfield  {author} {\bibinfo {author} {\bibfnamefont {G.}~\bibnamefont
  {Bellini}} \emph {et~al.} (\bibinfo {collaboration} {Borexino}),\ }\href
  {\doibase 10.1140/epja/i2013-13092-9} {\bibfield  {journal} {\bibinfo
  {journal} {Eur. Phys. J.}\ }\textbf {\bibinfo {volume} {A49}},\ \bibinfo
  {pages} {92} (\bibinfo {year} {2013}{\natexlab{d}})},\ \Eprint
  {http://arxiv.org/abs/1212.1332} {arXiv:1212.1332 [nucl-ex]} \BibitemShut
  {NoStop}%
\bibitem [{\citenamefont {Vogel}\ and\ \citenamefont
  {Beacom}(1999)}]{PhysRevD.60.053003}%
  \BibitemOpen
  \bibfield  {author} {\bibinfo {author} {\bibfnamefont {P.}~\bibnamefont
  {Vogel}}\ and\ \bibinfo {author} {\bibfnamefont {J.~F.}\ \bibnamefont
  {Beacom}},\ }\href {\doibase 10.1103/PhysRevD.60.053003} {\bibfield
  {journal} {\bibinfo  {journal} {Phys. Rev.}\ }\textbf {\bibinfo {volume}
  {D60}},\ \bibinfo {pages} {053003} (\bibinfo {year} {1999})},\ \Eprint
  {http://arxiv.org/abs/hep-ph/9903554} {arXiv:hep-ph/9903554 [hep-ph]}
  \BibitemShut {NoStop}%
\bibitem [{\citenamefont {Prezado}\ \emph {et~al.}(2005)\citenamefont {Prezado}
  \emph {et~al.}}]{PREZADO200543}%
  \BibitemOpen
  \bibfield  {author} {\bibinfo {author} {\bibfnamefont {Y.}~\bibnamefont
  {Prezado}} \emph {et~al.},\ }\href {\doibase 10.1016/j.physletb.2005.05.030}
  {\bibfield  {journal} {\bibinfo  {journal} {Phys. Lett.}\ }\textbf {\bibinfo
  {volume} {B618}},\ \bibinfo {pages} {43} (\bibinfo {year}
  {2005})}\BibitemShut {NoStop}%
\bibitem [{\citenamefont {Mei}\ and\ \citenamefont
  {Hime}(2006)}]{PhysRevD.73.053004}%
  \BibitemOpen
  \bibfield  {author} {\bibinfo {author} {\bibfnamefont {D.~M.}\ \bibnamefont
  {Mei}}\ and\ \bibinfo {author} {\bibfnamefont {A.}~\bibnamefont {Hime}},\
  }\href {\doibase 10.1103/PhysRevD.73.053004} {\bibfield  {journal} {\bibinfo
  {journal} {Phys. Rev.}\ }\textbf {\bibinfo {volume} {D73}},\ \bibinfo {pages}
  {053004} (\bibinfo {year} {2006})},\ \Eprint
  {http://arxiv.org/abs/astro-ph/0512125} {arXiv:astro-ph/0512125 [astro-ph]}
  \BibitemShut {NoStop}%
\bibitem [{\citenamefont {Westerdale}()}]{Westerdale:NeuCBOT}%
  \BibitemOpen
  \bibfield  {author} {\bibinfo {author} {\bibfnamefont {S.}~\bibnamefont
  {Westerdale}},\ }\href@noop {} {\enquote {\bibinfo {title} {Neu{CBOT}
  ({N}eutron {C}alculator {B}ased {O}n {TALYS})},}\ }\bibinfo {note}
  {\href{https://github.com/shawest/neucbot}{GitHub}}\BibitemShut {NoStop}%
\bibitem [{\citenamefont {Westerdale}\ and\ \citenamefont
  {Meyers}(2017)}]{Westerdale:2017kml}%
  \BibitemOpen
  \bibfield  {author} {\bibinfo {author} {\bibfnamefont {S.}~\bibnamefont
  {Westerdale}}\ and\ \bibinfo {author} {\bibfnamefont {P.~D.}\ \bibnamefont
  {Meyers}},\ }\href {\doibase 10.1016/j.nima.2017.09.007} {\bibfield
  {journal} {\bibinfo  {journal} {Nucl. Instrum. Meth.}\ }\textbf {\bibinfo
  {volume} {A875}},\ \bibinfo {pages} {57} (\bibinfo {year} {2017})},\ \Eprint
  {http://arxiv.org/abs/1702.02465} {arXiv:1702.02465 [physics.ins-det]}
  \BibitemShut {NoStop}%
\bibitem [{\citenamefont {Westerdale}(2018)}]{Westerdale:2018hck}%
  \BibitemOpen
  \bibfield  {author} {\bibinfo {author} {\bibfnamefont {S.}~\bibnamefont
  {Westerdale}} (\bibinfo {collaboration} {DEAP-3600}),\ }\bibfield
  {booktitle} {\emph {\bibinfo {booktitle} {{Proceedings, 6th Topical Workshop
  on Low Radioactivity Techniques (LRT 2017): Seoul, Korea, May 24-26,
  2017}}},\ }\href {\doibase 10.1063/1.5018998} {\bibfield  {journal} {\bibinfo
   {journal} {AIP Conf. Proc.}\ }\textbf {\bibinfo {volume} {1921}},\ \bibinfo
  {pages} {060002} (\bibinfo {year} {2018})}\BibitemShut {NoStop}%
\bibitem [{\citenamefont {Westerdale}(2016)}]{Westerdale:2016}%
  \BibitemOpen
  \bibfield  {author} {\bibinfo {author} {\bibfnamefont {S.}~\bibnamefont
  {Westerdale}},\ }\emph {\bibinfo {title} {A Study of Nuclear Recoil
  Backgrounds in Dark Matter Detectors}},\ \href@noop {} {Ph.D. thesis},\
  \bibinfo  {school} {Princeton University} (\bibinfo {year} {2016}),\ \bibinfo
  {note}
  {\href{http://web.archive.org/web/20161216231000/https://www.princeton.edu/physics/graduate-program/theses/Westerdalethesis.pdf}{online
  version}}\BibitemShut {NoStop}%
\bibitem [{\citenamefont {Koning}\ \emph {et~al.}()\citenamefont {Koning},
  \citenamefont {Hilaire},\ and\ \citenamefont {Duijvestijn}}]{Talys:code}%
  \BibitemOpen
  \bibfield  {author} {\bibinfo {author} {\bibfnamefont {A.}~\bibnamefont
  {Koning}}, \bibinfo {author} {\bibfnamefont {S.}~\bibnamefont {Hilaire}}, \
  and\ \bibinfo {author} {\bibfnamefont {M.}~\bibnamefont {Duijvestijn}},\
  }\href@noop {} {\enquote {\bibinfo {title} {Talys-1.9},}\ }\bibinfo {note}
  {\href{http://www.talys.eu/download-talys/}{TALYS website}}\BibitemShut
  {NoStop}%
\bibitem [{\citenamefont {Bersillon}\ \emph {et~al.}(2008)\citenamefont
  {Bersillon}, \citenamefont {Gunsing}, \citenamefont {Bauge}, \citenamefont
  {Jacqmin},\ and\ \citenamefont {Leray}}]{Talys:2008}%
  \BibitemOpen
  \bibinfo {editor} {\bibfnamefont {O.}~\bibnamefont {Bersillon}}, \bibinfo
  {editor} {\bibfnamefont {F.}~\bibnamefont {Gunsing}}, \bibinfo {editor}
  {\bibfnamefont {E.}~\bibnamefont {Bauge}}, \bibinfo {editor} {\bibfnamefont
  {R.}~\bibnamefont {Jacqmin}}, \ and\ \bibinfo {editor} {\bibfnamefont
  {S.}~\bibnamefont {Leray}},\ eds.,\ \href@noop {} {\emph {\bibinfo {title}
  {{TALYS-1.0}}}}\ (\bibinfo  {publisher} {EDP Sciences},\ \bibinfo {address}
  {Nice, France},\ \bibinfo {year} {2008})\ \bibinfo {note} {proceedings of the
  International Conference on Nuclear Data for Science and Technology -
  ND2007}\BibitemShut {NoStop}%
\bibitem [{\citenamefont {McKee}\ \emph {et~al.}(2008)\citenamefont {McKee},
  \citenamefont {Busenitz},\ and\ \citenamefont {Ostrovskiy}}]{McKee:2007bk}%
  \BibitemOpen
  \bibfield  {author} {\bibinfo {author} {\bibfnamefont {D.~W.}\ \bibnamefont
  {McKee}}, \bibinfo {author} {\bibfnamefont {J.~K.}\ \bibnamefont {Busenitz}},
  \ and\ \bibinfo {author} {\bibfnamefont {I.}~\bibnamefont {Ostrovskiy}},\
  }\href {\doibase 10.1016/j.nima.2007.12.002} {\bibfield  {journal} {\bibinfo
  {journal} {Nucl. Instrum. Meth.}\ }\textbf {\bibinfo {volume} {A587}},\
  \bibinfo {pages} {272} (\bibinfo {year} {2008})},\ \Eprint
  {http://arxiv.org/abs/0711.3624} {arXiv:0711.3624 [physics.ins-det]}
  \BibitemShut {NoStop}%
\bibitem [{\citenamefont {Alimonti}\ \emph {et~al.}(1998)\citenamefont
  {Alimonti} \emph {et~al.}}]{ALIMONTI1998411}%
  \BibitemOpen
  \bibfield  {author} {\bibinfo {author} {\bibfnamefont {G.}~\bibnamefont
  {Alimonti}} \emph {et~al.},\ }\href {\doibase 10.1016/S0168-9002(98)00018-7}
  {\bibfield  {journal} {\bibinfo  {journal} {Nucl. Instrum. Meth.}\ }\textbf
  {\bibinfo {volume} {A406}},\ \bibinfo {pages} {411} (\bibinfo {year}
  {1998})}\BibitemShut {NoStop}%
\bibitem [{\citenamefont {Cowan}\ \emph {et~al.}(2011)\citenamefont {Cowan},
  \citenamefont {Cranmer}, \citenamefont {Gross},\ and\ \citenamefont
  {Vitells}}]{Cowan2011}%
  \BibitemOpen
  \bibfield  {author} {\bibinfo {author} {\bibfnamefont {G.}~\bibnamefont
  {Cowan}}, \bibinfo {author} {\bibfnamefont {K.}~\bibnamefont {Cranmer}},
  \bibinfo {author} {\bibfnamefont {E.}~\bibnamefont {Gross}}, \ and\ \bibinfo
  {author} {\bibfnamefont {O.}~\bibnamefont {Vitells}},\ }\href {\doibase
  10.1140/epjc/s10052-011-1554-0, 10.1140/epjc/s10052-013-2501-z} {\bibfield
  {journal} {\bibinfo  {journal} {Eur. Phys. J.}\ }\textbf {\bibinfo {volume}
  {C71}},\ \bibinfo {pages} {1554} (\bibinfo {year} {2011})},\ \bibinfo {note}
  {[Erratum: Eur. Phys. J.C73,2501(2013)]},\ \Eprint
  {http://arxiv.org/abs/1007.1727} {arXiv:1007.1727 [physics.data-an]}
  \BibitemShut {NoStop}%
\bibitem [{\citenamefont {Tanabashi}\ \emph {et~al.}(2018)\citenamefont
  {Tanabashi} \emph {et~al.}}]{PhysRevD.98.030001}%
  \BibitemOpen
  \bibfield  {author} {\bibinfo {author} {\bibfnamefont {M.}~\bibnamefont
  {Tanabashi}} \emph {et~al.} (\bibinfo {collaboration} {Particle Data
  Group}),\ }\href {\doibase 10.1103/PhysRevD.98.030001} {\bibfield  {journal}
  {\bibinfo  {journal} {Phys. Rev.}\ }\textbf {\bibinfo {volume} {D98}},\
  \bibinfo {pages} {030001} (\bibinfo {year} {2018})}\BibitemShut {NoStop}%
\end{thebibliography}%

\end{document}